\documentclass[preprint,12pt]{elsarticle}

\usepackage{caption}
\usepackage{amsmath}
\usepackage{amssymb}
\usepackage{amsthm}
\usepackage{mathrsfs}

\usepackage{epstopdf}

\usepackage{threeparttable}
\usepackage{subfig}

\usepackage{subcaption}

\usepackage{booktabs} 
\usepackage{multirow} 

\usepackage{url}
\usepackage{hyperref}
\usepackage{xcolor}
\usepackage{cleveref} 

\usepackage{lineno}

\usepackage{color}
\usepackage{epstopdf}
\usepackage{float}
\usepackage[top=2.5cm, bottom=2.5cm, left=2cm, right=2cm]{geometry}
\renewcommand{\vec}[1]{\boldsymbol{#1}}

\biboptions{numbers,sort&compress}

\journal{AIAA Journal}

\begin{document}


\begin{frontmatter}



\title{Nonequilibrium flow simulations using unified gas-kinetic wave-particle method}


\author[a]{Wenpei Long}
\author[a]{Yufeng Wei}
\author[a,b,c]{Kun Xu\corref{cor1}}

\cortext[cor1] {Corresponding author.}
\ead{makxu@ust.hk}

\address[a]{Department of Mathematics, Hong Kong University of Science and Technology, Hong Kong, China}
\address[b]{Department of Mechanical and Aerospace Engineering, Hong Kong University of Science and Technology, Hong Kong, China}
\address[c]{HKUST Shenzhen Research Institute, Shenzhen, 518057, China}
\begin{abstract}
Nonequilibrium flows are commonly encountered in aerospace engineering applications, such as spacecraft re-entry, rocket launch, and satellite attitude control.
Numerical simulations help greatly in the understanding of non-equilibrium flow and the dynamics in the spacecraft flight.
Based on the direct modeling of the cell size and time step scales, the unified gas-kinetic wave-particle (UGKWP) method has been developed for the multi-scale flow simulation. The core of the method is the modeling of the coupled particle transport and collision within a numerical time in the flux evaluation across a cell interface.
The coupled transport releases the restriction on the cell size and numerical time step being less than the particle mean free path and collision time. Additionally, the weights in the wave-particle decompositions in UGKWP vary automatically to cope with the degrees of freedom in an accurate description of flow physics in different regimes and archive a smooth transition from the particle tracking in the highly rarefied regime to the macroscopic wave evolution in the continuum Navier--Stokes (NS) regime.
Consequently, the UGKWP emerges as a prime strategy that strikes a balance between precision and efficiency in multiscale flow simulations, particularly in the study of high-speed flow.
In the present study, the UGKWP method is utilized to simulate a few challenge flow problems with a large variation of Knudsen numbers, which include supersonic flow around a sphere,  hypersonic flow around a space vehicle, nozzle plume into vacuum, and side-jet impingement on the hypersonic flow.
In the case of an unsteady nozzle plume, the UGKWP is capable of accurately rendering the Navier--Stokes (NS) solution within the nozzle, extending to the free molecular flow in the external environment, all within a singular computation.
For the side-jet impingement on the hypersonic flow, complicated structures in the flow interaction have been captured by the UGKWP method.
All obtained simulation results have been verified either through experimental measurements or via solutions obtained from Direct Simulation Monte Carlo (DSMC) methods.
In terms of computational cost, the UGKWP method requires only $60$ GiB of memory to simulate the three-dimensional space vehicle with $560,593$ cells under different flow conditions.  This is manageable even on personal workstations. Given its excellent efficiency and accuracy, the UGKWP method displays its substantial benefits in simulating multiscale flow for aerospace engineering applications.

\end{abstract}

\begin{keyword}
	Hypersonic rarefied flow \sep
	Nozzle plume flow \sep
	Side jet \sep
	Multiple flow regimes\sep
	Unified gas kinetic wave-particle method
\end{keyword}

\end{frontmatter}



\section{Introduction}\label{sec:intro}

Multiscale flows are frequently encountered in a variety of aerospace applications, including spacecraft re-entry, rocket launches, and aircraft attitude control using jet flows. Experimental investigations of such problems pose challenges due to the difficulty and expense of establishing high-fidelity experimental environments, especially when attempting to replicate high-altitude atmospheric conditions.
Computational Fluid Dynamics (CFD), a crucial tool in fluid mechanics research, can surmount these experimental limitations. It offers precise and comprehensive representations of various physical quantities in evolving flow fields.

For rarefied flow simulation, the methods are mainly categorized as the deterministic method with the discretization of particle velocity space and the stochastic method with particles. Both deterministic methods \cite{broadwell1964study,lizhihui20140302,zhang2023conservative,jiangdw,michael2004scaling} and stochastic methods \cite{Padilla,loth2008compressibility,loth2021supersonic} have been widely used in aerodynamic applications.
However, the deterministic discrete velocity methods (DVM) consume a significant amount of computing resources.
For high-speed rarefied flow, direct simulation Monte Carlo (DSMC)\cite{bird1963approach} method based on stochastic particles is commonly used,
such as the calculations of shuttle orbiters, capsules, and aerospace planes in re-entry problems, and has presented reliable results \cite{rault1994aerodymics,LEBEAU2001595,justiz1994dsmc,Gimelshein2002,zhang2010spacecraft,ivanovdsmc,
	ivanovich2013aerodynamic,titarev2020comparison,zuppardi2016aerodynamic,zuppardi2014,moss2005dsmc}. In particular, for the X38-like spacecraft, Lebeau et al. used the DSMC-based DAC software to simulate the surface pressure distribution \cite{LEBEAU2001595} in the transitional flow regime.
Li \cite{li2021kinetic} and Jiang \cite{jiangdw} used the deterministic unified gas-kinetic scheme (UGKS) and stochastic DSMC to study the surface force and thermal properties of X38. It was realized that local Knudsen numbers can vary significantly around the surface of X38.
In comparison with DSMC, the UGKS is a multiscale method, but its memory consumption is significantly high in high-speed flow simulation.
On the other hand, the DSMC is valid in the particle mean free path and collision time scales and has difficulties in computing flow in the near-continuum regime due to the constraints on the cell size and time step. The unified gas-kinetic wave-particle (UGKWP) method takes advantage of both UGKS and DSMC and optimizes the computational efficiency and accuracy in flow studies with the co-existence of multiple flow regimes.

The UGKWP method follows the direct modeling methodology of UGKS \cite{xu2015}. The UGKS is a DVM method and its time evolution of particle distribution function is based on the coupled particle free transport and collision. With the variation of the particle collision frequency within a numerical time step, the UGKS can capture different flow physics from the free molecular flow to the NS evolution solution.
The UGKS releases the restrictions on the cell size and time step being less than particle mean free path and mean collision time. 
In order to improve the efficiency of UGKS in the high flow simulation, the UGKWP method replaces the discrete velocity space in UGKS with stochastic particles, overcoming the huge computational load and memory consumption of UGKS \cite{liu2020unified,zhu2019unified}. The wave-particle decomposition in the UGKWP method makes it possible to adjust the degrees of freedom in the computation, such as tracking all particles in the highly rarefied flow regime and
evolving macroscopic flow variables in the Navier--Stokes (NS) regime. In the continuum flow regime, the UGKWP will precisely recover the 
Gas-kinetic scheme (GKS) in the continuum flow limit \cite{xu2001gas}.
Recently, the UGKWP method has been used to study the gas effect with rotational and vibrational nonequilibrium \cite{wei2022unified}.
At the same time, the methodology has been used to construct multiscale methods for phonon transport \cite{li2020unified}, plasma \cite{liu2021unified}, and gas-solid particle flow \cite{yang2023unified}. 
The current study uses the UGKWP method to simulate supersonic flow around a sphere and hypersonic flow around a space vehicle, as well as nozzle plume to vacuum and transversal jet interactions with the hypersonic flow.

This article is organized as follows. In Section \ref{sec:method}, the UGKWP method based on the finite volume framework will be briefly described. In Section \ref{sec:result} various multi-scale flow problems will be simulated and analyzed, and the outcomes will be compared with existing experimental and simulation data to demonstrate the accuracy and efficiency of the UGKWP method. Section \ref{sec:conclusion} will summarize the characteristics and advantages of the UGKWP method in multi-scale flows in large-scale aerodynamic applications.

\section{UGKWP method}\label{sec:method}

The unified gas kinetic wave-particle (UGKWP) method is used for numerical simulations in this study. The method is constructed under the finite-volume framework. 
Within a discrete finite volume cell $i$ and a discrete time step $\Delta t = t^{n+1} - t^n$, the macroscopic quantities of the flow field $\vec{W}=(\rho, \rho \textbf{U}, \rho E)$, i.e., densities of mass, momentum, energy, are updated by the fluxes through the cell interfaces
\begin{equation*}
\vec{W}_i^{n+1}=\vec{W}_i^n-\frac{\Delta t}{{\Omega}_i}\sum_{j\in N(i)}\vec{F}_{ij}\mathcal{A}_{ij},
\end{equation*}
where $\Omega_i$ denotes the volume of cell $i$, $N(i)$ is the set of all interface-adjacent neighboring cells of the control volume $i$, and $j$ is one of the neighboring cells with the interface between them labeled as $ij$ and the area denoted as $\mathcal{A}_{ij}$. $\vec{F}_{ij}$ is the time-averaged macroscopic flux crossing the interface $ij$.

Under the framework of the finite volume method, the key of a numerical method lies in the evolution model of the flux function through the cell interface. In the gas kinetic theory, the macroscopic flux function is obtained by taking moments of the microscopic distribution function
\begin{equation*}
\vec{F}_{ij}=\int  \mathcal{F}_{ij}\psi\mathrm{d}\vec{u},
\end{equation*}
where $\psi=\left(1,\vec{u},\frac12\vec{u}^2\right)^T$, and $\vec{u}$ represents the microscopic particle velocity, and the time-averaged flux of the microscopic distribution function, denoted as $\mathcal{F}_{ij}$, is expressed as
\begin{equation*}
\mathcal{F}_{ij}=\frac1{\Delta t}{\int_0^{\Delta t}\vec{u}\cdot\vec{n}_{ij}f_{ij}(t)\mathrm{d}t},
\end{equation*}
where $\vec{n}_{ij}$ is the unit normal vector to the interface, and $f_{ij}(t)$ represents the time-dependent distribution function on the interface. In the UGKWP method, the distribution function is described by the BGK model
\begin{equation*}\label{eq:BGK}
\frac{\partial f}{ \partial t} + \vec{u} \cdot \frac{\partial f}{\partial \vec{r}} = \frac{ g - f}{\tau }.
\end{equation*}
Along the characteristic line, its integral solution gives the local evolution of the gas distribution function
\begin{equation*}
f(\vec{r},t)=\frac1\tau\int_0^te^{-\left(t-t^\prime\right)/\tau} g \left(\vec{r}^\prime,t^\prime\right)\mathrm{d}t^\prime
+e^{-t/\tau} f_0(\vec{r}-\vec{u}t),
\end{equation*}
where $\tau$ is the relaxation time, $\vec{r}$ is the spatial location, $f_0(\vec{r}-\vec{u}t)$ is the initial distribution function at the beginning of each step $t_n$, and $g(\vec{r}^\prime, t^\prime)$ represents the local equilibrium following the Maxwell distribution
\begin{equation}\label{eq:g}
g= \rho \left(\frac{\lambda}{\pi}\right)^{\frac{D}{2}} e^{-\lambda {\vec{c}}^2},
\end{equation}
where $D$ is the degrees of freedom of gas, and $\lambda$ is related to the temperature $T$ by $\lambda = m_0/2k_BT$. Here, $m_0$ and $k_B$ are the molecular mass and Boltzmann constant, respectively. $\vec{c} = \vec{u} - \vec{U}$ denotes the peculiar velocity.

\begin{figure}[H]
\centering
\subfloat[]{\includegraphics[width=0.24\textwidth]
	{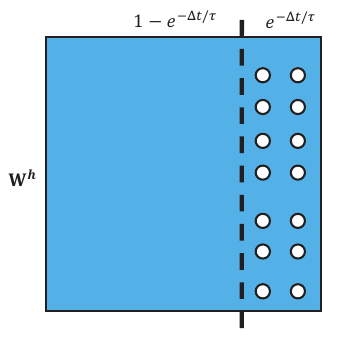}}
\subfloat[]{\includegraphics[width=0.24\textwidth]
	{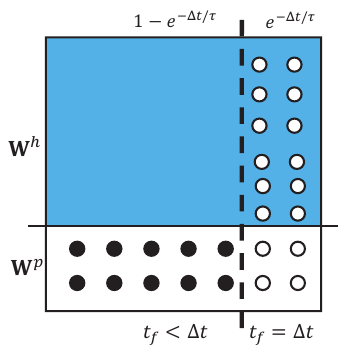}}
\subfloat[]{\includegraphics[width=0.24\textwidth]
	{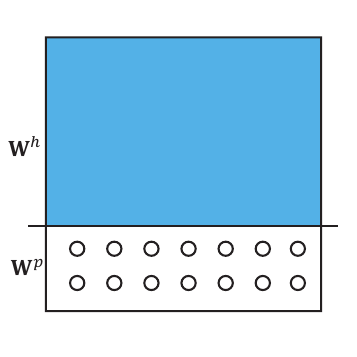}}
\subfloat[]{\includegraphics[width=0.24\textwidth]
	{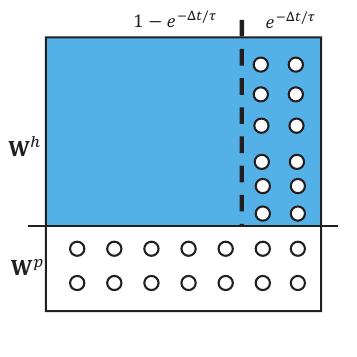}}
\caption{Diagram to illustrate the composition of the particles during time evolution in the UGKWP method. (a) Initial field, (b) classification of the collisionless and collisional particles for $\vec{W}^p_i$, (c) update on the macroscopic level, and (d) update on the microscopic level.}
\label{fig:ugkwp}
\end{figure}

For the numerical description of the integral solution, the UGKWP method introduces stochastic particles $\phi_k=\left(m_k,m_k\vec{u}_k,\frac12m_k\vec{u}_k^2\right)^T$ and analytical wave $\vec{W}^h = \left(\rho^h, (\rho \vec{U})^h, (\rho E)^h \right)$ to represent the initial distribution function $f_0$, enhancing its ability to accurately capture non-equilibrium flows. Additionally, its adaptive wave-particle decomposition ensures efficient solutions for high-speed, multi-scale flow across different flow regimes. The macroscopic flux function of the UGKWP method consists of three parts. The first part is contributed by local equilibrium state $g(\vec{r}^\prime, t^\prime)$ in the integral solution can be directly obtained through integration
\begin{equation*}
\vec{F}_{ij}^{eq}=\int\vec{u}\cdot\vec{n}_{ij}\left[C_1 g_0+ C_2\vec{u}\cdot\frac{\partial g}{\partial \vec{r}}
+C_3\frac{\partial g}{\partial t}\right]\psi\mathrm{d}\vec{u}.
\end{equation*}
The second part is derived from the initial distribution function where part of free streaming flux can be calculated analytically
\begin{equation*}
\vec{F}_{ij}^{fr,h}=\int\vec{u}\cdot\vec{n}_{ij}\Bigg[C_4 g_0+C_5\vec{u}\cdot\frac{\partial g}{\partial\vec{r}}\Bigg]\psi\mathrm{d}\vec{u}.
\end{equation*}
Therefore, $\vec{F}_{ij}^{eq}$ and $F_{ij}^{fr,h}$ are regarded as analytical waves, where $C_1$, $C_2$, $C_3$, $C_4$, and $C_5$ are corresponding time
\begin{equation*}
\begin{aligned}
	C_1 &= 1 - \frac{\tau}{\Delta t} \left( 1 - e^{-\Delta t / \tau} \right) , \\
	C_2 &= -\tau + \frac{2\tau^2}{\Delta t} - e^{-\Delta t / \tau} \left( \frac{2\tau^2}{\Delta t} + \tau\right) ,\\
	C_3 &=  \frac12 \Delta t - \tau + \frac{\tau^2}{\Delta t} \left( 1 - e^{-\Delta t / \tau} \right) , \\
	C_4 &= \frac{\tau}{\Delta t} \left(1 - e^{-\Delta t / \tau}\right) -  e^{-\Delta t / \tau}, \\
	C_5 &= \tau  e^{-\Delta t / \tau} - \frac{\tau^2}{\Delta t}(1 -  e^{-\Delta t / \tau}) + \frac12\Delta t e^{-\Delta t / \tau}.
\end{aligned}
\end{equation*}
The third part of macroscopic flux is contributed by particle free transport $\vec{W}_i^{^{fr}}$, which is obtained by tracking the discrete particles crossing the interface
\begin{equation*}
\vec{W}_i^{^{fr}}=\frac1{\Delta t}\Bigg(\sum_{\vec{x}_k\in\Omega_i}\phi_k-\sum_{\vec{x}_k^n\in\Omega_i}\phi_k\Bigg),
\end{equation*}
where $\vec{x}_k^n$ denotes the position of $k$-th particle at initial time $t^n$, and $\vec{x}_k$ is the position of the particle after free transport. After deriving macroscopic flux on the cell interface, the cell-averaged macroscopic variable $\vec{W}_i$ is updated by
\begin{equation*}
\begin{aligned}\vec{W}_i^{n+1}&=\vec{W}_i^n-\frac{\Delta t}{\Omega_i}\sum_{j\in N(i)}\vec{F}_{ij}^{eq}\mathcal{A}_{ij}-\frac{\Delta t}{\Omega_i}\sum_{j\in N(i)}\vec{F}_{ij}^{fr,h}\mathcal{A}_{ij}+\frac{\Delta t}{\Omega_i}\vec{W}_i^{fr,p}.\end{aligned}
\end{equation*}
The wave-particle decomposition process is reflected at the start and end of each time step in Fig.~\ref{fig:ugkwp}. At the beginning of the update, random particles are sampled to separate them from the macroscopic waves. After the update, particles with free transport time less than the time step $\Delta t$ are removed and absorbed into the macroscopic waves. For a detailed introduction to the UGKWP method, please refer to references \cite{liu2020unified,zhu2019unified}.

\section{Flow studies and discussion}\label{sec:result}

The UGKWP method can simulate flow in all flow regimes with great efficiency and accuracy. Its wave-particle
decomposition takes a significant advantage in three-dimensional flow, especially for hypersonic one around complex geometry, unsteady nozzle plume flow to background vacuum, and hypersonic side-jet interaction. The method can greatly reduce memory consumption and improve computation efficiency. This section will show the numerical simulation results of the UGKWP method in many cases. Detailed comparisons of the key flow structures will be presented, such as the aerodynamic force and aerothermal distribution of the high-speed flow around the space vehicle, the axial pressure and temperature of the nozzle plume, as well as the flow separation, reattachment, and penetration phenomena in the side-jet interaction. Combined with computational time, memory occupation, and other computational performance, this section will fully demonstrate the great potential of the UGKWP method in simulating the multi-scale flow in large-scale engineering applications.

The Knudsen number is used to describe the degree of rarefaction in multi-scale flows. In the non-equilibrium flow, the gradient-length-dependent  local Knudsen number ${\rm Kn}_{Gll}$ is used to reveal the local non-equilibrium 
\begin{equation*}
	{\rm Kn}_{Gll} = \frac{l}{\rho/|\nabla \rho|},
\end{equation*}
where $l$ is the local average molecular free path.

For the initial condition, the Knudsen number of freestream flow ${\rm Kn}_\infty$ is adopted to define a case
\begin{equation*}
	{\rm Kn}_\infty = \frac{l_\infty}{L_{ref}},
\end{equation*}
where $l_{\infty}$ is the mean free path of molecules of the freestream flow, and $L_{ref}$ is the reference length for defining the Knudsen number ${\rm Kn}_\infty$. The density of the specific gas can be obtained based on the Knudsen number of the freestream flow
\begin{equation*}
	\rho_\infty=\frac{4 \alpha(5-2 \omega)(7-2 \omega)}{5(\alpha+1)(\alpha+2)} \sqrt{\frac{m}{2 \pi k_B T_\infty}} \frac{\mu_\infty}{L_{r e f} \mathrm{Kn}_\infty},
\end{equation*}
where $m$ represents the molecular mass. The dynamic viscosity is determined by a power law
\begin{equation*}
	\mu = \mu_{ref}\left(\frac{T}{T_{ref}}\right)^\omega.
\end{equation*}
The gas internal degrees of freedom in the Maxwell distribution Eq.~\eqref{eq:g} are calculated based on the specific heat ratio
\begin{equation*}
	D = \frac{2}{\gamma-1}.
\end{equation*}
Table~\ref{tab:gas} provides the gas parameters used in this study. Note that the reference dynamic viscosity is defined at a reference temperature $T_{ref} = 273$ K.
\begin{table}[H]
	\centering
	\caption{The gas and physical properties parameters used in this study}
	\begin{tabular}{ccccccc}
		\toprule
		Gas & Case & $m$, kg & $\mu_{ref}$, $\rm{N}\rm{s}\rm{m}^{-2}$ & $\omega$ & $\alpha$ & $\gamma$
		\\ \midrule
		${\rm N}_2$  & Sphere,~Side jet & $4.65\times 10^{-26}$ & $1.65\times 10^{-5}$ & 0.74 & 1.0  & 1.4\\
		Ar           & X38             & $6.63\times 10^{-26}$ & $2.12\times 10^{-5}$ & 0.81 & 1.0 & 1.67 \\
		${\rm CO}_2$ & Nozzle           & $7.31\times 10^{-26}$ & $1.38\times 10^{-5}$ & 0.67 & 1.0 & 1.4 \\ \bottomrule
	\end{tabular}
	\label{tab:gas}
\end{table}
In addition, all cases in this work are simulated by the explicit UGKWP method with the CFL number 0.95. The spatial reconstruction of macroscopic flow variables is carried out by the least-square method with Venkatarishnan limiter\cite{venkatakrishnan1995convergence}.

\subsection{Supersonic flow around a sphere}
\begin{figure}[H]
	\centering
	\subfloat[]{\includegraphics[width=0.4\textwidth]
		{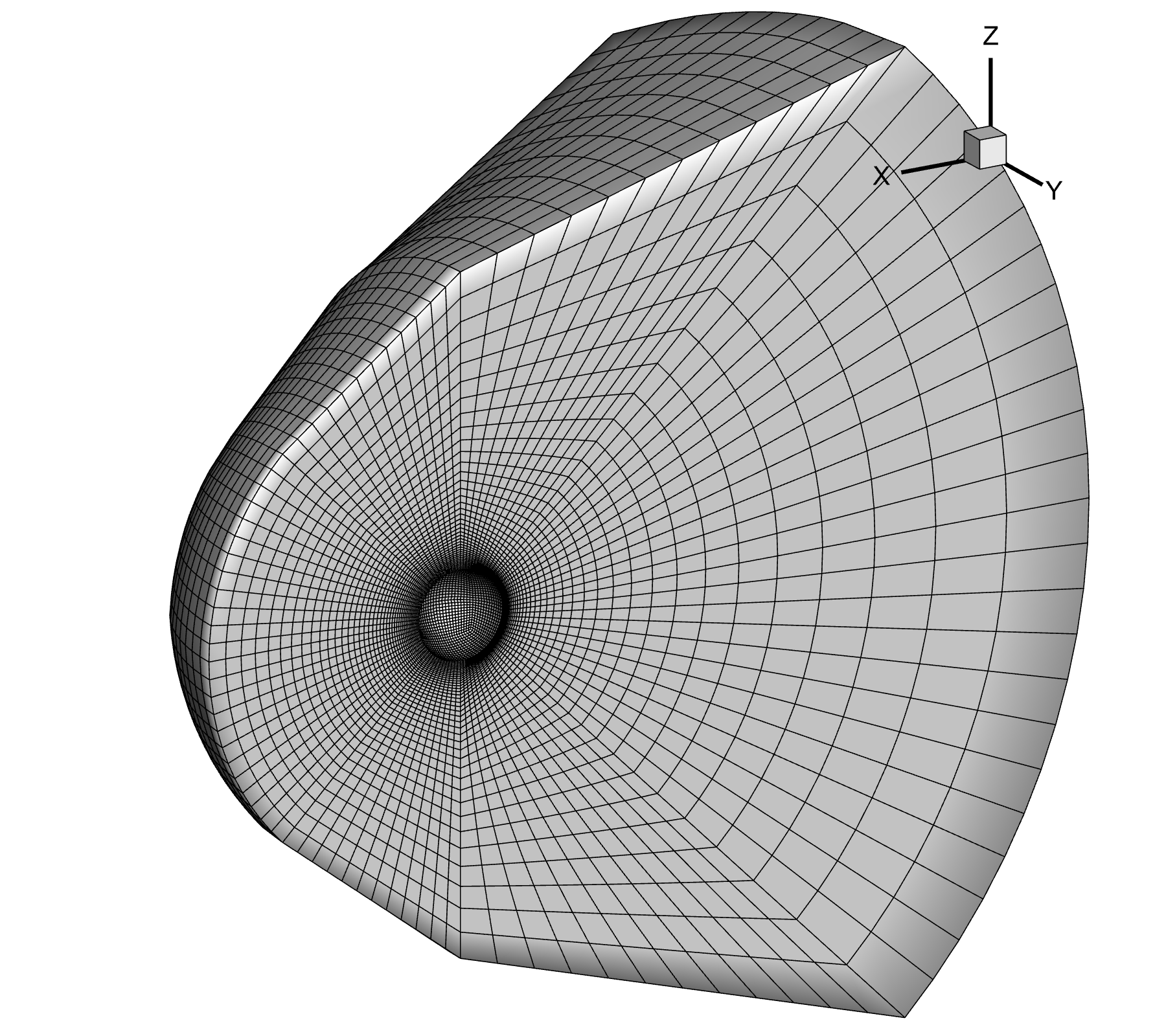}}
	\subfloat[]{\includegraphics[width=0.4\textwidth]
		{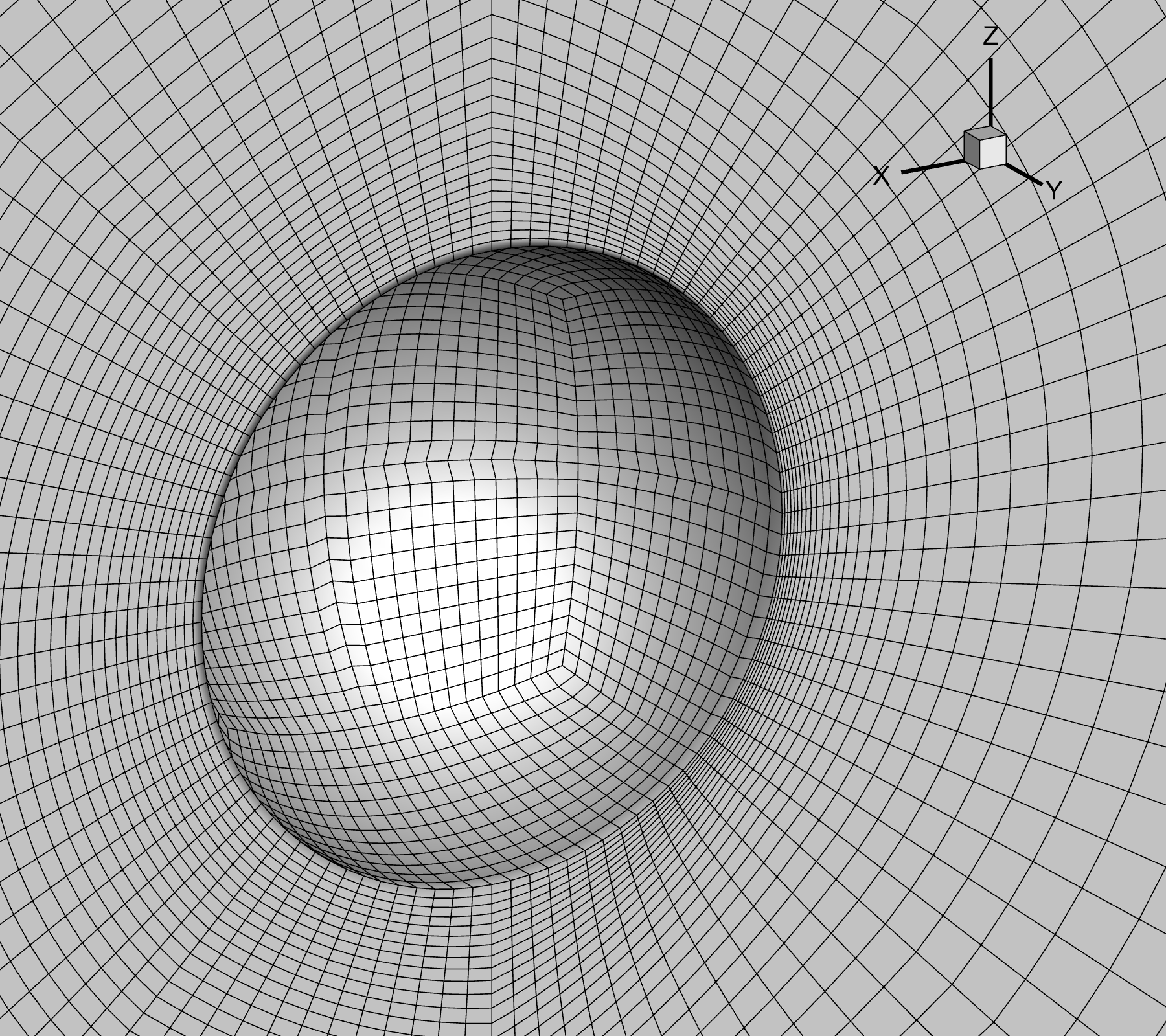}} \\
	\caption{Physical domain of supersonic flow around a sphere which consists 138240 hexahedral cells. (a) The entire domain, and (b) detailed mesh distribution in the vicinity of the solid surface.}
	\label{fig:spheremesh}
\end{figure}

A supersonic flow passing over a three-dimensional sphere at Mach number 4.25 is simulated for various Knudsen numbers using nitrogen gas.  To define the Knudsen number, the reference length is chosen as the diameter of the sphere, i.e., $d=0.002$ m. The physical domain, consisting of $3456 \times 40$ hexahedral cells, is illustrated in Fig.~\ref{fig:spheremesh}, together with a detailed view of the mesh near the solid surface, where the height of the first layer cell is $0.005d$. Tab.~\ref{table:spherecondition} presents the initial conditions of cases tested, including Knudsen number, density, and temperature of the freestream flow, as well as the altitude estimated by density. The simulation adopts supersonic far-field inflow and outflow conditions, and an isothermal wall at the sphere surface with a temperature of the free stream flow. 
The reference number of particles sampled per cell in the UGKWP method is set to 100 for all cases.

\begin{table}[H]
	\caption{Freestream flow parameters of supersonic flow at ${\rm Ma}_\infty=4.25$ around a sphere.}
	\centering
	\begin{tabular}{cccccc}
		\toprule
		${\rm Ma}_\infty$  & ${\rm Kn}_\infty$ & Altitude, km & $\rho_\infty$, kg/($\rm{m}^3 \cdot \rm{s}$) & $T_\infty$, K & $T_w$, K \\
		\midrule
		$4.25$  & 0.672 & 89.1 & $3.173\times 10^{-5}$ & $65.04$ & $302$  \\
		$4.25$  & 0.338 & 83.3 & $6.313\times 10^{-5}$ & $65.04$ & $302$  \\
		$4.25$  & 0.121 & 74.7 & $1.761\times 10^{-4}$ & $65.04$ & $302$  \\
		$4.25$  & 0.080 & 71.1 & $2.675\times 10^{-4}$ & $65.04$ & $302$  \\
		$4.25$  & 0.031 & 63.2 & $6.879\times 10^{-4}$ & $65.04$ & $302$  \\
		\bottomrule
	\end{tabular}
	\label{table:spherecondition}
\end{table}
Table~\ref{table:spherecd} presents the drag coefficients computed by the UGKWP method, along with the corresponding errors compared to experimental results. All errors are less than or nearly $1\%$. As the free stream flow becomes more rarefied, the drag force on the sphere increases.
\begin{table}[H]
	\caption{Comparison of the drag coefficients for supersonic flow at ${\rm Ma}_\infty=4.25$ over a sphere.}
	\centering
	\begin{tabular}{ccccc}
		\toprule
		\multirow{2}{*}{${\rm Ma}_\infty$} & \multirow{2}{*}{${\rm Kn}_\infty$} & \multicolumn{3}{c}{Drag Coefficient (Error)} \\
		\cline{3-5} &  &
		\begin{tabular}[c]{@{}c@{}}Experiment (Air)\end{tabular} &
		\begin{tabular}[c]{@{}c@{}}UGKS      (${\rm N}_2$)\end{tabular} &
		\begin{tabular}[c]{@{}c@{}}UGKWP (${\rm N}_2$)\end{tabular}
		\\ \midrule
		4.25 & 0.672 & 2.42 & 2.356 (-2.64\%) & 2.423 (0.13\%)  \\
		4.25 & 0.338 & 2.12 & 2.101 (-0.87\%) & 2.114 (-0.30\%)	\\
		4.25 & 0.121 & 1.69 & 1.694 (-0.23\%) & 1.671 (-1.12\%)	\\
		4.25 & 0.080 & 1.53 & 1.558 (1.80\%) & 1.527 (-0.23\%) \\
		4.25 & 0.031 & 1.35 & 1.355 (0.39\%) & 1.352 (0.12\%)\\
		\bottomrule
	\end{tabular}
	\label{table:spherecd}
\end{table}

\begin{figure}[H]
	\centering
	\subfloat[]{\includegraphics[width=0.4\textwidth]
		{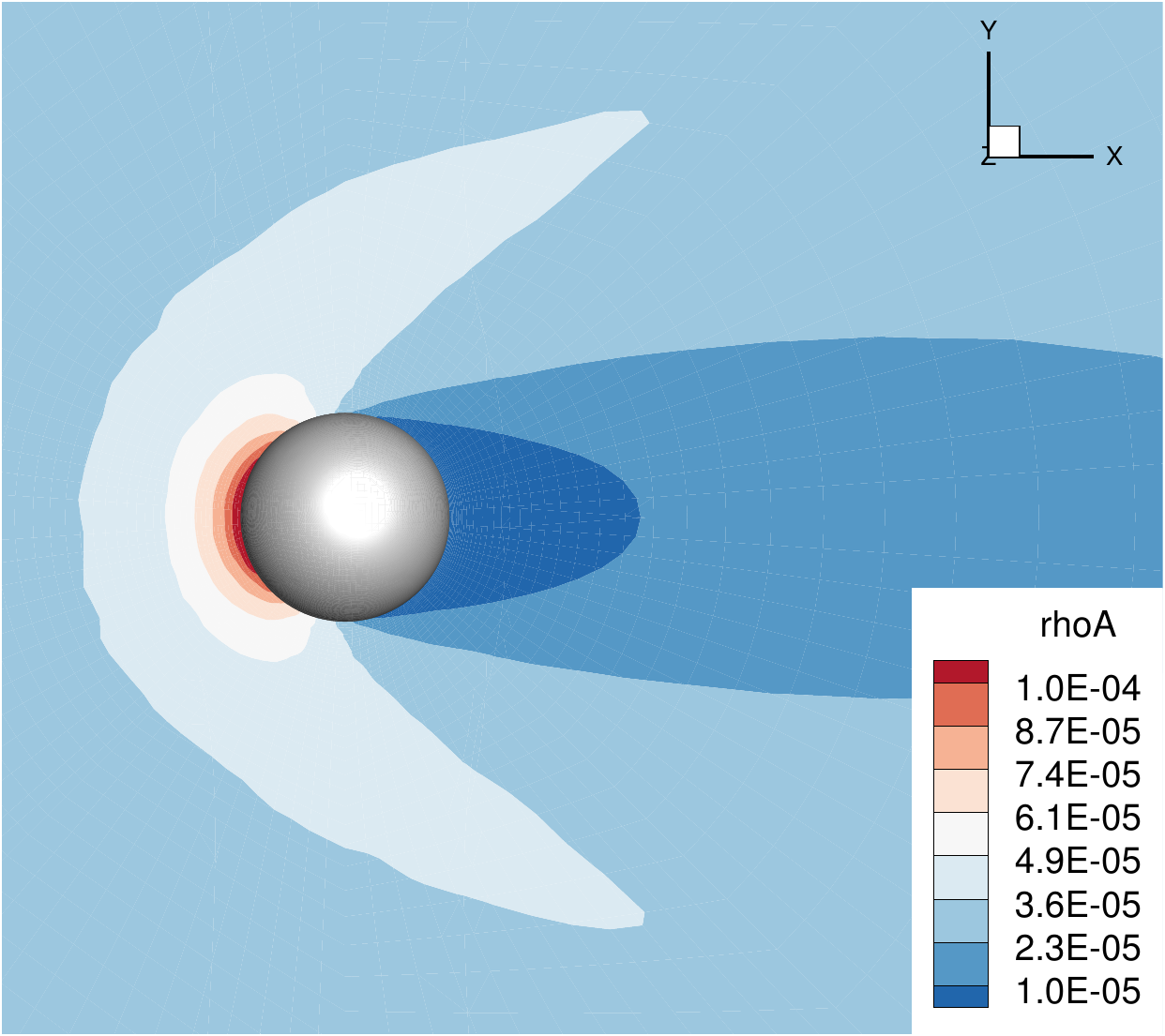}}
	\subfloat[]{\includegraphics[width=0.4\textwidth]
		{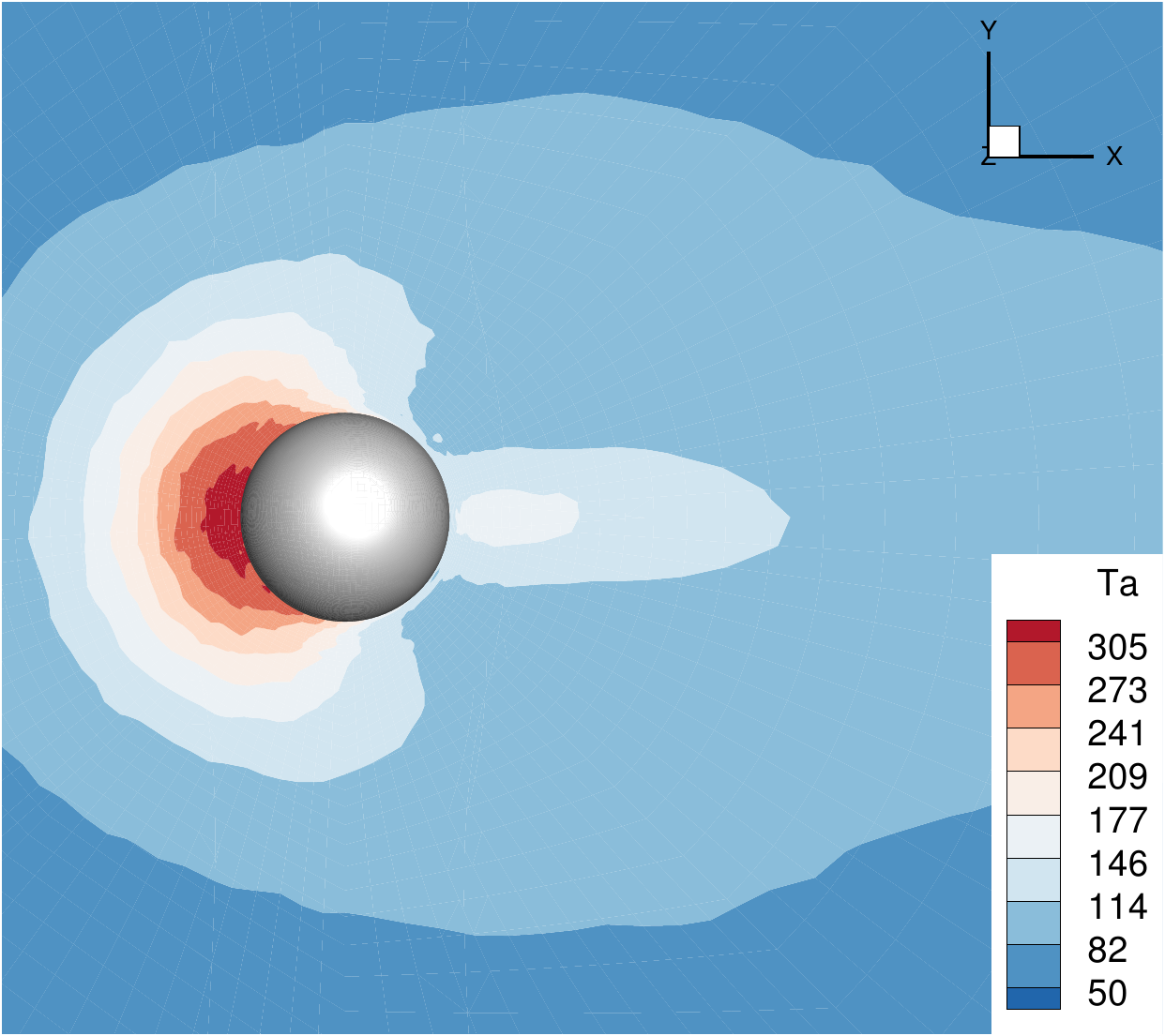}} \\
	\subfloat[]{\includegraphics[width=0.4\textwidth]
		{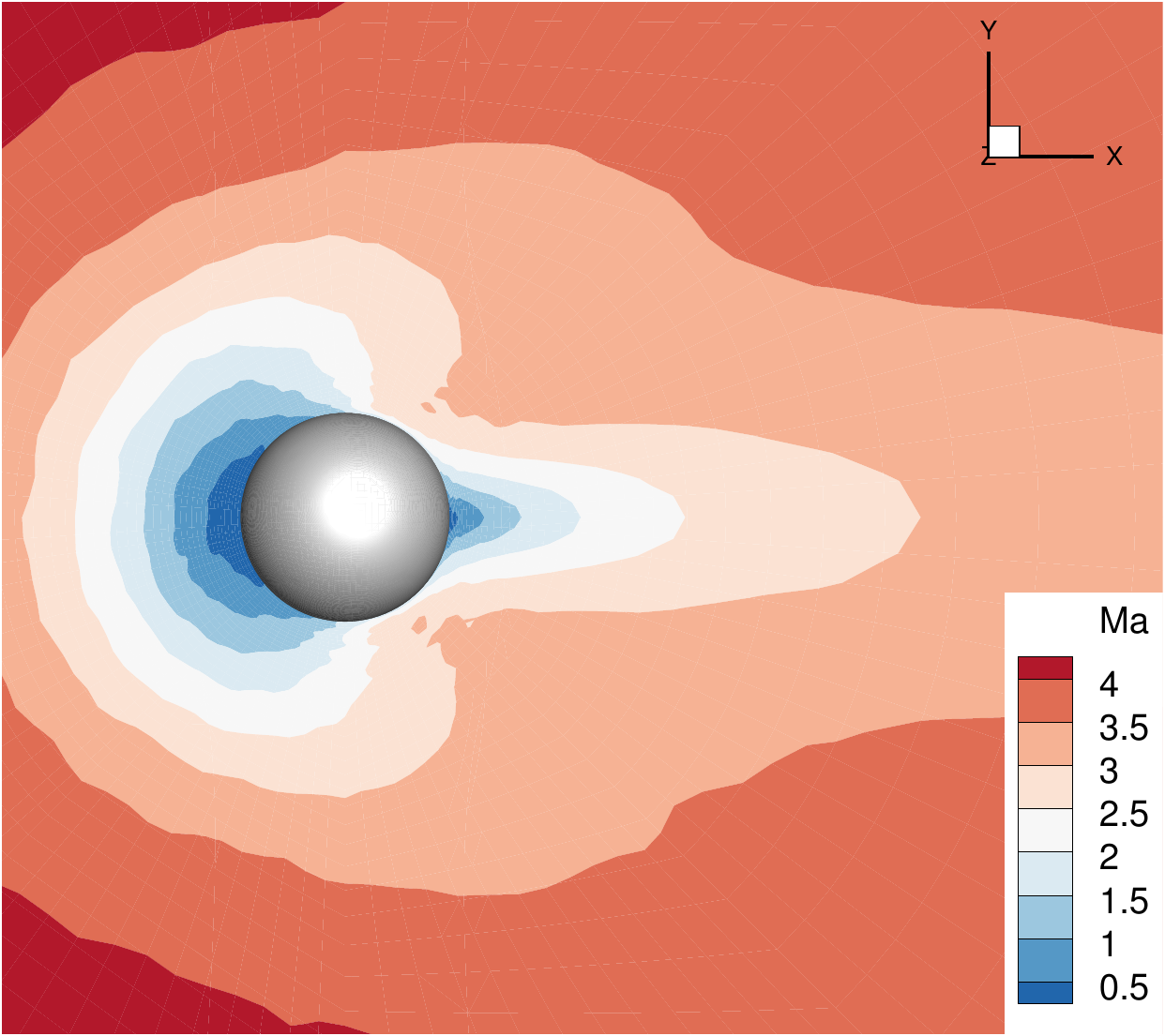}}
	\subfloat[]{\includegraphics[width=0.4\textwidth]
		{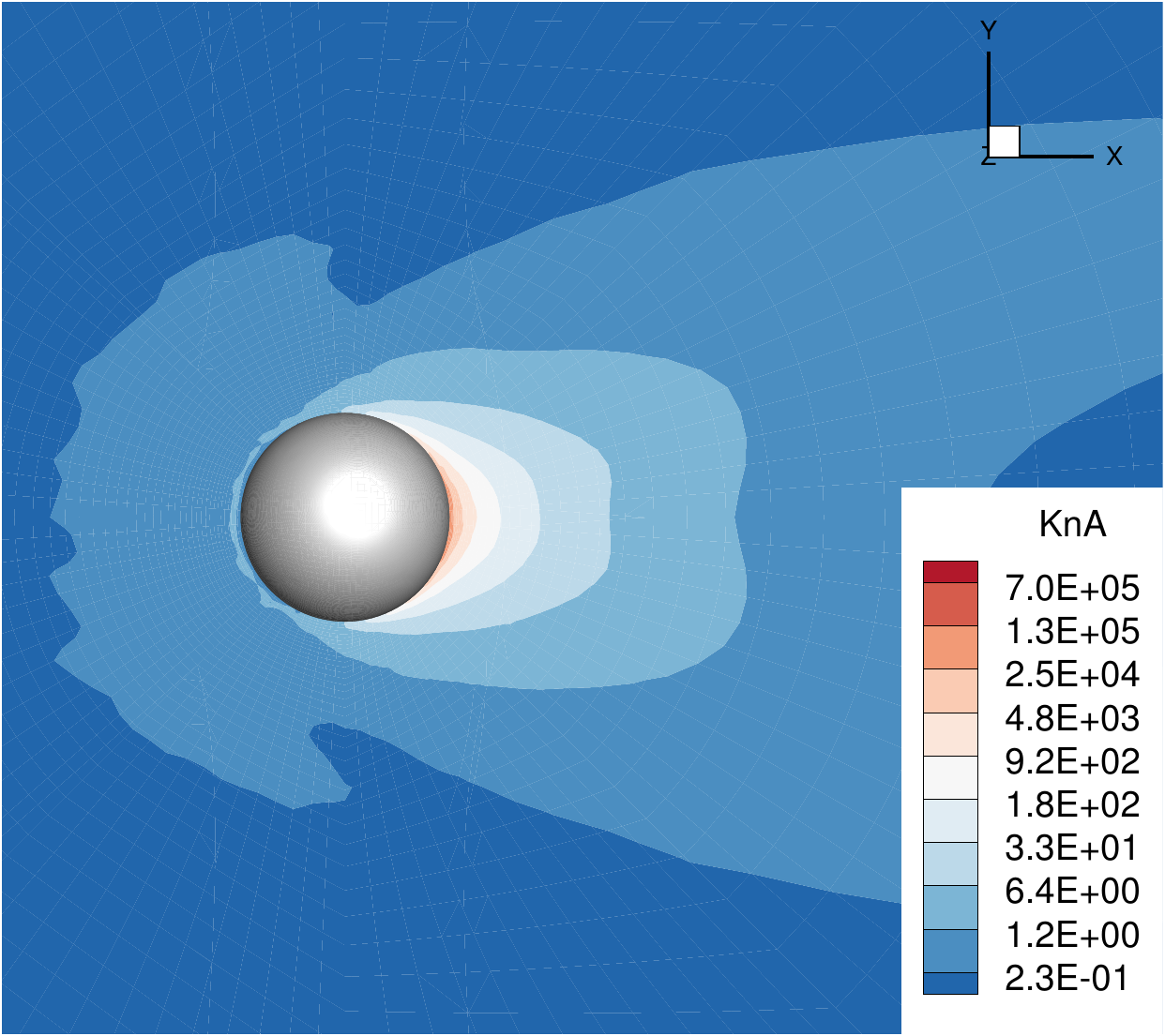}} \\
	\caption{Supersonic flow
		around a sphere at ${\rm Ma}_\infty = 4.25$ and ${\rm Kn}_\infty = 0.672$ by the UGKWP method. (a) Density, (b)temperature, (c) Mach number,
		and (d) local Knudsen number contours.}
	\label{fig:sphere-Ma4.25-1}
\end{figure}

\begin{figure}[H]
	\centering
	\subfloat[]{\includegraphics[width=0.4\textwidth]
		{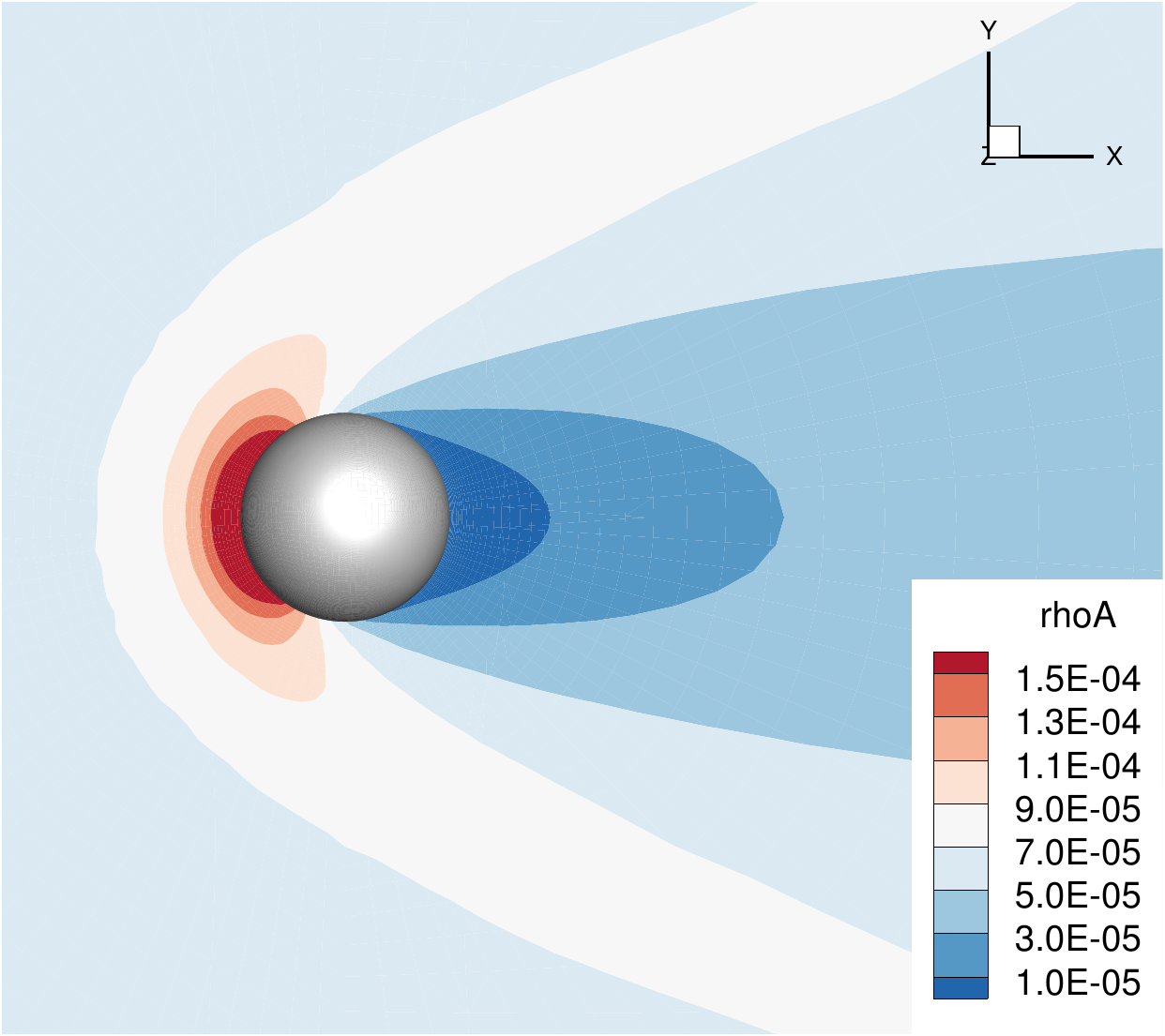}}
	\subfloat[]{\includegraphics[width=0.4\textwidth]
		{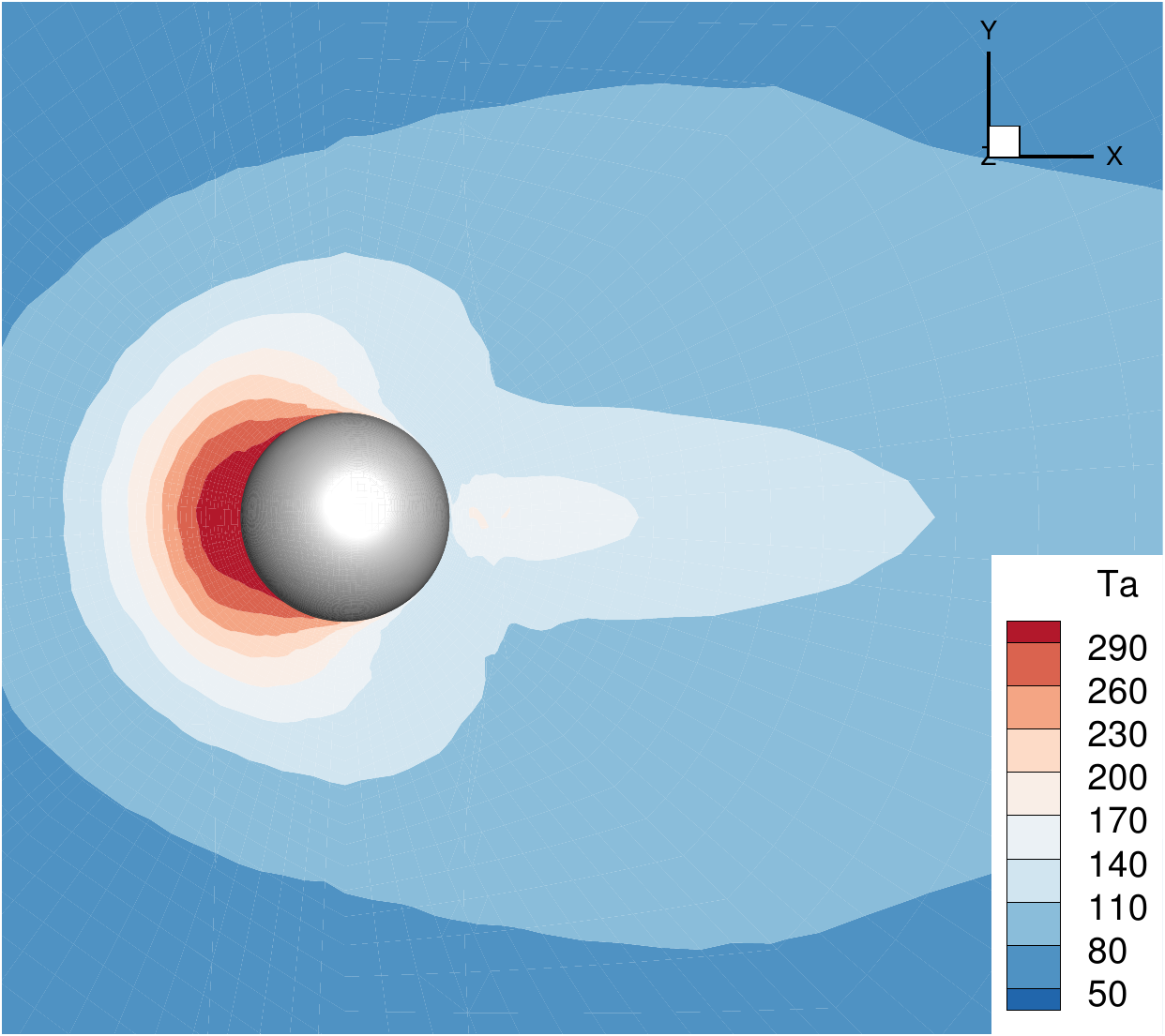}} \\
	\subfloat[]{\includegraphics[width=0.4\textwidth]
		{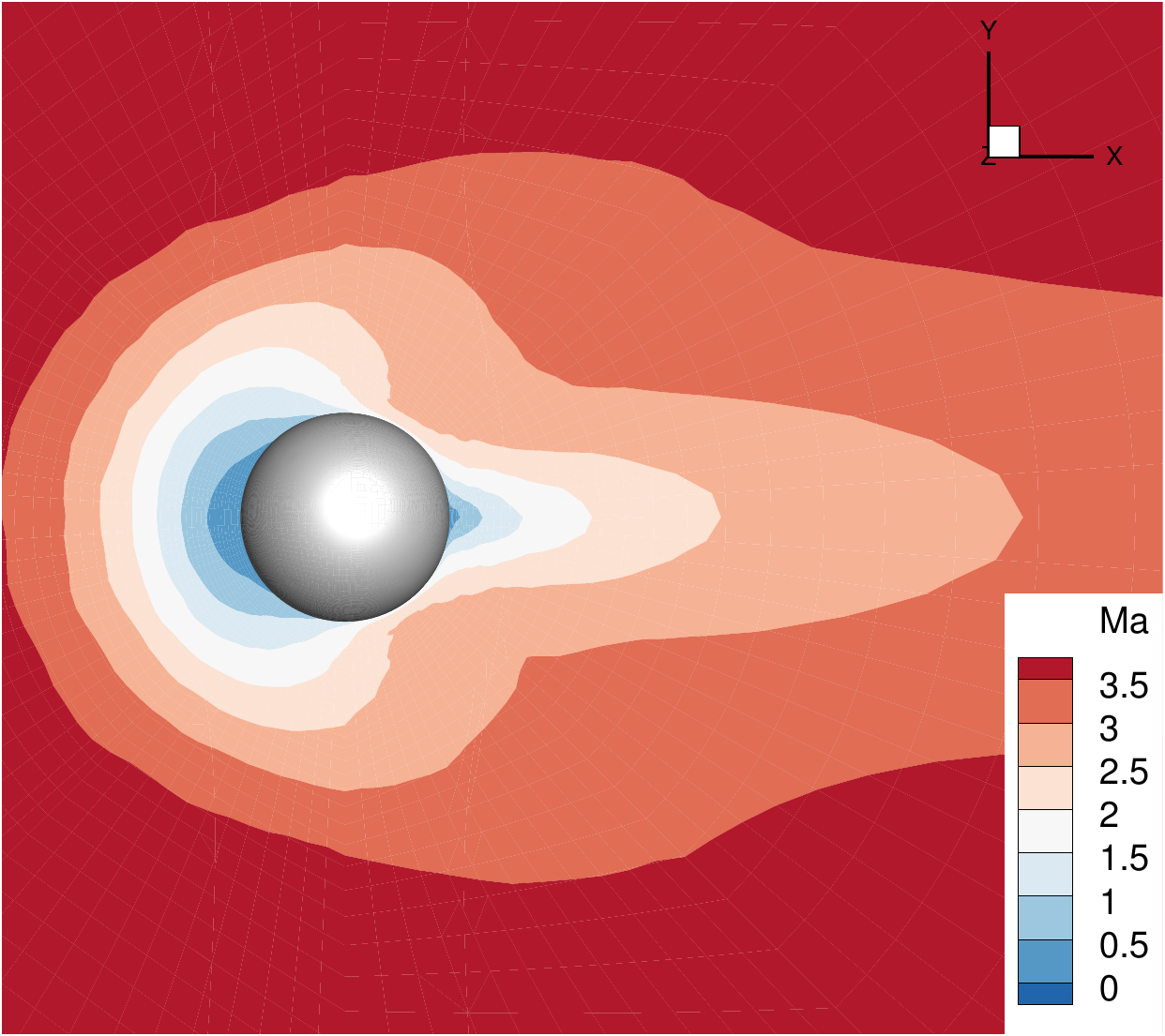}}
	\subfloat[]{\includegraphics[width=0.4\textwidth]
		{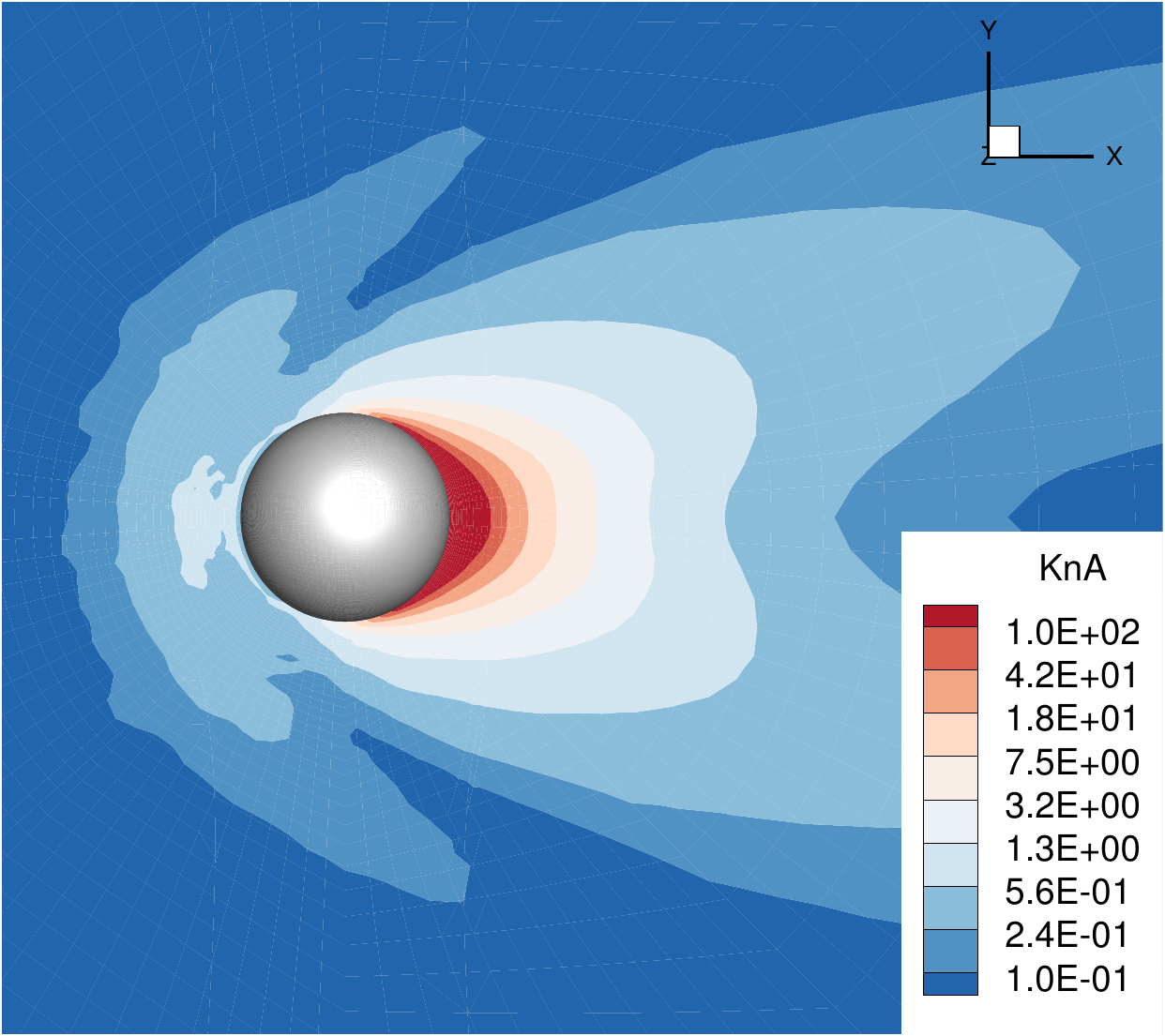}} \\
	\caption{Supersonic flow
		around a sphere at ${\rm Ma}_\infty = 4.25$ and ${\rm Kn}_\infty = 0.338$ by the UGKWP method. (a) Density, (b) temperature, (c) Mach number,
		and (d) local Knudsen number contours.}
	\label{fig:sphere-Ma4.25-2}
\end{figure}

\begin{figure}[H]
	\centering
	\subfloat[]{\includegraphics[width=0.4\textwidth]
		{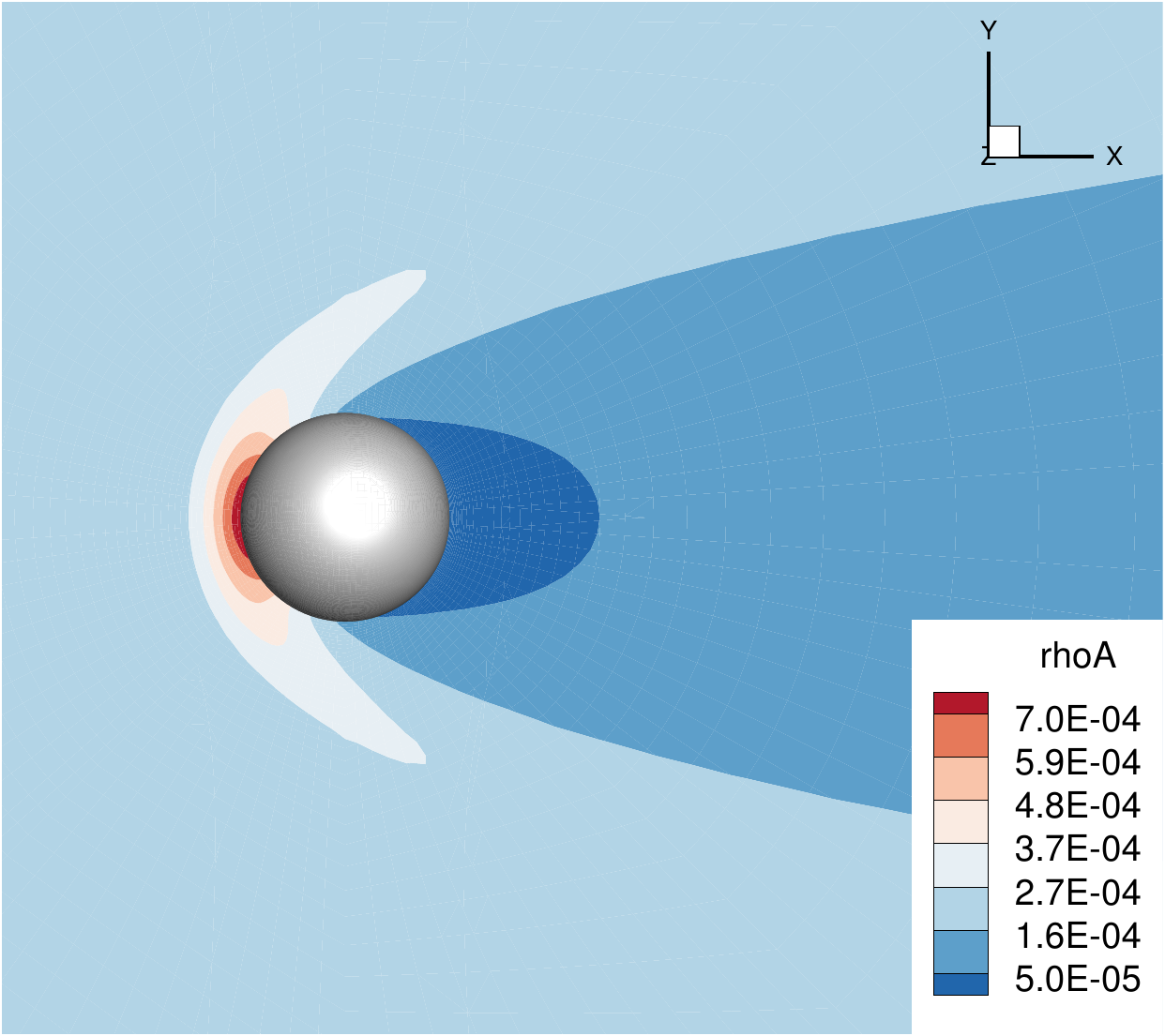}}
	\subfloat[]{\includegraphics[width=0.4\textwidth]
		{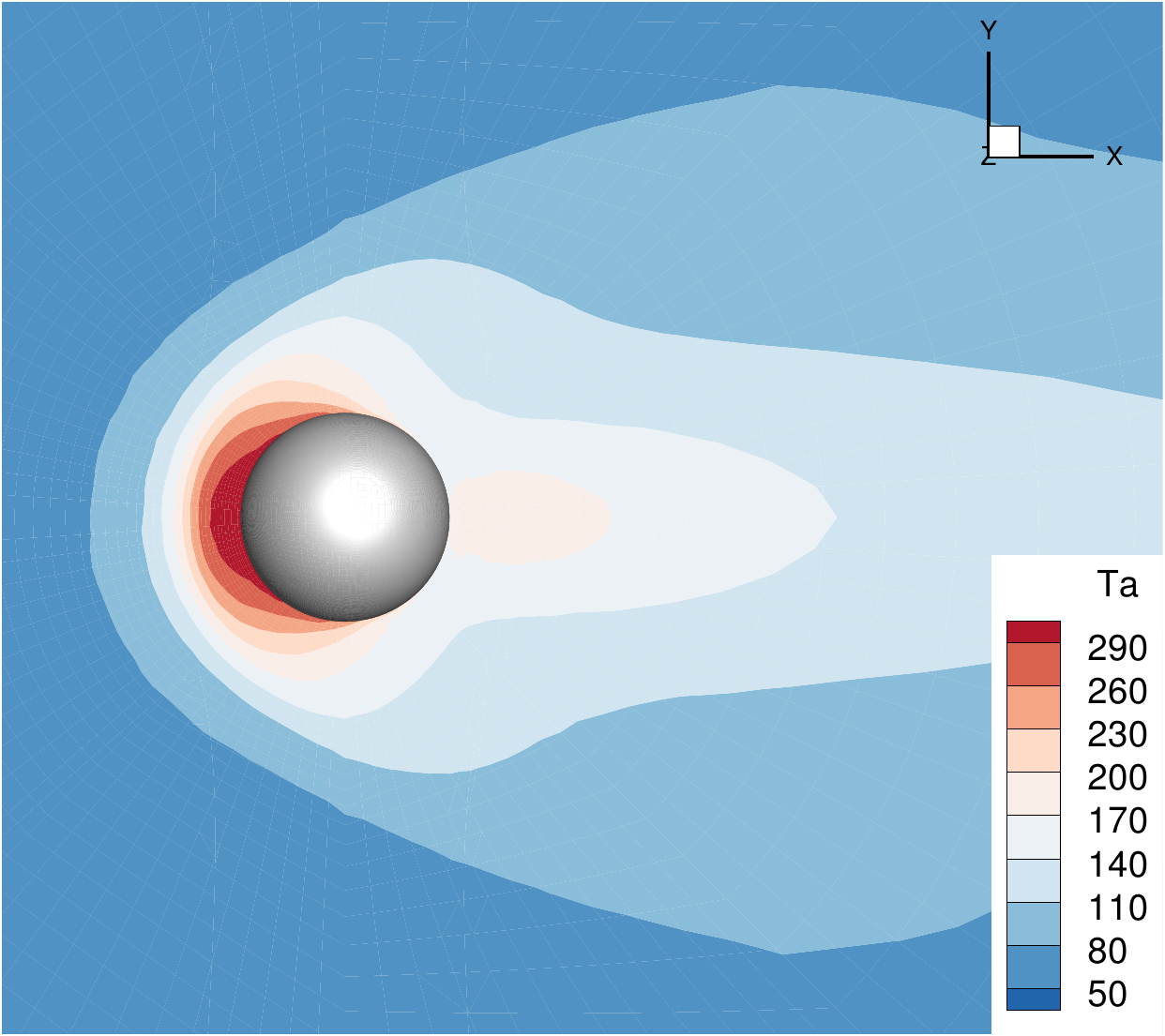}} \\
	\subfloat[]{\includegraphics[width=0.4\textwidth]
		{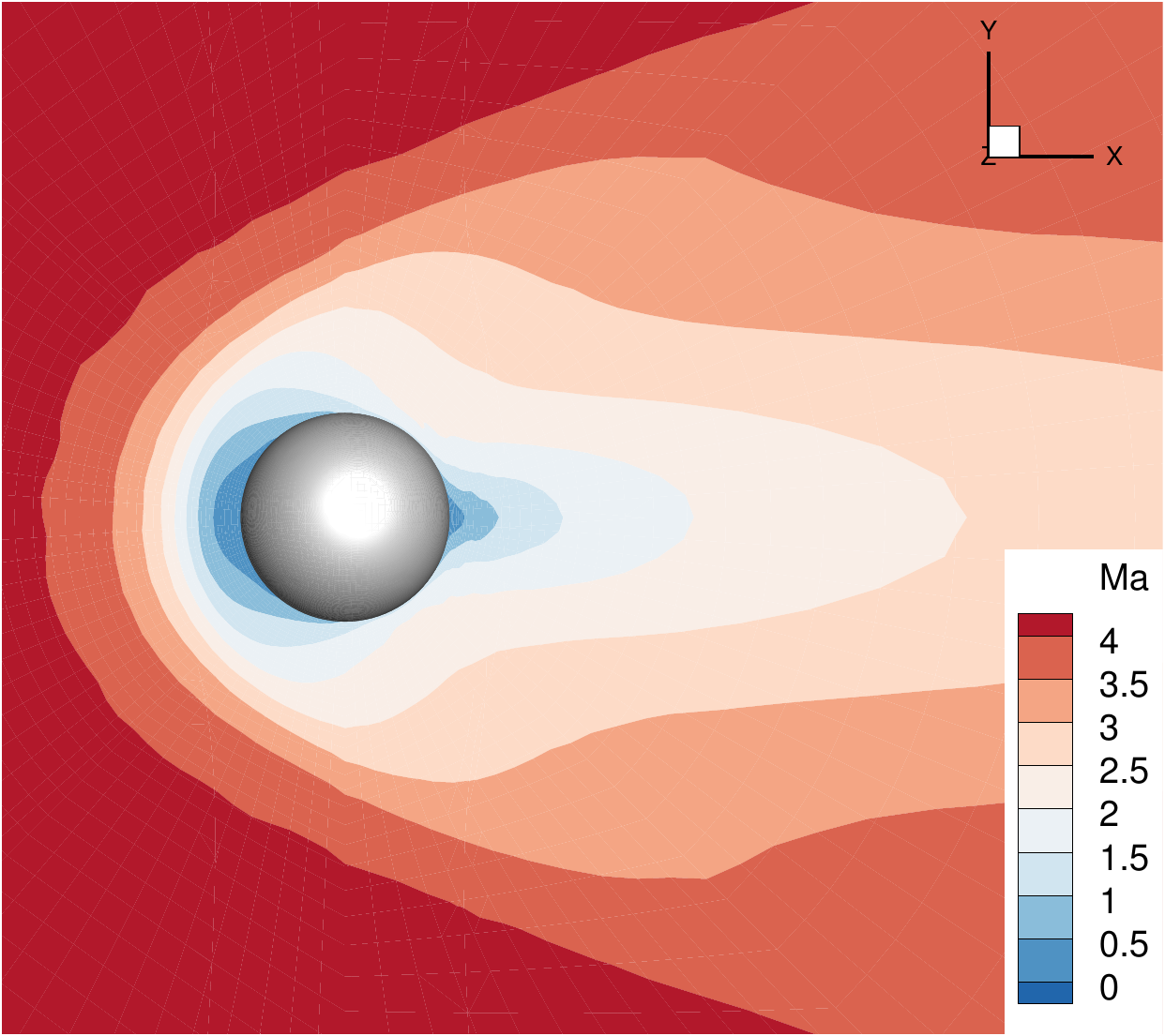}}
	\subfloat[]{\includegraphics[width=0.4\textwidth]
		{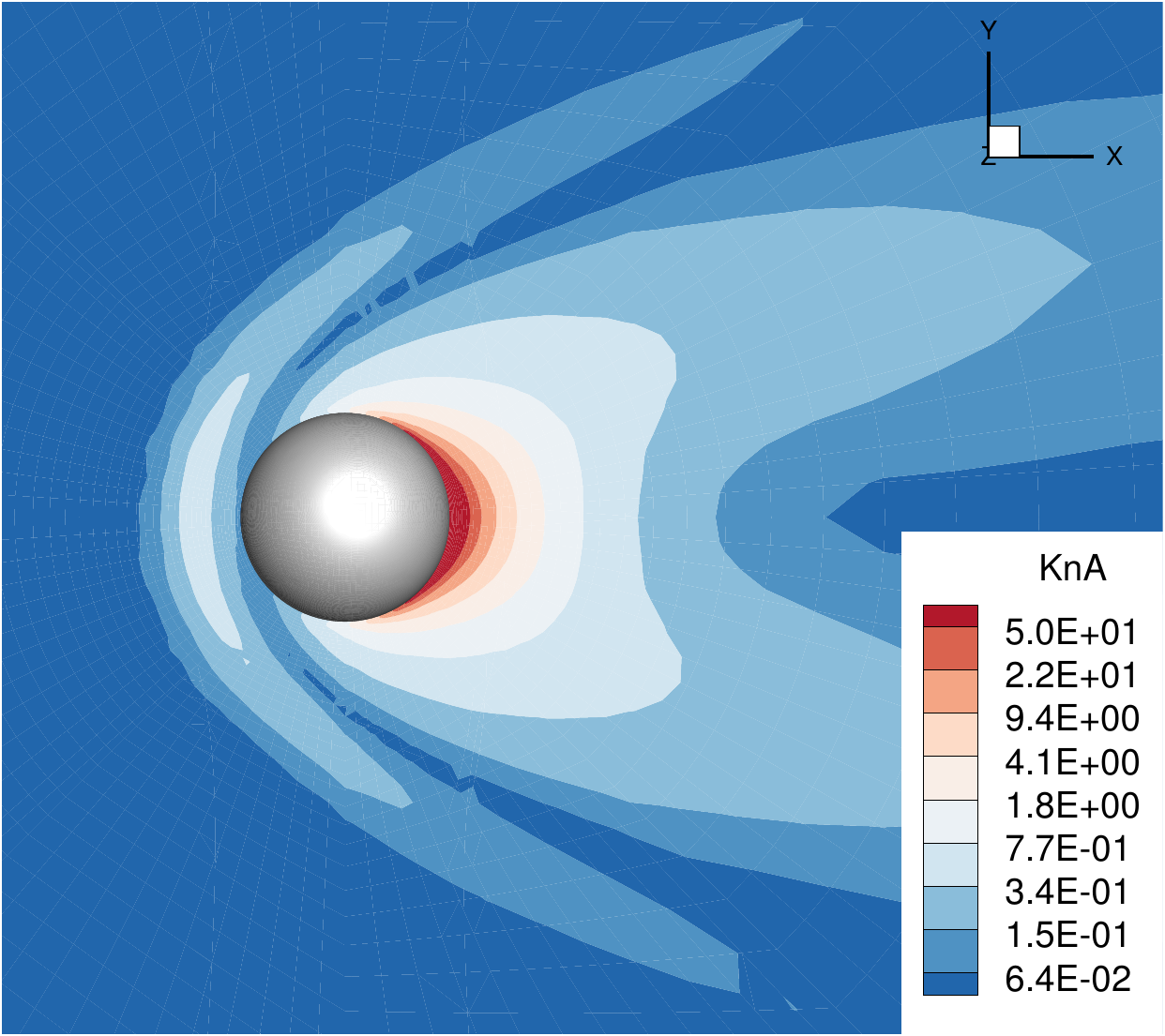}} \\
	\caption{Supersonic flow
		around a sphere at ${\rm Ma}_\infty = 4.25$ and ${\rm Kn}_\infty = 0.121$ by the UGKWP method. (a) Density, (b) temperature, (c) Mach number,
		and (d) local Knudsen number contours.}
	\label{fig:sphere-Ma4.25-3}
\end{figure}

\begin{figure}[H]
	\centering
	\subfloat[]{\includegraphics[width=0.4\textwidth]
		{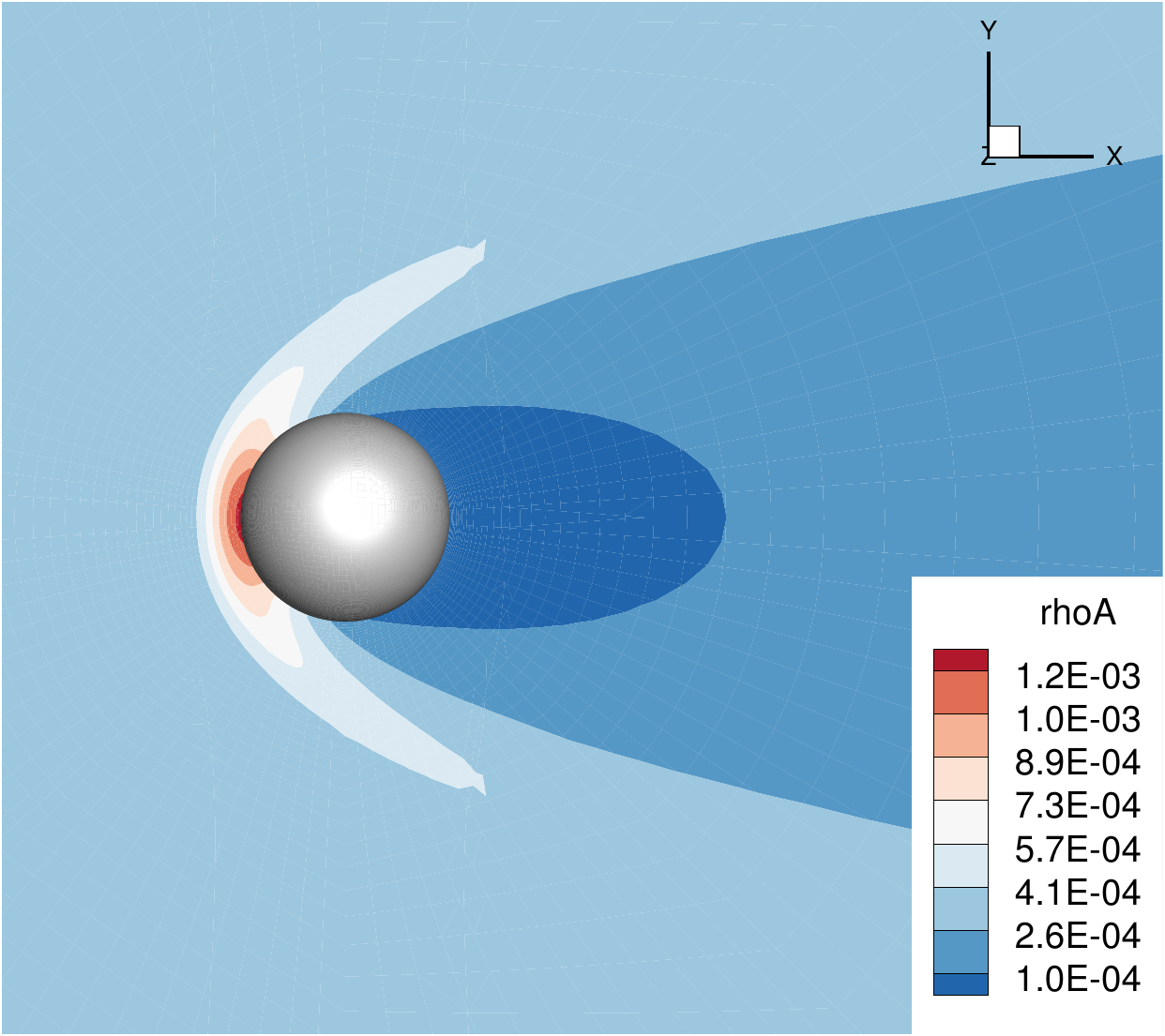}}
	\subfloat[]{\includegraphics[width=0.4\textwidth]
		{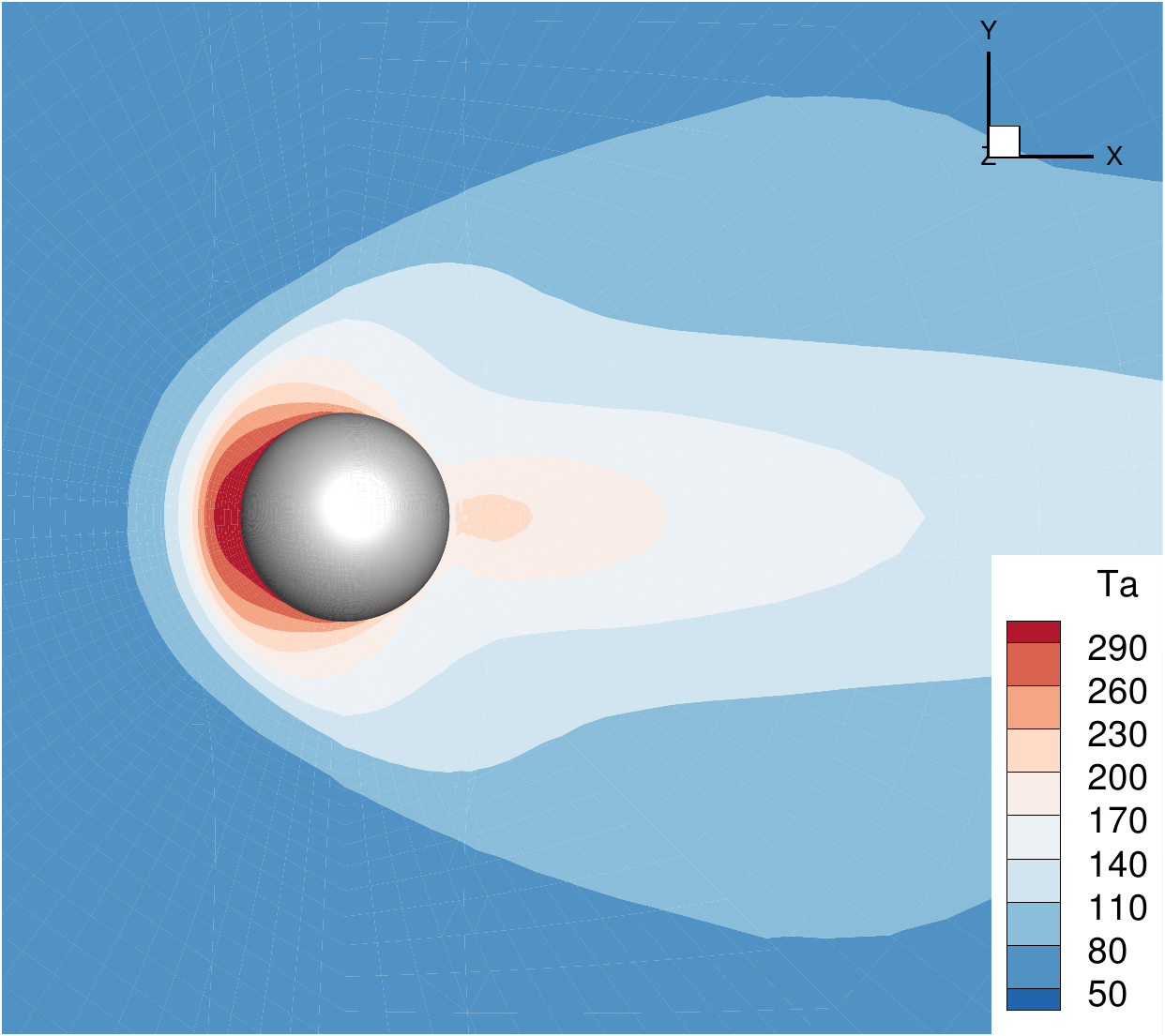}} \\
	\subfloat[]{\includegraphics[width=0.4\textwidth]
		{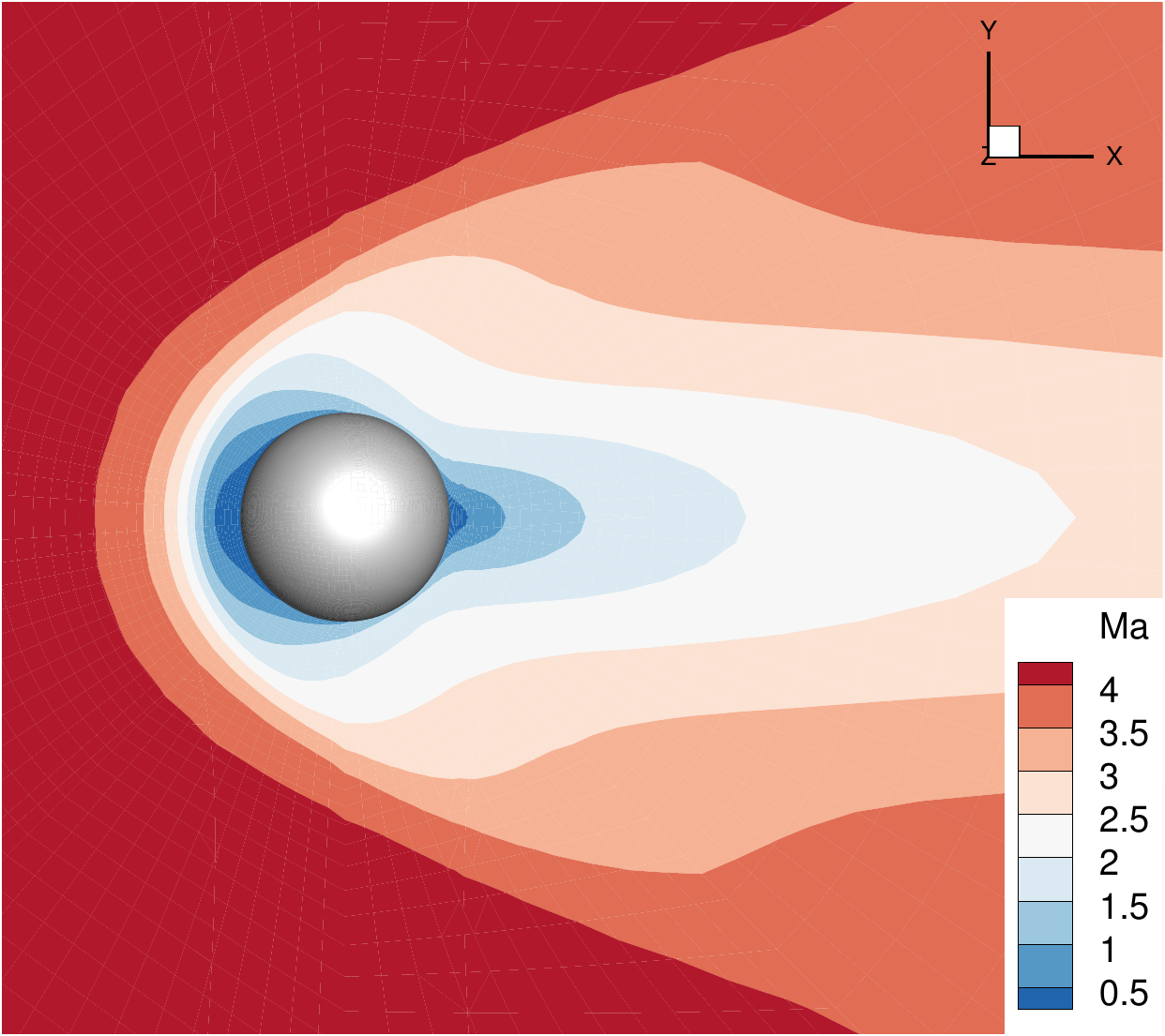}}
	\subfloat[]{\includegraphics[width=0.4\textwidth]
		{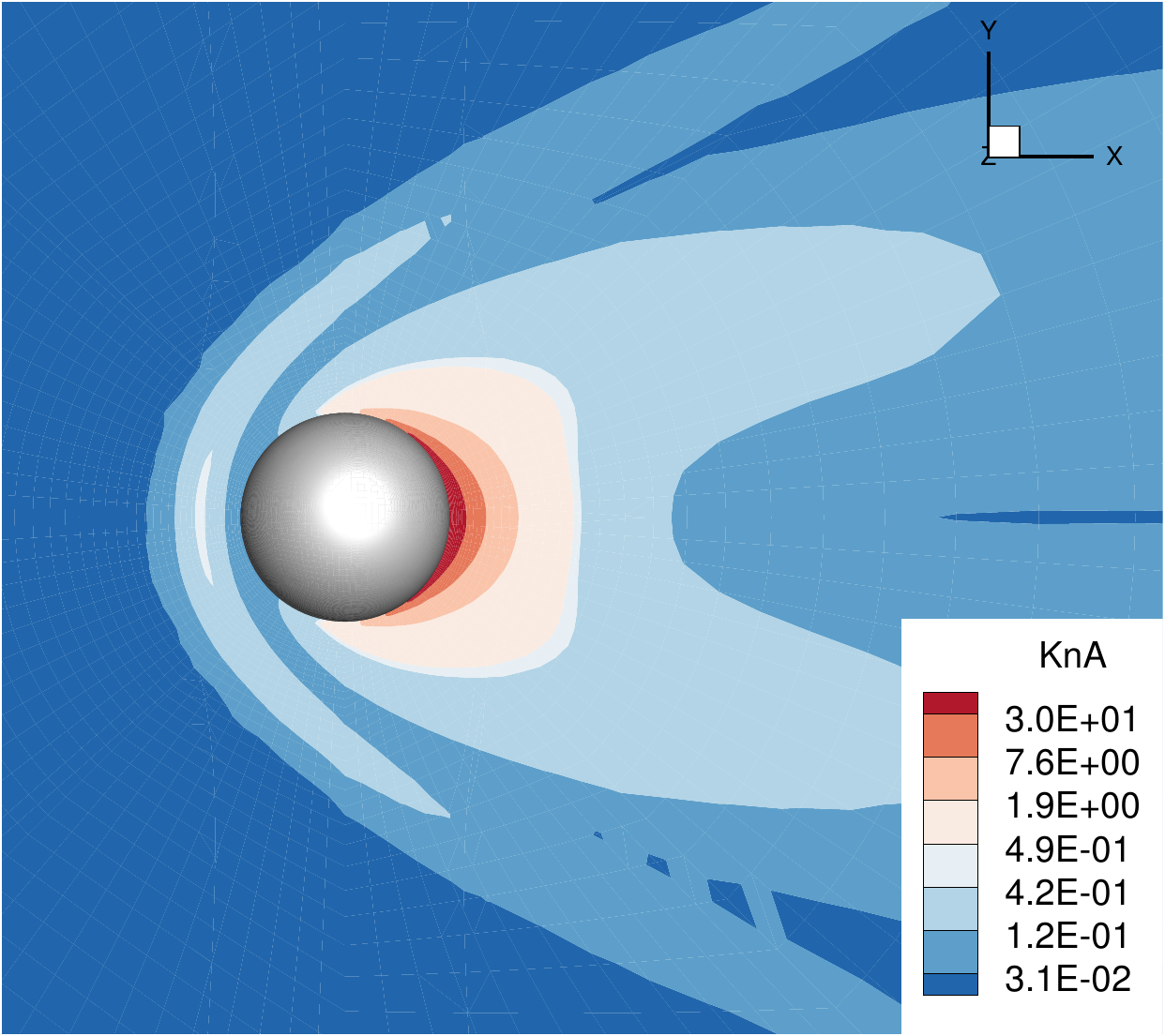}} \\
	\caption{Supersonic flow
		around a sphere at ${\rm Ma}_\infty = 4.25$ and ${\rm Kn}_\infty = 0.080$ by the UGKWP method. (a) Density, (b) temperature, (c) Mach number, and (d) local Knudsen number contours.}
	\label{fig:sphere-Ma4.25-4}
\end{figure}

\begin{figure}[H]
	\centering
	\subfloat[]{\includegraphics[width=0.4\textwidth]
		{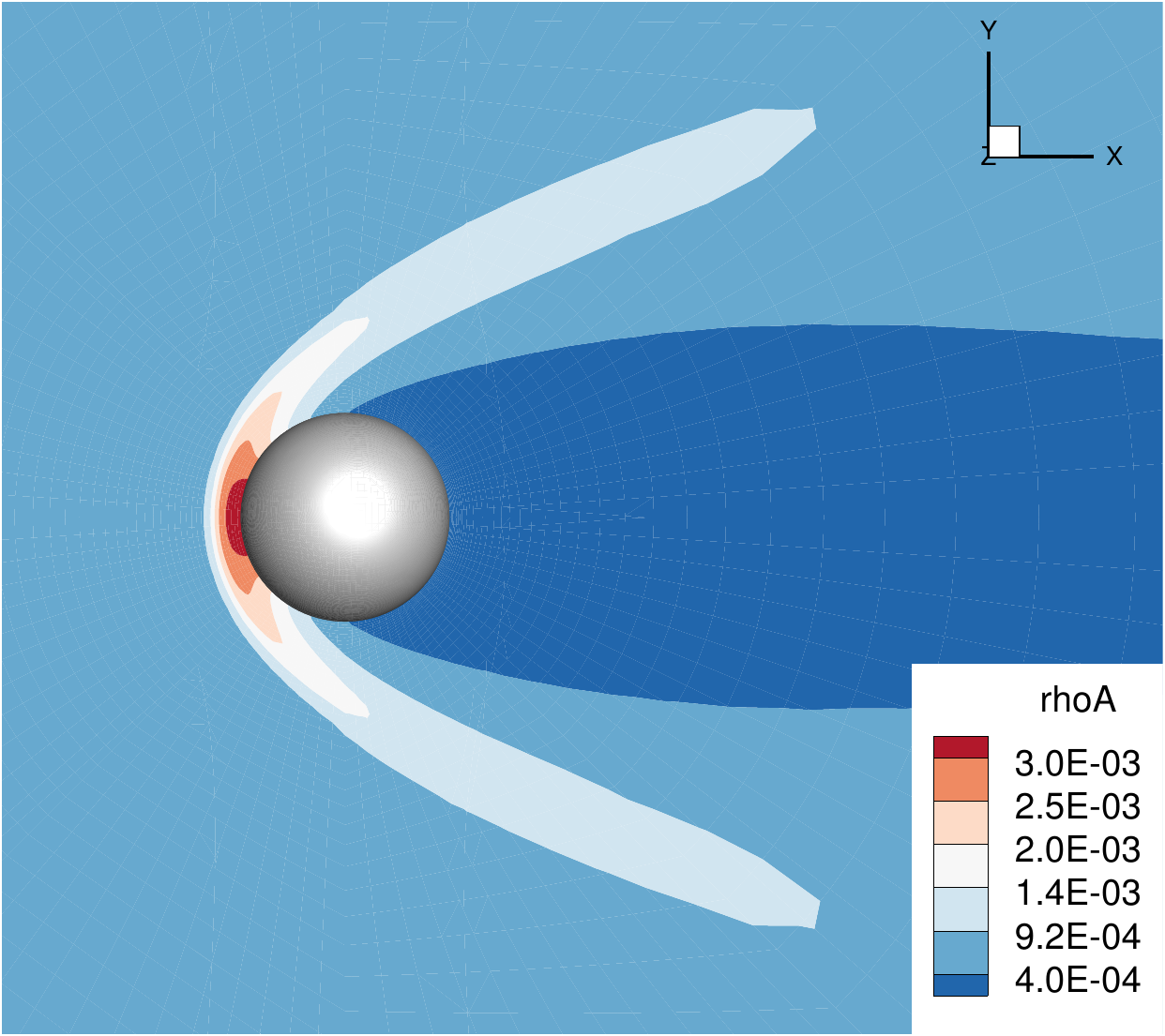}}
	\subfloat[]{\includegraphics[width=0.4\textwidth]
		{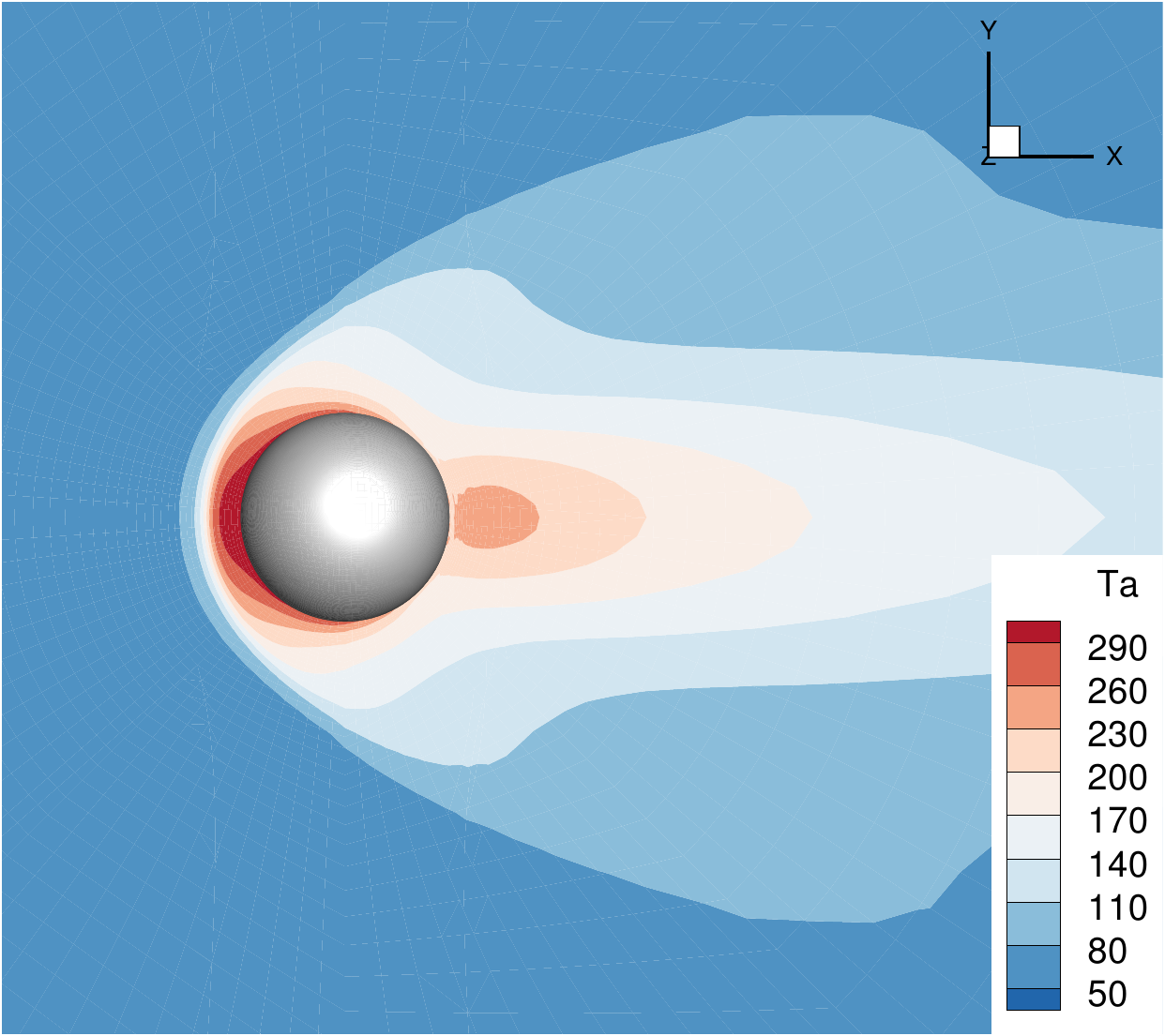}} \\
	\subfloat[]{\includegraphics[width=0.4\textwidth]
		{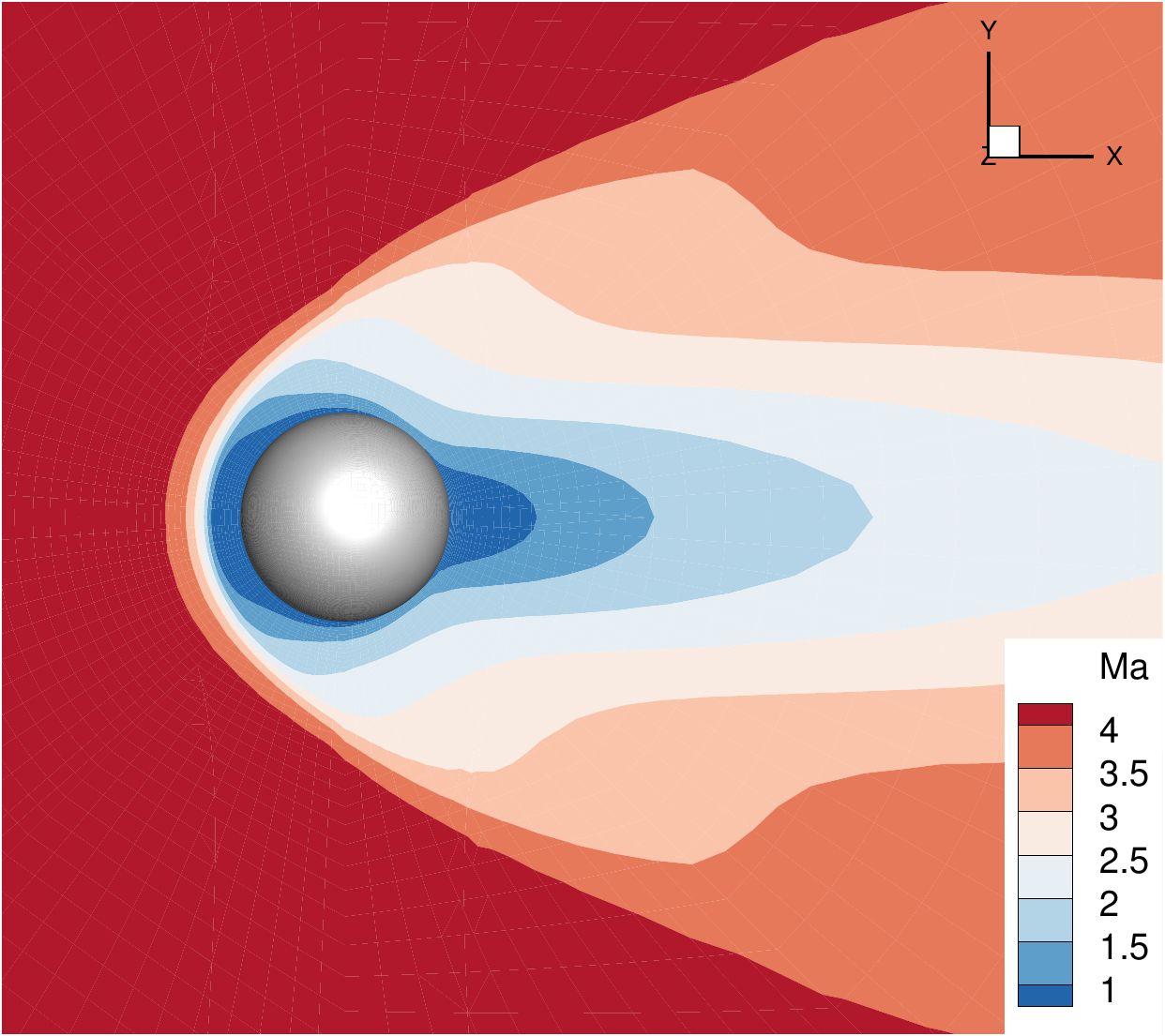}}
	\subfloat[]{\includegraphics[width=0.4\textwidth]
		{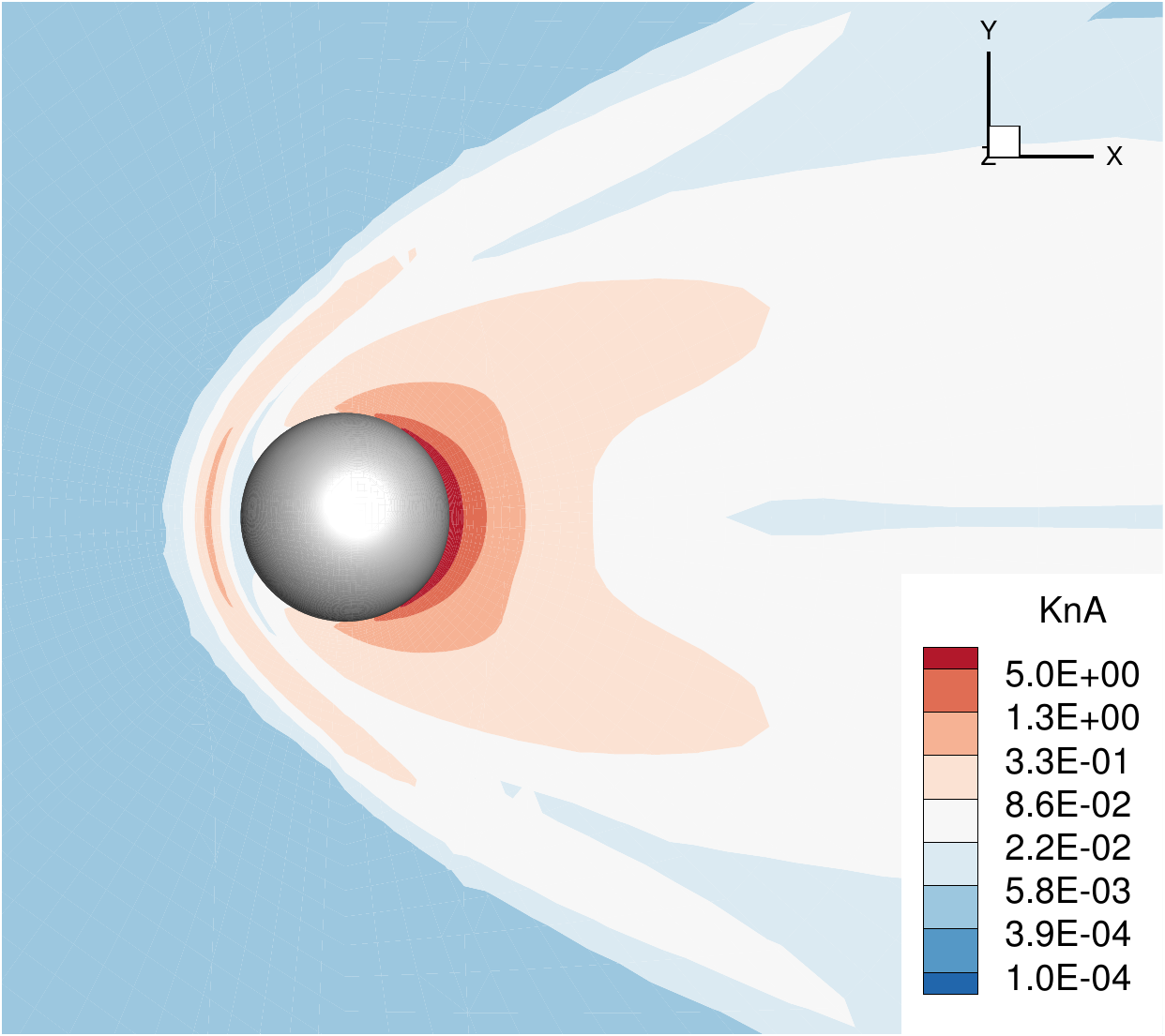}} \\
	\caption{Supersonic flow
		around a sphere at ${\rm Ma}_\infty = 4.25$ and ${\rm Kn}_\infty = 0.031$ by the UGKWP method. (a) Density, (b) temperature, (c) Mach number,
		and (d) local Knudsen number contours.}
	\label{fig:sphere-Ma4.25-5}
\end{figure}

Figure~\ref{fig:sphere-Ma4.25-1} to~\ref{fig:sphere-Ma4.25-5} depict the distribution of density, temperature, Mach number, and gradient-length-dependent local Knudsen number in the flow field simulated by UGKWP. An increase in the thickness of the bow shock in front of the sphere is observed as the Knudsen number increases. The local Knudsen number distribution reveals strong non-equilibrium effects in the leeward area of the sphere and within the shock.

\begin{figure}[H]
	\centering
	\subfloat[]{\includegraphics[width=0.33\textwidth]
		{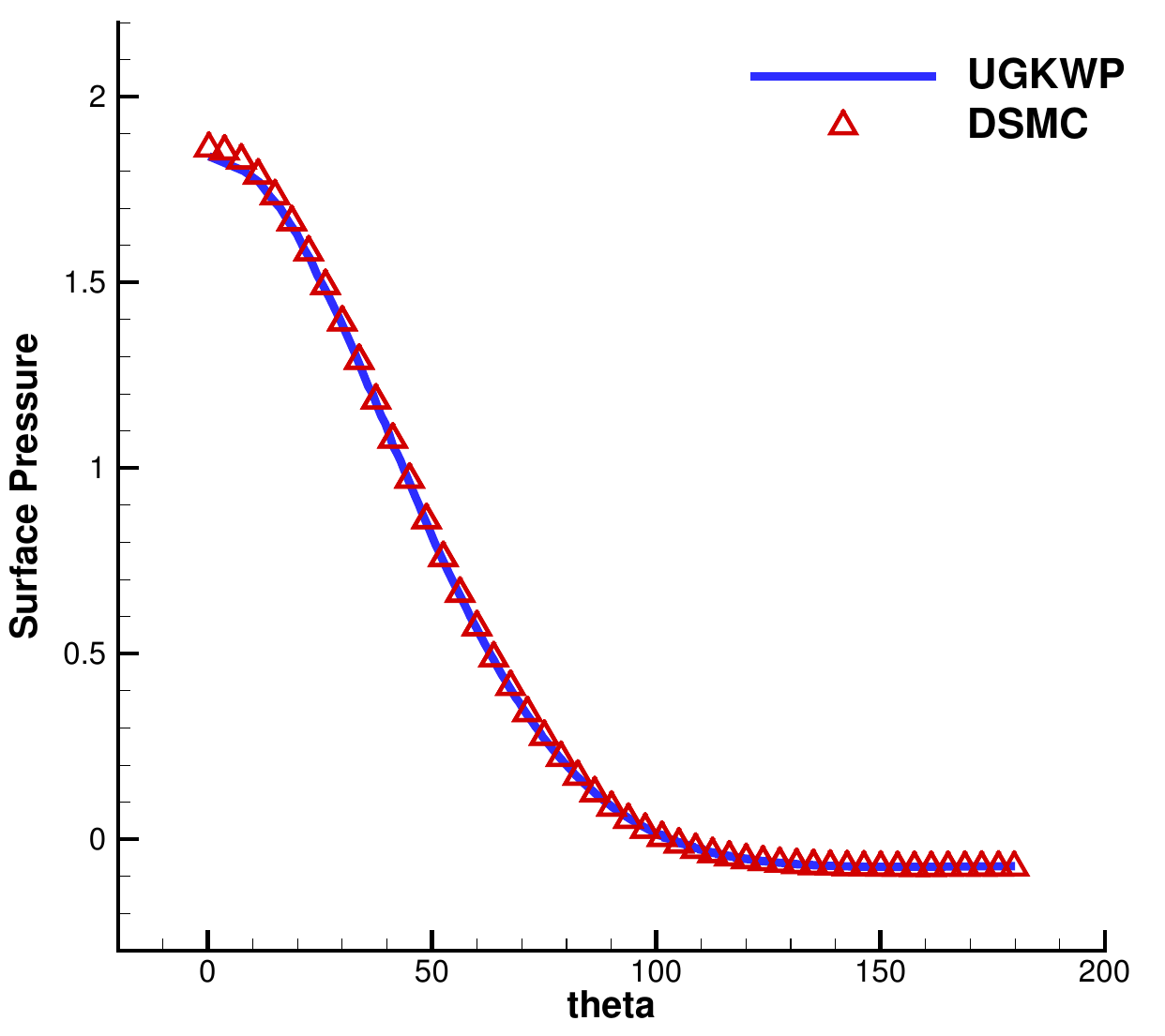}}
	\subfloat[]{\includegraphics[width=0.33\textwidth]
		{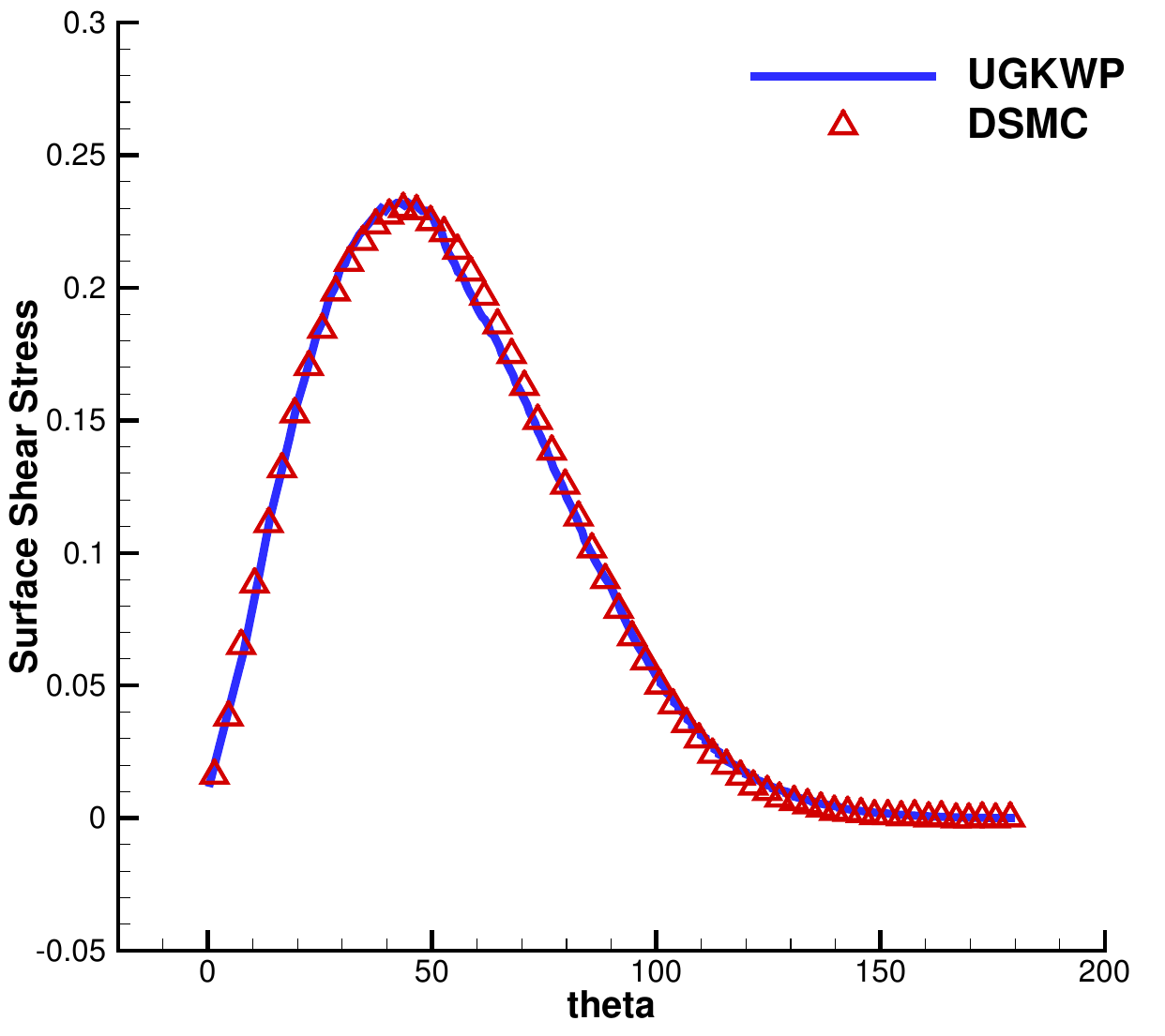}}
	\subfloat[]{\includegraphics[width=0.33\textwidth]
		{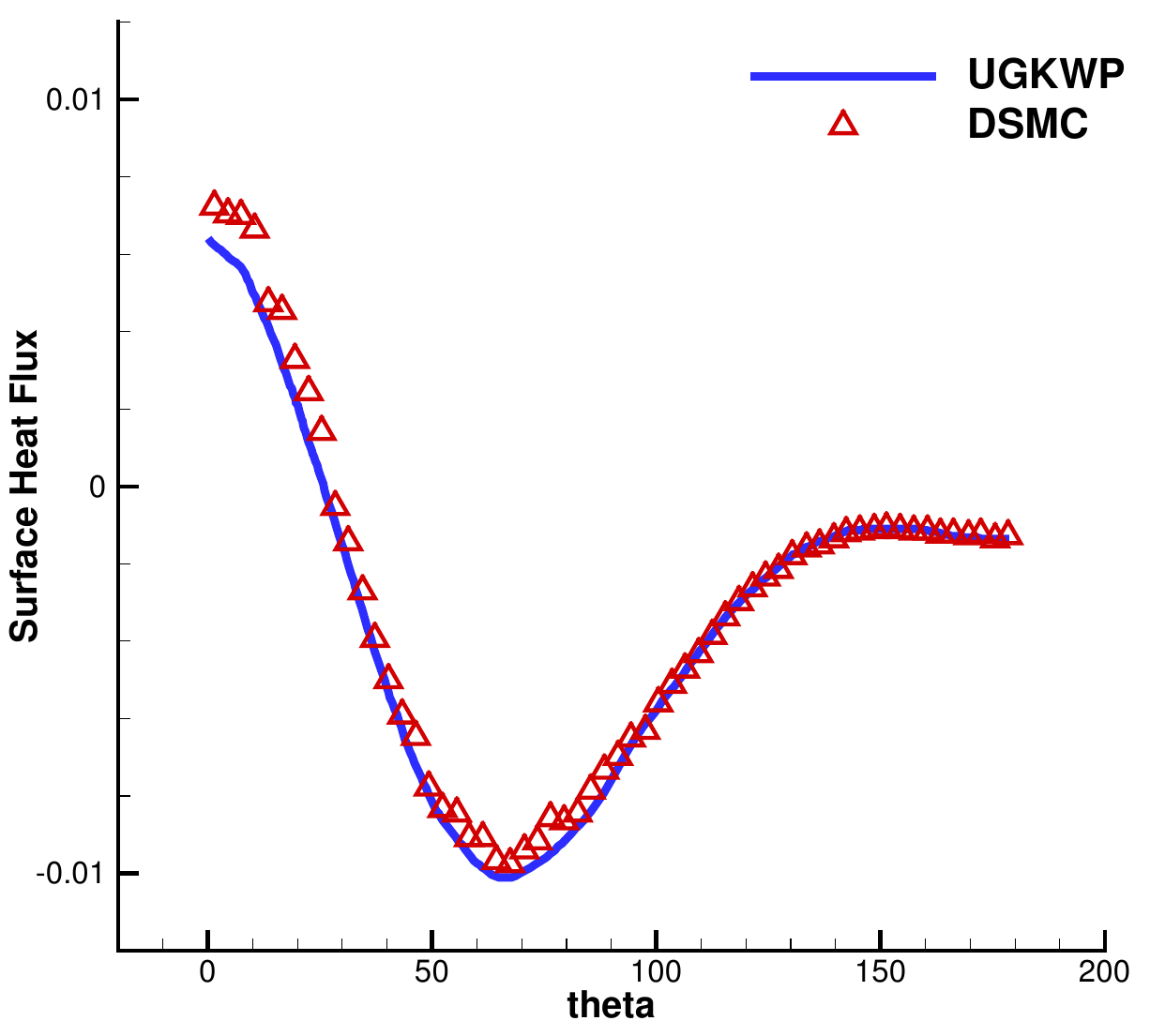}}  \\
	\caption{Surface quantities on a sphere at ${\rm Ma_\infty} = 4.25$ and ${\rm Kn_\infty} = 0.031$ by the UGKWP method. (a) Surface pressure coefficient $C_p$, (b) surface shear stress coefficient $C_{\tau}$, and (c) surface heat flux coefficient $C_{h}$.}
	\label{fig:sphere-surface-0.031}
\end{figure}

Figure~\ref{fig:sphere-surface-0.031} illustrates the distribution of pressure, shear stress, and heat flux coefficients on the solid surface. The pressure and shear stress coefficients exhibit excellent agreement with DSMC data from the study by Zhang et al.\cite{zhang2023conservative}. However, the heat flux on the windward side of the sphere deviates from the reference data, most likely due to differences in the Prandtl number. In this case, the surface coefficients are non-dimensionalized as follows
\begin{equation*}
	C_p=\dfrac{p_s-p_{\infty}}{\frac{1}{2} \rho_{\infty}U_{\infty}^{2}},~
	C_{\tau}=\dfrac{f_s}{\frac{1}{2} \rho_{\infty}U_{\infty}^{2}},~
	C_h=\dfrac{h_s}{\frac{1}{2} \rho_{\infty}U_{\infty}^{3}},
\end{equation*}
where $p_s$ is the surface pressure, $p_{\infty}$ is the pressure in free stream flow, $f_s$ is the surface friction and $h_s$ is the surface heat flux.

The computational efficiency and resource consumption of the simulations are shown in Tab.~\ref{table:spheretime} which are conducted on the SUGON computation platform with a CPU model of 7285 32C 2.0GHz. Time-averaging begins after 12000 steps, with a variation of averaging steps. 
Steady states are achieved with fewer averaging steps in rarefied cases. In the near-continuous cases, 
few particles will emerge in the wave-particle decomposition. The case at Knudsen number of $0.121$ has the highest computational cost due to the large number of particles sampled and removed in a single time step with the dynamic wave-particle decomposition. 
In the current study, UGKWP exhibits high computational efficiency and consumes minimal memory resources in comparison with DVM methods, 
and has particular advantages in approaching the limiting solutions of continuum and free molecular flow in a single computation.
\begin{table}[H]
	\caption{The computational cost for simulations of the supersonic flow around a sphere at ${\rm Ma_\infty} = 4.25$ by the UGKWP method. The physical domain consists of 138240 cells, and the reference number of particles per cell in the UGKWP method is set as $N_r = 100$.}
	\centering
	\begin{threeparttable}
		\begin{tabular}{ccccc}
			\toprule
			${\rm Kn}_\infty$ & Computation Steps & Wall Clock Time, h  & Cores & Estimated Memory, GiB  \\
			\midrule
			$0.672$ & $12000+500\tnote{1}$ & 4.64  & 128  & 2.12  \\
			$0.338$ & $12000+1000\tnote{1}$ & 5.75 & 128  & 2.01   \\
			$0.121$ & $12000+3500\tnote{1}$ & 6.08 & 128  & 1.74  \\
			$0.080$ & $12000+4000\tnote{1}$ & 4.63  & 128  & 1.56  \\
			$0.031$ & $12000+13000\tnote{1}$ & 3.46 & 128  & 1.40  \\
			\bottomrule
		\end{tabular}
		
		\begin{tablenotes}
			\item[1] Steps of time-averaging process in the UGKWP simulations.
		\end{tablenotes}
	\end{threeparttable}
	\label{table:spheretime}
\end{table}

\subsection{Hypersonic flow around X38-like space vehicle }

\begin{figure}[H]
	\centering
	\subfloat[]{\includegraphics[height=3.5cm]{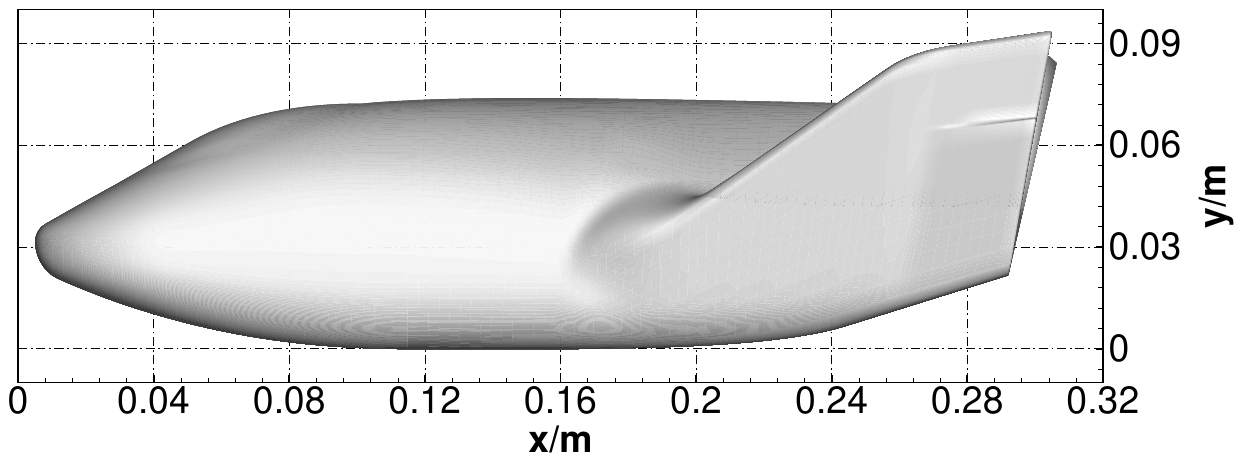}} \\
	\subfloat[]{\includegraphics[height=3.5cm]{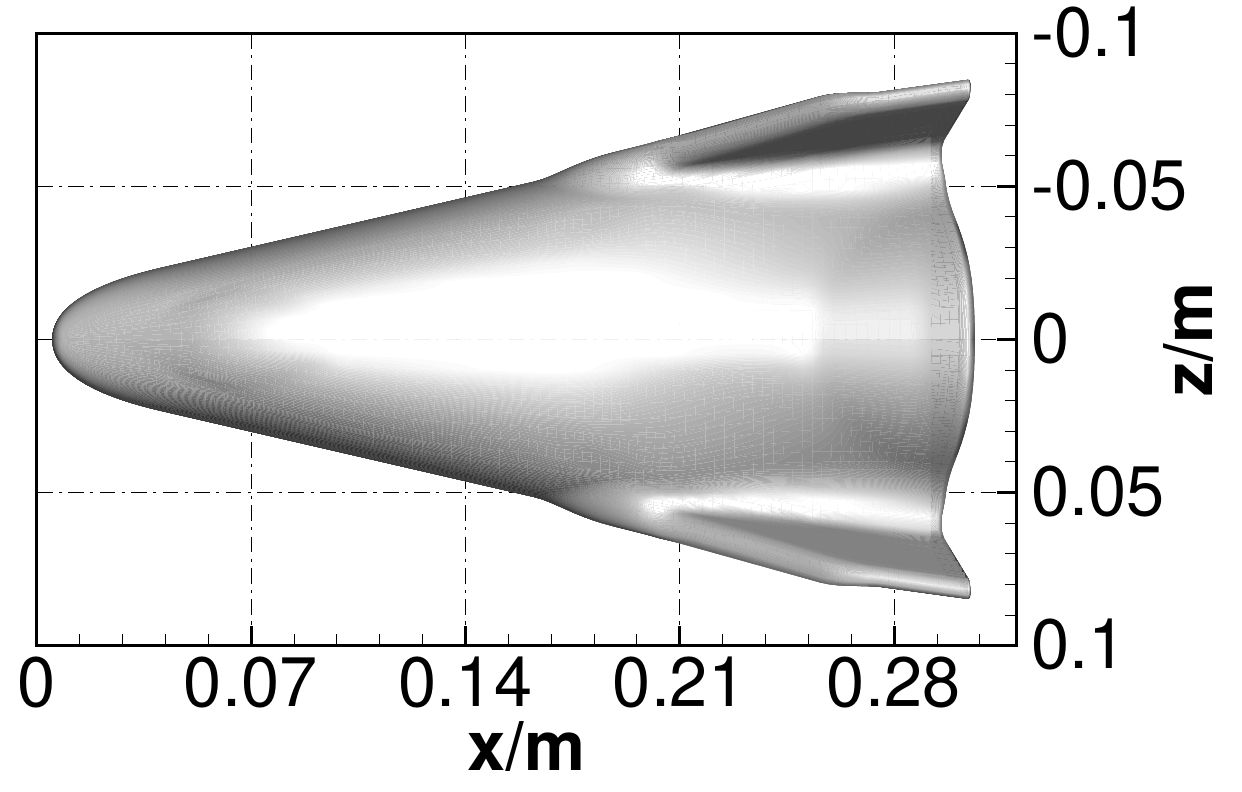}}
	\subfloat[]{\includegraphics[height=3.5cm]{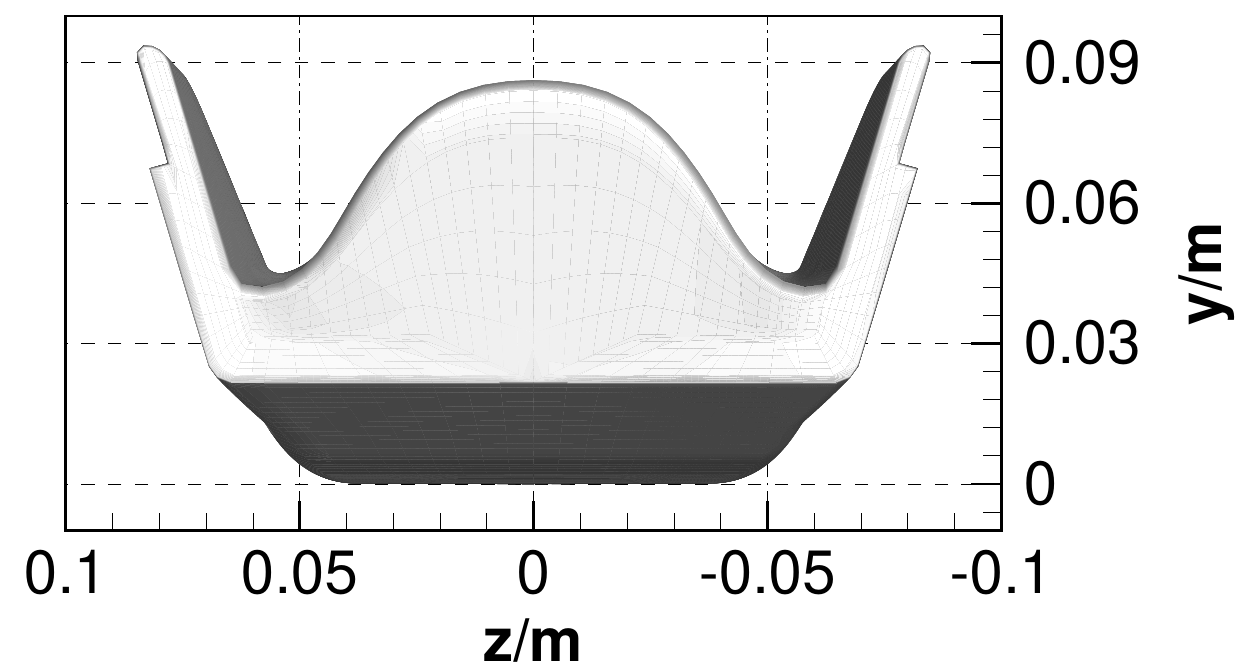}}
	\caption{Sketch of the X38-like space vehicle.}
	\label{fig:x38-geo}
\end{figure}

\begin{figure}[H]
	\centering
	\subfloat[]{\includegraphics[width=0.4\textwidth]
		{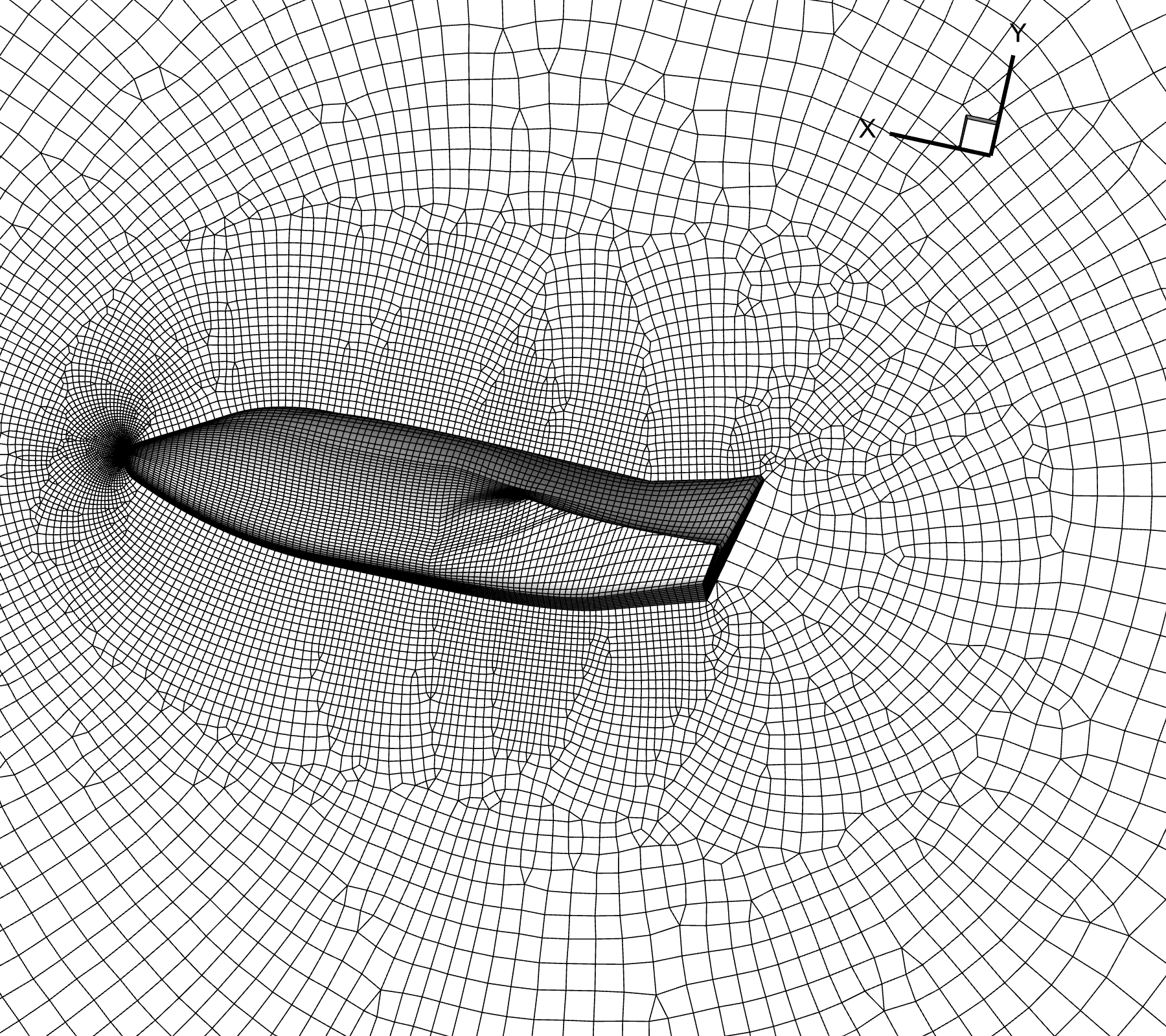}}
	\subfloat[]{\includegraphics[width=0.4\textwidth]
		{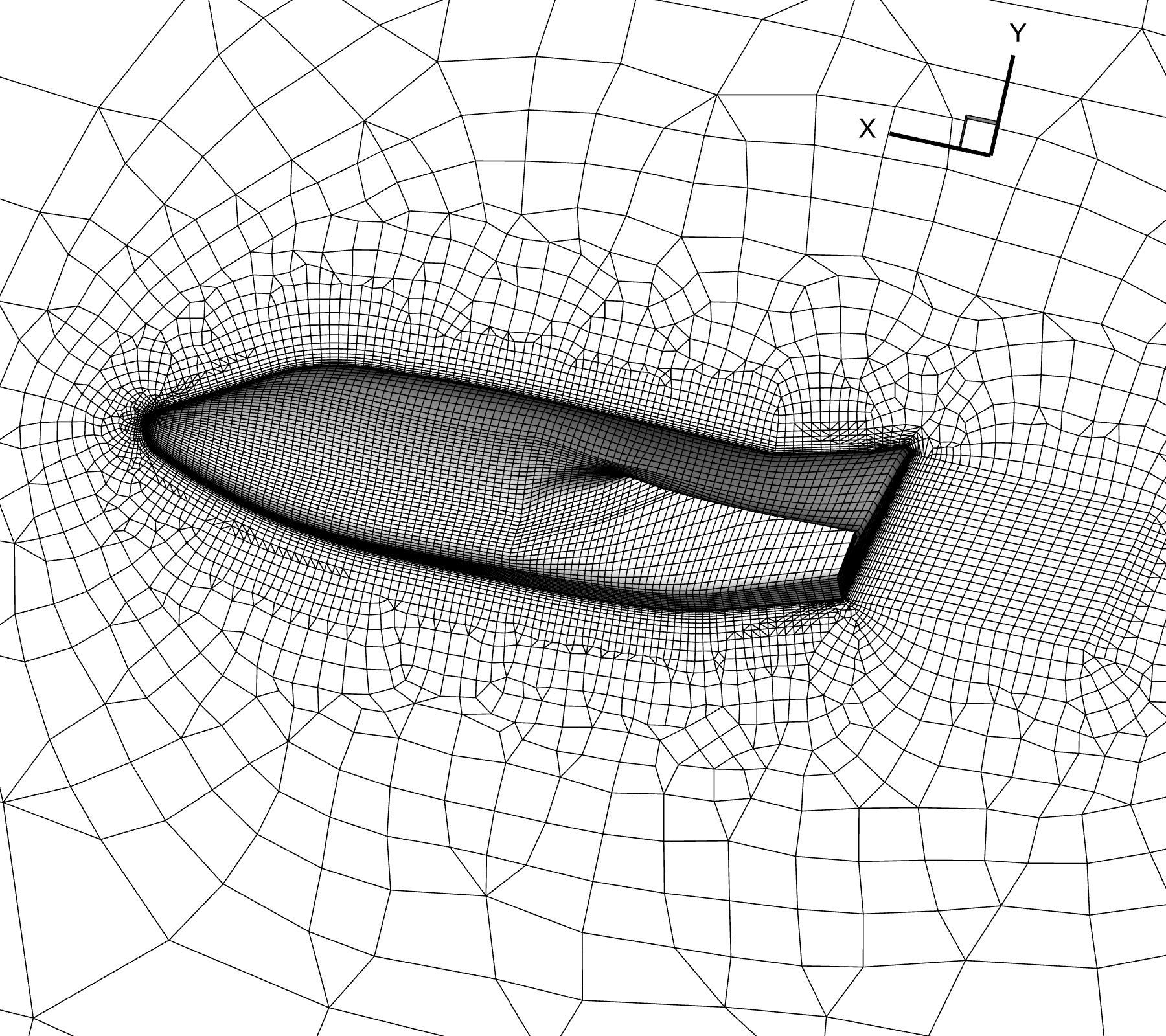}}
	\caption{Three-dimensional physical meshes of hypersonic flow at $Ma_\infty = 8.0$ around a X38-like space vehicle. Domain consists of (a) 560593 cells, and (b) 246558 cells.}
	\label{fig:x38-mesh}
\end{figure}

Simulations of hypersonic flows at ${\rm Ma_\infty}=8.0$ in all flow regimes passing over an X38-like space vehicle at angles of attack of $\rm{AoA}=0^\circ$ and $\rm{AoA}=20^\circ$, are conducted with Argon gas. The geometry of the space vehicle used in the simulations is depicted in Fig.~\ref{fig:x38-geo}. The unstructured symmetry mesh used is shown in Fig.~\ref{fig:x38-mesh}, consisting of 560593 cells with no surface refinement for most cases. For the case of Knudsen number $0.00275$, a refined mesh in the vicinity of the solid surface is adopted to obtain more accurate results of surface quantities, with a reduction of the cell height of the first layer to $0.1$ of the original cell height.

Flow conditions of all cases are listed in Tab.~\ref{table:x38condition}. To determine the Knudsen number of the freestream flow and for the non-dimensionalization procedure, the characteristic length and area of the vehicle are chosen as $0.28$ m and $0.012$ $\rm{m}^2$, respectively. The Knudsen number is calculated from the free stream flow density by the hard-sphere model, varying from $0.00275$ to $2.75$, which covers flow regimes from near continuum flow to free molecule one. The third column is the corresponding altitude estimated by the free-steam flow density.

\begin{table}[H]
	\caption{Freestream flow parameters of hypersonic flow around an X38-like vehicle.}
	\centering
	\begin{tabular}{ccccccc}
		\toprule
		$Ma_\infty$ & ${\rm Kn}_\infty$ & AoA & Altitude, km & $\rho, \rm{kg}/(\rm{m}^3 \cdot \rm{s})$ & $T_\infty$, K & $T_w$, K \\
		\midrule
		$8.0$ & 2.75    & $0^{\circ}$ & 136.8 & $1.11\times 10^{-7}$ & $56$ & $300$  \\
		$8.0$ & 0.275  &$0^{\circ}$ & 117.4 & $1.11\times 10^{-6}$ & $56$ & $300$  \\
		$8.0$ & 0.0275  & $0^{\circ}$ & 98.0 & $1.11\times 10^{-5}$ & $56$ & $300$  \\
		$8.0$ & 0.00275 & $0^{\circ}$& 78.6 & $1.11\times 10^{-4}$ & $56$ & $300$  \\
		$8.0$ & 2.75    & $20^{\circ}$ & 136.8 & $1.11\times 10^{-7}$ & $56$ & $300$  \\
		$8.0$ & 0.275  &$20^{\circ}$ & 117.4 & $1.11\times 10^{-6}$ & $56$ & $300$  \\
		$8.0$ & 0.0275  & $20^{\circ}$ & 98.0 & $1.11\times 10^{-5}$ & $56$ & $300$  \\
		$8.0$ & 0.00275 & $20^{\circ}$& 78.6 & $1.11\times 10^{-4}$ & $56$ & $300$  \\
		\bottomrule
	\end{tabular}
	\label{table:x38condition}
\end{table}

Drag coefficients at $\rm{AoA} = 0^{\circ}$ and $\rm{AoA} = 20^{\circ}$ are presented in Tab.~\ref{table:x38cd}. The drag coefficients increase dramatically for all cases as the flow becomes rarefied. The UGKWP method simulated results agree well with DSMC and UGKS data in reference \cite{li2021kinetic}.

\begin{table}[H]
	\caption{Comparison of the drag coefficients for hypersonic flow at ${\rm Ma}_\infty = 8.0$ over a X38-like vehicle.}
	\centering
	\begin{tabular}{cccccc}
		\toprule
		\multirow{2}{*}{$Ma_\infty$} & \multirow{2}{*}{$Kn_\infty$} & \multirow{2}{*}{AoA} & \multicolumn{3}{c}{Drag Coefficient (Error)} \\
		\cline{4-6} &  &  &
		\begin{tabular}[c]{@{}c@{}}DSMC \end{tabular} &
		\begin{tabular}[c]{@{}c@{}}UGKS \end{tabular} &
		\begin{tabular}[c]{@{}c@{}}UGKWP \end{tabular}
		\\ \midrule
		8.0 & 0.00275 & 20° & 0.297 & 0.289 (-2.64\%) & 0.294 (-1.28\%)  \\
		8.0 & 0.0275  & 20° & 0.536 & 0.556 (3.73\%)  & 0.520 (-2.94\%)	\\
		8.0 & 0.275   & 20° & 0.979 & 1.002 (2.35\%)  & 0.989 (1.00\%)	\\
		8.0 & 2.75    & 20° & 1.187 & 1.209 (0.02\%)  & 1.209 (0.02\%)   \\
		8.0 & 0.00275 & 0° & 0.227 & - & 0.229 (0.90\%)  \\
		8.0 & 0.0275  & 0° & 0.463 & - &  0.157 (-1.30\%) \\
		8.0 & 0.275   & 0° & 0.833 & - &  0.803 (-3.58\%) \\
		8.0 & 2.75    & 0° & 0.937 & - &  0.943 (0.61\%)  \\
		\bottomrule
	\end{tabular}
	\label{table:x38cd}
\end{table}

\begin{figure}[H]
	\centering
	\subfloat[]{\includegraphics[width=0.4\textwidth]
		{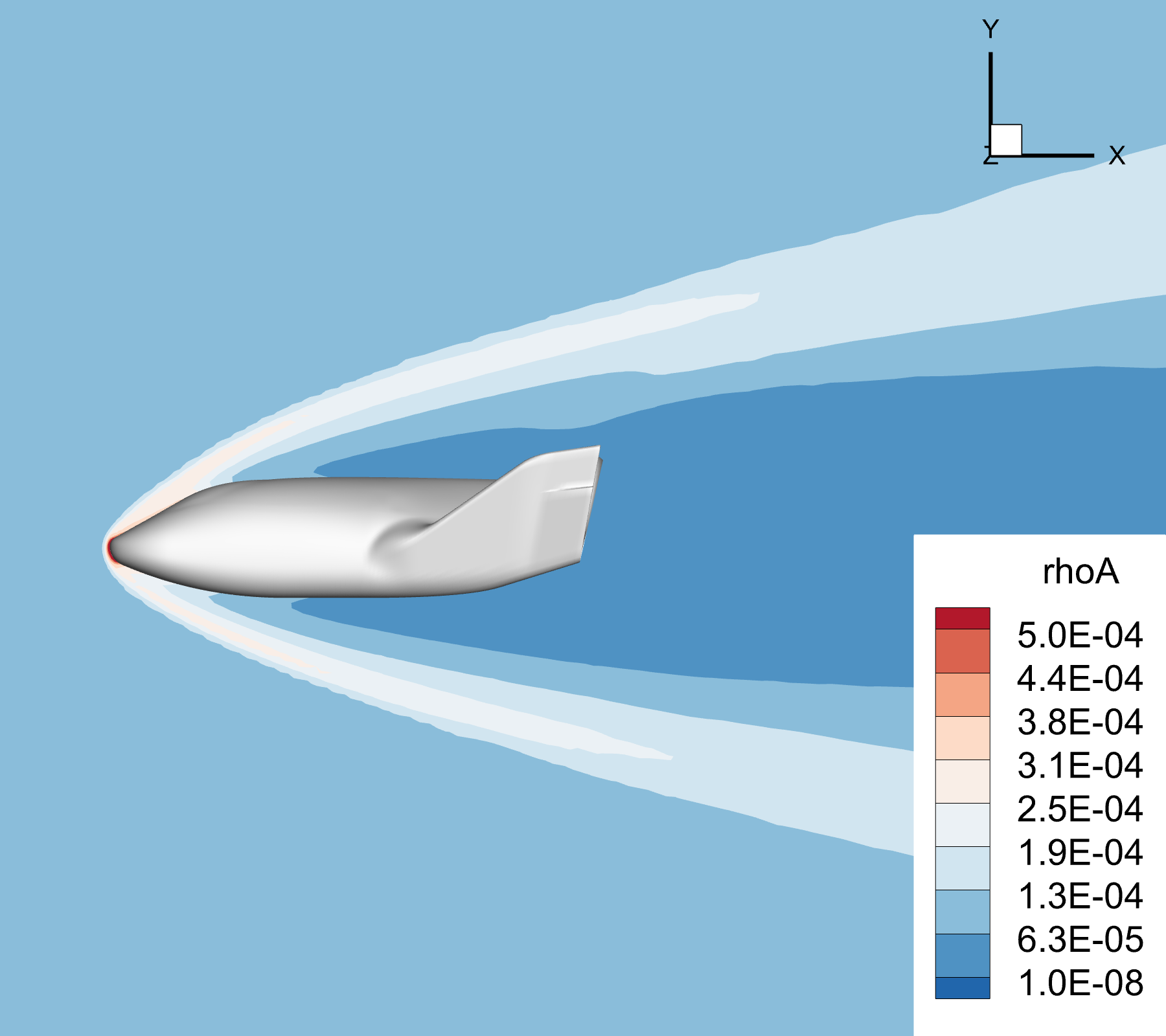}}
	\subfloat[]{\includegraphics[width=0.4\textwidth]
		{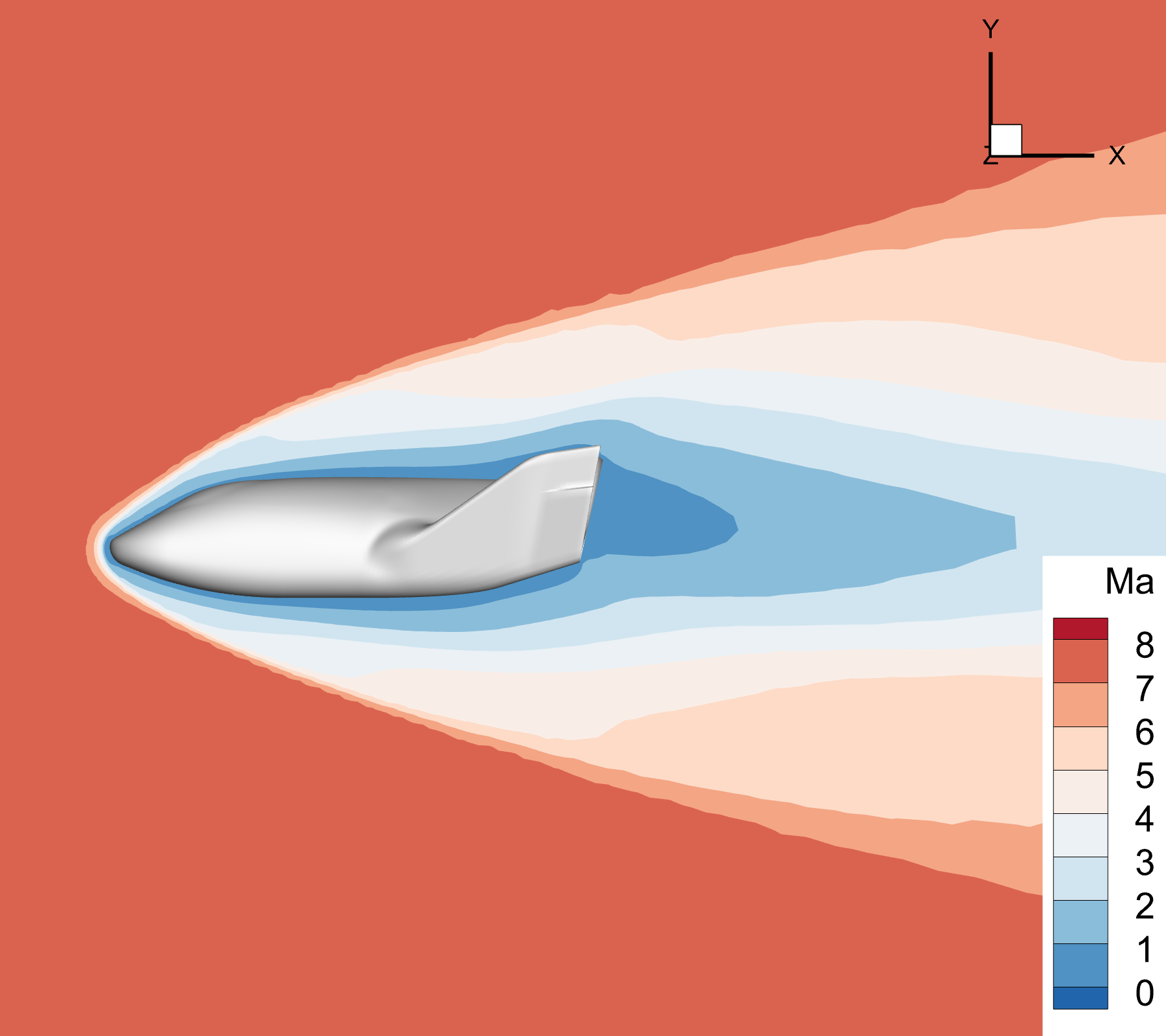}} \\
	\subfloat[]{\includegraphics[width=0.4\textwidth]
		{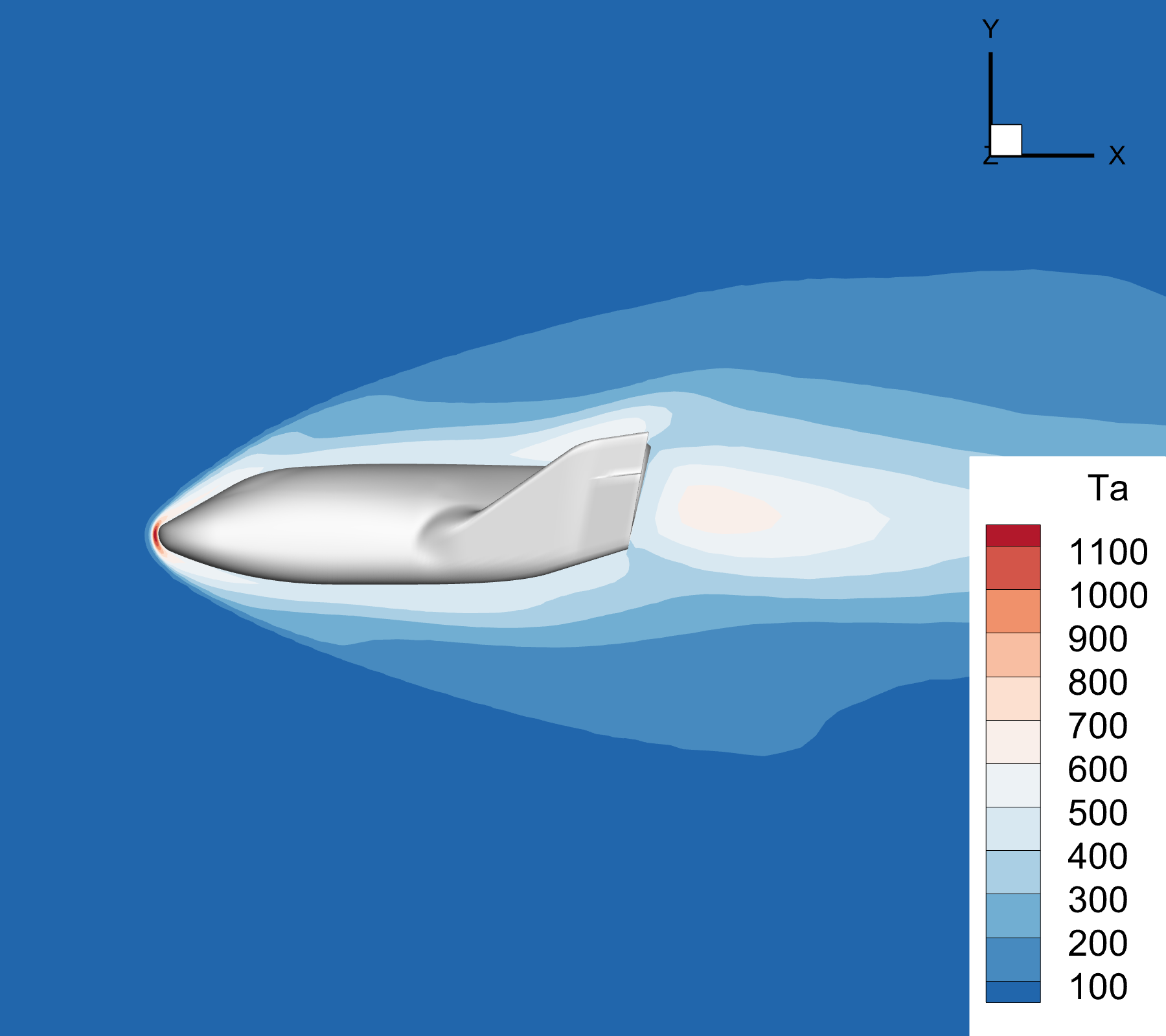}}
	\subfloat[]{\includegraphics[width=0.4\textwidth]
		{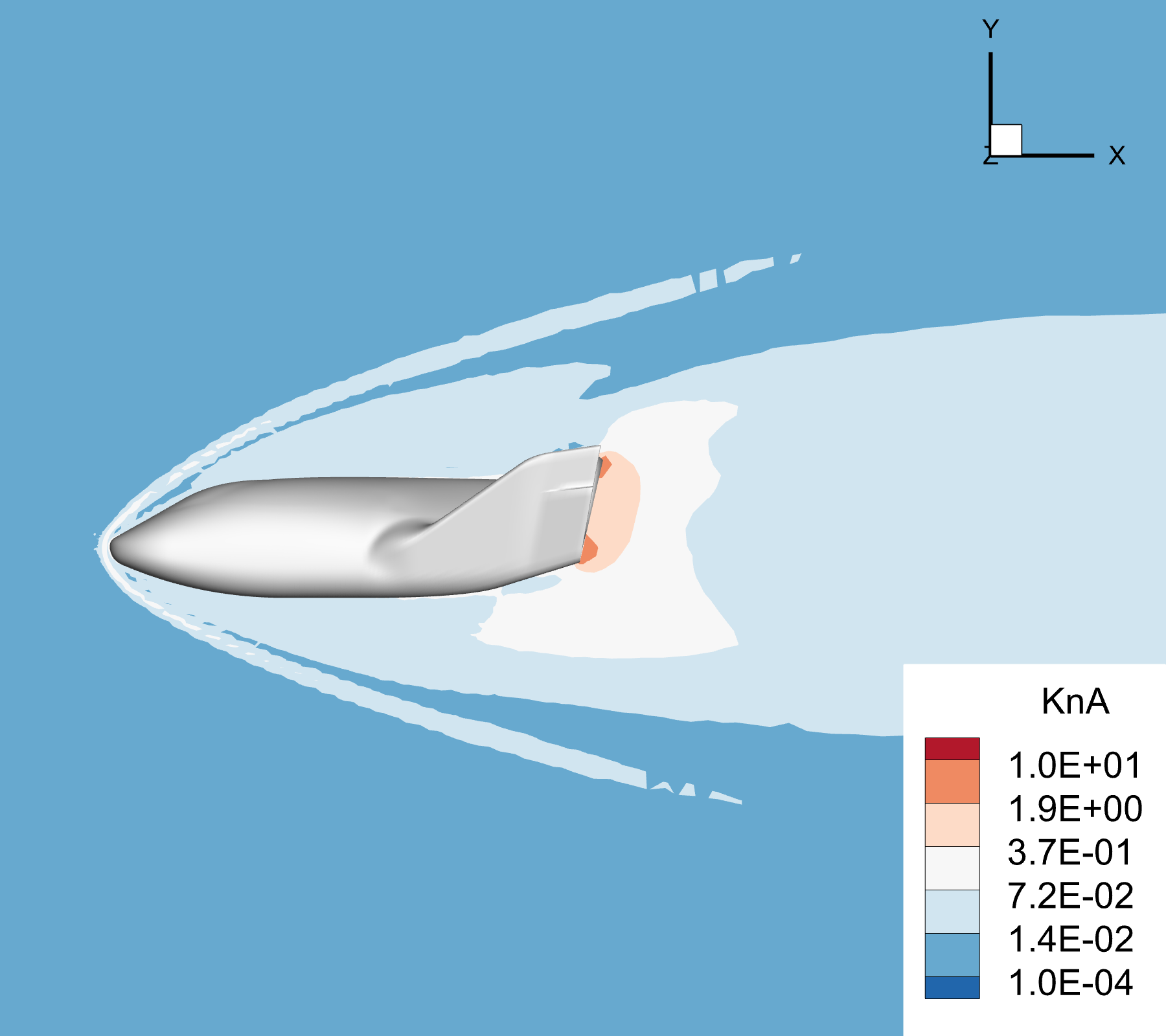}} \\
	\caption{Hypersonic flow
		around a X38-like vehicle at ${\rm Ma}_\infty = 8.0$ with ${\rm AoA} = 0^{\circ}$ and ${\rm Kn}_\infty = 0.00275$ by the UGKWP method. (a) Density, (b) temperature, (c) Mach number, and
		(d) local Knudsen number contours.}
	\label{fig:x38contour1}
\end{figure}

\begin{figure}[H]
	\centering
	\subfloat[]{\includegraphics[width=0.4\textwidth]
		{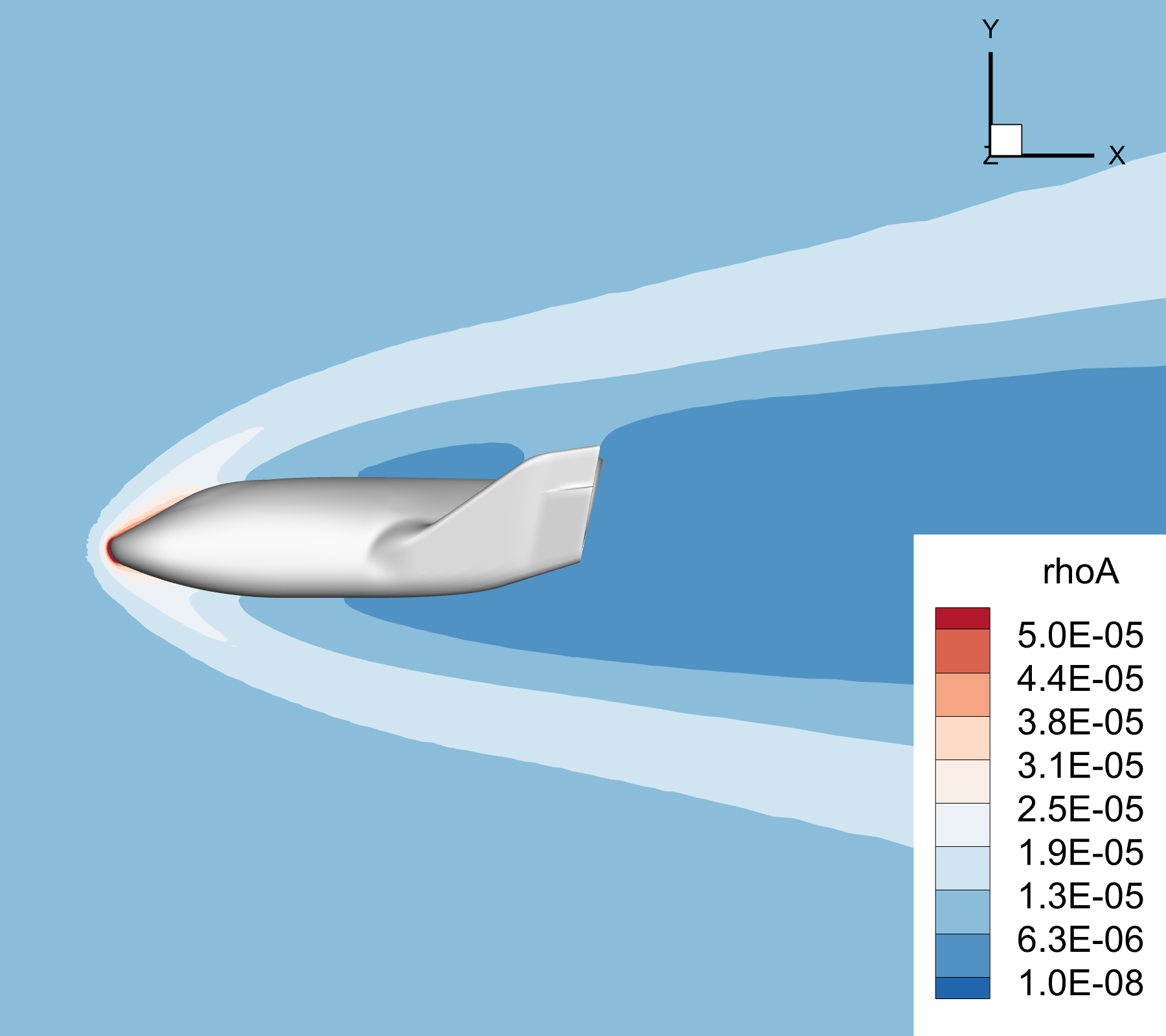}}
	\subfloat[]{\includegraphics[width=0.4\textwidth]
		{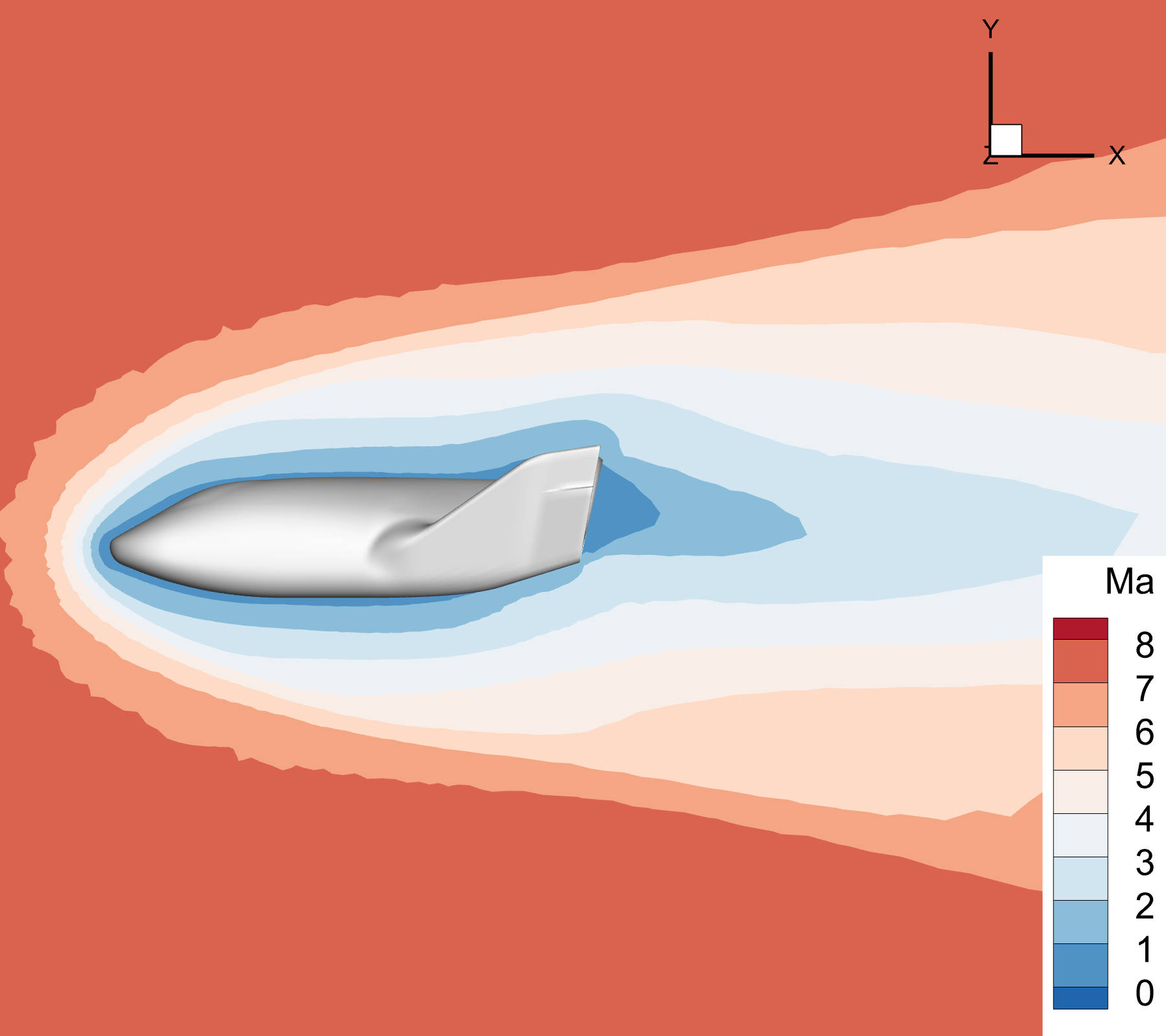}} \\
	\subfloat[]{\includegraphics[width=0.4\textwidth]
		{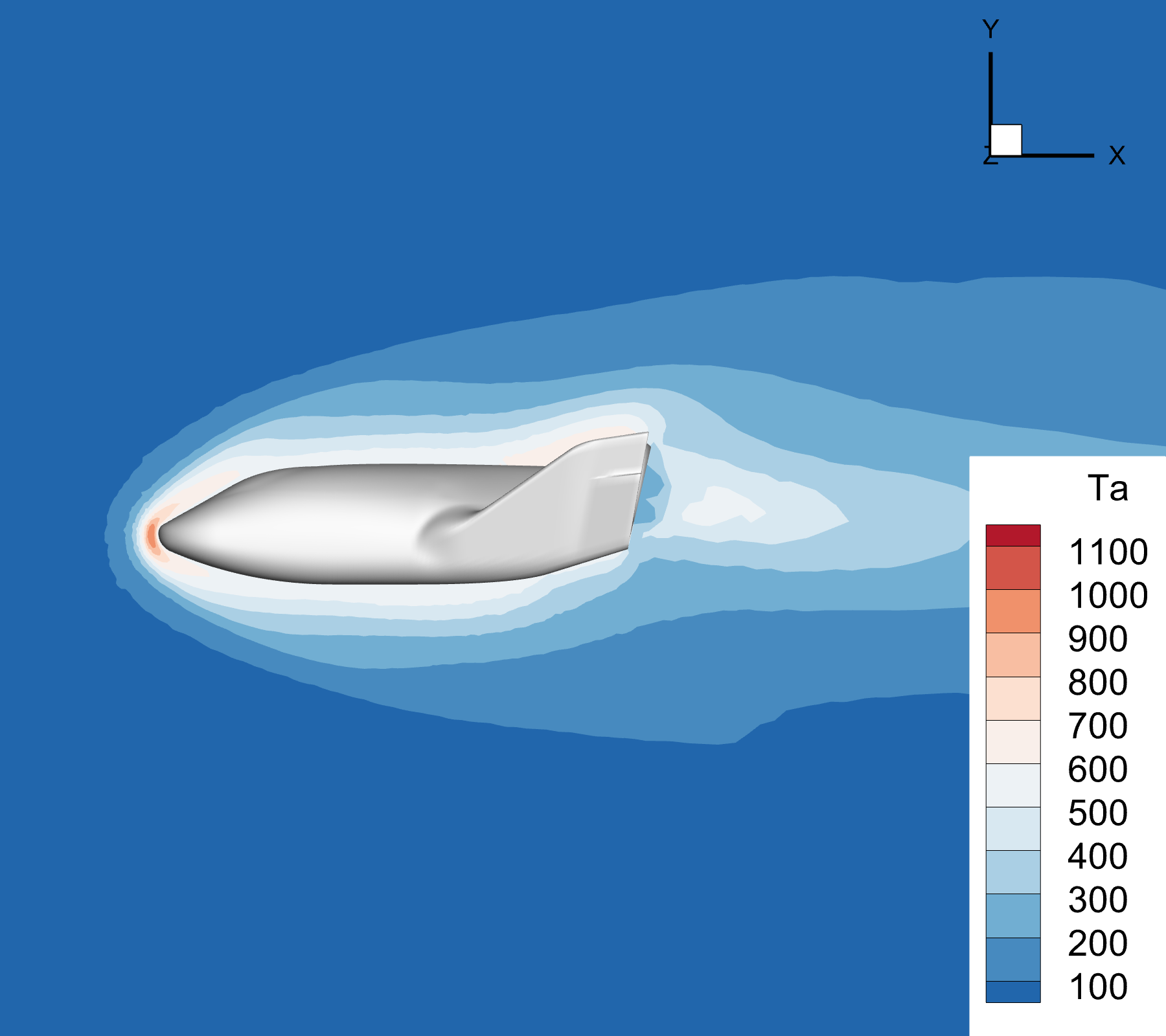}}
	\subfloat[]{\includegraphics[width=0.4\textwidth]
		{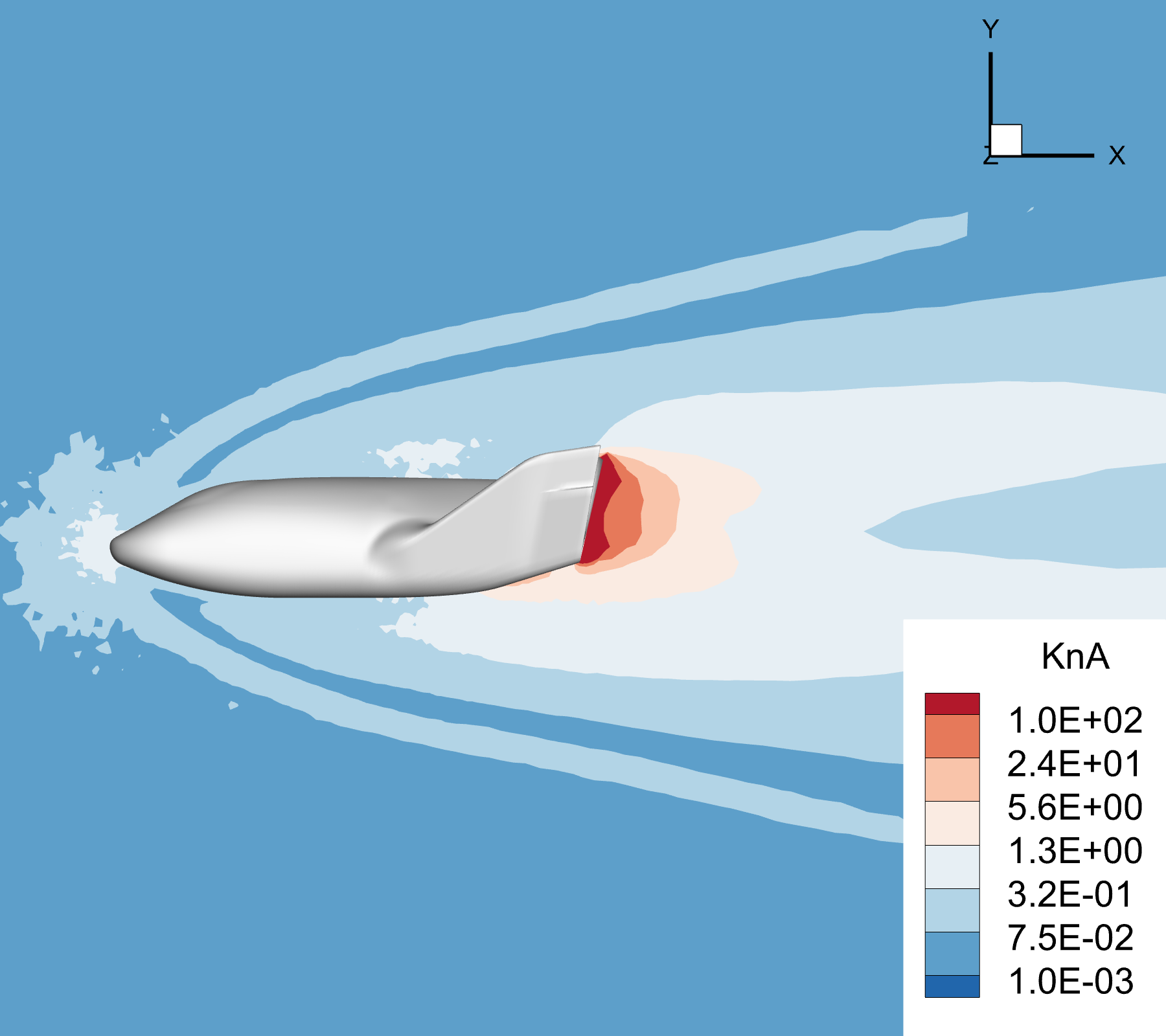}} \\
	\caption{Hypersonic flow
		around a X38-like vehicle at ${\rm Ma}_\infty = 8.0$ with $\rm{AoA} = 0^{\circ}$ and ${\rm Kn}_\infty = 0.0275$ by the UGKWP method.  (a) Density, (b) temperature, (c) Mach number, and
		(d) local Knudsen number contours.}
	\label{fig:x38contour3}
\end{figure}

\begin{figure}[H]
	\centering
	\subfloat[]{\includegraphics[width=0.4\textwidth]
		{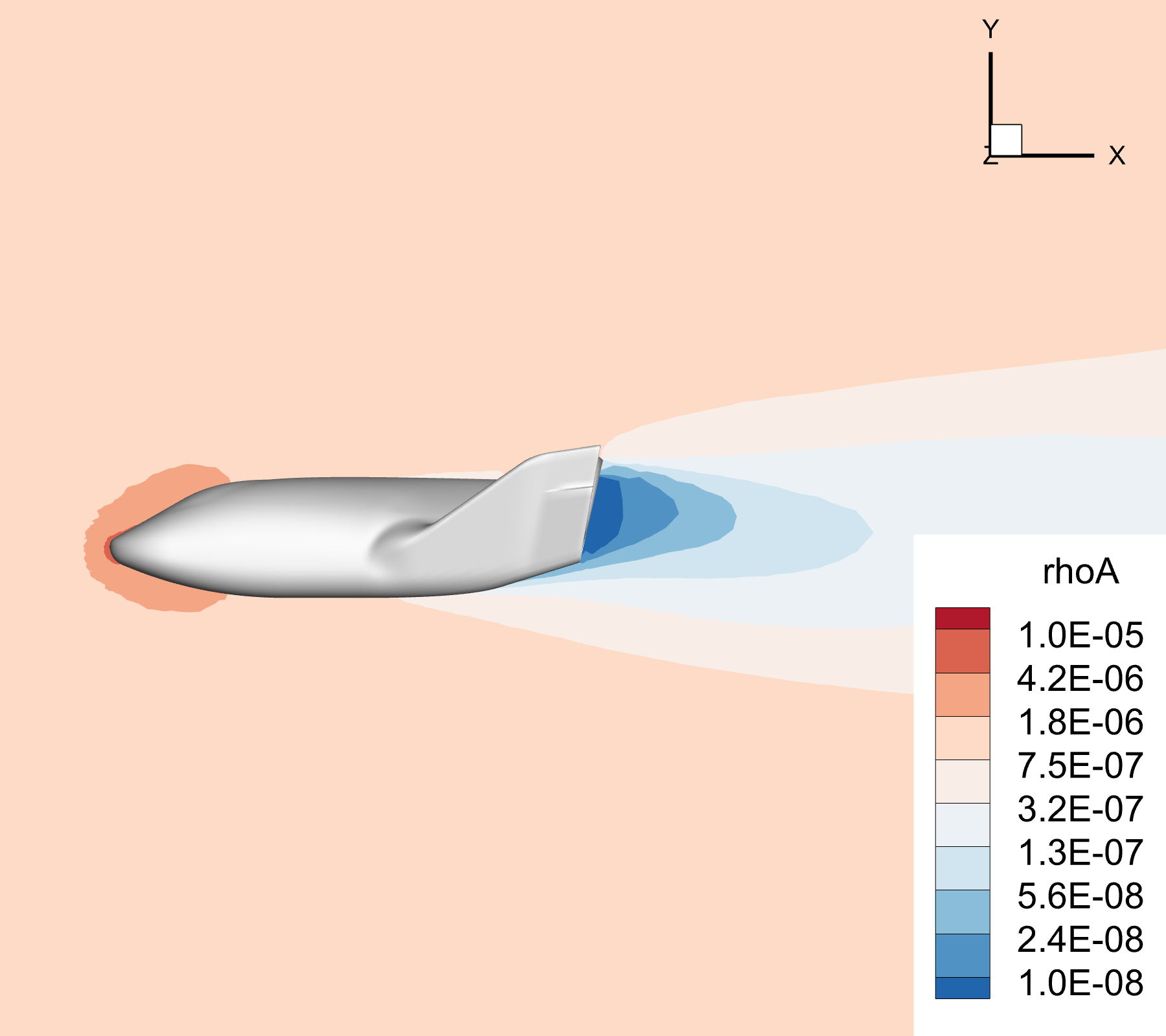}}
	\subfloat[]{\includegraphics[width=0.4\textwidth]
		{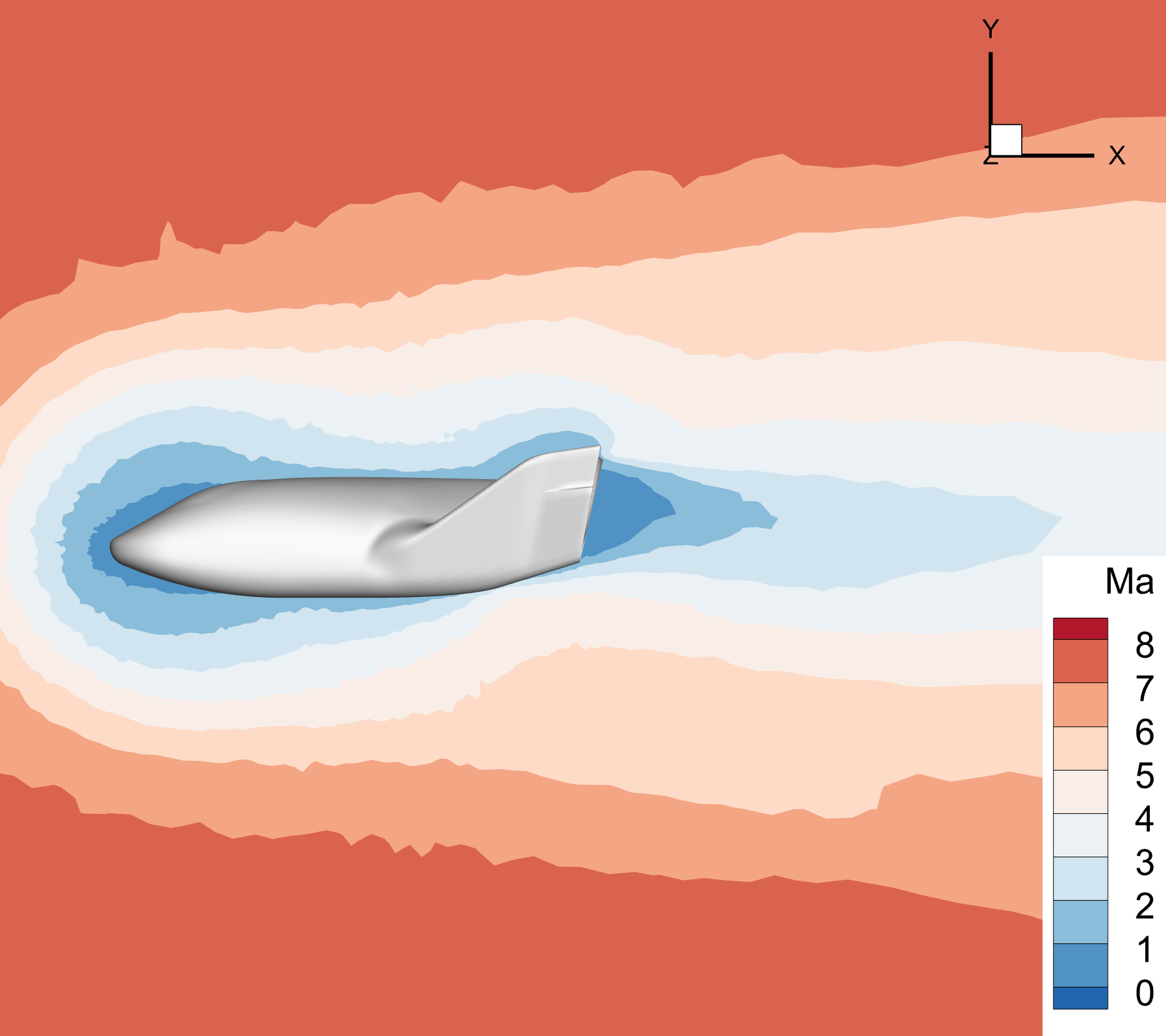}} \\
	\subfloat[]{\includegraphics[width=0.4\textwidth]
		{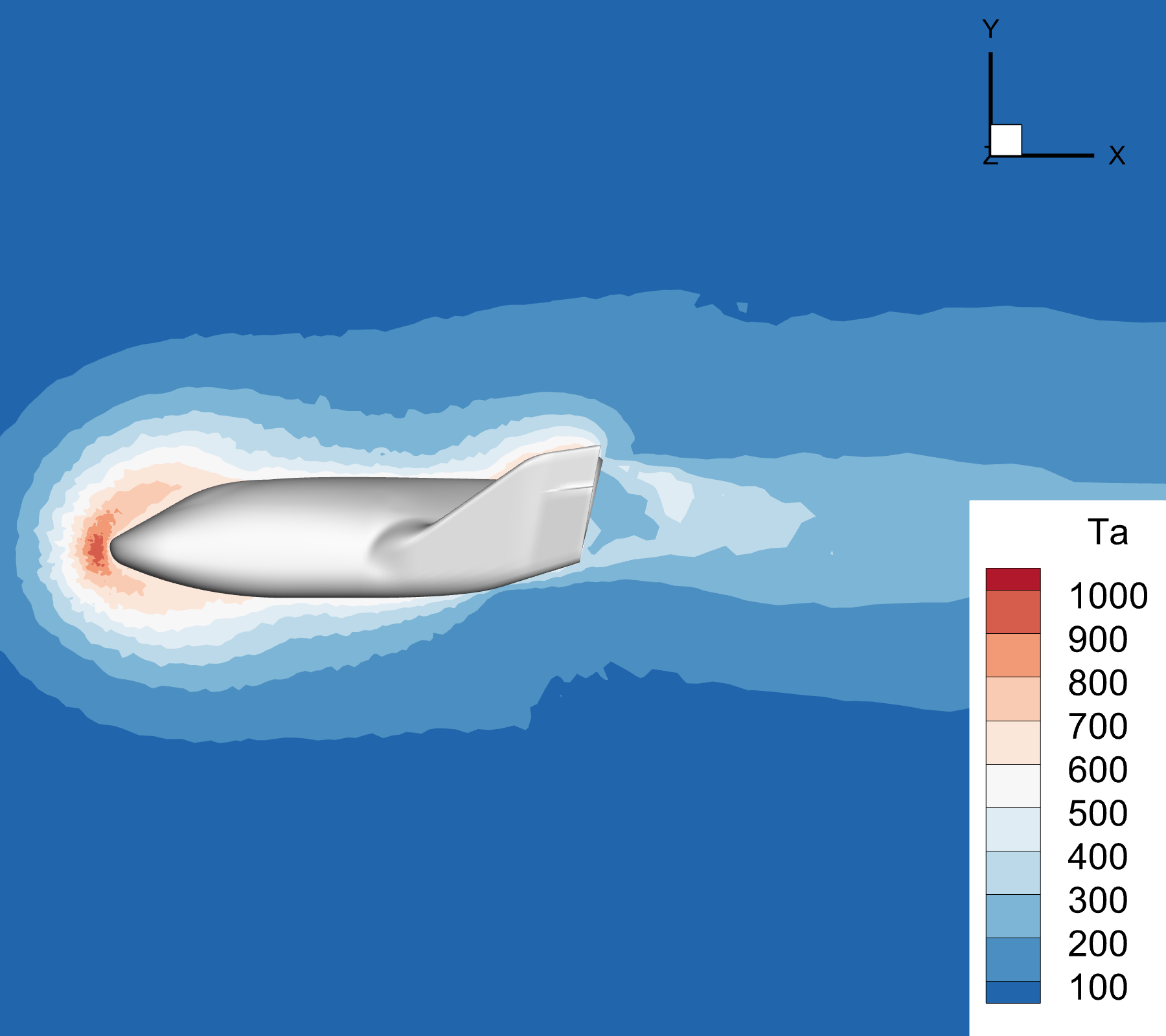}}
	\subfloat[]{\includegraphics[width=0.4\textwidth]
		{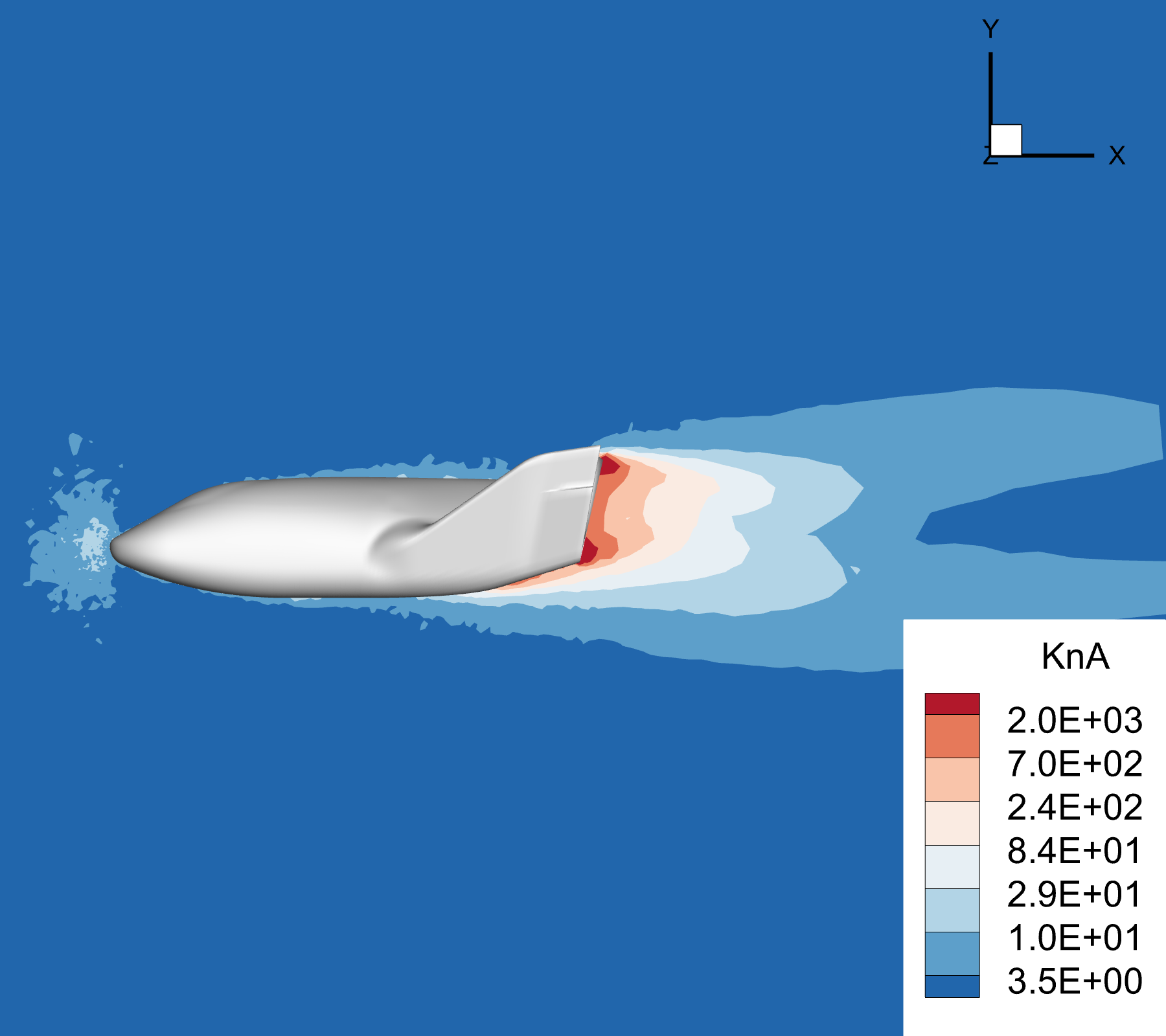}} \\
	\caption{Hypersonic flow
		around a X38-like vehicle at ${\rm Ma}_\infty = 8.0$ with $\rm{AoA} = 0^{\circ}$ and ${\rm Kn}_\infty = 0.275$ by the UGKWP method. (a) Density, (b) temperature, (c) Mach number, and
		(d) local Knudsen number contours.}
	\label{fig:x38contour5}
\end{figure}

\begin{figure}[H]
	\centering
	\subfloat[]{\includegraphics[width=0.4\textwidth]
		{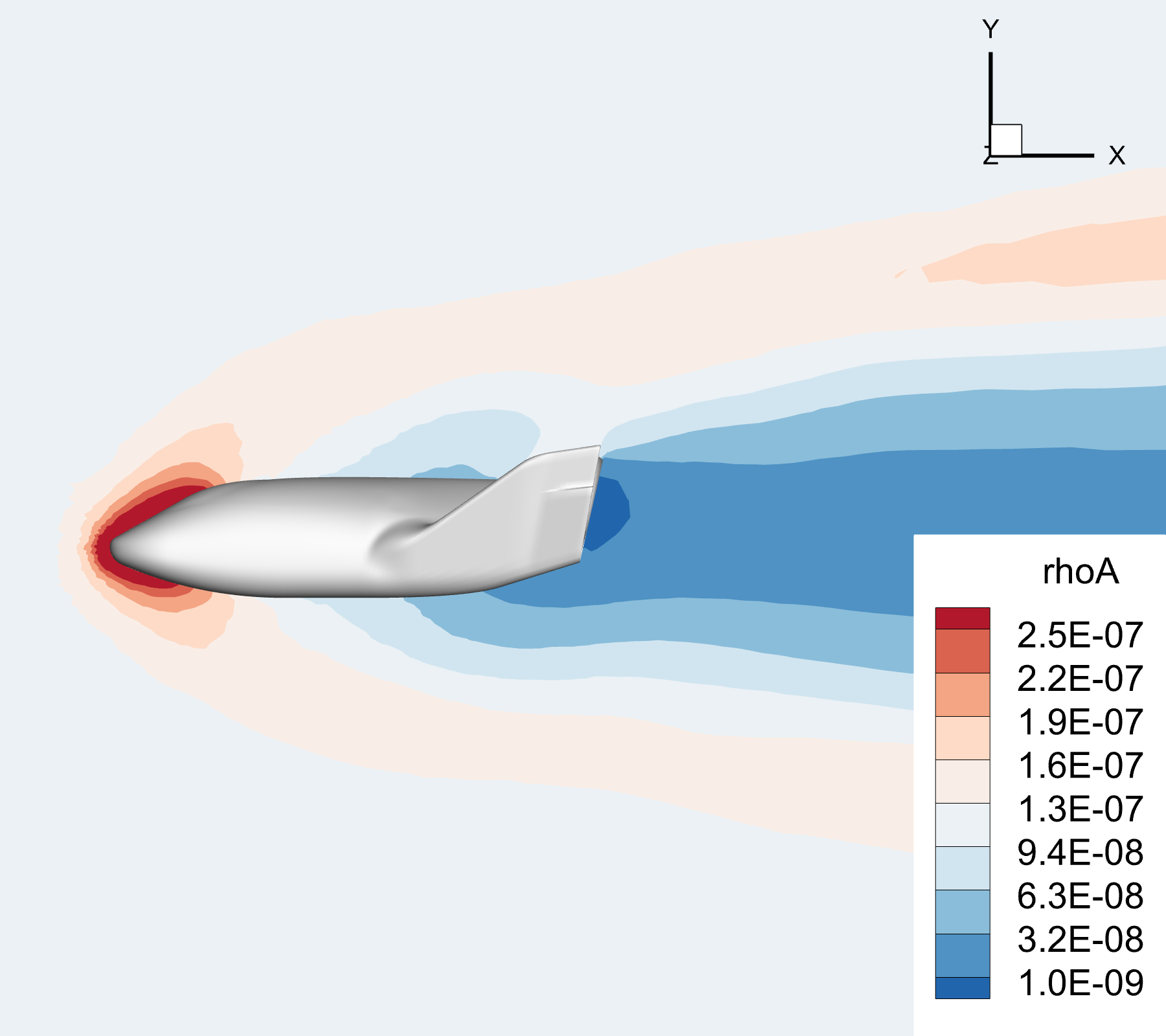}}
	\subfloat[]{\includegraphics[width=0.4\textwidth]
		{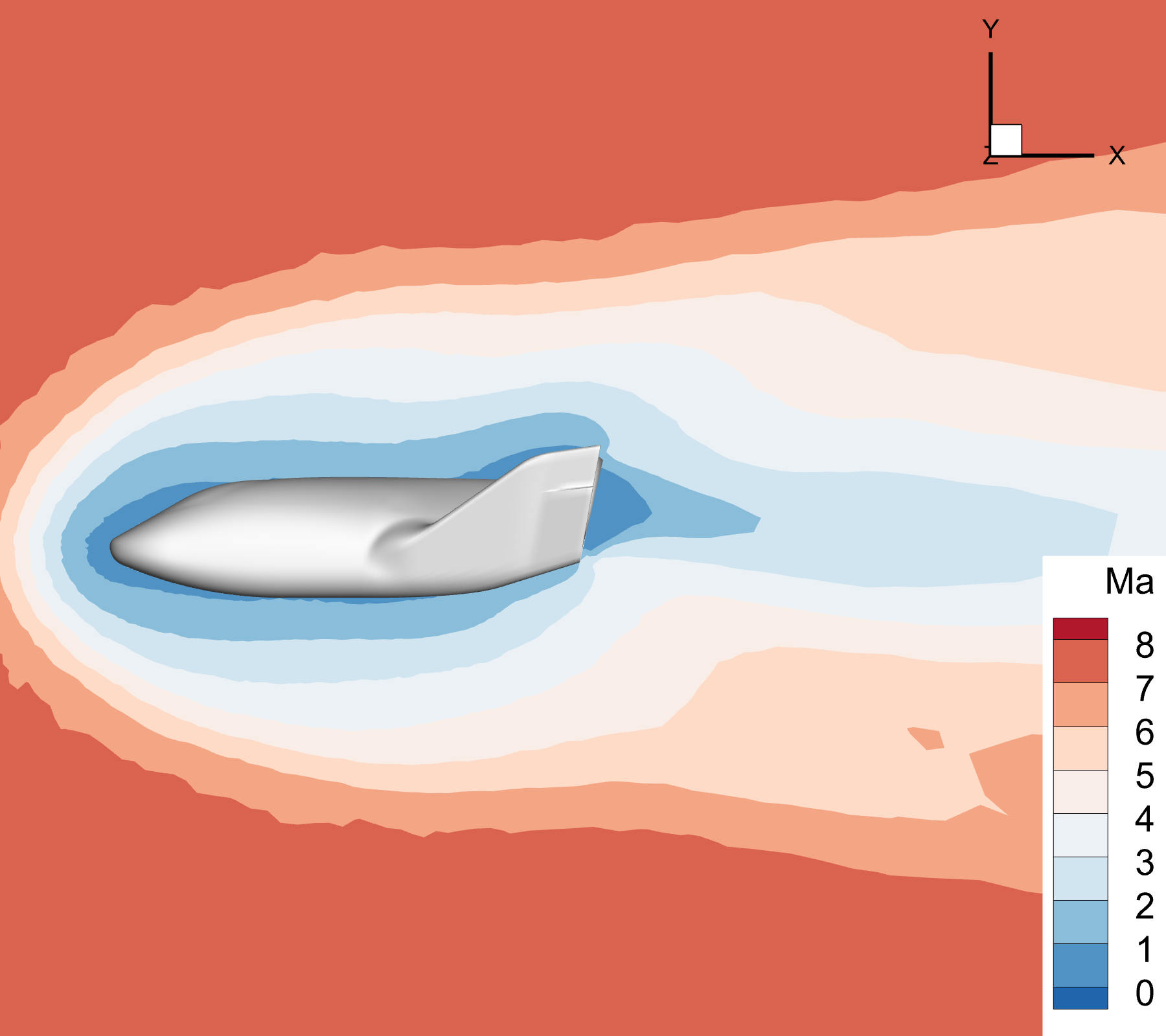}} \\
	\subfloat[]{\includegraphics[width=0.4\textwidth]
		{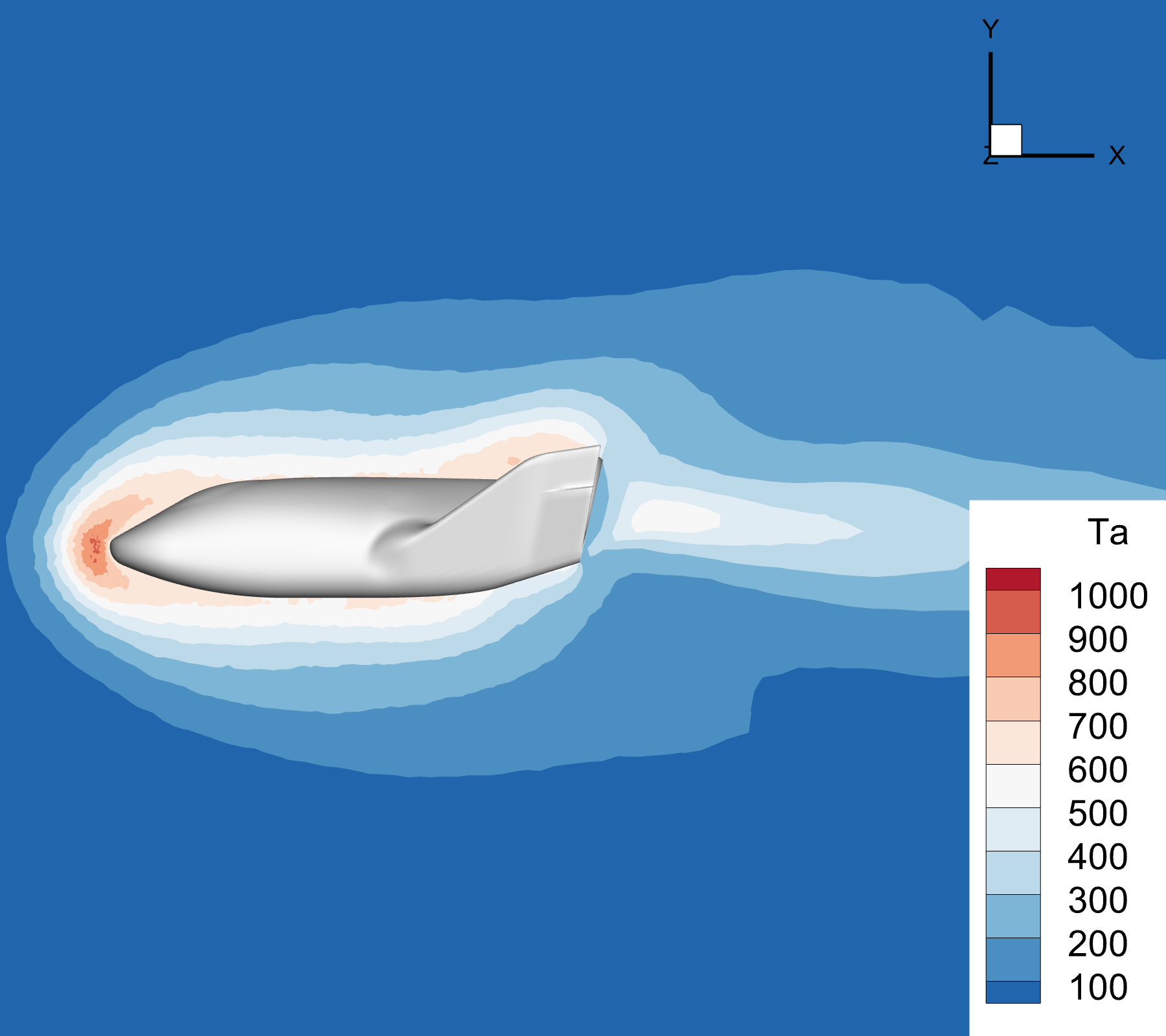}}
	\subfloat[]{\includegraphics[width=0.4\textwidth]
		{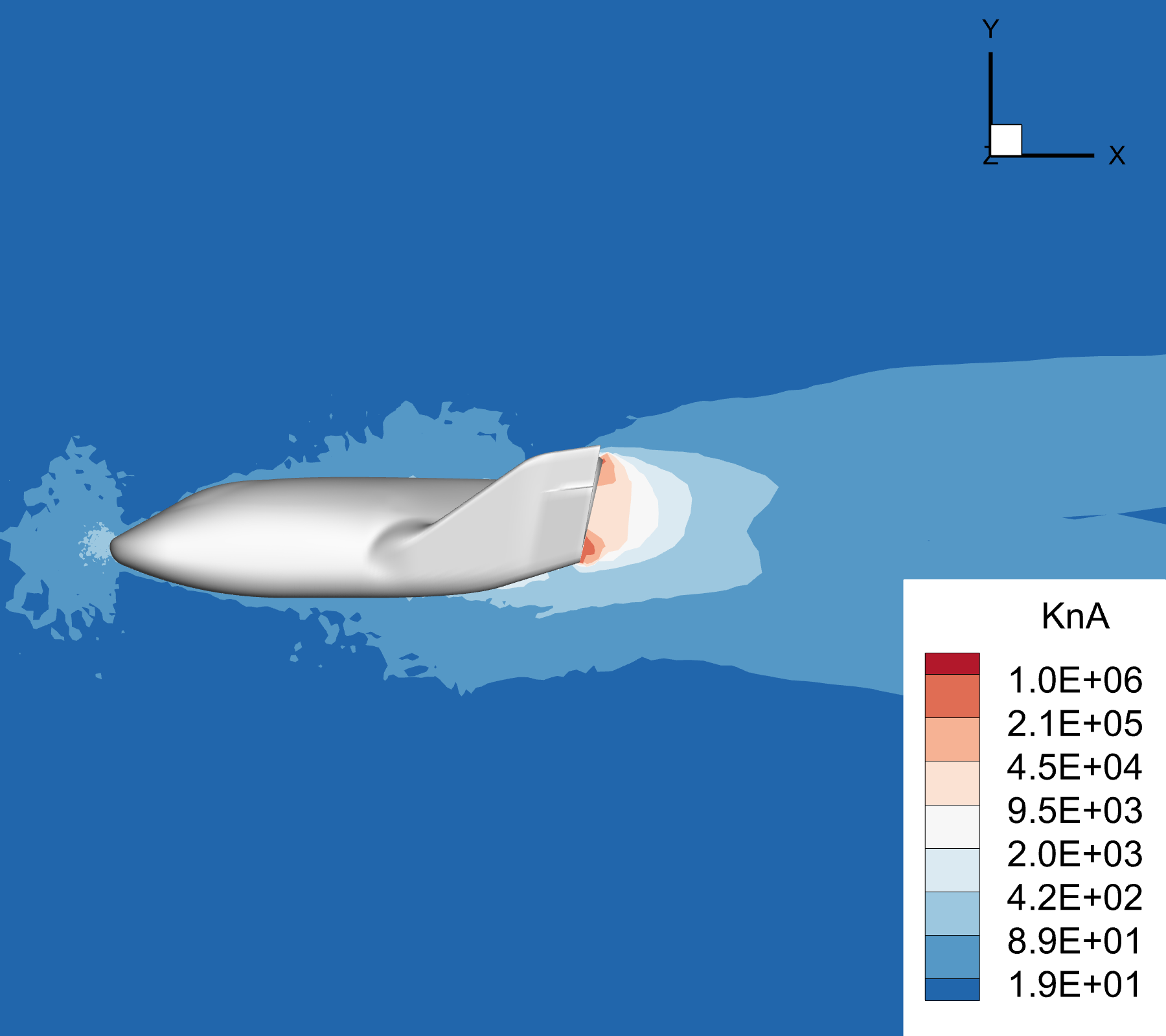}} \\
	\caption{Hypersonic flow
		around a X38-like vehicle at ${\rm Ma}_\infty = 8.0$ with $\rm{AoA} = 0^{\circ}$ and ${\rm Kn}_\infty = 2.75$ by the UGKWP method. (a) Density, (b) temperature, (c) Mach number, and
		(d) local Knudsen number contours.}
	\label{fig:x38contour7}
\end{figure}

\begin{figure}[H]
	\centering
	\subfloat[]{\includegraphics[width=0.4\textwidth]
		{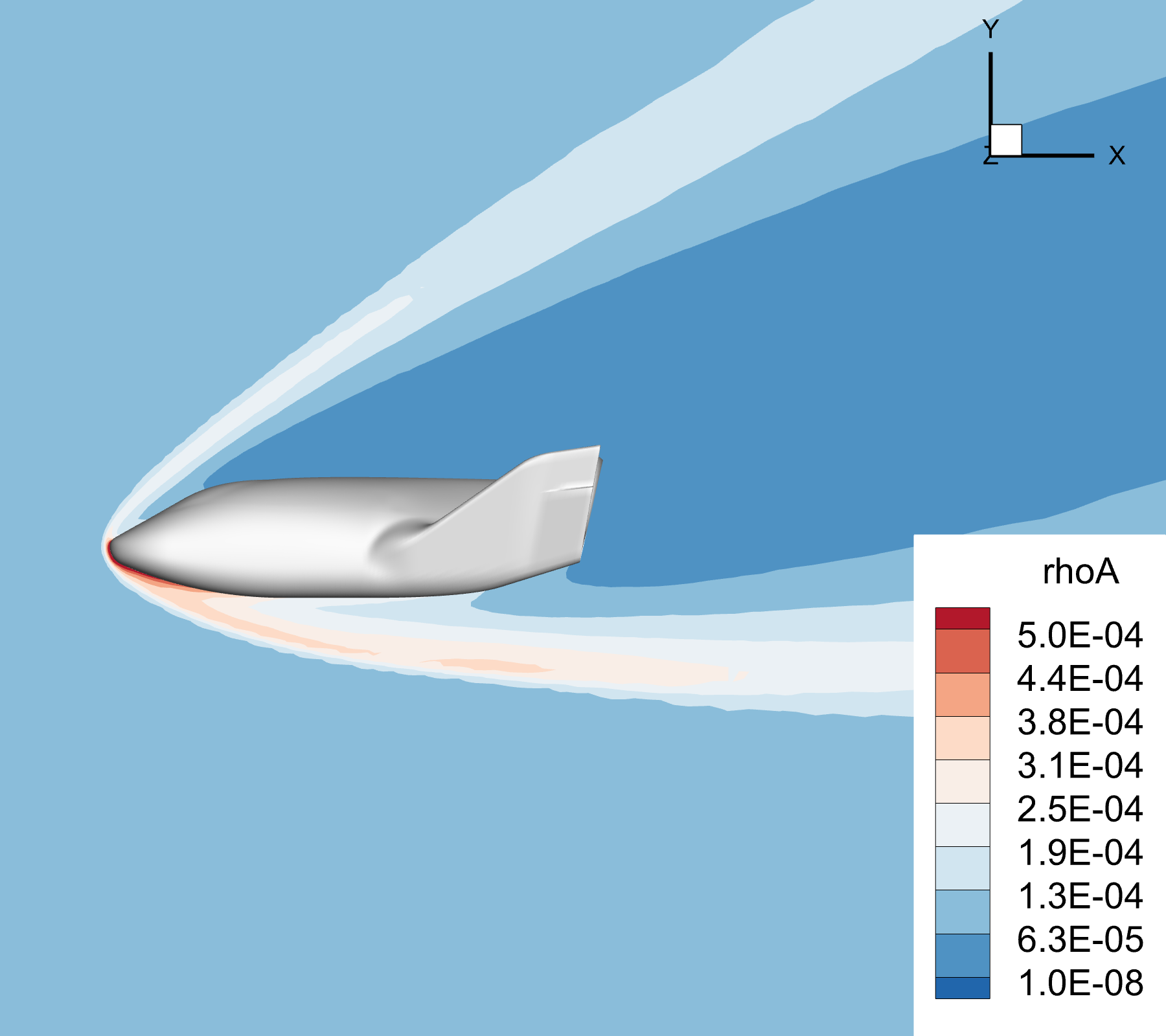}}
	\subfloat[]{\includegraphics[width=0.4\textwidth]
		{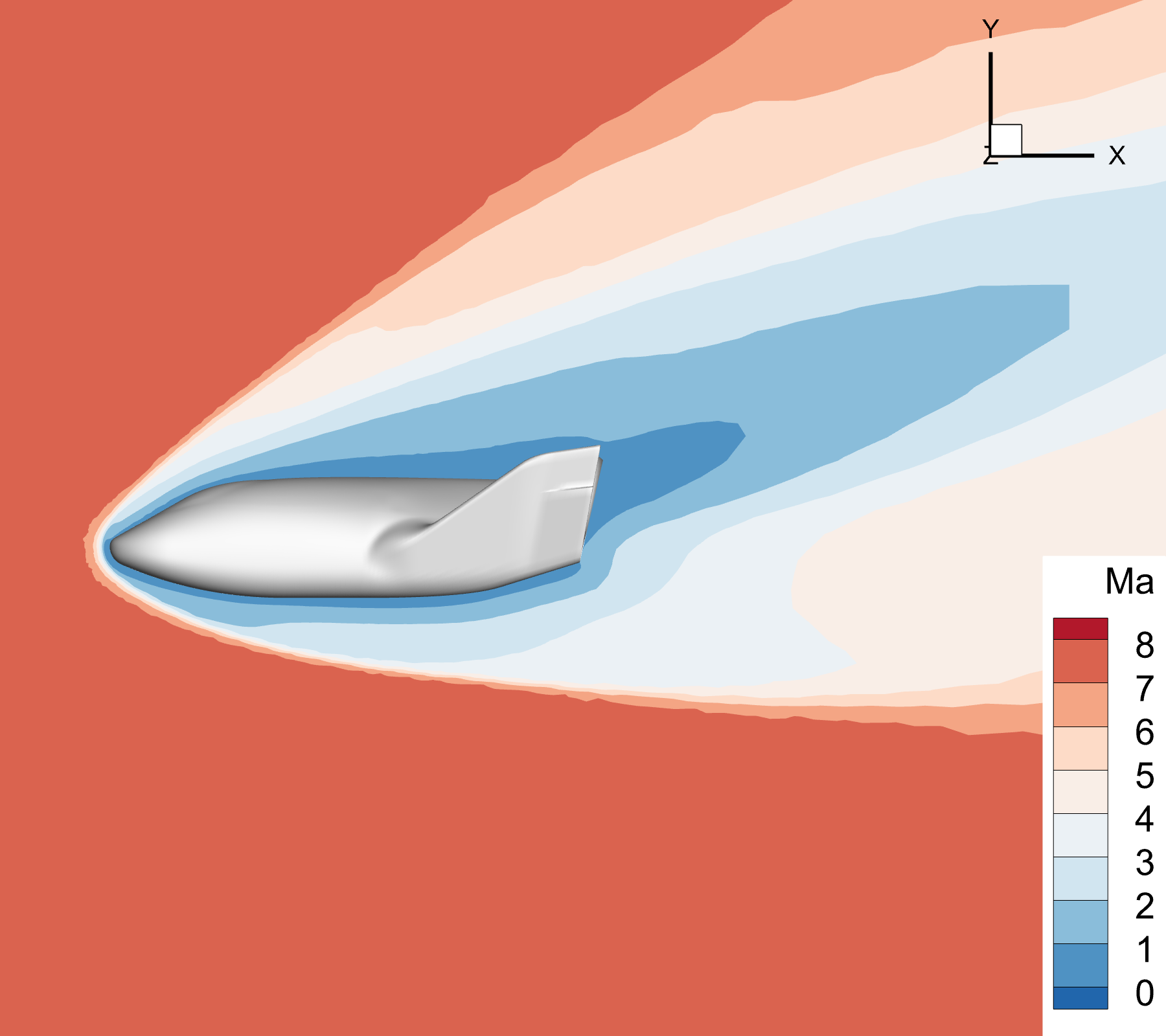}} \\
	\subfloat[]{\includegraphics[width=0.4\textwidth]
		{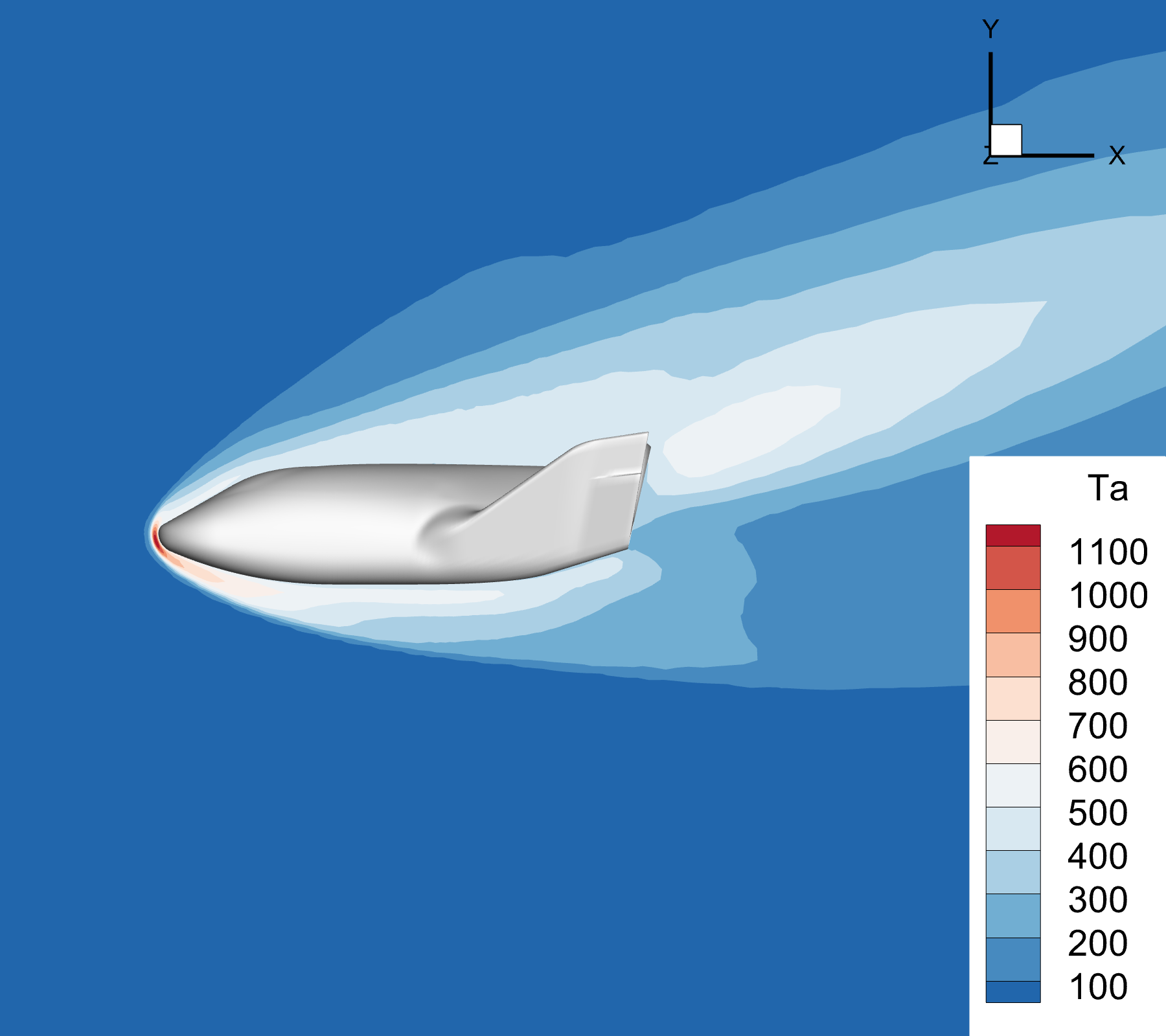}}
	\subfloat[]{\includegraphics[width=0.4\textwidth]
		{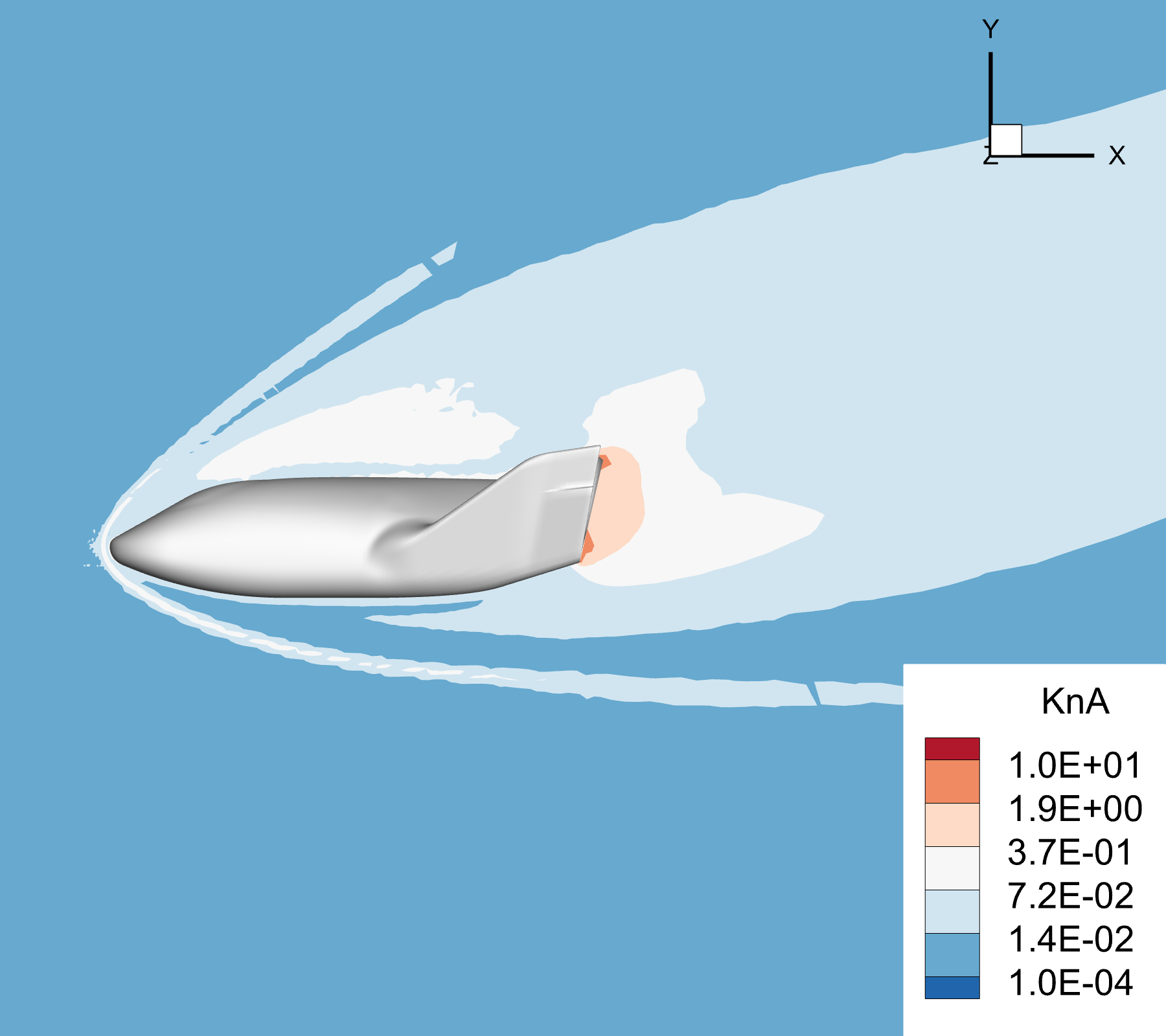}} \\
	\caption{Hypersonic flow
		around a X38-like vehicle at ${\rm Ma}_\infty = 8.0$ with $\rm{AoA} = 20^{\circ}$ and ${\rm Kn}_\infty = 0.00275$ by the UGKWP method. (a) Density, (b) temperature, (c) Mach number, and
		(d) local Knudsen number contours.}
	\label{fig:x38contour2}
\end{figure}

\begin{figure}[H]
	\centering
	\subfloat[]{\includegraphics[width=0.4\textwidth]
		{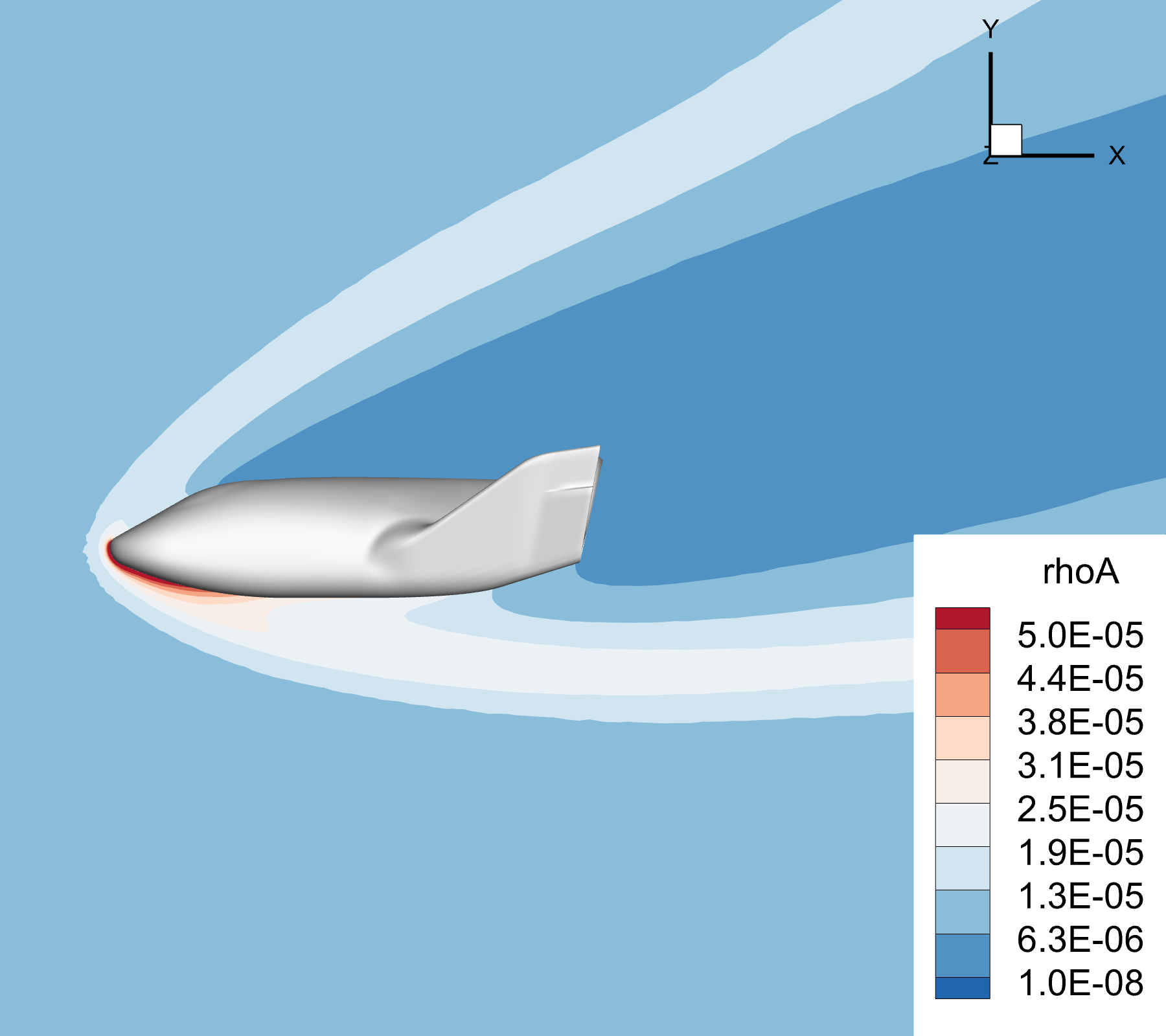}}
	\subfloat[]{\includegraphics[width=0.4\textwidth]
		{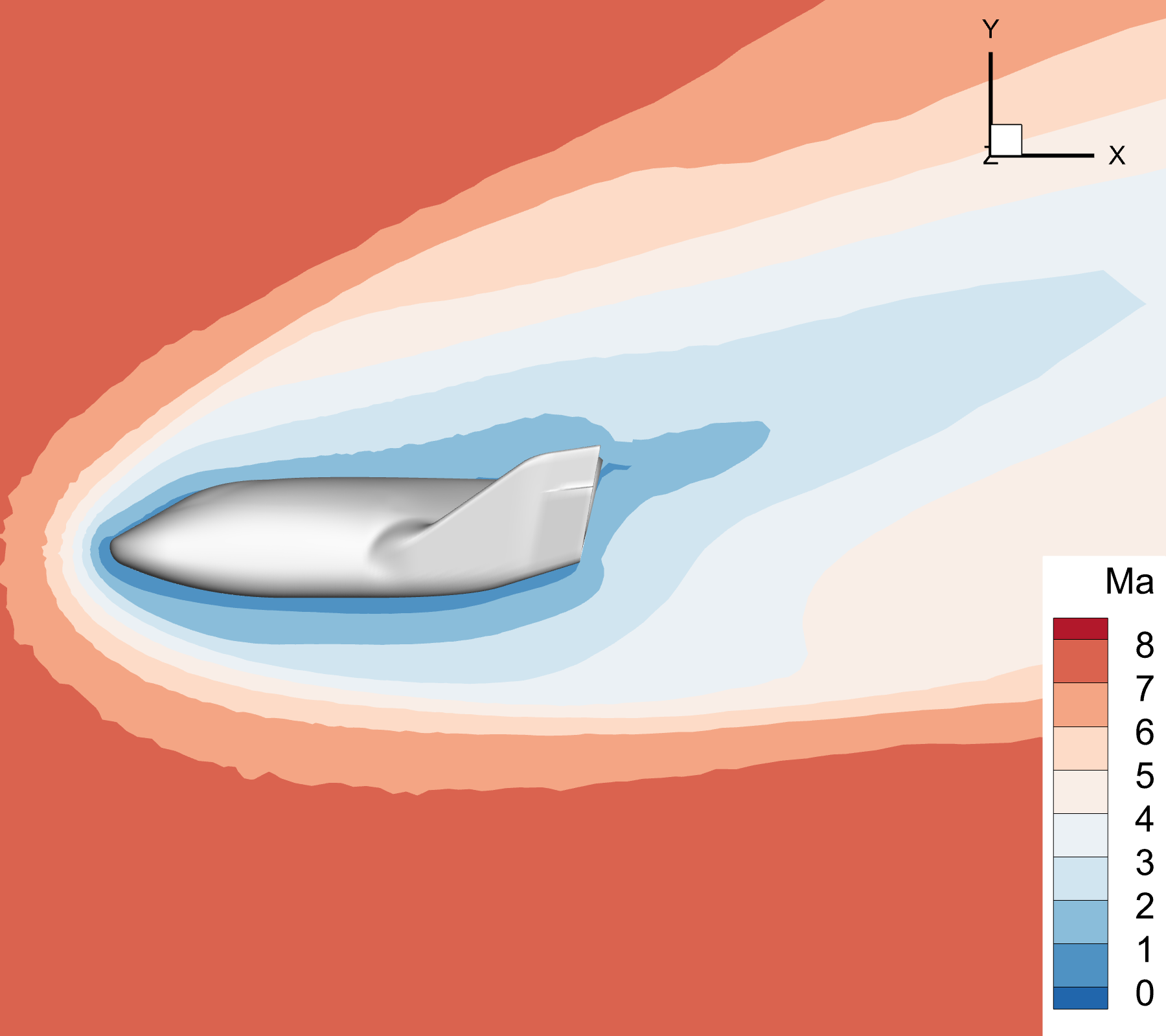}} \\
	\subfloat[]{\includegraphics[width=0.4\textwidth]
		{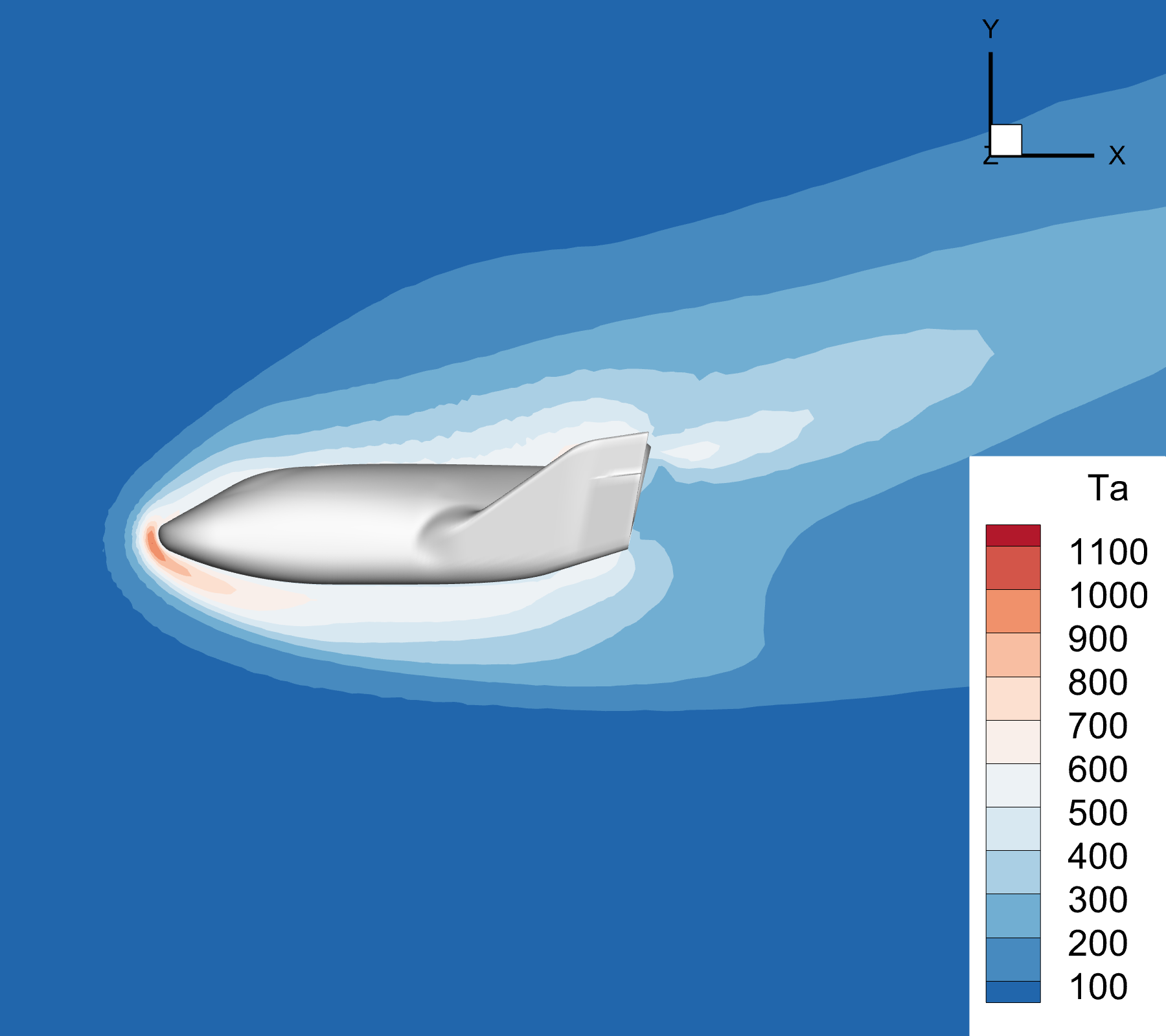}}
	\subfloat[]{\includegraphics[width=0.4\textwidth]
		{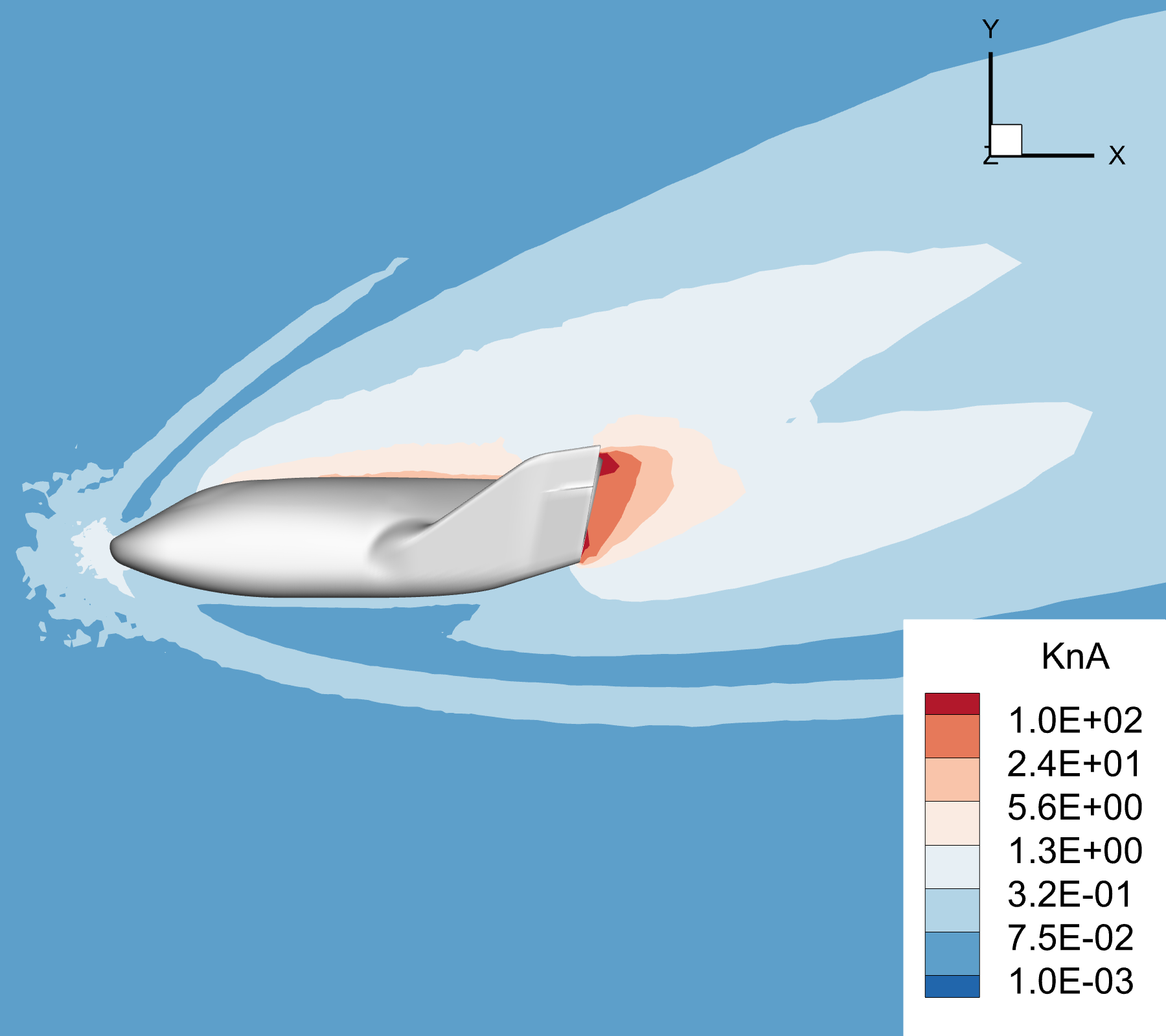}} \\
	\caption{Hypersonic flow
		around a X38-like vehicle at ${\rm Ma}_\infty = 8.0$ with $\rm{AoA} = 20^{\circ}$ and ${\rm Kn}_\infty = 0.0275$ by the UGKWP method. (a) Density, (b) temperature, (c) Mach number, and
		(d) local Knudsen number contours.}
	\label{fig:x38contour4}
\end{figure}

\begin{figure}[H]
	\centering
	\subfloat[]{\includegraphics[width=0.4\textwidth]
		{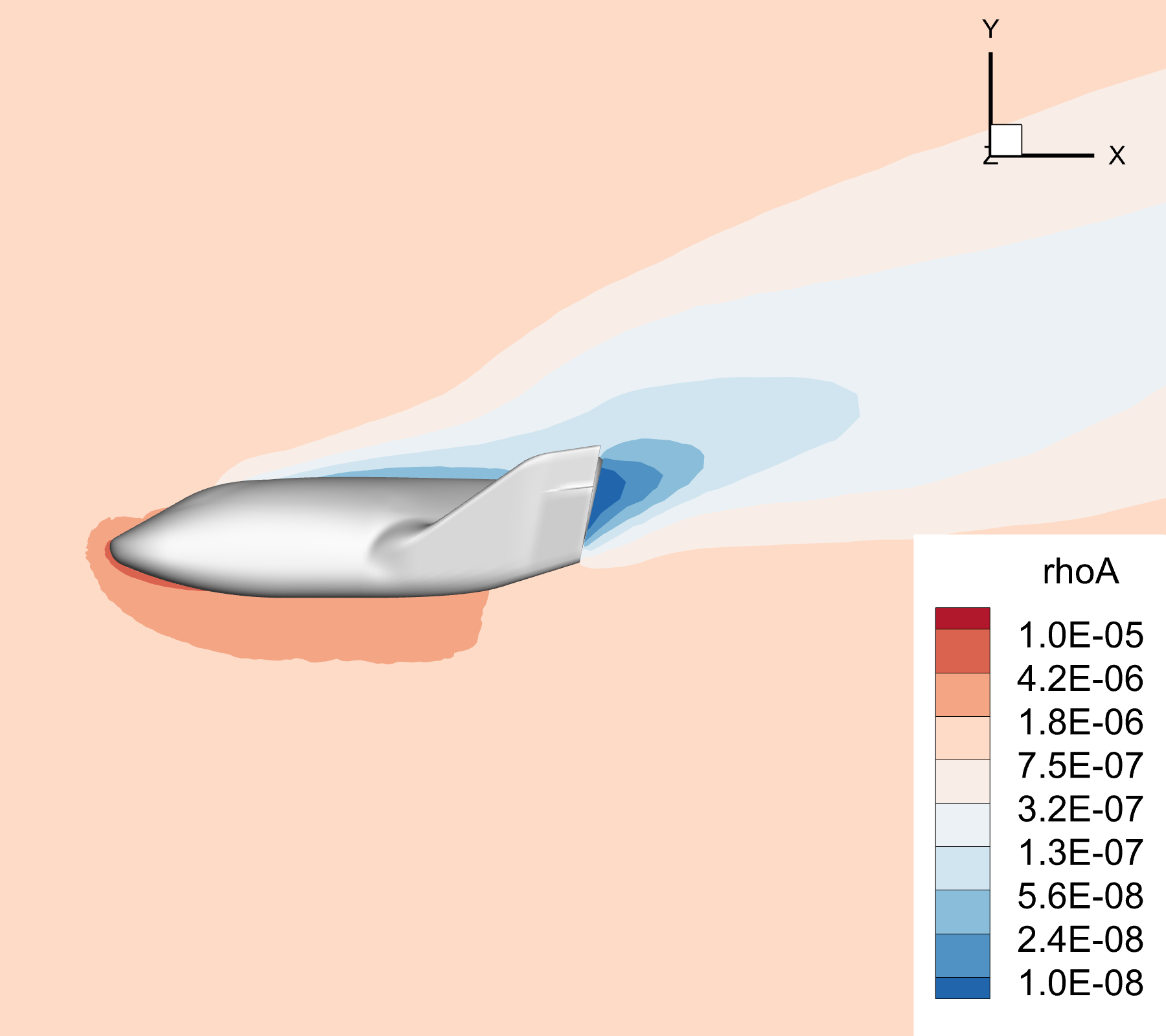}}
	\subfloat[]{\includegraphics[width=0.4\textwidth]
		{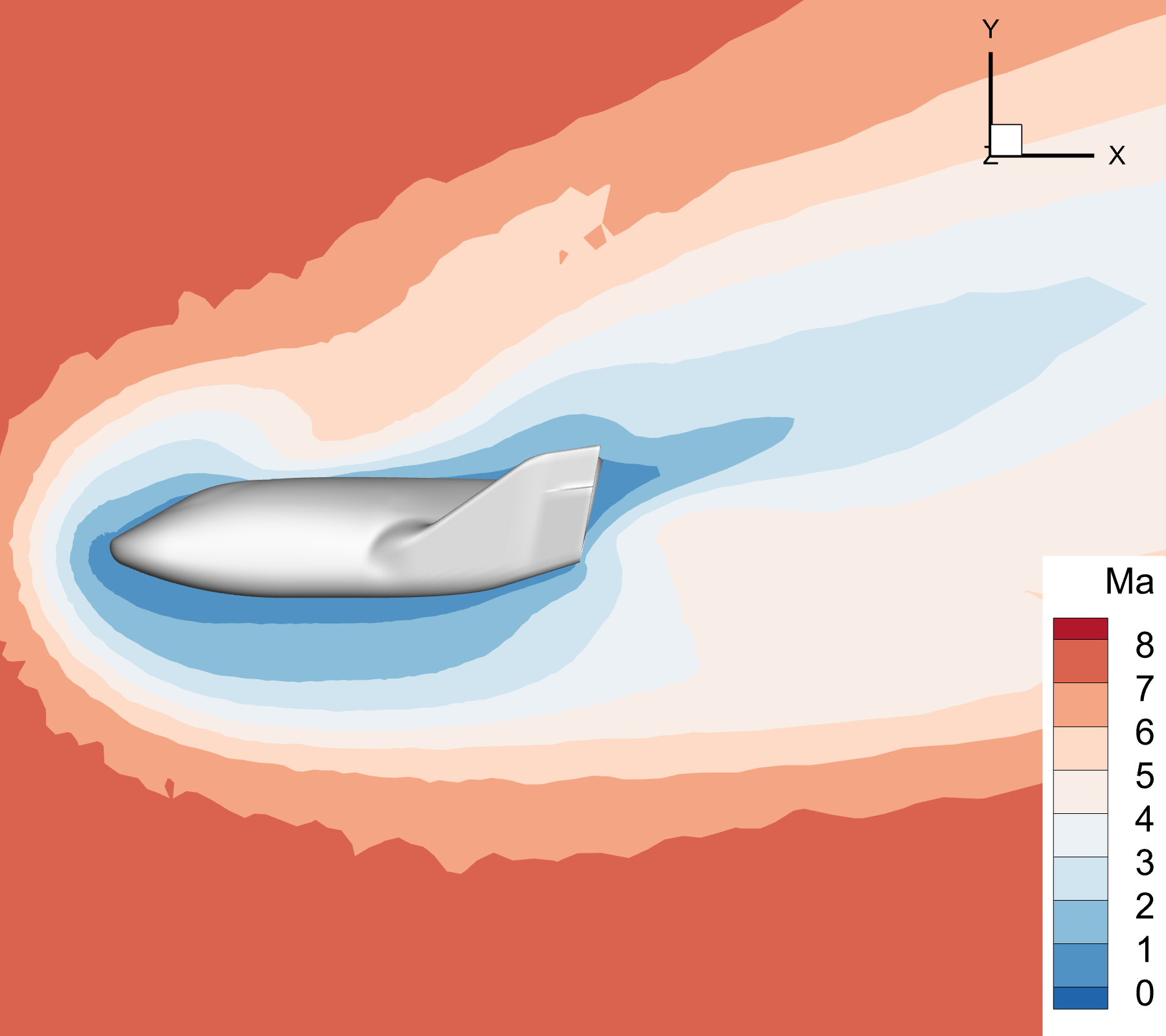}} \\
	\subfloat[]{\includegraphics[width=0.4\textwidth]
		{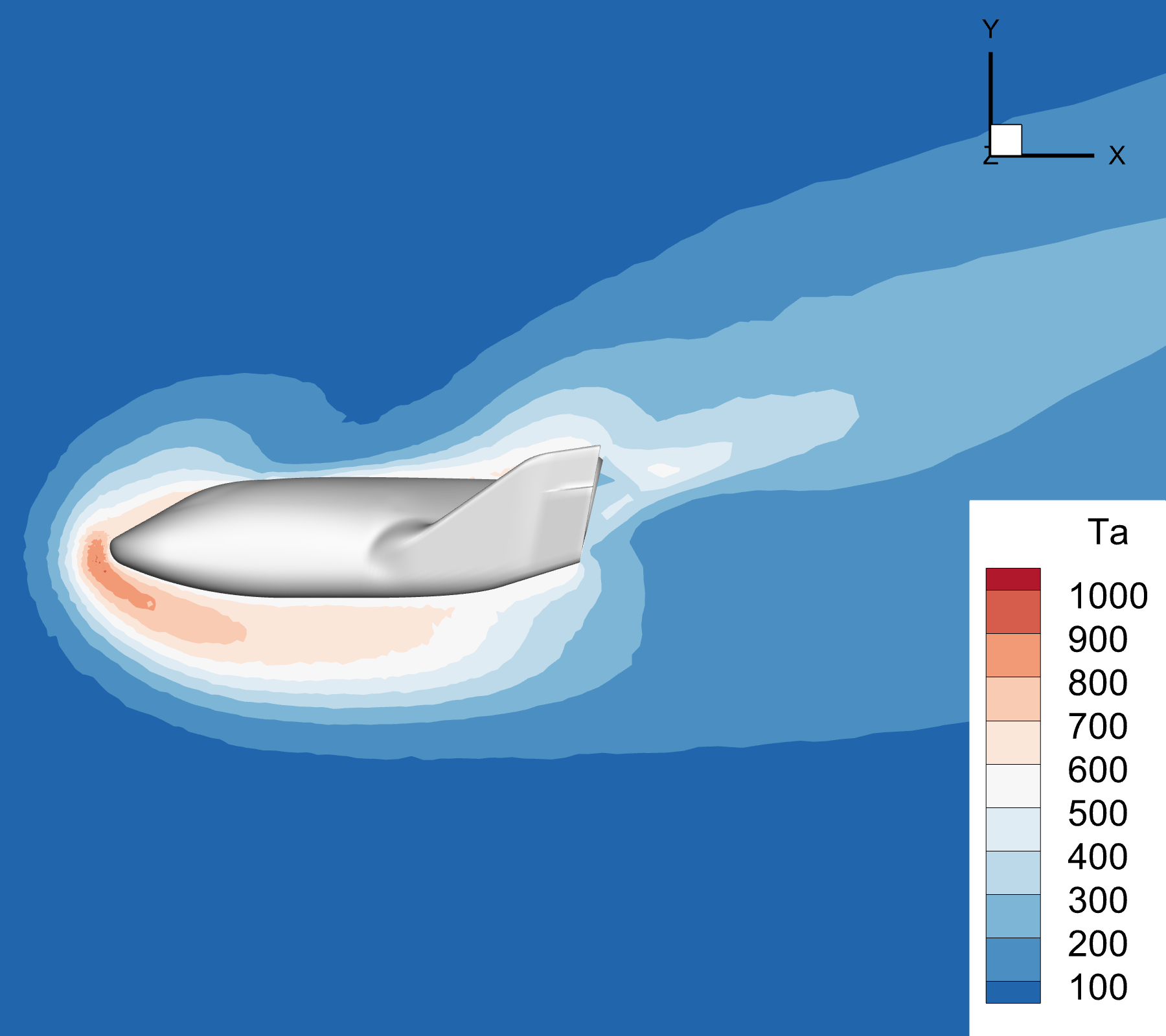}}
	\subfloat[]{\includegraphics[width=0.4\textwidth]
		{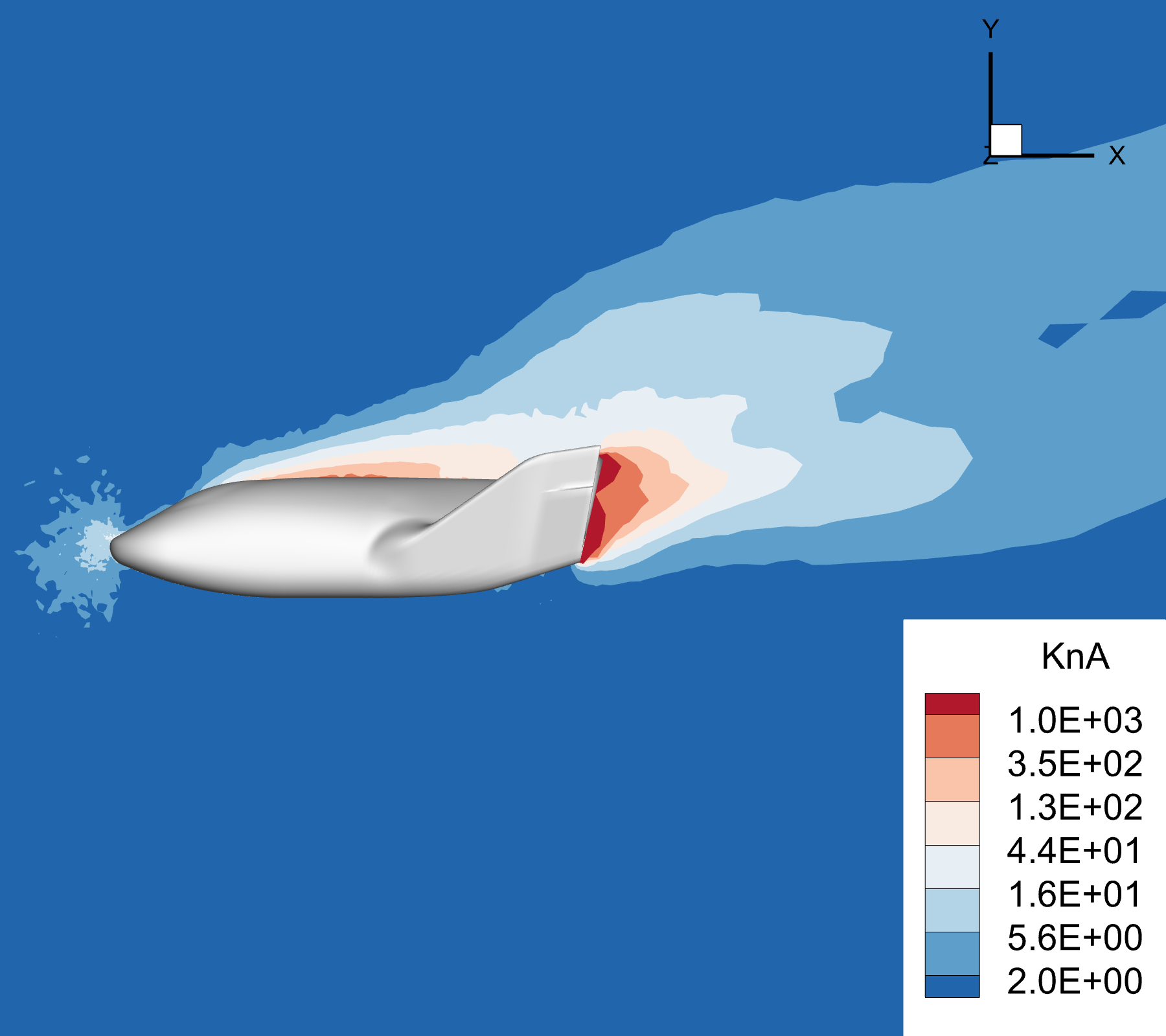}} \\
	\caption{Hypersonic flow
		around a X38-like vehicle at ${\rm Ma}_\infty = 8.0$ with $\rm{AoA} = 20^{\circ}$ and ${\rm Kn}_\infty = 0.275$ by the UGKWP method. (a) Density, (b) temperature, (c) Mach number, and
		(d) local Knudsen number contours.}
	\label{fig:x38contour6}
\end{figure}

\begin{figure}[H]
	\centering
	\subfloat[]{\includegraphics[width=0.4\textwidth]
		{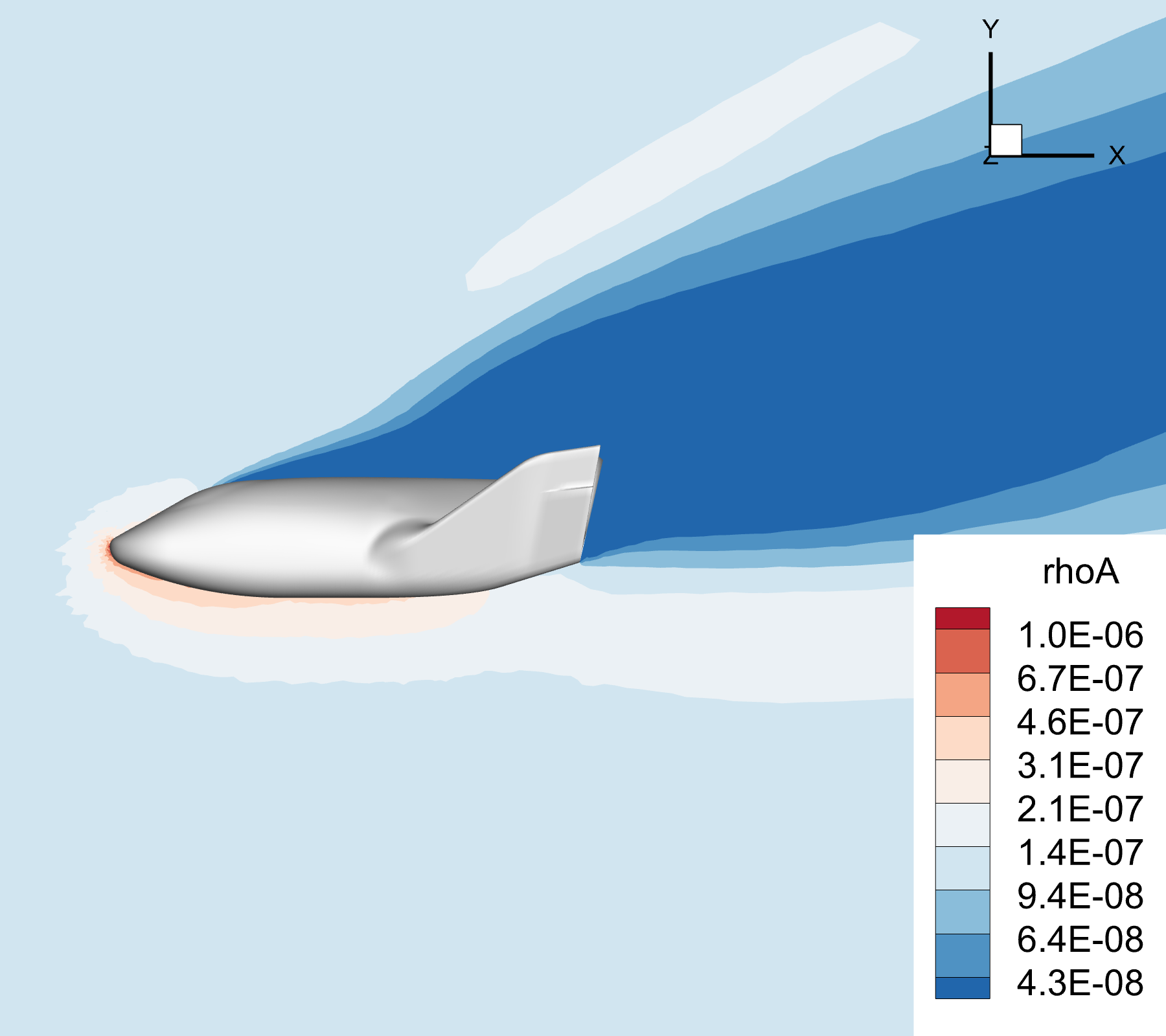}}
	\subfloat[]{\includegraphics[width=0.4\textwidth]
		{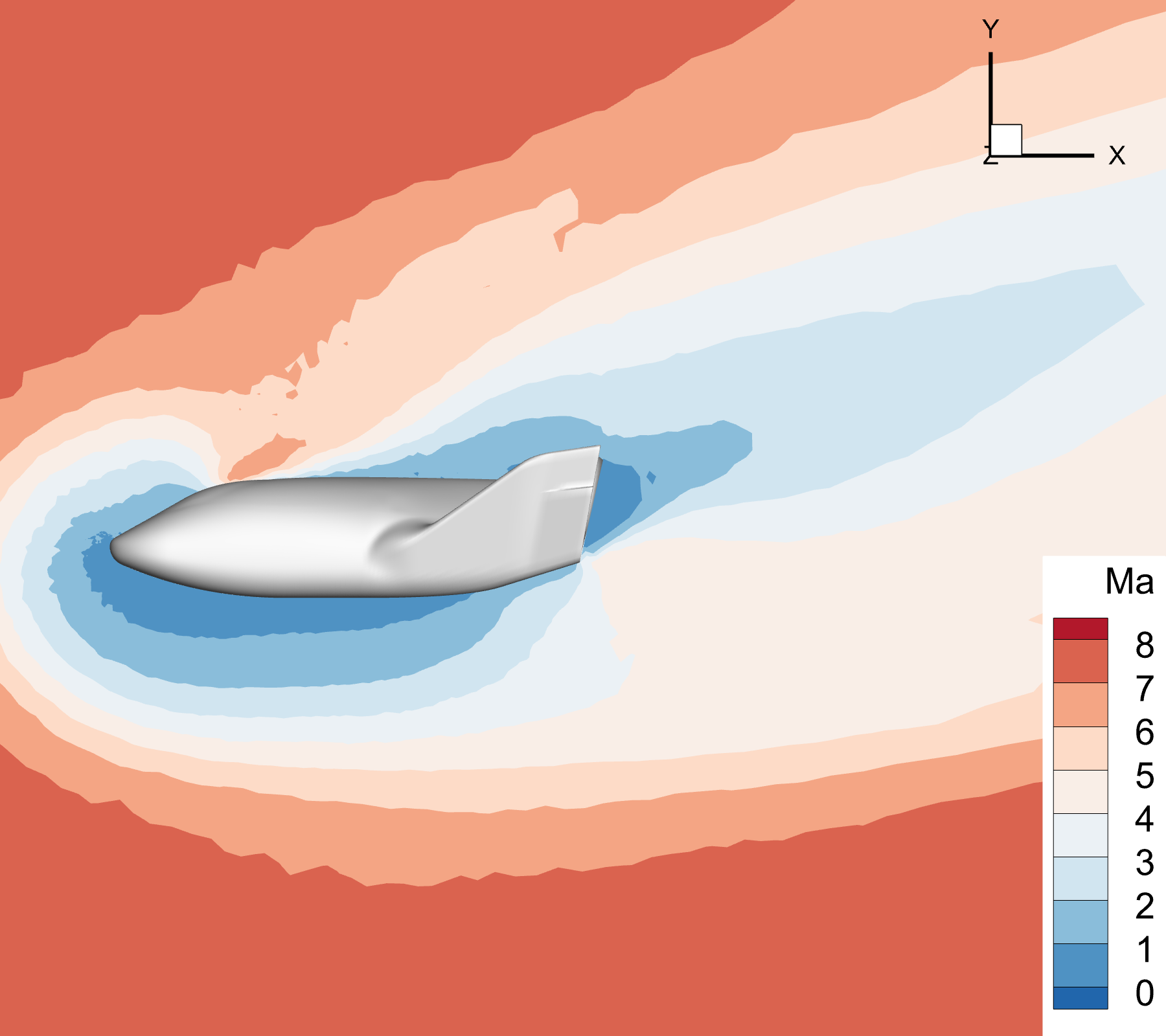}} \\
	\subfloat[]{\includegraphics[width=0.4\textwidth]
		{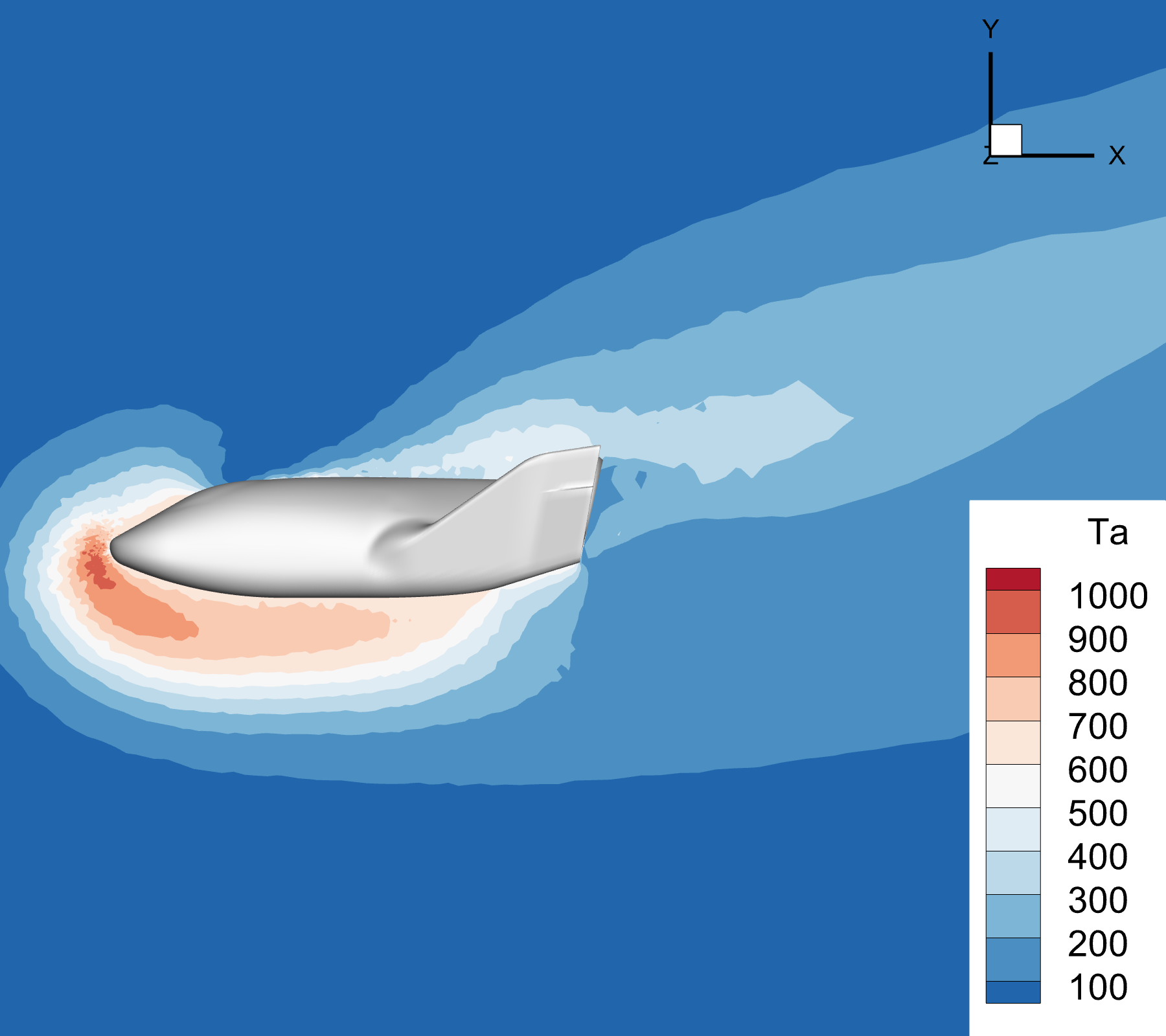}}
	\subfloat[]{\includegraphics[width=0.4\textwidth]
		{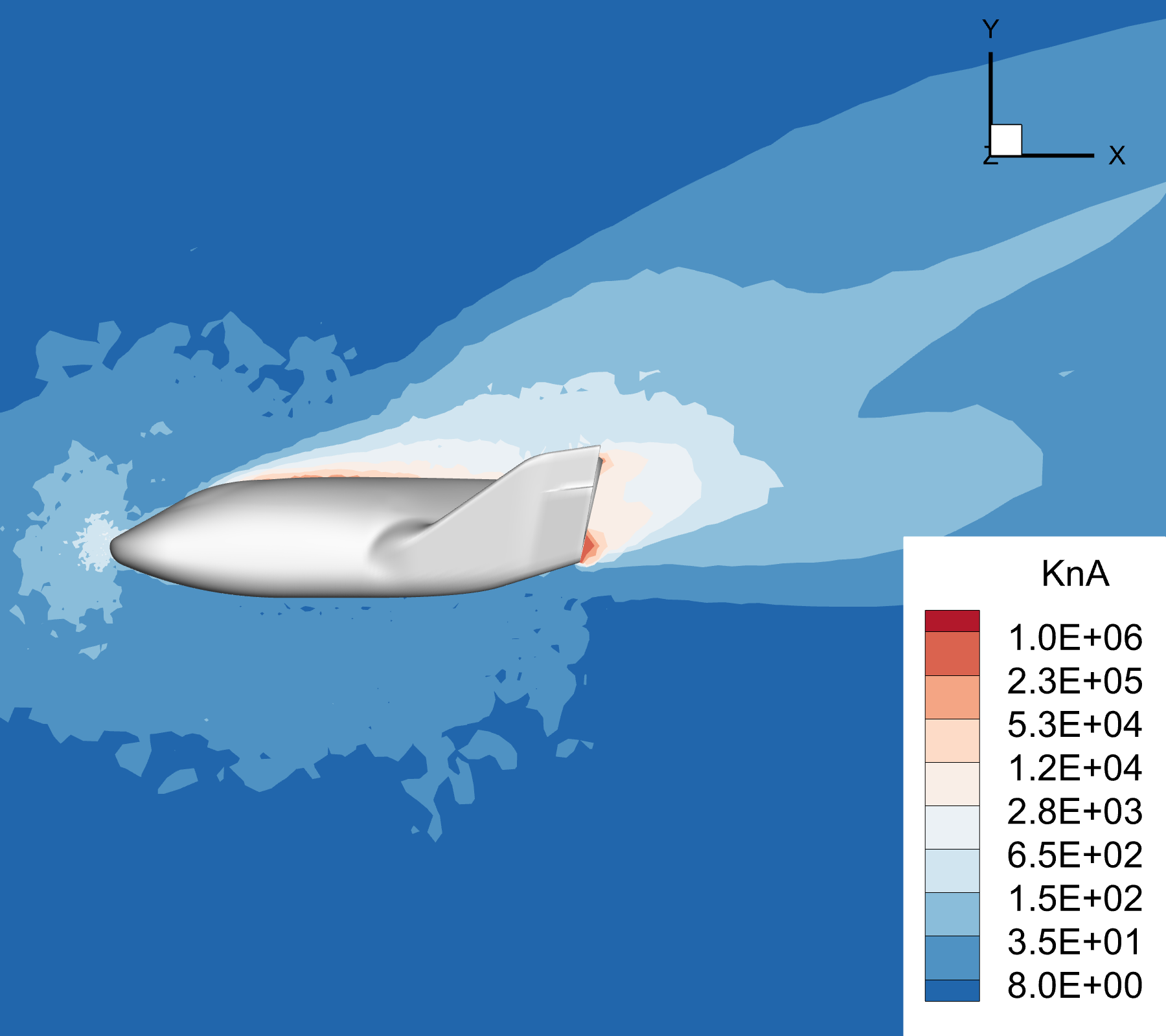}} \\
	\caption{Hypersonic flow
		around a X38-like vehicle at ${\rm Ma}_\infty = 8.0$ with $\rm{AoA} = 20^{\circ}$ and ${\rm Kn}_\infty = 2.75$ by the UGKWP method. (a) Density, (b) temperature, (c) Mach number, and
		(d) local Knudsen number contours.}
	\label{fig:x38contour8}
\end{figure}

Distributions of density, Mach number, temperature, and local Knudsen number of all cases are illustrated. Fig.~\ref{fig:x38contour1} to \ref{fig:x38contour7} show the simulation results at $\rm{AoA} = 0^{\circ}$ in different flow regimes, while Fig.~\ref{fig:x38contour2} to \ref{fig:x38contour8} give contours at $\rm{AoA} = 20^{\circ}$. The density contours at different Knudsen numbers show the high density in the windward region and low density in the leeward region and present the multiple-scale flow structure in all cases. 
The local Knudsen number distributions show strong non-equilibrium effects in the vicinity of the vehicle tail. 
From the temperature and Mach number distributions, shock and discontinuous flow distributions can be observed in Fig.~\ref{fig:x38contour1}(b) and Fig.~\ref{fig:x38contour1}(c), i.e., the cases at ${\rm Kn}_\infty = 0.00275$, while the shock wave gets thicker and weaker as the Knudsen number rises. At ${\rm Kn}_\infty = 2.75$ shown in Fig.~\ref{fig:x38contour7}(d), there is no clear discontinuity in the windward region, and the shock merges together with the boundary layer to form a diffusive structure.

\begin{figure}[H]
	\centering
	\subfloat[]{\includegraphics[width=0.4\textwidth]
		{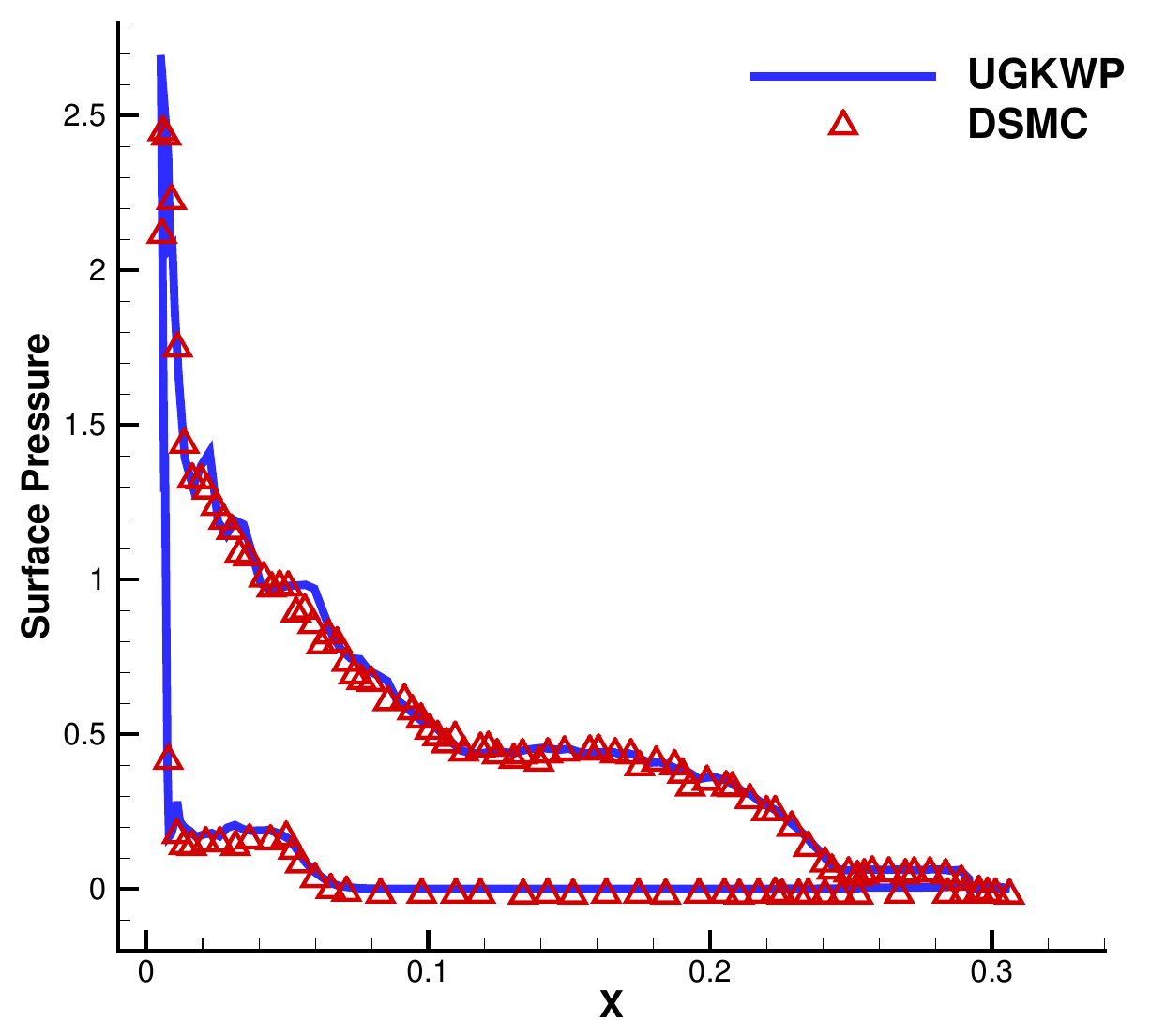}}
	\subfloat[]{\includegraphics[width=0.4\textwidth]
		{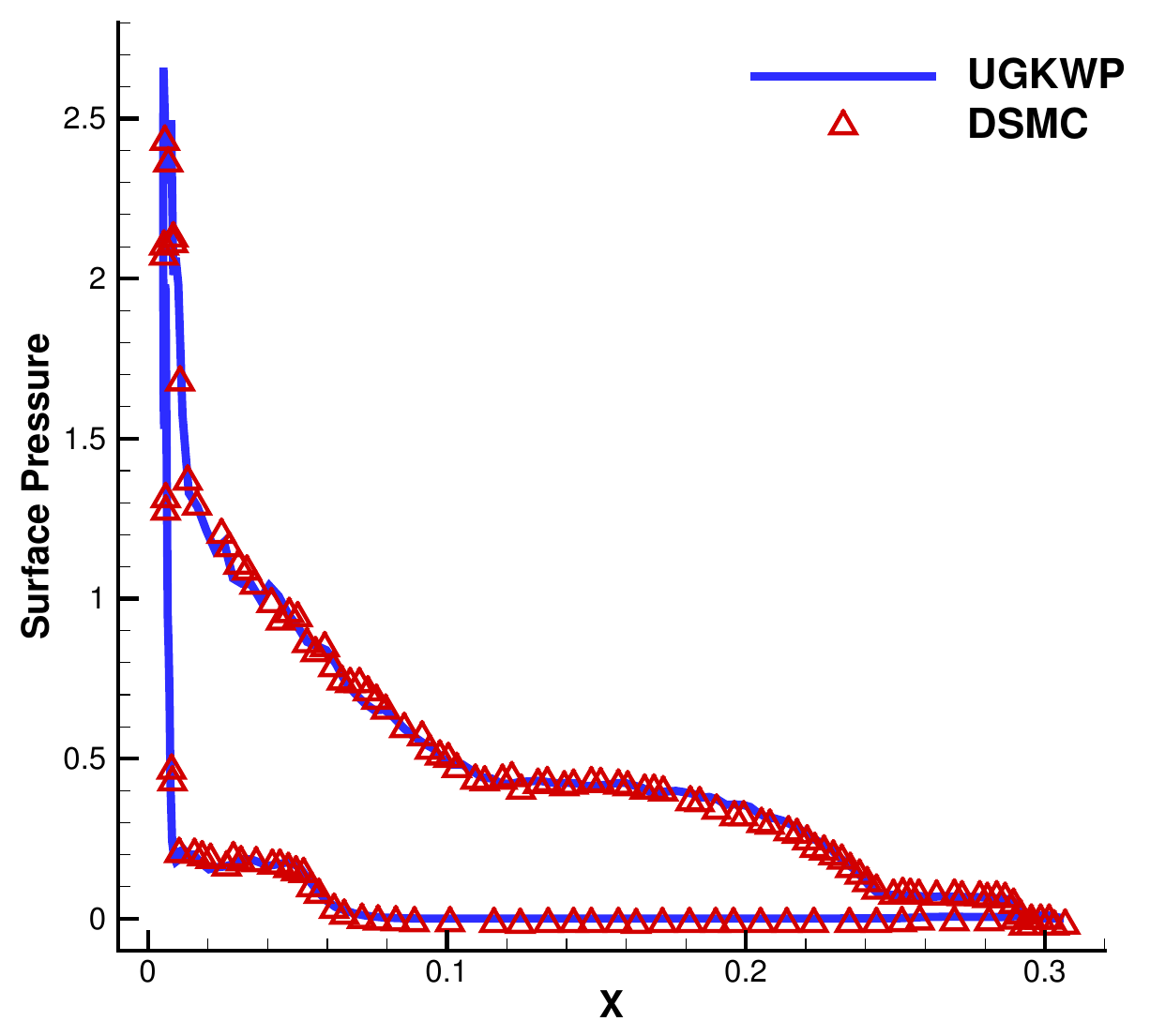}} \\
	\subfloat[]{\includegraphics[width=0.4\textwidth]
		{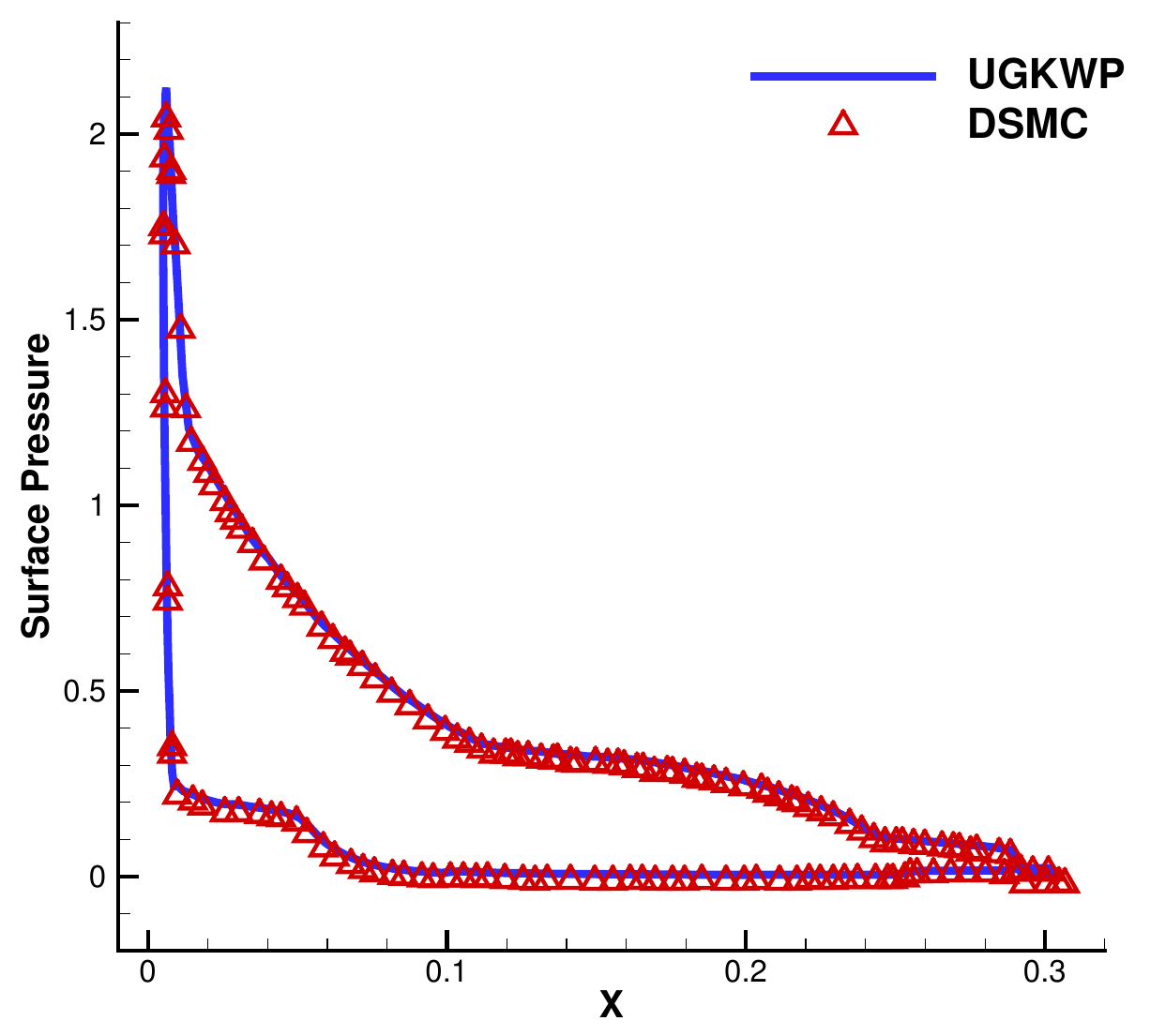}}
	\subfloat[]{\includegraphics[width=0.4\textwidth]
		{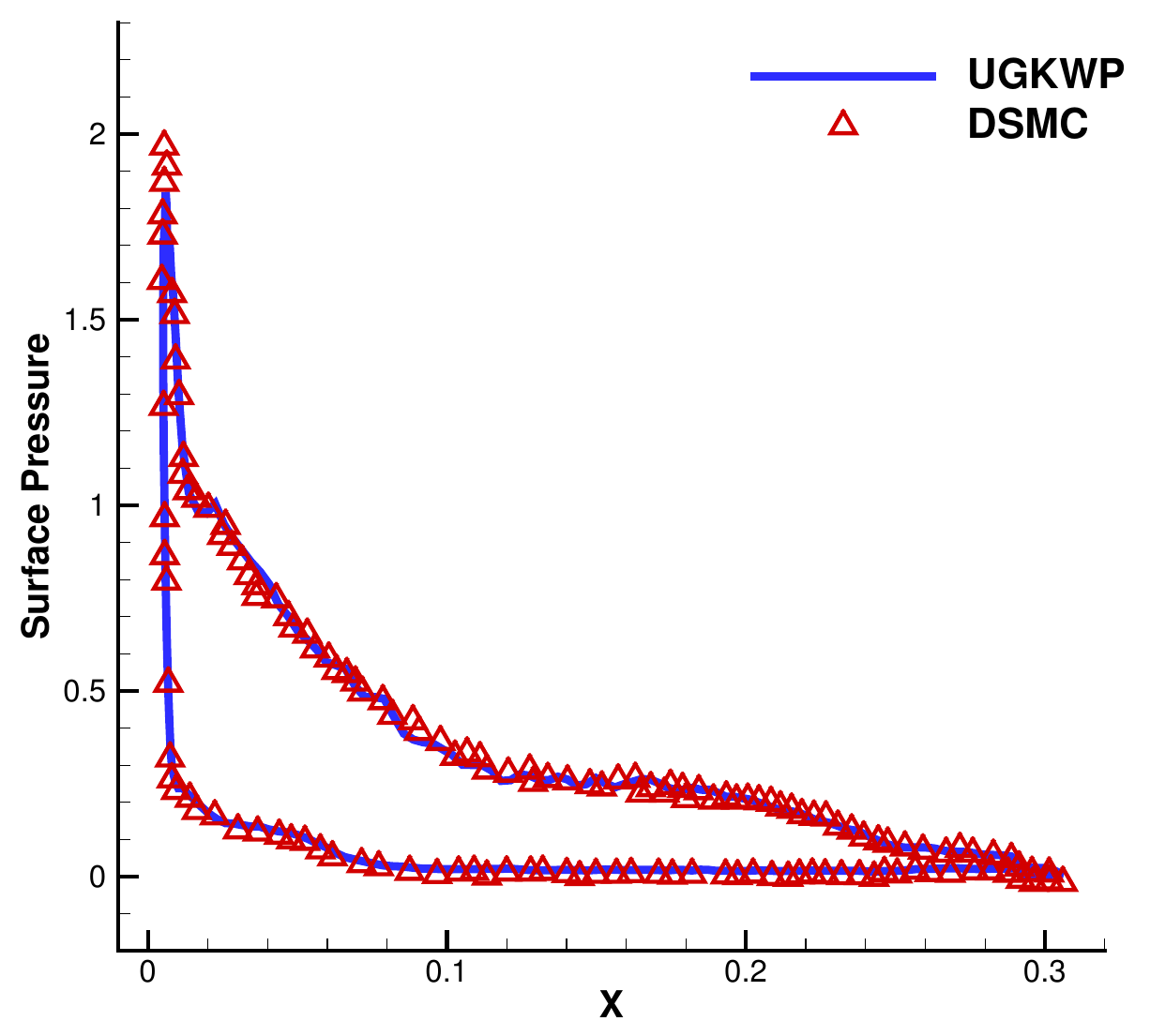}} \\
	\caption{Pressure coefficients on the axial surface of X38-like vehicle at ${\rm Ma}_\infty = 8.0$ with $\rm{AoA} = 20^{\circ}$ by the UGKWP method. (a)~${\rm Kn}_\infty = 2.75$ ,(b)~${\rm Kn}_\infty = 0.275$, (c)~${\rm Kn}_\infty = 0.0275$, and (d)~${\rm Kn}_\infty = 0.00275$.}
	\label{fig:x38Cp}
\end{figure}

\begin{figure}[H]
	\centering
	\subfloat[]{\includegraphics[width=0.4\textwidth]
		{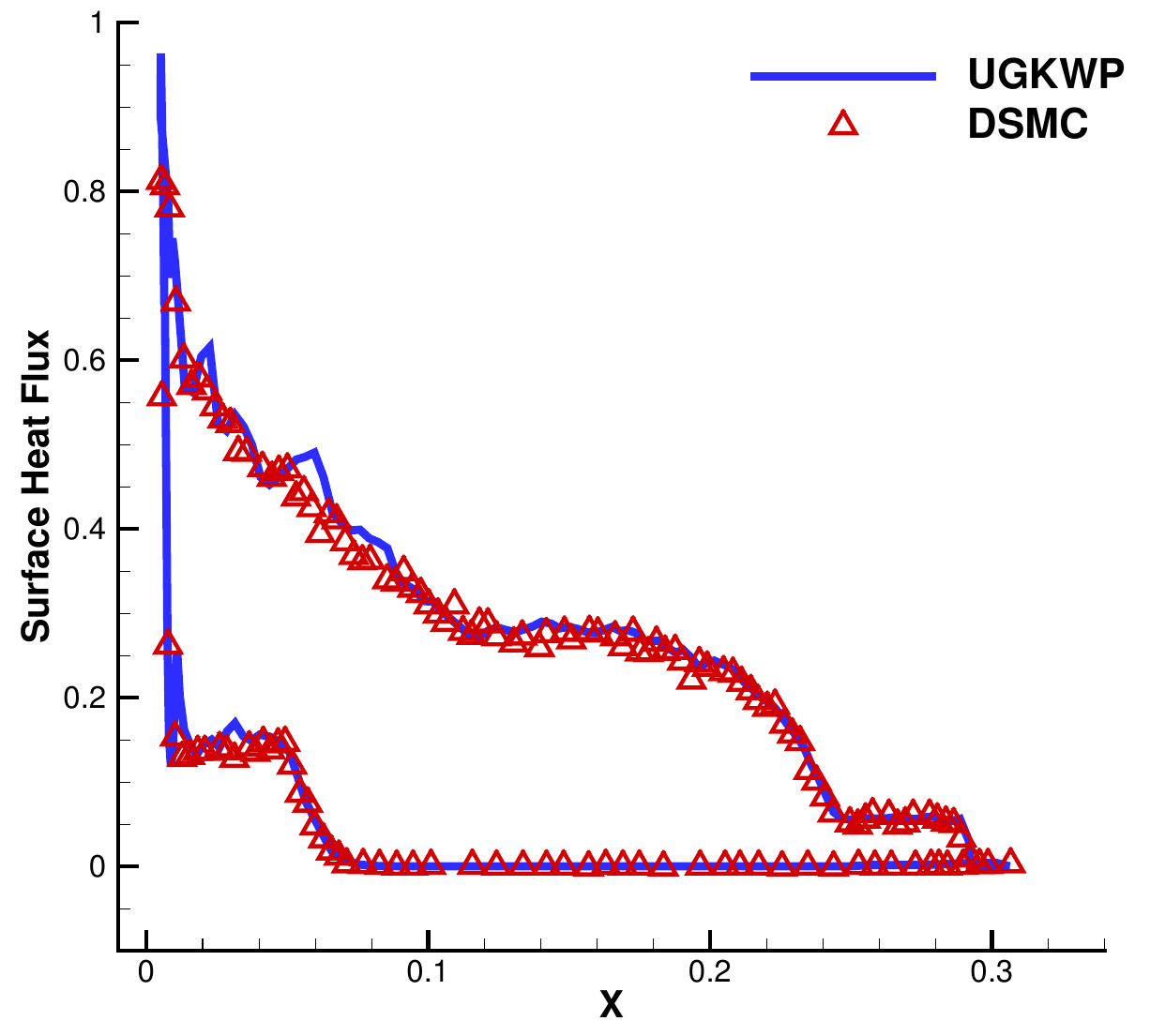}}
	\subfloat[]{\includegraphics[width=0.4\textwidth]
		{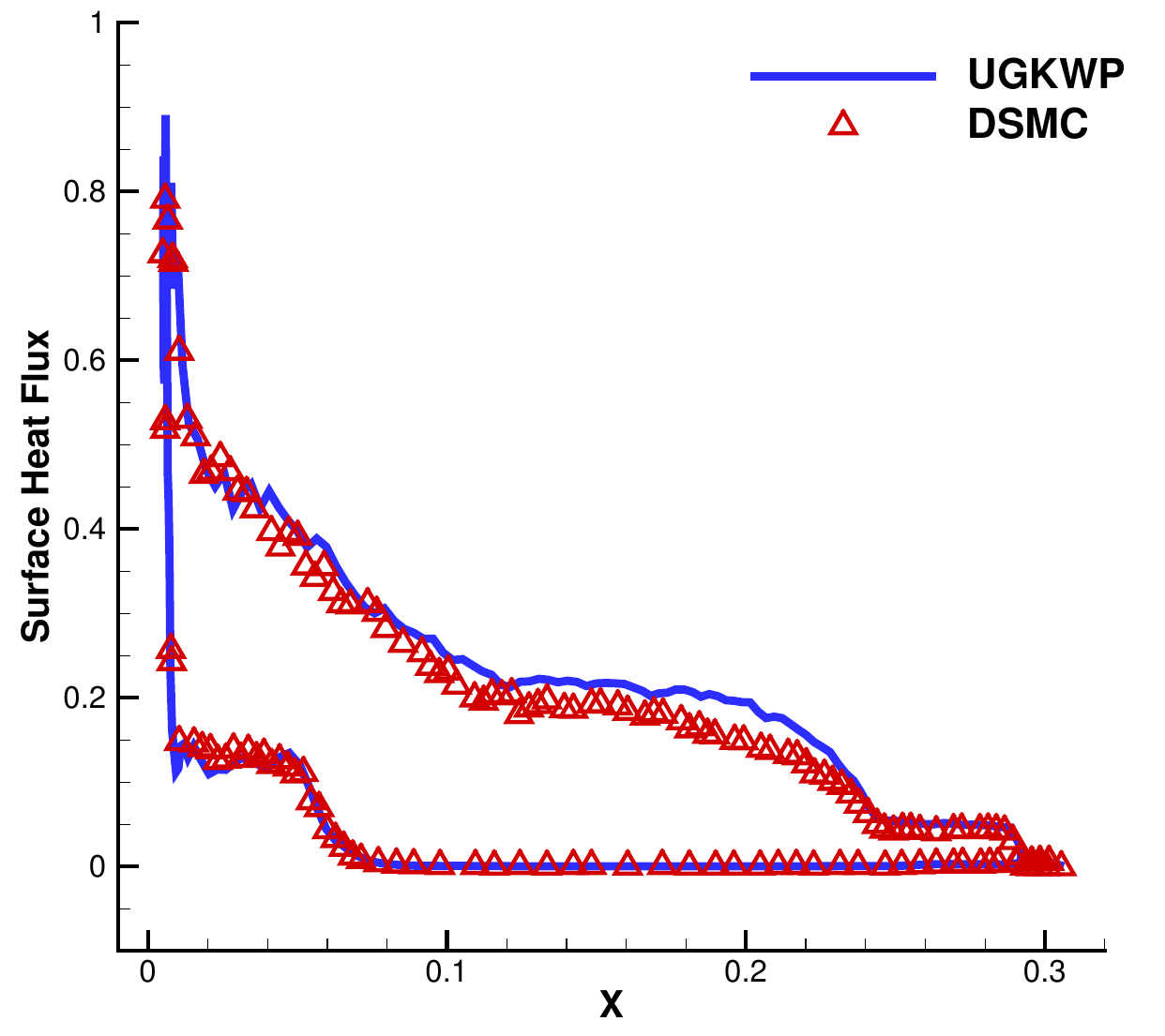}} \\
	\subfloat[]{\includegraphics[width=0.4\textwidth]
		{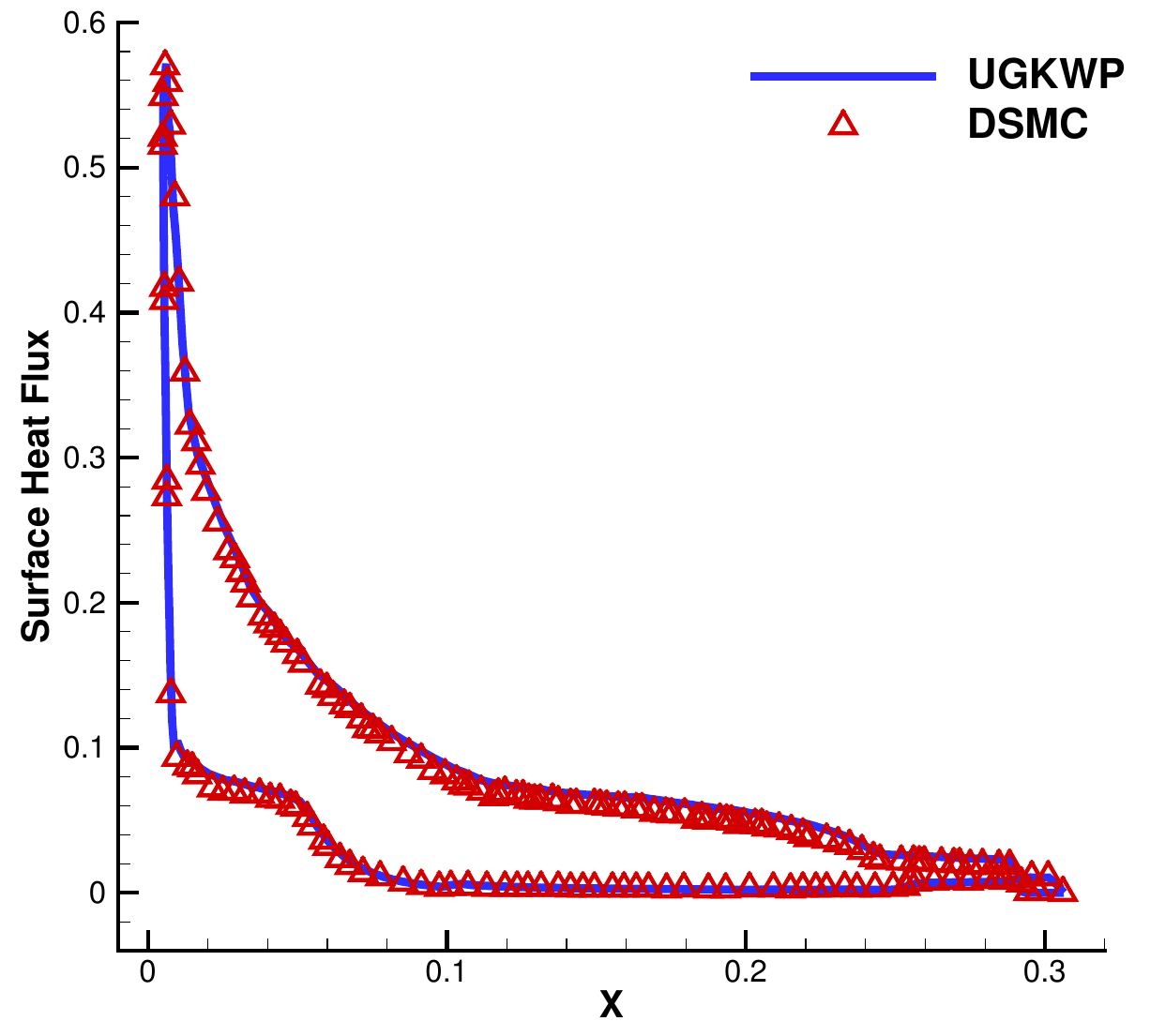}}
	\subfloat[]{\includegraphics[width=0.4\textwidth]
		{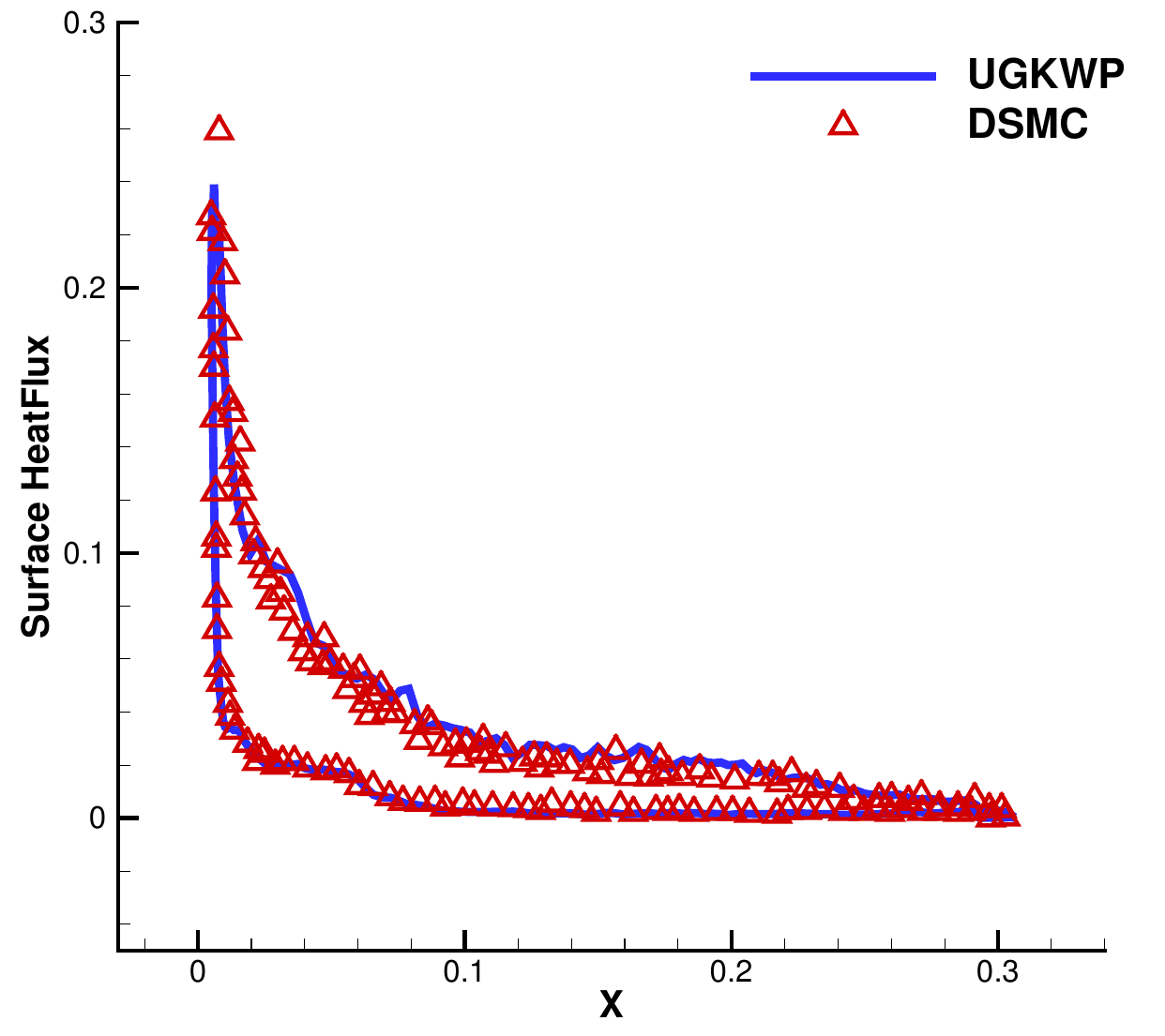}} \\
	\caption{Heat transfer coefficients on the axial surface of X38-like vehicle at ${\rm Ma}_\infty = 8.0$ with $\rm{AoA} = 20^{\circ}$ by the UGKWP method. (a)~${\rm Kn}_\infty = 2.75$, (b) ${\rm Kn}_\infty = 0.275$, (c) ${\rm Kn}_\infty = 0.0275$, and (d) ${\rm Kn}_\infty = 0.00275$.}
	\label{fig:x38Ch}
\end{figure}

\begin{figure}[H]
	\centering
	\subfloat[]{\includegraphics[width=0.4\textwidth]
		{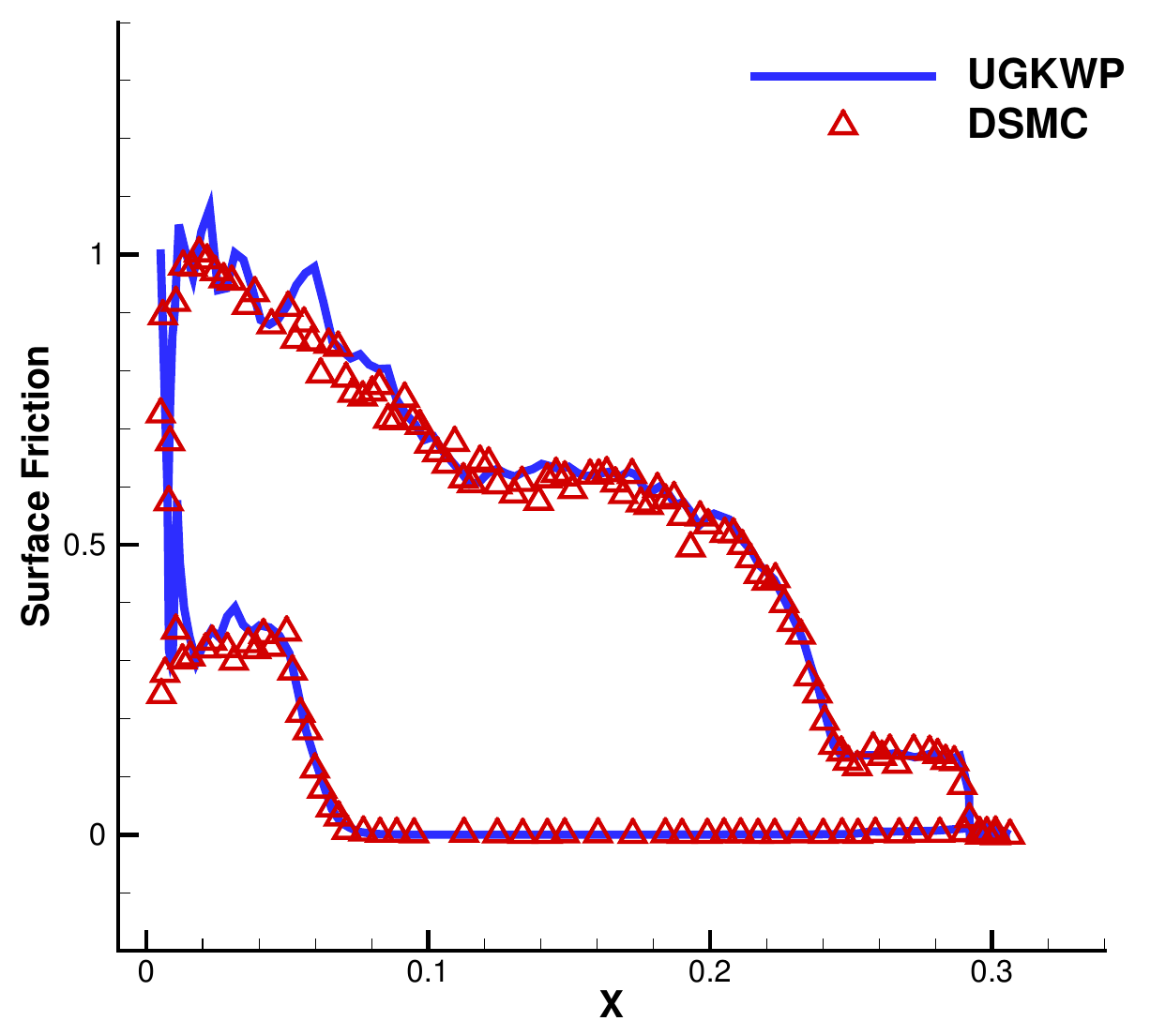}}
	\subfloat[]{\includegraphics[width=0.4\textwidth]
		{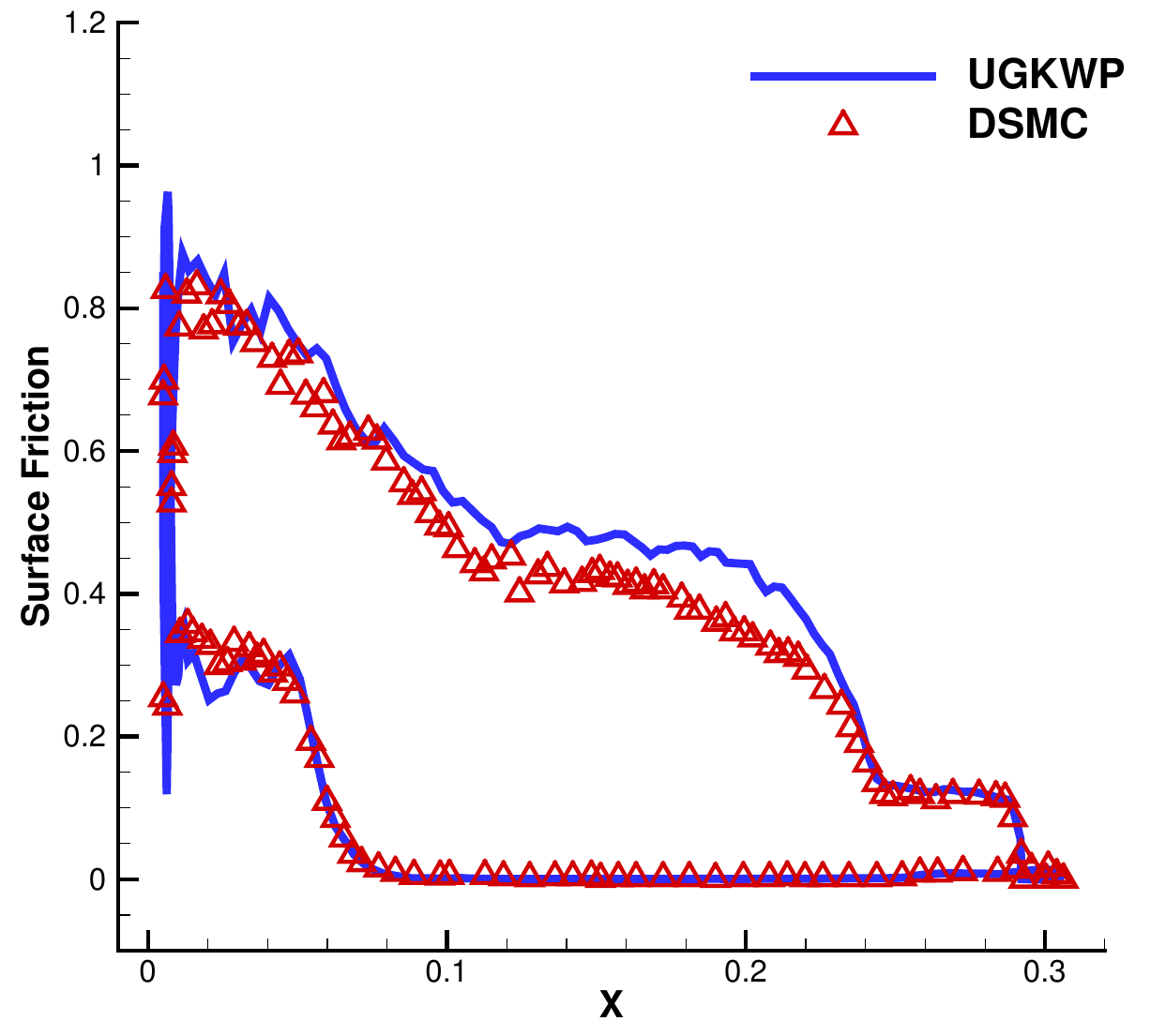}} \\
	\subfloat[]{\includegraphics[width=0.4\textwidth]
		{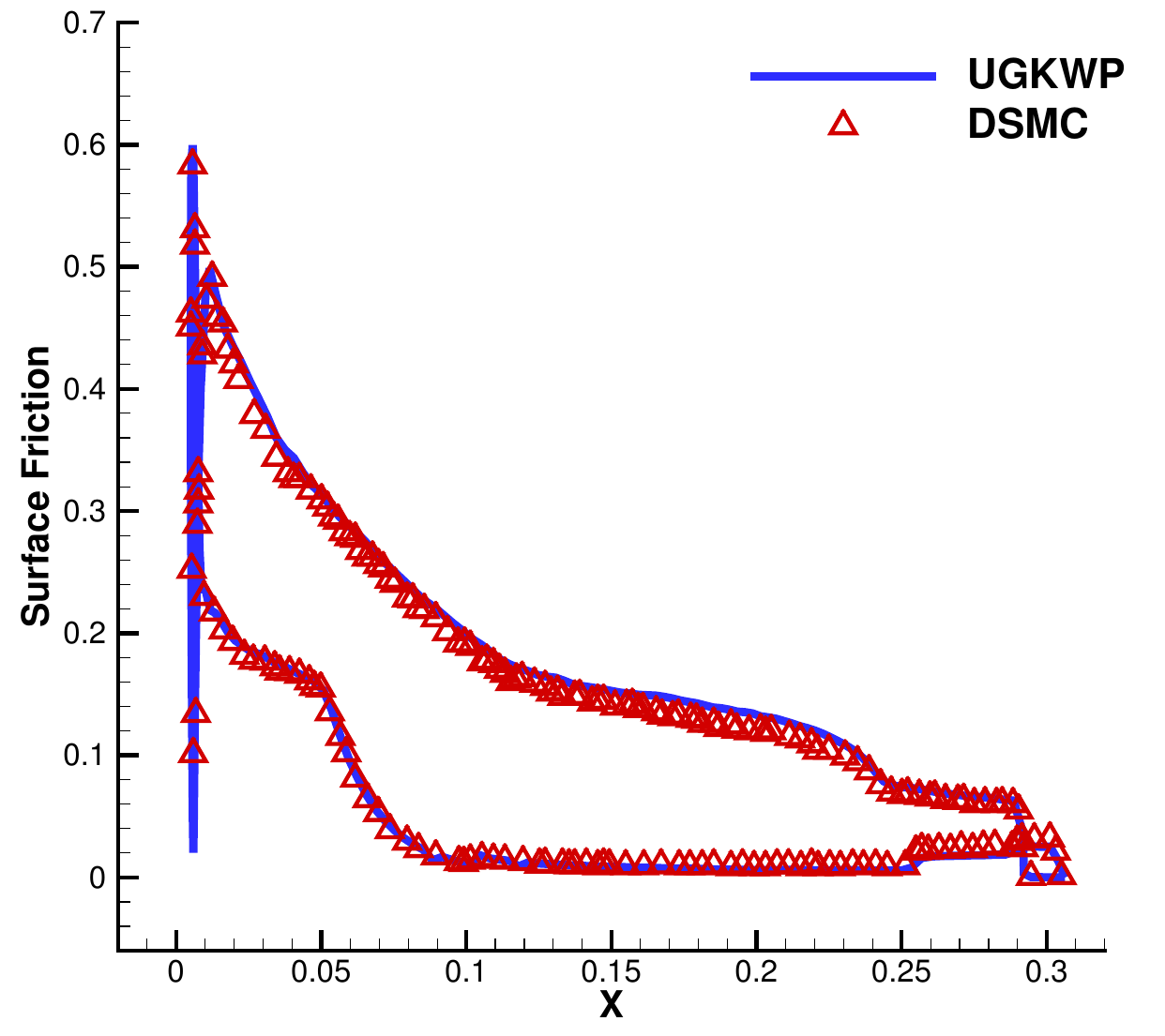}}
	\subfloat[]{\includegraphics[width=0.4\textwidth]
		{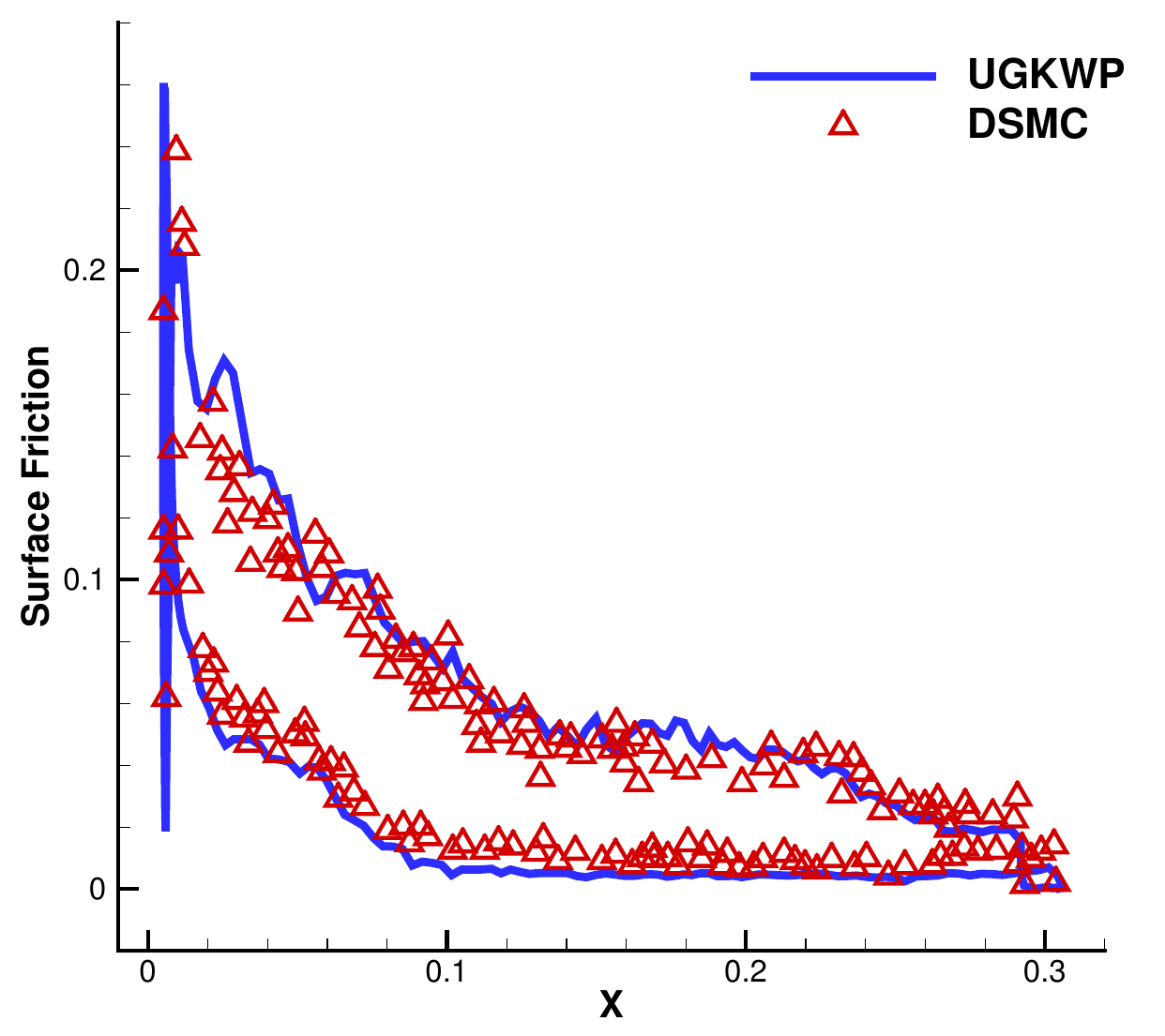}} \\
	\caption{Shear stress coefficients on the axial surface of X38-like vehicle at ${\rm Ma}_\infty = 8.0$ with $\rm{AoA} = 20^{\circ}$ by the UGKWP method. (a) ${\rm Kn}_\infty = 2.75$,(b) ${\rm Kn}_\infty = 0.275$, (c) ${\rm Kn}_\infty = 0.0275$, and (d) ${\rm Kn}_\infty = 0.00275$.}
	\label{fig:x38Ct}
\end{figure}

Figure~\ref{fig:x38Cp} to \ref{fig:x38Ct} show the distributions of surface physical quantities at the symmetrical section for all cases with $\rm{AoA} = 20^{\circ}$. The pressure coefficients $C_p$, heat transfer coefficients $C_h$ and shear stress coefficients $C_{\tau}$ are normalized by density $\rho_{\infty}$ and velocity $U_{\infty}$ of free stream flow:
\begin{equation*}
	C_p=\dfrac{p_s}{\frac{1}{2} \rho_{\infty}U_{\infty}^{2}},
	C_{\tau}=\dfrac{f_s}{\frac{1}{2} \rho_{\infty}U_{\infty}^{2}},
	C_h=\dfrac{h_s}{\frac{1}{2} \rho_{\infty}U_{\infty}^{3}}.
\end{equation*}
The peak values are concentrated near the nose of the vehicle, and decrease as the Knudsen number falls, which indicates the significant non-equilibrium in the rarefied regime. Comparing the maximum value of surface quantities at ${\rm Kn}_\infty = 0.00275$ and ${\rm Kn}_\infty = 2.75$, surface pressure decreases from near 3 to 2, and heat transfer diminishes from over 1 to less than 1/3, while surface friction is reduced to nearly 1/5.  Due to flat-plate-like geometry, coefficients on the lower surface of the model are almost constant at 0. Although statistical scatter can be observed in the case at ${\rm Kn}_\infty = 2.75$, all coefficients agree well with DSMC data.

Table~\ref{table:x38time} gives the computational efficiency and the consumption of computational resources for  $\rm{AoA} = 0^{\circ}$ and $\rm{AoA} = 20^{\circ}$. All the calculations are taken on the SUGON computation platform, CPU model used is 7285H 32C 2.5GHz. Simulations start averaging after 40000 steps and reach a steady state after varying averaging steps. For cases at ${\rm Kn}_\infty=0.00275$, it consumes dramatically less memory and costs due to the reduction of particles in the wave-particle decomposition. 
Cases at  ${\rm Kn}_\infty=0.0275$ converge slowly due to the competitive decomposition of wave-particle and a large number of particles sampled and removed in each step. The tendency of the simulation costs in all cases agrees with that observed in the sphere case.

\begin{table}[H]
	\caption{The computational cost for simulations of hypersonic flow around a X38-like vehicle at ${\rm Ma}_\infty = 8.0$ by the UGKWP method. The physical domain consists of 246558 cells for cases at ${\rm Kn}_\infty = 0.00275$ and 560593 cells for others, and the reference number of particles per cell in the UGKWP method is set as $N_r = 150$. }
	\centering
	\begin{threeparttable}
		\begin{tabular}{cccccc}
			\toprule
			${\rm Kn}_\infty$ & AoA & Computation steps & Wall Clock Time, h  & Cores & Estimated Memory, GiB  \\
			\midrule
			$0.00275$ & 20° & $40000+10000\tnote{1}$ & 6.25  & 640  &  9.28 \\
			$0.0275$ & 20° & $40000+14000\tnote{1}$ & 13.6 & 640  & 57.7   \\
			$0.275$ & 20° & $40000+12000\tnote{1}$ & 8.15 & 640  & 29.5  \\
			$2.75$ & 20° & $40000+12000\tnote{1}$ & 11.1  & 640  & 35.4  \\		
			$0.00275$ & 0° & $40000+10000\tnote{1}$ & 6.58  & 640  & 9.61  \\		
			$0.0275$ & 0° & $40000+14000\tnote{1}$ & 15.1 & 640  & 60.0   \\		
			$0.275$ & 0° & $40000+12000\tnote{1}$ & 8.22 & 640  & 29.8  \\		
			$2.75$ & 0° & $40000+12000\tnote{1}$ & 12.3  & 640  & 36.6  \\	
			\bottomrule
		\end{tabular}
		\begin{tablenotes}
			\item[1] Steps of averaging process in UGKWP simulation.
		\end{tablenotes}
	\end{threeparttable}
	\label{table:x38time}
\end{table}

\subsection{Nozzle plume expanding into a vacuum}

\begin{figure}[H]
	\centering
	\subfloat[]{\includegraphics[width=0.4\textwidth]
		{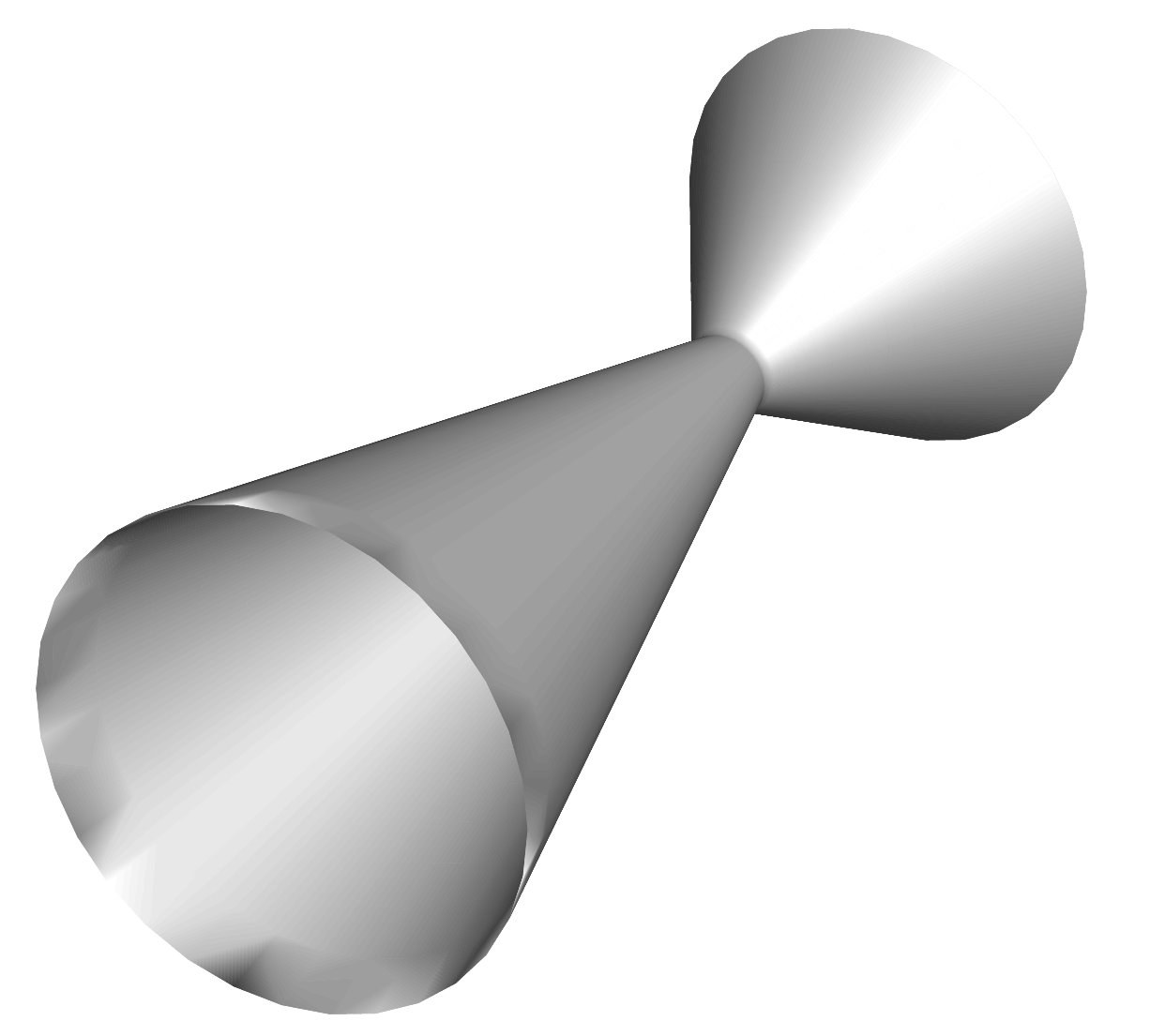}}
	\subfloat[]{\includegraphics[width=0.4\textwidth]
		{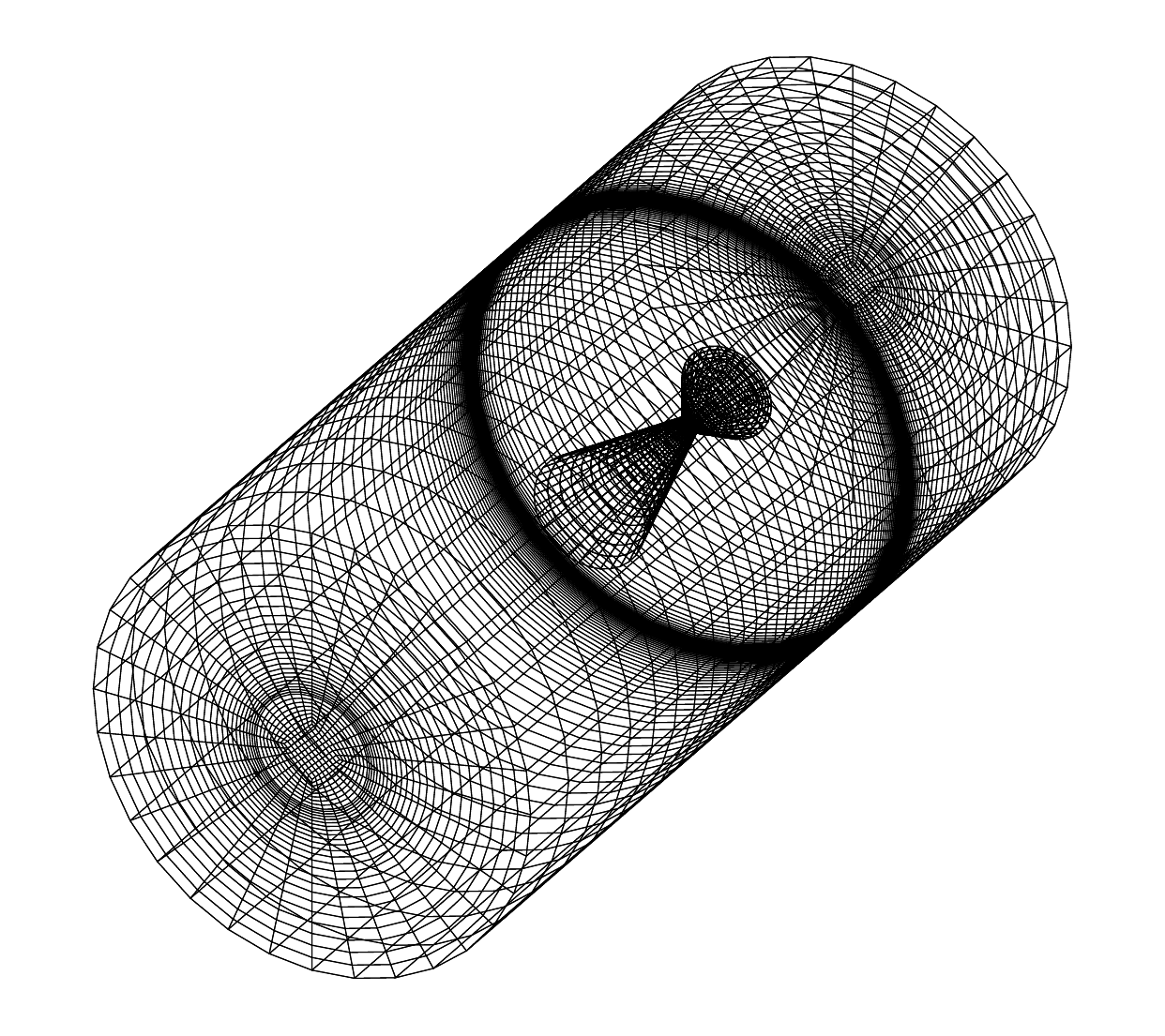}}  \\
	\caption{Geometry shape and the physical domain of nozzle plume expanding into a vacuum. (a) Geometry shape, and (b) physical mesh with 83776 cells.}
	\label{fig:nozzleplume-mesh}
\end{figure}

In this section, a carbon dioxide nozzle plume expanding to an extreme vacuum background is simulated. 
The geometric shape of the nozzle is based on the model used in Boyd et al.\cite{george1999simulation}, as shown in Fig.~\ref{fig:nozzleplume-mesh}. Inside the nozzle, the Knudsen number of the flow is small enough to approach the continuum flow regime. 
In the outer back-flow and plume region, the Knudsen number gets to the free molecule regime. 
This multi-scale problem poses great challenges to traditional CFD methods. The DSMC method encounters excessive computational cost in the near-continuum regime and makes the simulation of the whole expansion process unacceptable. 
The UGKWP method is perfectly suitable for this case and yields a reliable multi-scale solution with a reasonable computational cost.

The mesh adopted in this case is shown in Fig.~\ref{fig:nozzleplume-mesh}. The physical mesh of the entire calculation domain has 83776 cells, and the maximum number of particles in each cell is 200. Simulation results by the UGKWP method are compared with those of experiments and DSMC-NS method\cite{george1999simulation}. In this section, the ambient pressure outside the nozzle is $P_{\infty}$ = 0.01 Pa, and the ambient temperature is $T_{\infty}$ = 300K; At the inlet of the nozzle, the stagnation temperature is $T_s$ = 710 K, and the stagnation pressure is $T_s$ = 4866.18 Pa; The isothermal wall boundary condition is used for the nozzle wall with a temperature of $T_w$ = 500 K.

\begin{figure}[H]
	\centering
	\subfloat[]{\includegraphics[width=0.45\textwidth]
		{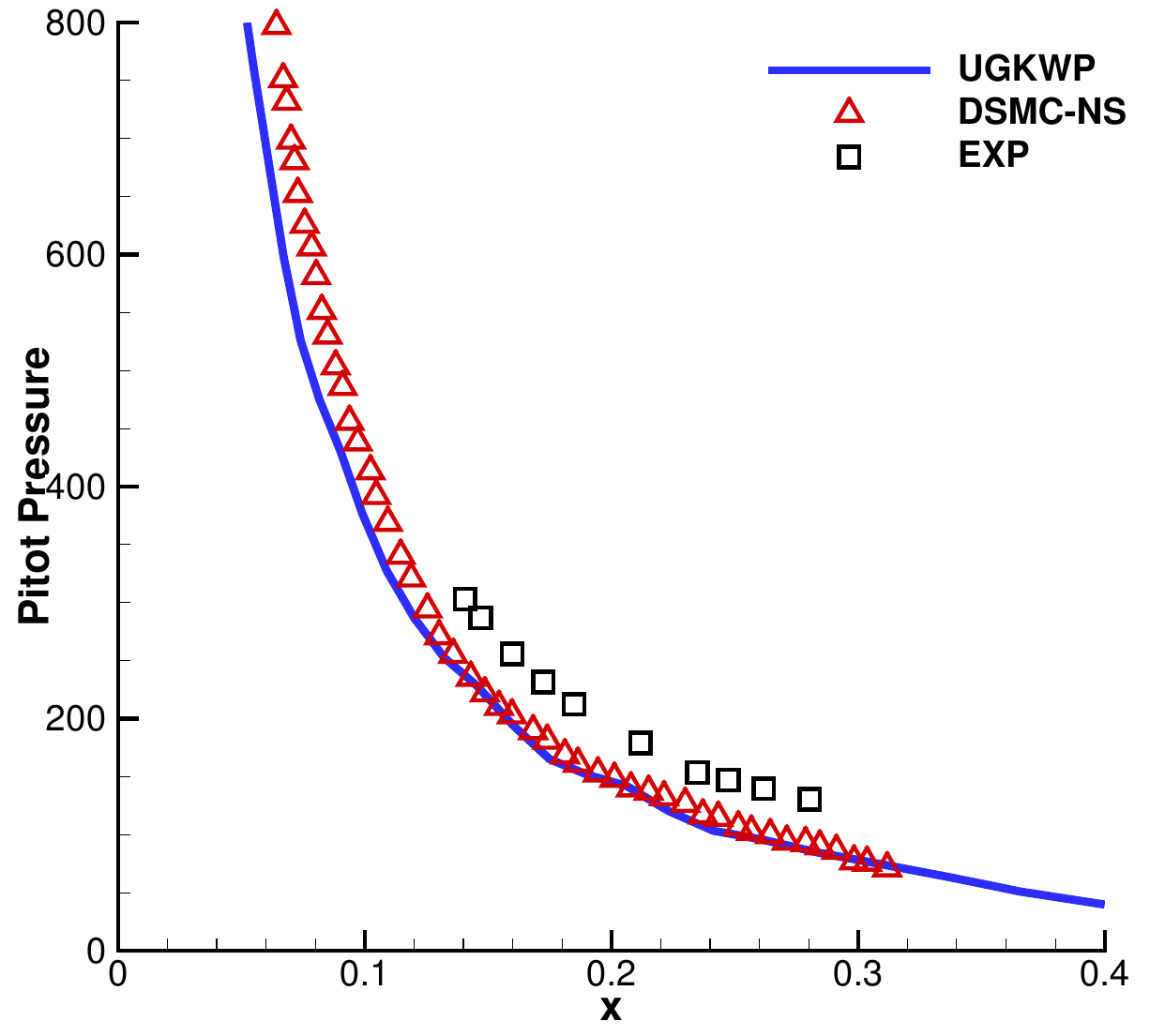}}
	\subfloat[]{\includegraphics[width=0.45\textwidth]
		{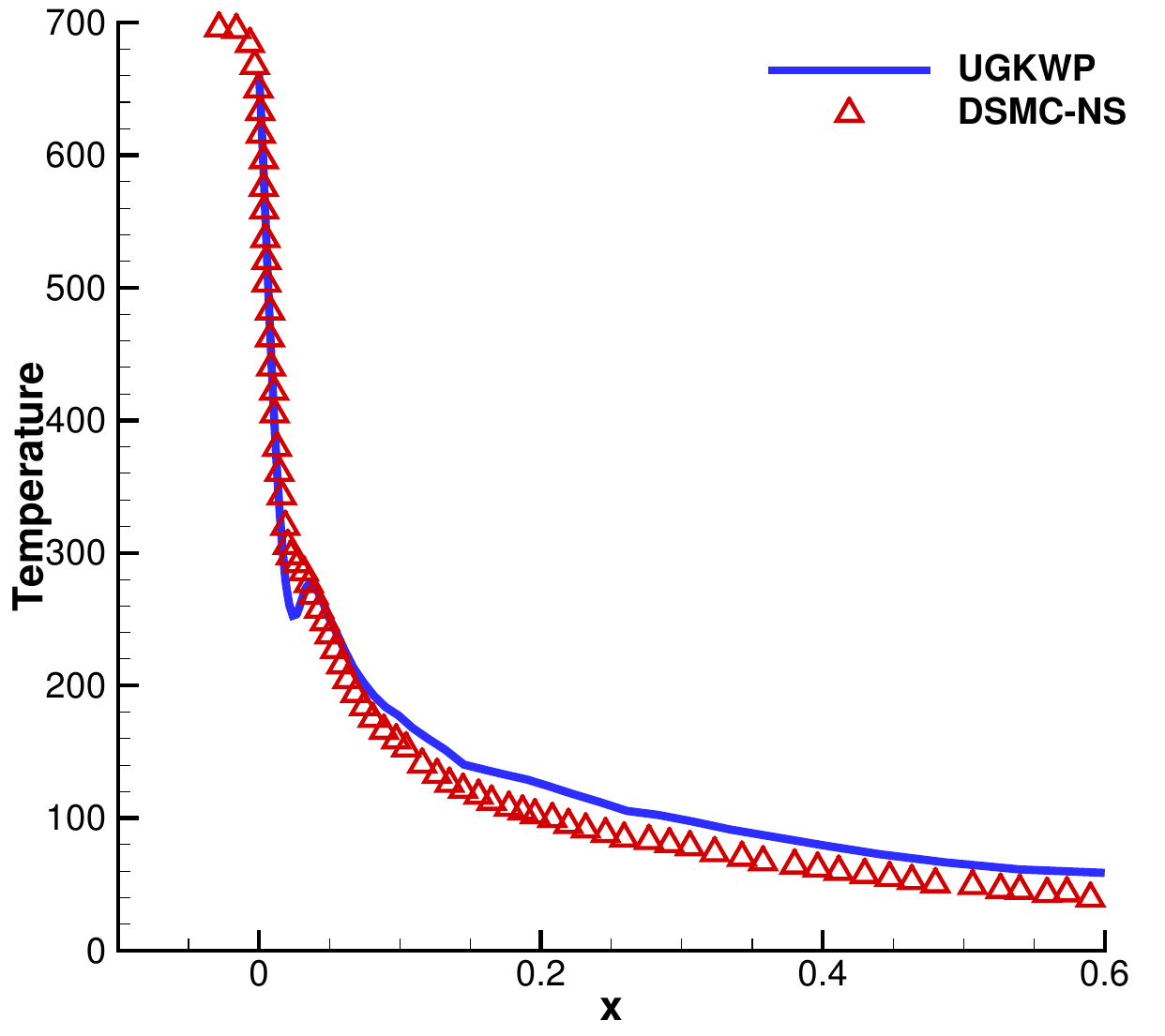}} \\
	\caption{Pitot pressure and temperature profiles along the axis of the nozzle by the UGKWP method. (a) Pitot Pressure, and (b) temperature.}
	\label{fig:nozzleplume-linecompare}
\end{figure}

To validate the accuracy of the simulation, the Pitot pressure and temperature along the central axis of the nozzle are shown in Fig.~\ref{fig:nozzleplume-linecompare}. A comparison is made among the solutions of UGKWP, experimental data, and DSMC-NS method \cite{george1999simulation}. Reasonable agreements have been observed. A temperature jump at the nozzle throat is captured by UGKWP while the DSMC-NS method is unable to get it due to the loss of accuracy.

\begin{figure}[H]
	\centering
	\subfloat[]{\includegraphics[width=0.4\textwidth]
		{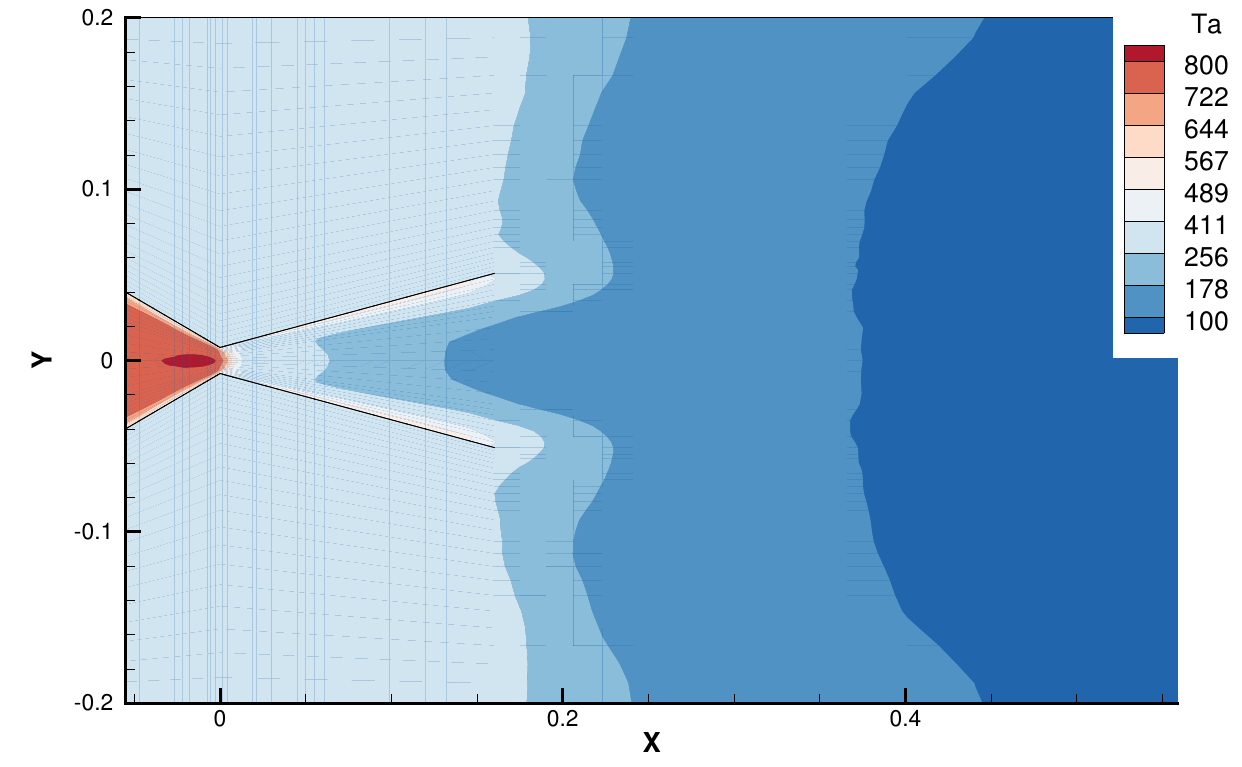}}
	\subfloat[]{\includegraphics[width=0.4\textwidth]
		{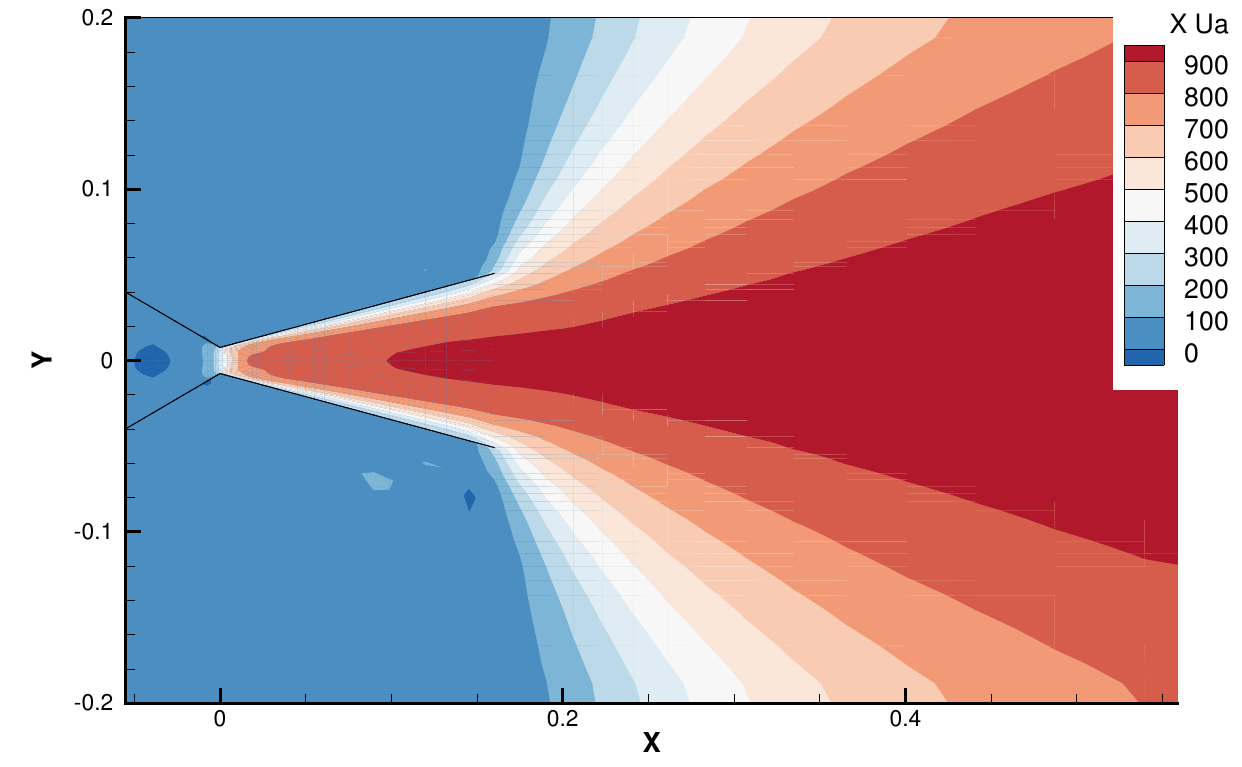}}  \\
	\subfloat[]{\includegraphics[width=0.4\textwidth]
		{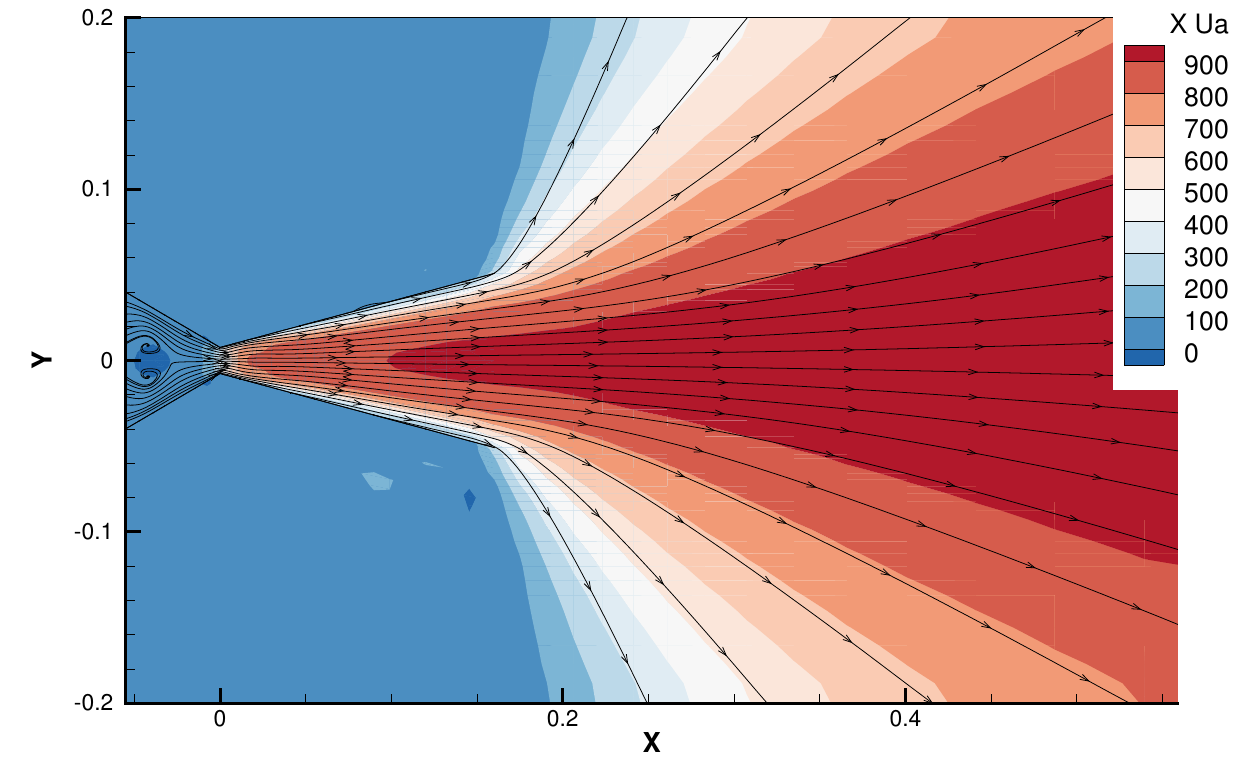}} \\
	\caption{Nozzle plume expanding to vacuum by the UGKWP method. (a) Temperature, (b) $x$-component velocity contours, and (c) streamlines.}
	\label{fig:nozzleplume-contour}
\end{figure}

Figure~\ref{fig:nozzleplume-contour} shows the temperature, $x$-component velocity distribution, and streamlines on the symmetry plane of the computational domain. In Fig.~\ref{fig:nozzleplume-contour}(a), the temperature field shows that due to the influence of high-temperature gas at the inlet and contraction effect at the nozzle throat, the inlet section temperature is significantly higher than other regions, reaching up to a maximum value of 1100K. The temperature near the solid wall of the nozzle is affected by the isothermal wall condition and is around 500 K; In the expansion section of the nozzle, the expansion effect causes the temperature to drop rapidly, reaching below 50 K far from the exit plane. The back-flow region is less affected, with a temperature of around 300 K. In Fig.~\ref{fig:nozzleplume-contour}(b), a sharp increase in velocity can be observed in the exhaust expansion section, and the velocity continues to rise rapidly after leaving the exit plane. In the back-flow region, the velocity is generally below 50 m/s, indicating that the back-flow is not intense. Streamlines in Fig.~\ref{fig:nozzleplume-contour}(c) demonstrate the vortex structure within the inlet section, as well as the expansion process after leaving the exit plane.

\begin{figure}[H]
	\centering
	\subfloat[t=0.000100025 temperature contour]{\includegraphics[width=0.4\textwidth]
		{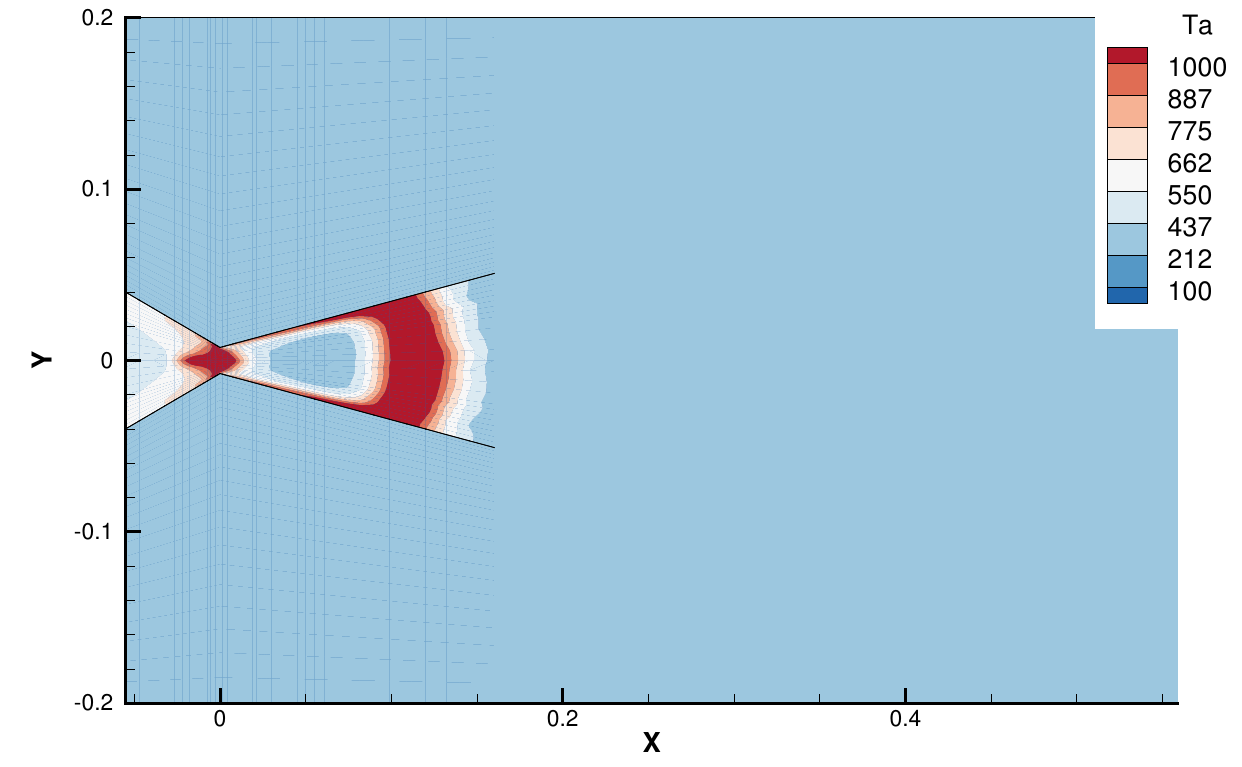}}
	\subfloat[t=0.000199976 temperature contour]{\includegraphics[width=0.4\textwidth]
		{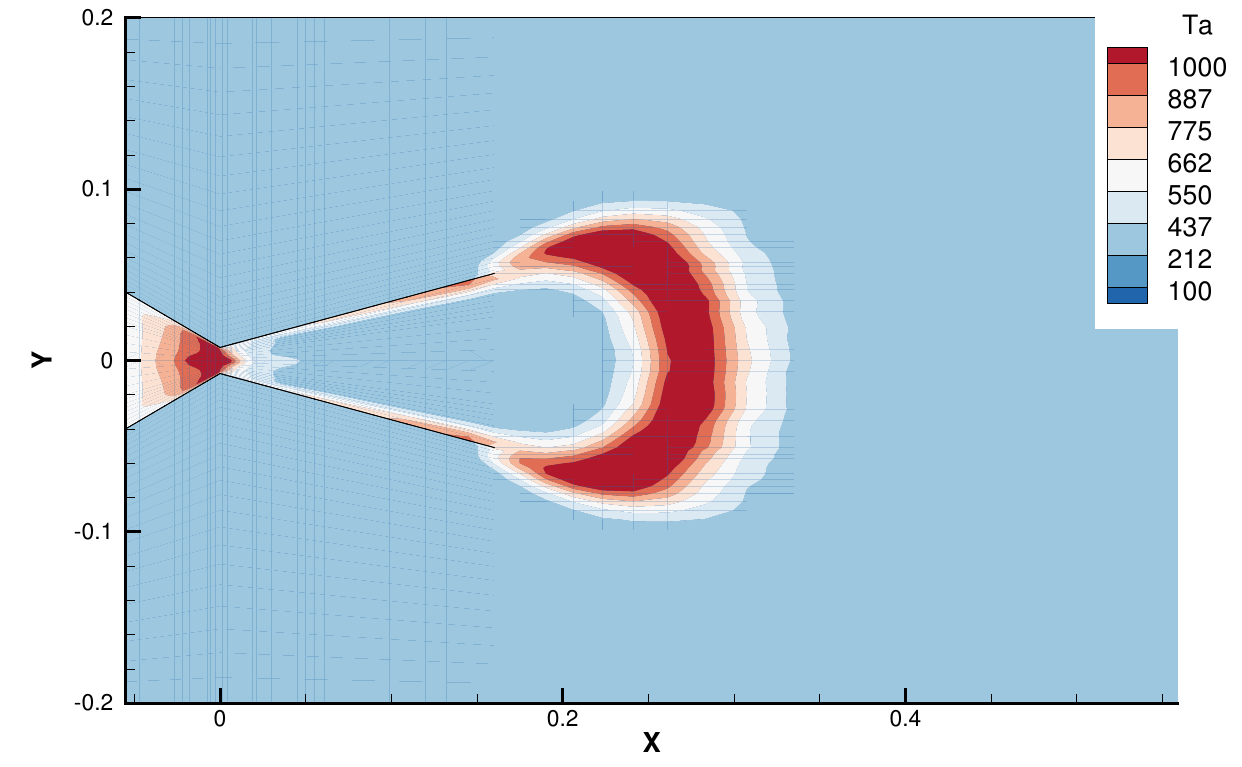}}  \\
	\subfloat[t=0.000100025 ${\rm Kn}_{Gll}$ distribution line]{\includegraphics[width=0.4\textwidth]
		{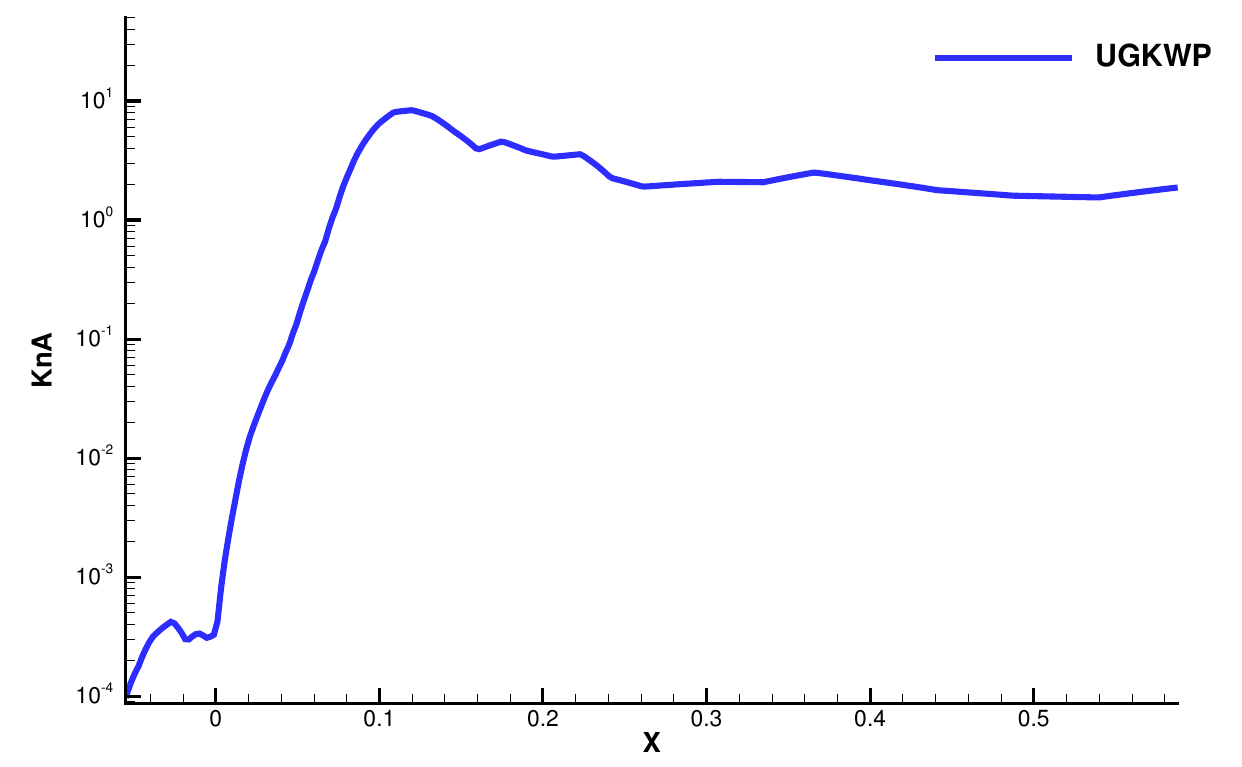}}
	\subfloat[t=0.000199976 ${\rm Kn}_{Gll}$ distribution line]{\includegraphics[width=0.4\textwidth]
		{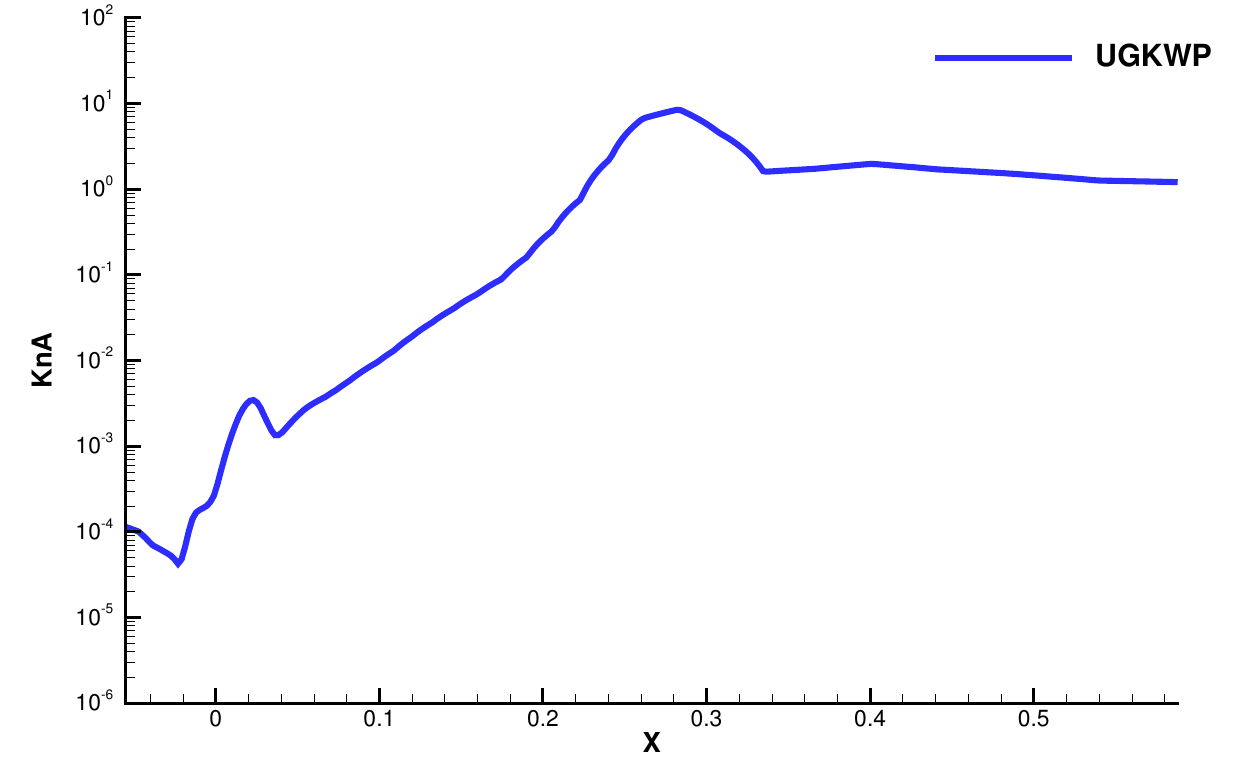}}  \\
	\subfloat[t=0.000299987 temperature contour]{\includegraphics[width=0.4\textwidth]
		{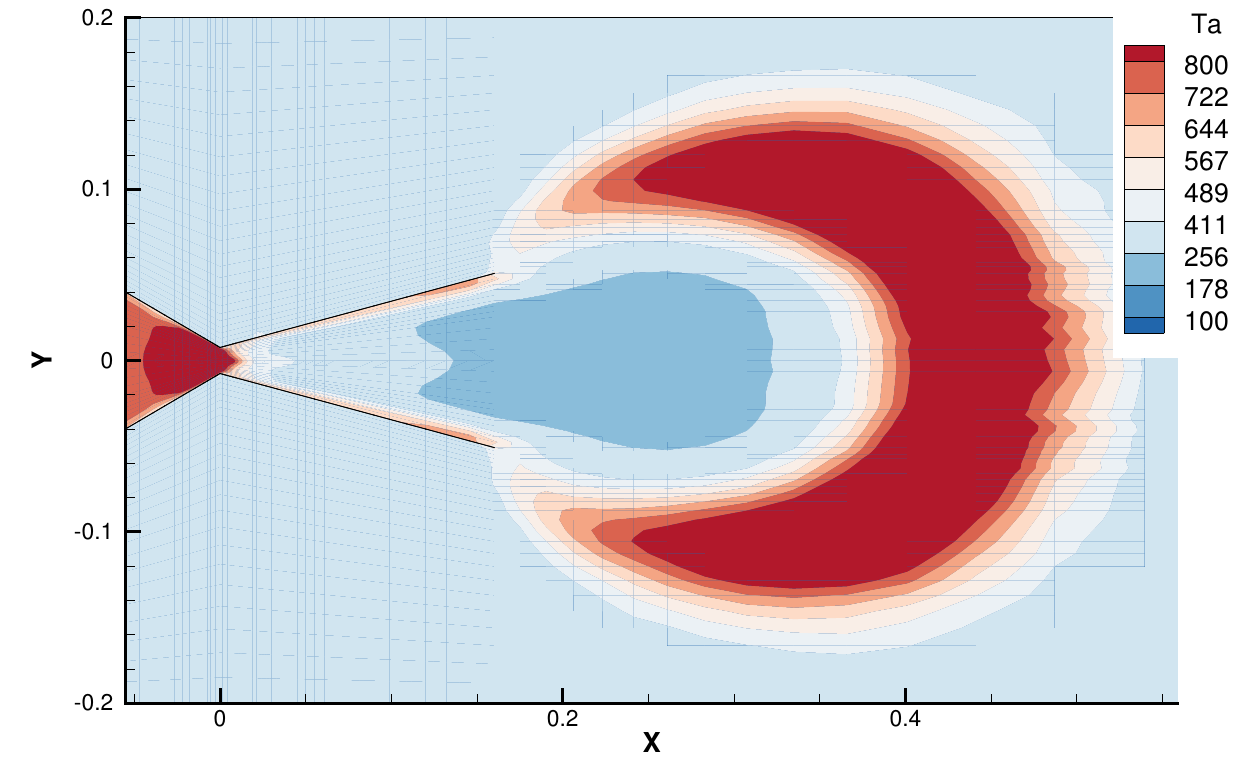}}
	\subfloat[t=0.00295003 temperature contour]{\includegraphics[width=0.4\textwidth]
		{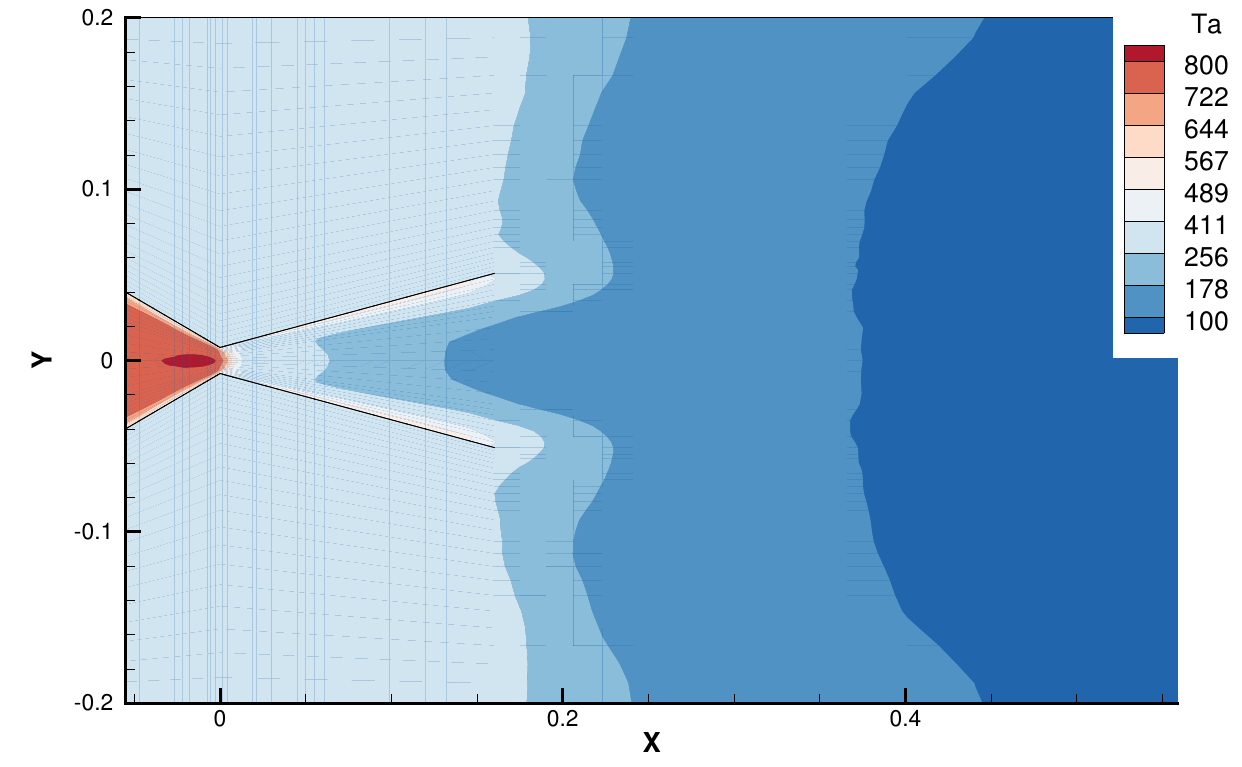}}  \\
	\subfloat[t=0.000299987 ${\rm Kn}_{Gll}$ distribution line]{\includegraphics[width=0.4\textwidth]
		{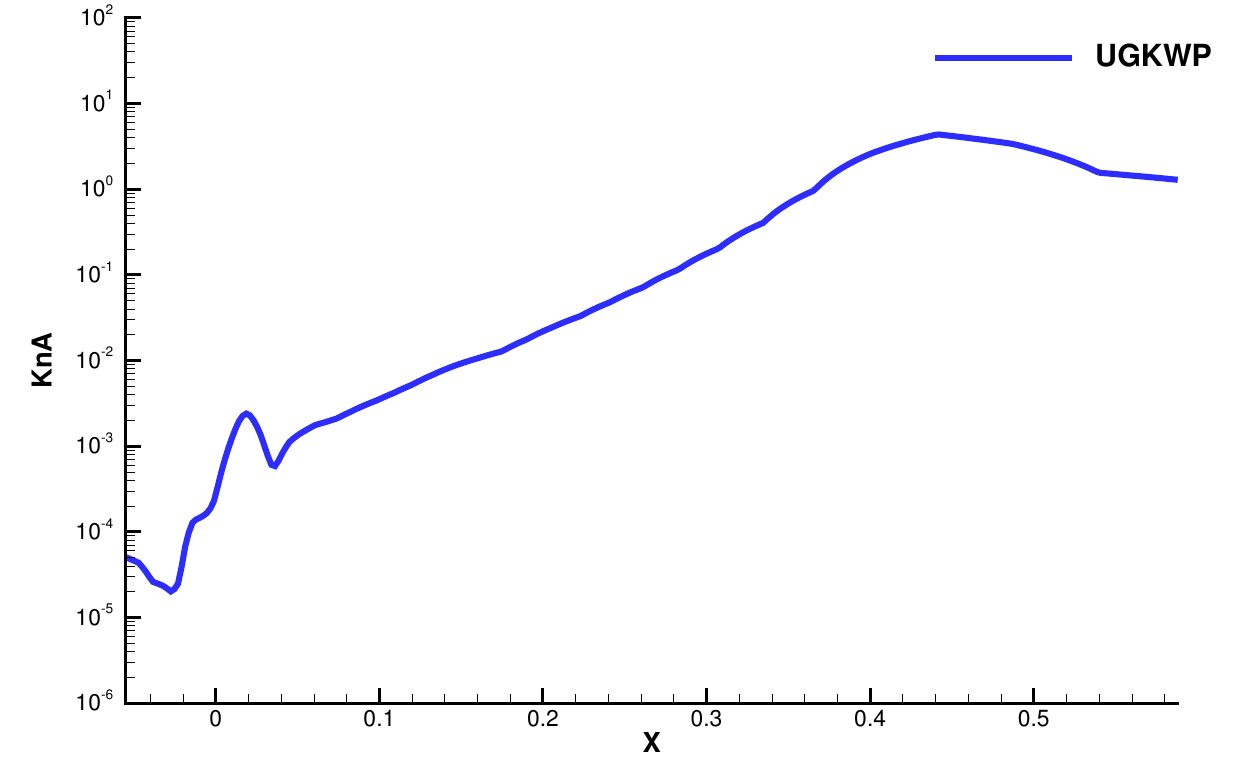}}
	\subfloat[t=0.00295003 ${\rm Kn}_{Gll}$ distribution line]{\includegraphics[width=0.4\textwidth]
		{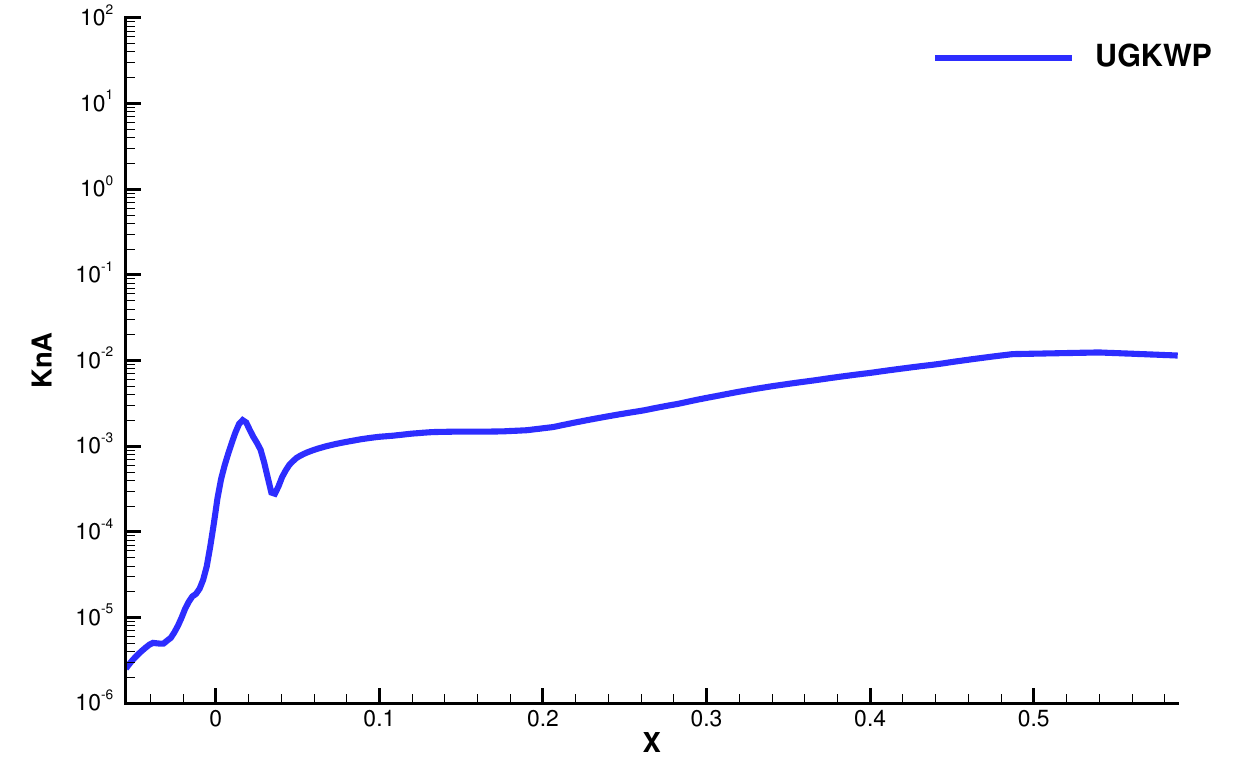}}  \\		
	\caption{Distribution of gradient-length local Knudsen number ${\rm Kn}_{Gll}$ and temperature on the symmetry plane of nozzle by the UGKWP method. }
	\label{fig:nozzleplume-Knline}
\end{figure}

Figure~\ref{fig:nozzleplume-Knline} shows the unsteady process of nozzle plume expansion. The contours and corresponding line plots show the temperature distribution and the gradient-length-dependent local Knudsen number ${\rm Kn}_{Gll}$ along the central axis at different simulation times. As shown in Fig.~\ref{fig:nozzleplume-Knline}, ${\rm Kn}_{Gll}$ exhibits a variation across seven orders of magnitude., indicating the existence of multi-scale flow throughout the entire transport process.

Table~\ref{table:nozzletime} presents the computational cost of the UGKWP method in the current study. 
Simulation is taken on the SUGON computation platform and the CPU model used is 7285H 32C 2.5GHz.

\begin{table}[H]
	\caption{The computational cost for the simulation of nozzle plume expansion into vacuum by the UGKWP method. The physical domain consists of 83776 cells, and the reference number of particles per cell in the UGKWP method is set as $N_r = 200$.}
	\centering
	\begin{threeparttable}
		\begin{tabular}{cccc}
			\toprule
			Computation Steps & Wall Clock Time, h  & Cores & Estimated Memory, GiB  \\
			\midrule
			$6000+15000\tnote{1}$ & 2.94  & 128  & 0.92  \\
			\bottomrule
		\end{tabular}
		\begin{tablenotes}
			\item[1] Steps of time-averaging process in the UGKWP simulation.
		\end{tablenotes}
	\end{threeparttable}
	\label{table:nozzletime}
\end{table}

\subsection{Side-jet impingement on hypersonic cone flow}

\begin{figure}[H]
	\centering
	\includegraphics[width=0.7\textwidth]
	{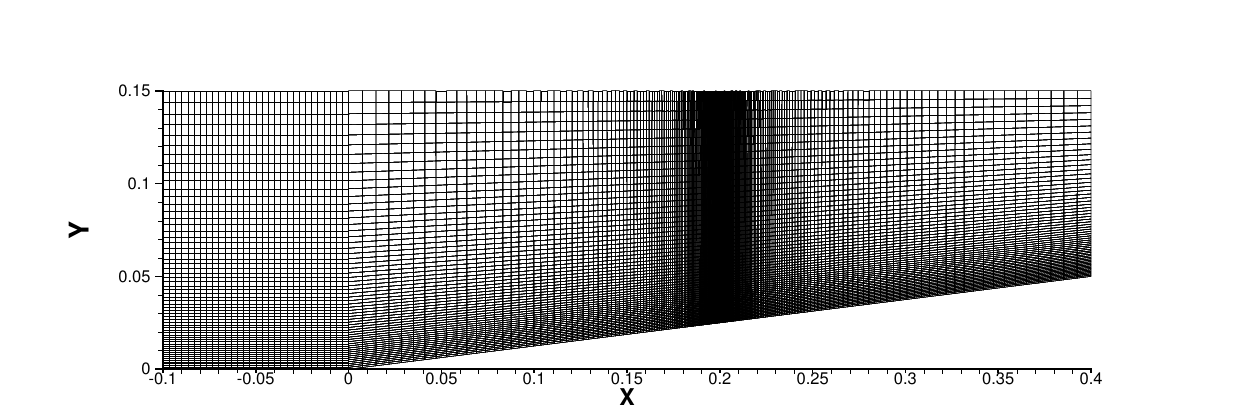}
	\caption{Physical domain of the side-jet interaction with hypersonic cone flow. }
	\label{fig:sidejet-mesh}
\end{figure}

The complex interaction between the jet flow and shock waves generates unique flow structures. In this section, we focus on the flow structure of the interactions between jet flow and hypersonic flow over a 7° cone simulated by the UGKWP method.  Referring to the configuration of previous study\cite{karpuzcu2022jet,karpuzcu2023study}, a two-dimensional mesh is generated as shown in Fig.~\ref{fig:sidejet-mesh}. The cone's base diameter is  0.1 m, while its height reaches 0.4 m. Positioned at a height of 0.2 m, the side jet nozzle has a diameter of 1 mm. The simulation employs a total of 13980 cells, with the reference number of particles per cell in the UGKWP method set as $N_r = 200$.

The simulation parameters follow the setup in Karpuzcu's work\cite{karpuzcu2023study} and are presented in Tab.~\ref{table:sidejet-parameter}.
\begin{table}[H]
	\caption{Freestream flow and jet flow parameters. The gas used for the simulation is nitrogen. }
	\centering
	\begin{tabular}{ccc}
		\toprule
		Flow Region & Free Stream & Side Jet \\
		\midrule
		Molecule Number Density $(m^{-3})$ & $0.5\times 10^{22}$ & $8.25\times 10^{23}$ \\
		Mach Number & $12.0$ & $1.0$ \\
		Temperature (K) & $50.0$ & $250.0$ \\
		Pressure (Pa) & $3.45$ & $2847.4$ \\
		Knudsen Number & $0.15$ & $0.0013$ \\
		\bottomrule
	\end{tabular}
	\label{table:sidejet-parameter}
\end{table}

\begin{figure}[H]
	\centering
	\subfloat[]{\includegraphics[width=0.4\textwidth]
		{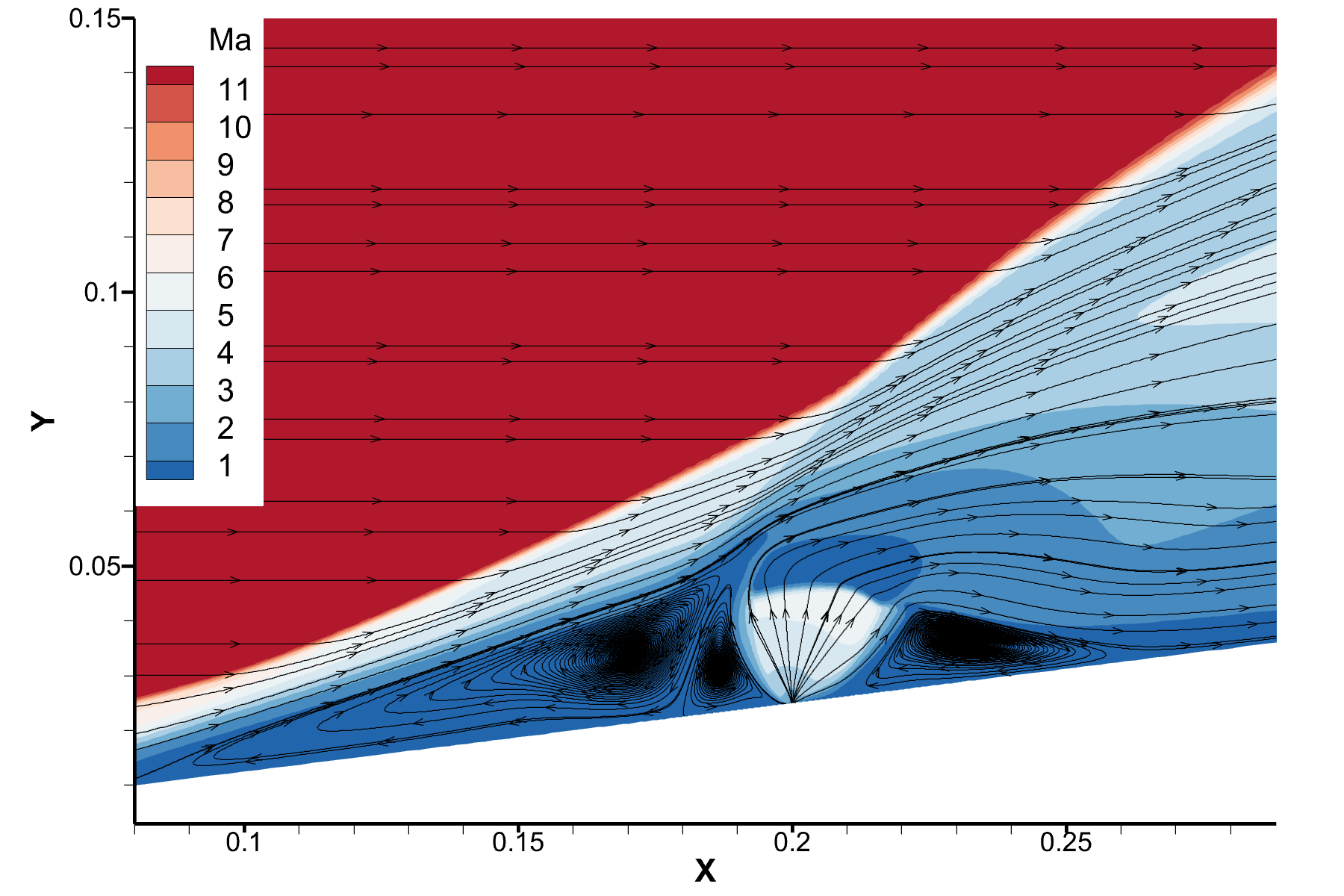}}
	\subfloat[]{\includegraphics[width=0.4\textwidth]
		{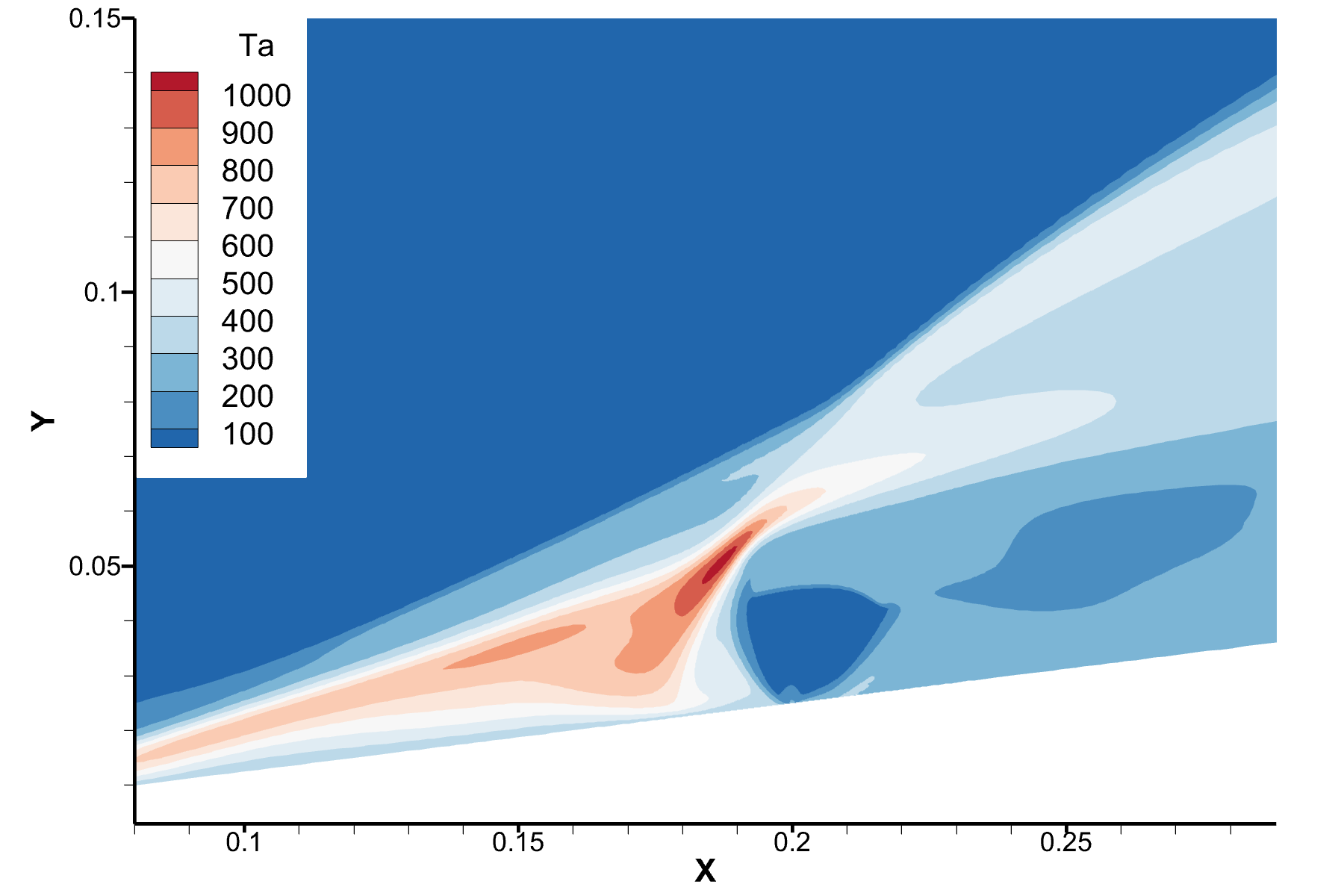}} \\
	\subfloat[]{\includegraphics[width=0.4\textwidth]
		{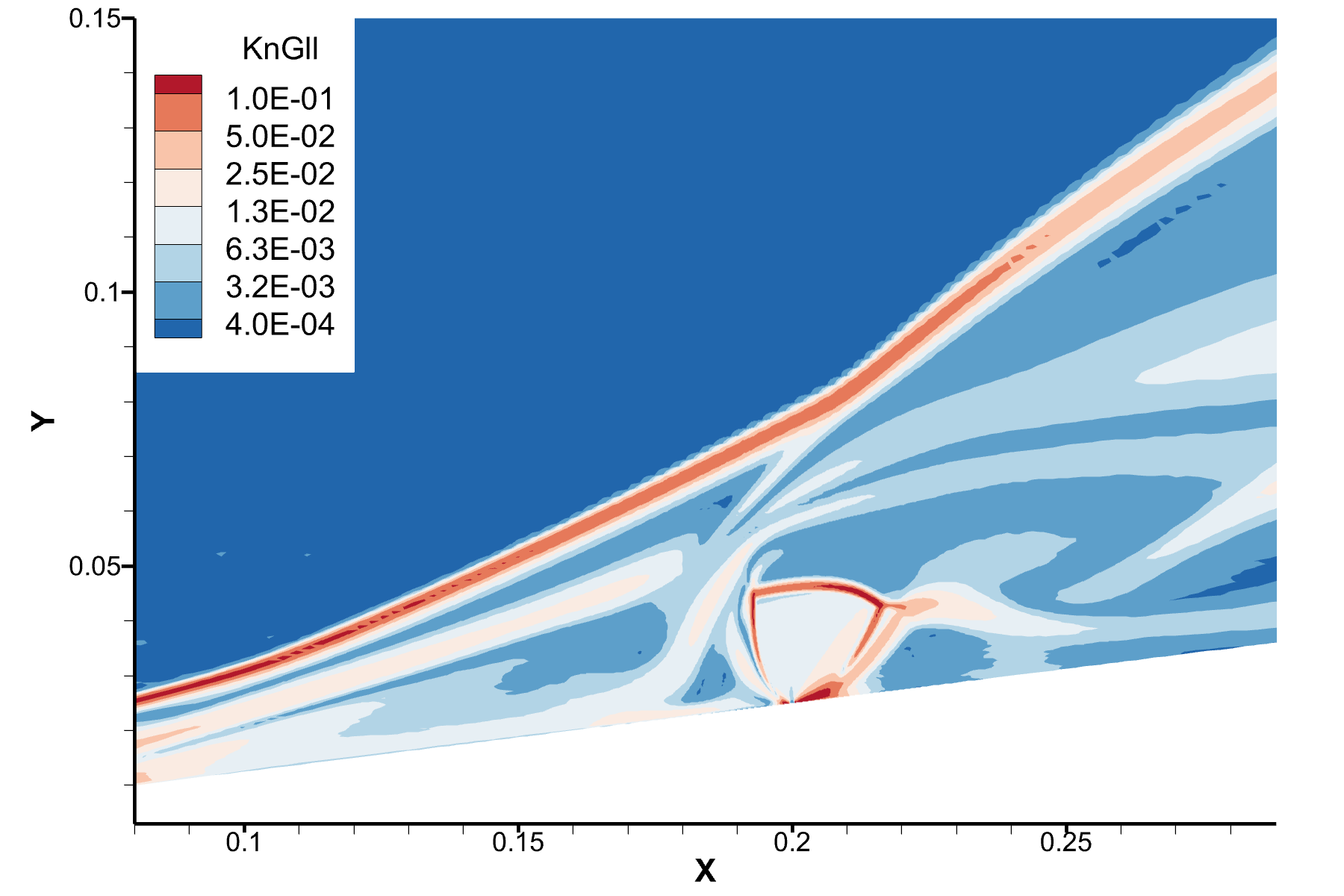}} \\
	\caption{Flow structures of Hypersonic freestream flow at ${\rm Ma}_\infty=12.0$ and ${\rm Kn}_\infty=0.15$ interacting with a side-jet at ${\rm Ma}_\infty=1.0$ and ${\rm Kn}_\infty=0.0013$ by the UGKWP method. (a) Mach Number with streamlines, (b) temperature, and (c) local Knudsen number contours. }
	\label{fig:sidejet-contour}
\end{figure}

Figure~\ref{fig:sidejet-contour} illustrates the streamline, Mach number, temperature, and local Knudsen number distribution ${\rm Kn}_{Gll}$. Two upstream recirculation zones and one downstream recirculation zone can be identified from the streamline plot. The expansion of the jet flow is restricted to a small area by a strong leading edge shock, as the streamline from the nozzle changes the direction sharply. The gradient-length-dependent local Knudsen number distribution shows regions with shock and strong shear layers and their interaction.

\begin{figure}[H]
	\centering
	\subfloat[]{\includegraphics[width=0.7\textwidth]
		{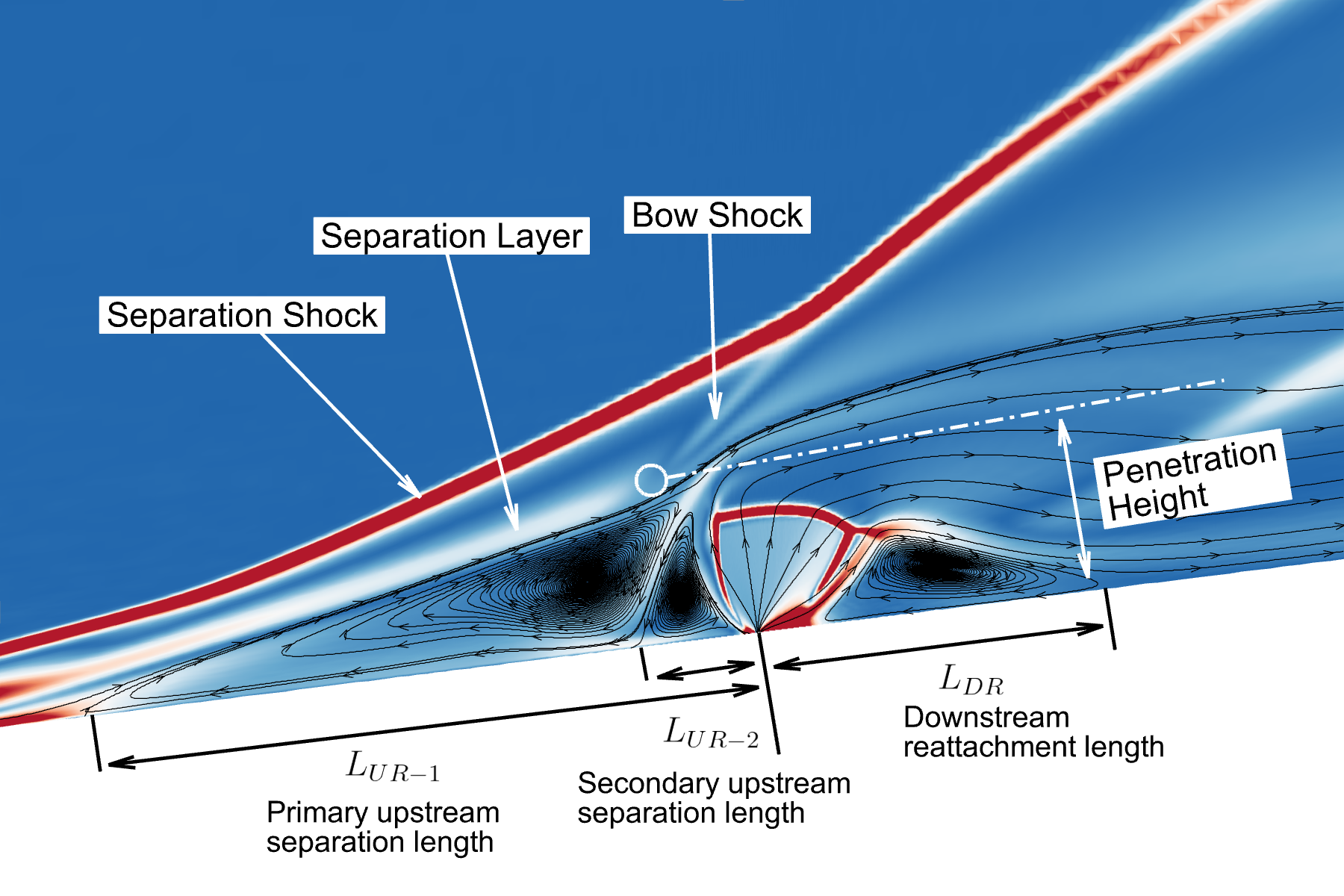}}\\
	\caption{A schematic diagram depicting several main flow structures of the hypersonic flow-jet interaction. }
	\label{fig:sidejet-shiyitu}
\end{figure}

A schematic diagram (see Fig.~\ref{fig:sidejet-shiyitu}) is drawn on a continuous contour of ${\rm Kn}_{Gll}$ to further describe the complex flow structure. Due to the presence of the jet, a separation shock forms and causes a strong separation layer, together with two recirculation zones. The jet expansion interacts with the separation shock, thus creating a strong bow shock. The hypersonic flow forces jet flow to reattach on the solid surface, creating the downstream recirculation zone.

A detailed comparison of flow structures is conducted. Specifically, the penetration height of the side jet, the length of primary upstream separation, secondary upstream separation, and downstream reattachment zones, and the angle between the bow shock, separation shear layer, and the cone surface respectively are compared as characteristic features.
All flow structures are illustrated in Fig.\ref{fig:sidejet-shiyitu}. Table~\ref{table:sidejet-properties} shows the comparison of different flow characteristic parameters among the literature data, DSMC solution \cite{karpuzcu2023study}, and UGKWP simulation. The errors of the UGKWP method for each parameter are close to those of the DSMC method, and the overall errors compared to the literature data are less than 10$\%$.
\begin{table}[H]
	\caption{Comparison of the parameters in hypersonic flow-jet interaction by the UGKWP method with literature and DSMC data.}
	\centering
	\begin{tabular}{ccccccc}
		\toprule
		Source & $h/r_b$ & $L_{UR-1}/h$ & $L_{UR-2}/h$ & $L_{DR}/h$ & $\theta_{BS}/^{\circ}$ & $\theta_{SL}/^{\circ}$  \\
		\midrule
		Literature & 0.523 & $-$ & $-$ & $-$ & 13  & 33 \\
		
		DSMC & $0.5$ & $3.32$ & $0.55$ & $1.81$ & $11$ & $33$  \\
		Error(With Literature) & $-4.40\%$ & $-$ & $-$ & $-$ & $-15.4\%$ & $0\%$  \\
		UGKWP & $0.564$ & $3.26$ & $0.658$ & $1.71$ & $13.6$ & $33.6$  \\
		Error(With Literature/DSMC) & $7.84\%$ & $-1.80\%$ & $-19.63\% $ & $5.52\%$ & $4.61\%$ & $1.82\%$  \\
		\bottomrule
	\end{tabular}
	\label{table:sidejet-properties}
\end{table}

Table~\ref{table:sidejet-time} resents the computational cost using the UGKWP method. 
Simulation is taken on the SUGON computation platform with the CPU model 7285H 32C 2.5GHz.

\begin{table}[H]
	\caption{The computational cost for the simulation of side-jet interaction with hypersonic cone flow by the UGKWP method. The physical domain consists of 13980 cells, and the reference number of particles per cell in the UGKWP method is set as $N_r = 200$.}
	\centering
	\begin{threeparttable}
		\begin{tabular}{cccc}
			\toprule
			Computation Steps & Wall Clock Time, h  & Cores & Estimated Memory, GiB  \\
			\midrule
			$10000+20000\tnote{1}$ & 0.53  & 128  & 0.189  \\
			\bottomrule
		\end{tabular}
		
		\begin{tablenotes}
			\item[1] Steps of time-averaging process in the UGKWP simulation.
		\end{tablenotes}
	\end{threeparttable}
	\label{table:sidejet-time}
\end{table}

\subsection{Interaction of hypersonic flow with an annular jet over a three-dimensional cone}

\begin{figure}[H]
	\centering
	\includegraphics[width=0.45\textwidth]
	{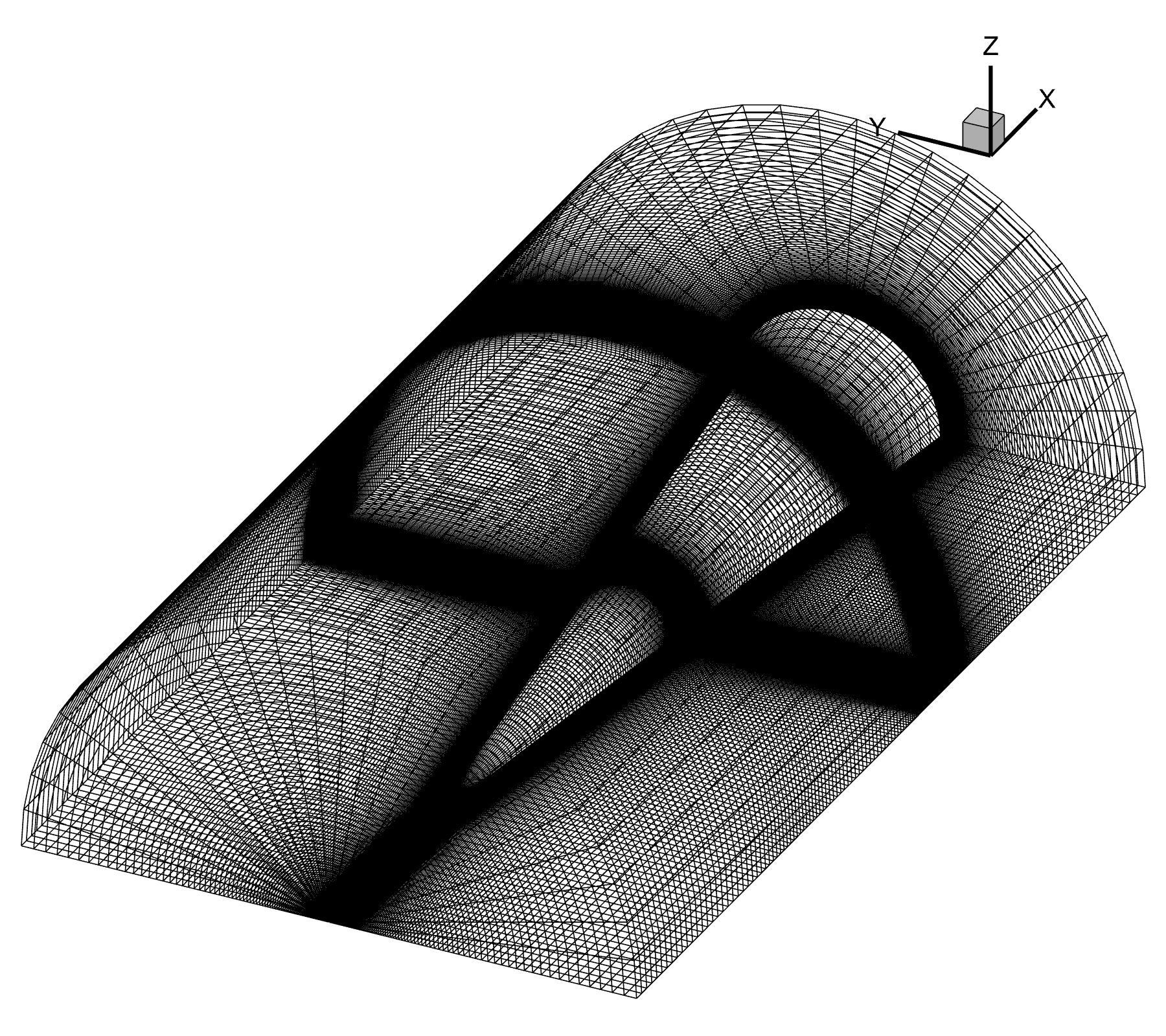}
	\caption{Three-dimensional mesh over a cone for the simulation of the interaction between hypersonic flow and annular jet.}
	\label{fig:3dsidejet-mesh}
\end{figure}

In this section, hypersonic flow passing over a three-dimensional cone interacting with a ring of transverse jets is simulated. 
The geometric model used is the same as that in the previous section, with the cone's base diameter of 0.1 meters and its height of 0.4 meters. 
An annular jet with a width of 1 millimeter is positioned at a height of 0.2 meters. The UGKWP method is used in the simulation. 
The mesh with 612500 cells around the cone is shown in Fig.~\ref{fig:3dsidejet-mesh}. 
The computation is performed on a symmetric half-cone domain, with a maximum particle capacity of 200 per cell.

\begin{figure}[H]
	\centering
	\subfloat[]{\includegraphics[width=0.4\textwidth]
		{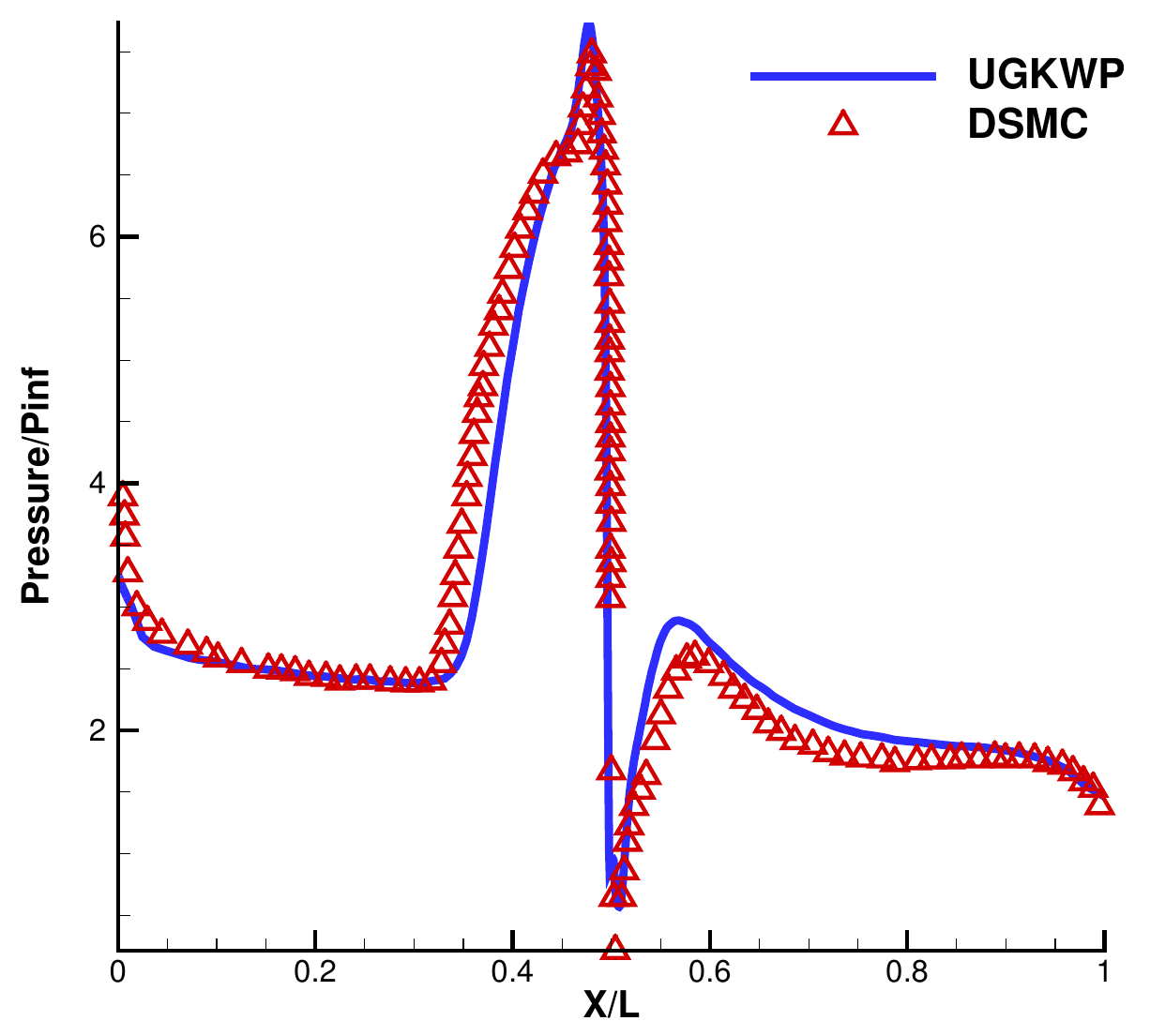}}
	\subfloat[]{\includegraphics[width=0.4\textwidth]
		{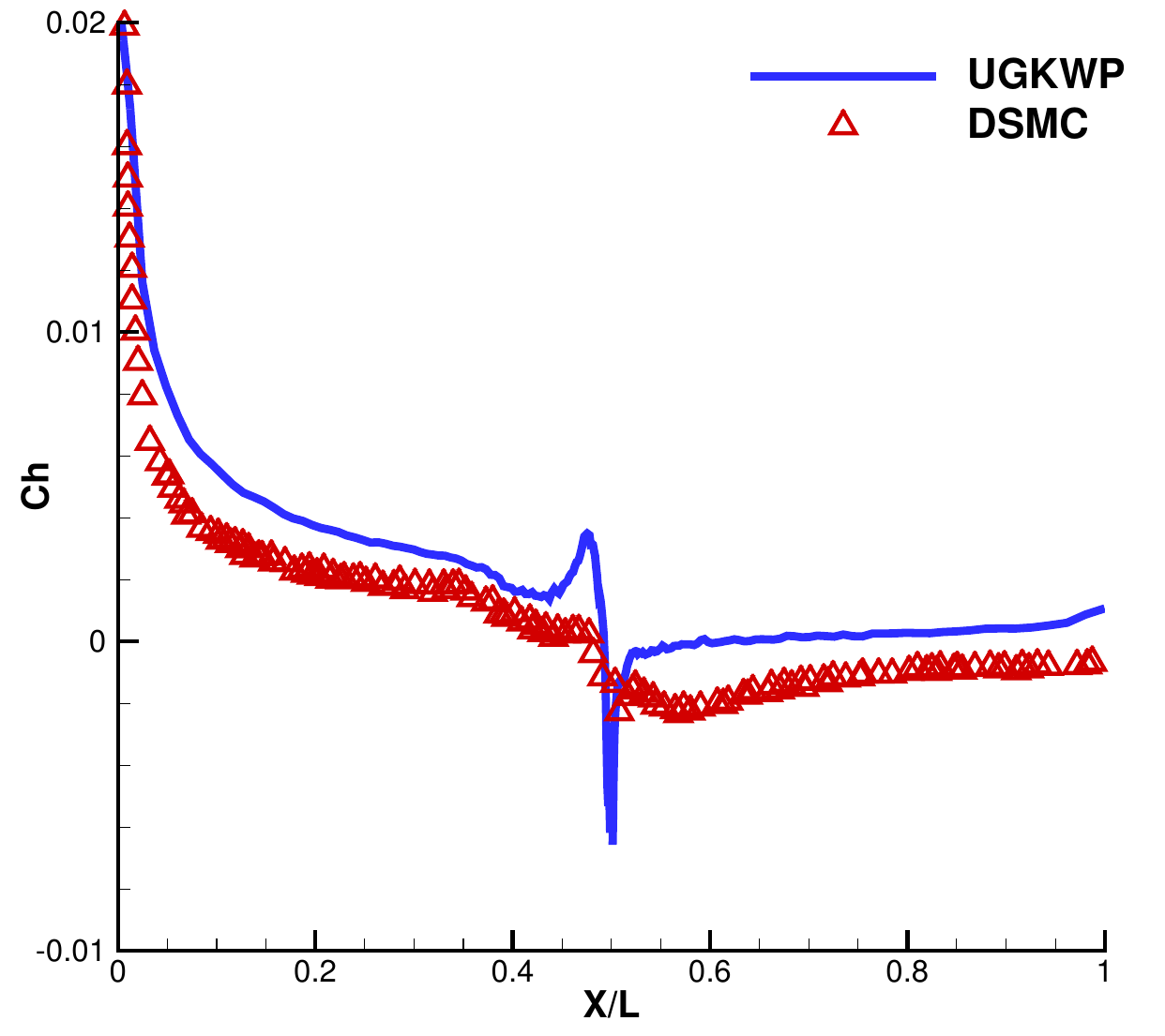}} \\
	\caption{Surface quantities of the three-dimensional cone by the UGKWP method. (a) Surface normalized pressure, and (b) surface heat transfer coefficient. }
	\label{fig:3dsidejet-surface}
\end{figure}

The simulation parameters follow the setup in Karpuzcu's work\cite{karpuzcu2023study} as presented in Tab.~\ref{table:3dsidejet-parameter}.
\begin{table}[H]
	\caption{Free stream flow and jet flow parameters. The gas used for the simulation is nitrogen.}
	\centering
	\begin{tabular}{ccc}
		\toprule
		Flow Region & Free Stream & Side Jet \\
		\midrule
		Molecule Number Density $(m^{-3})$ & $1.0\times 10^{22}$ & $1.65\times 10^{23}$ \\
		Mach Number & $6.0$ & $1.0$ \\
		Temperature (K) & $50.0$ & $250.0$ \\
		Pressure (Pa) & $6.9$ & $571.2$ \\
		Knudsen Number & $0.07$ & $0.0067$ \\
		\bottomrule
	\end{tabular}
	\label{table:3dsidejet-parameter}
\end{table}
The surface normalized pressure $p_{n}$ and heat transfer coefficient $C_h$ are calculated and shown in Fig.~\ref{fig:3dsidejet-surface} together with comparison with the DSMC
data \cite{karpuzcu2023study}. Good agreements have been observed. The normalization is taken below
\begin{equation*}
	p_{n}=\dfrac{p_s}{p_{\infty}},
	C_h=\dfrac{h_s}{\frac{1}{2} \rho_{\infty}C_{\infty}^{3}},
\end{equation*}
where $C_{\infty} = \sqrt{2RT_{\infty}}$, $T_{\infty},\rho_{\infty}$ are temperature and density of free stream flow.

\begin{figure}[H]
	\centering
	\subfloat[]{\includegraphics[width=0.4\textwidth]
		{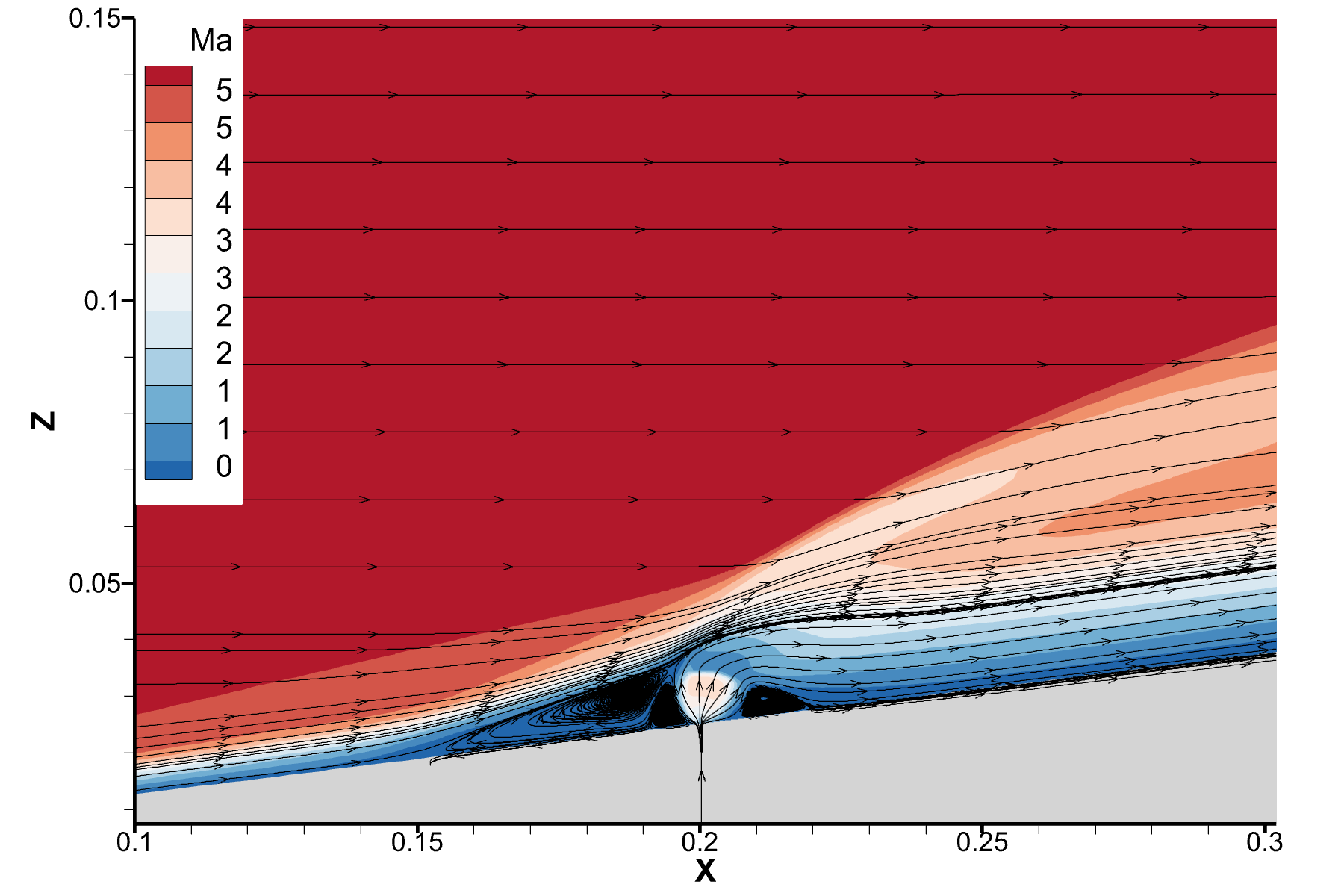}}
	\subfloat[]{\includegraphics[width=0.4\textwidth]
		{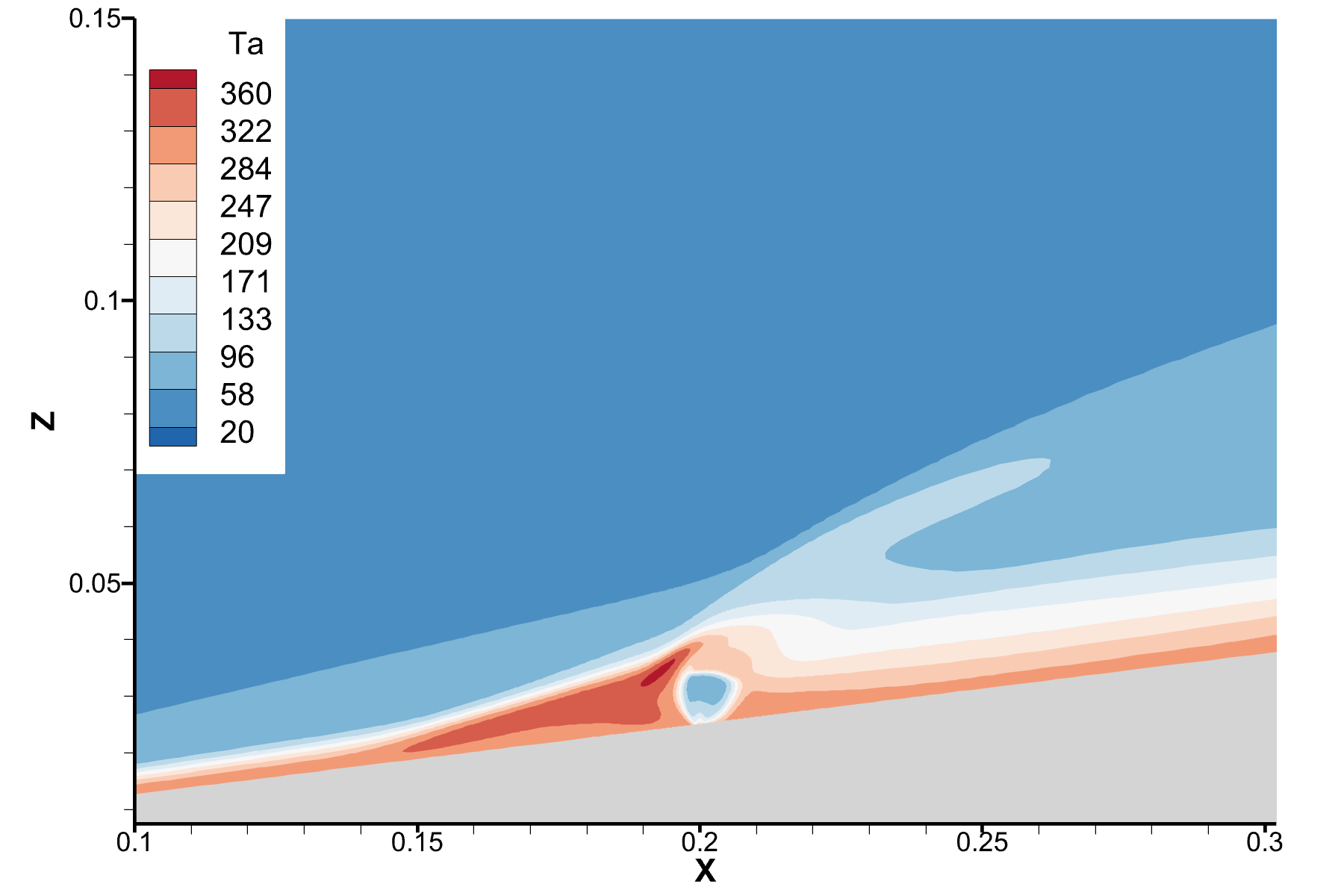}} \\
	\subfloat[]{\includegraphics[width=0.4\textwidth]
		{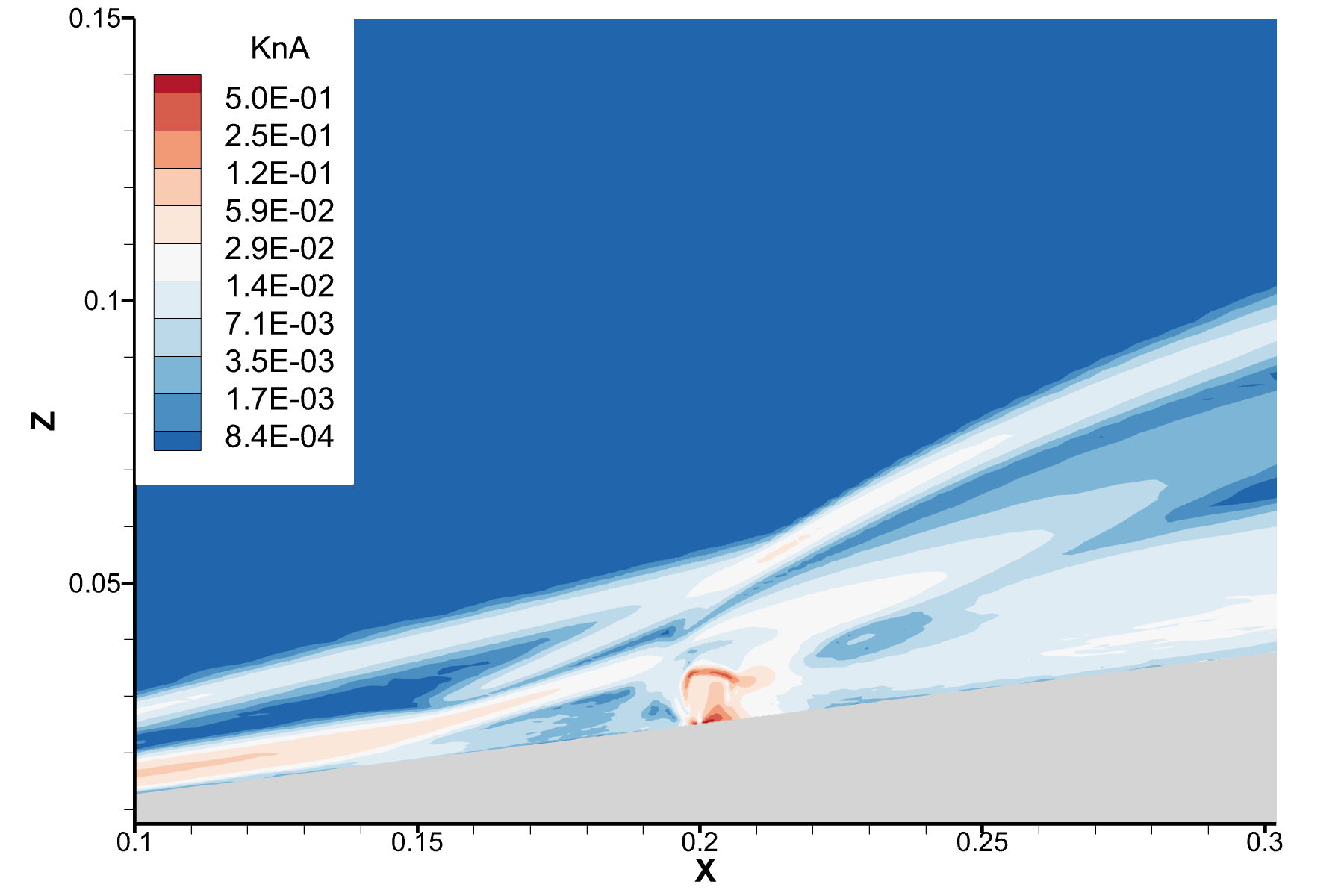}} \\
	\caption{Flow structures of Hypersonic freestream flow at ${\rm Ma}_\infty=6.0$ and ${\rm Kn}_\infty=0.07$ interacting with an annular jet at ${\rm Ma}_\infty=1.0$ and ${\rm Kn}_\infty=0.0067$ by the UGKWP method. (a) Mach Number with streamlines, (b) temperature, and (c) local Knudsen number contours. }
	\label{fig:3dsidejet-contour}
\end{figure}

Figure~\ref{fig:3dsidejet-contour} illustrates the streamline, Mach number, temperature, and local Knudsen number distribution ${\rm Kn}_{Gll}$, all of which demonstrate a flow structure similar to the two-dimensional case. The recirculation zones and expansion area of the jet are much reduced due to the less intensified jet flow, the three-dimensional effect, and a low Mach number of the free stream flow.

Table~\ref{table:3dsidejettime} presents the computational cost in the simulation which is taken on the SUGON computation platform with a CPU model of 7285H 32C 2.5GHz.

\begin{table}[H]
	\caption{The computational cost for the simulation of hypersonic flow over a three-dimensional cone with an annular jet by the UGKWP method. The physical domain consists of 612500 cells, and the reference number of particles per cell in the UGKWP method is set as $N_r = 200$.}
	\centering
	\begin{threeparttable}
		\begin{tabular}{cccc}
			\toprule
			Computation Steps & Wall Clock Time, h  & Cores & Estimated Memory, GiB  \\
			\midrule
			$28000+7000\tnote{1}$ & 4.73  & 640  & 8.29  \\
			\bottomrule
		\end{tabular}
		\begin{tablenotes}
			\item[1] Steps of time-averaging process in the UGKWP simulation.
		\end{tablenotes}
	\end{threeparttable}
	\label{table:3dsidejettime}
\end{table}

\section{Conclusion}\label{sec:conclusion}

In this study, we utilize the UGKWP method to examine several non-equilibrium flow issues. The adaptive wave-particle decomposition in UGKWP significantly aids in efficiently capturing multiscale solutions in all flow regimes. The computational aerodynamic and aerothermodynamic distributions of supersonic and hypersonic external flows over a sphere and a space vehicle agree well with experimental and DSMC data.
The UGKWP method is able to capture non-equilibrium transport in all flow regimes and is exemplified in the simulation of nozzle plume expansion into a background vacuum, where solutions ranging from the continuum Navier--Stokes to the free molecular flow are accurately obtained.
Complex flow structures resulting from the interaction of hypersonic flow with side-jet are accurately reproduced in the UGKWP simulation. For 3D simulations involving complex geometries, like the X38 vehicle with 560,593 cells in the computational domain, the UGKWP method consumes only 60GiB of memory and takes less than 15 hours to get the converged solutions. These calculations can be performed on personal workstations.
In conclusion, all the test cases in this study demonstrate the significant potential of the UGKWP method for engineering applications involving non-equilibrium multiscale transport.

\section*{Author's contributions}

All authors contributed equally to this work.

\section*{Acknowledgments}
This work was supported by National Key R$\&$D program of China 2022YFA1004500, National Natural Science Foundation of China (Grant Nos. 12172316),
and Hong Kong research grant council (16208021,16301222). 

\section*{Data Availability}

The data that support the findings of this study are available from the corresponding author upon reasonable request.

	




\bibliographystyle{elsarticle-num}
\bibliography{wp.bib}

\begin{thebibliography}{10}
\expandafter\ifx\csname url\endcsname\relax
  \def\url#1{\texttt{#1}}\fi
\expandafter\ifx\csname urlprefix\endcsname\relax\def\urlprefix{URL }\fi
\expandafter\ifx\csname href\endcsname\relax
  \def\href#1#2{#2} \def\path#1{#1}\fi

\bibitem{broadwell1964study}
J.~E. Broadwell, Study of rarefied shear flow by the discrete velocity method,
  Journal of Fluid Mechanics 19~(3) (1964) 401--414.

\bibitem{lizhihui20140302}
L.~Zhihui, J.~Xinyu, W.~Junlin, P.~Aoping, Gas-kinetic unified algorithm for
  {Boltzmann model equation} in rotational nonequilibrium and its application
  to the whole range flow regimes, Chinese Journal of Theoretical and Applied
  Mechanics 46~(3) (2014) 336--351.

\bibitem{zhang2023conservative}
R.~Zhang, S.~Liu, J.~Chen, C.~Zhong, C.~Zhuo, A conservative implicit scheme
  for three-dimensional steady flows of diatomic gases in all flow regimes
  using unstructured meshes in the physical and velocity spaces, arXiv preprint
  arXiv:2303.10846 (2023).

\bibitem{jiangdw}
J.~Dingwu, Study of the {Gas Kinetic Scheme Based on the Analytic Solution of
  Model Equations}, Phd thesis, China Aerodynamics Research and Development
  Center, Mianyang (May 2016).

\bibitem{michael2004scaling}
N.~Michael, Scaling parameters in rarefied flow and the breakdown of the
  {Navier-Stokes} equations (2004).

\bibitem{Padilla}
J.~Padilla, I.~Boyd, Assessment of rarefied hypersonic aerodynamics modeling
  and windtunnel data, in: 9th AIAA/ASME Joint Thermophysics and Heat Transfer
  Conference, 2006, p. 3390.

\bibitem{loth2008compressibility}
E.~Loth, Compressibility and rarefaction effects on drag of a spherical
  particle, AIAA journal 46~(9) (2008) 2219--2228.

\bibitem{loth2021supersonic}
E.~Loth, J.~Tyler~Daspit, M.~Jeong, T.~Nagata, T.~Nonomura, Supersonic and
  hypersonic drag coefficients for a sphere, AIAA journal 59~(8) (2021)
  3261--3274.

\bibitem{bird1963approach}
G.~Bird, Approach to translational equilibrium in a rigid sphere gas, Phys.
  fluids 6 (1963) 1518--1519.

\bibitem{rault1994aerodymics}
D.~F. Rault, Aerodymics of the {Shuttle Orbiter} at high altitudes, Journal of
  Spacecraft and Rockets 31~(6) (1994) 944--952.

\bibitem{LEBEAU2001595}
G.~LeBeau, F.~Lumpkin~Iii, Application highlights of the {DSMC Analysis Code
  (DAC)} software for simulating rarefied flows, Computer Methods in Applied
  Mechanics and Engineering 191~(6-7) (2001) 595--609.

\bibitem{justiz1994dsmc}
C.~R. Justiz, R.~M. Sega, C.~Dalton, A.~Ignatiev, {DSMC-and BGK-based}
  calculations for return flux contamination of an outgassing spacecraft,
  Journal of thermophysics and heat transfer 8~(4) (1994) 802--803.

\bibitem{Gimelshein2002}
S.~Gimelshein, N.~Gimelsheint, D.~Levin, M.~Ivmov, G.~Markelov, Modeling of
  rarefied hypersonic flows over spacecraft in martian atmosphere using the
  dsmc method, in: 8th AIAA/ASME Joint Thermophysics and Heat Transfer
  Conference 2002, 2002.

\bibitem{zhang2010spacecraft}
W.-p. Zhang, B.~Han, C.-y. Zhang, Spacecraft aerodynamics and trajectory
  simulation during aerobraking, Applied Mathematics and Mechanics 31~(9)
  (2010) 1063--1072.

\bibitem{ivanovdsmc}
M.~Ivanov, P.~Vashchenkov, G.~Markelov, A.~Kashkovsky, A.~Krylov, {DSMC Study
  of the Near-Continuum Flow Near the Nose Part of the
  Spacecraft``Progress-M''}, in: 37th AIAA Thermophysics Conference, p. 2688.

\bibitem{ivanovich2013aerodynamic}
K.~Y. Ivanovich, Z.~Y.~M. Myint, K.~A. Yurievich, {Aerodynamic Investigation
  for Prospective Aerospace Vehicle in the Transitional Regime}, International
  Journal of Aeronautical and Space Sciences 14~(3) (2013) 215--221.

\bibitem{titarev2020comparison}
V.~A. Titarev, A.~A. Frolova, V.~Rykov, P.~Vashchenkov, A.~Shevyrin, Y.~A.
  Bondar, Comparison of the {Shakhov kinetic equation and DSMC} method as
  applied to space vehicle aerothermodynamics, Journal of Computational and
  Applied Mathematics 364 (2020) 112354.

\bibitem{zuppardi2016aerodynamic}
G.~Zuppardi, R.~Savino, G.~Russo, L.~Spano’Cuomo, E.~Petrosino, Aerodynamic
  analysis of the aerospaceplane {HyPlane} in supersonic rarefied flow, Acta
  Astronautica 123 (2016) 229--238.

\bibitem{zuppardi2014}
G.~Zuppardi, L.~Morsa, M.~Sippel, T.~Schwanekamp, Aero-thermo-dynamic analysis
  of the {SpaceLiner-7.1} vehicle in high altitude flight, in: AIP Conference
  Proceedings, Vol. 1628, American Institute of Physics, 2014, pp. 1268--1276.

\bibitem{moss2005dsmc}
J.~Moss, Dsmc simulations of ballute aerothermodynamics under hypersonic
  rarefied conditions, in: 38th AIAA Thermophysics Conference, 2005, p. 4949.

\bibitem{li2021kinetic}
J.~Li, D.~Jiang, X.~Geng, J.~Chen, Kinetic comparative study on aerodynamic
  characteristics of hypersonic reentry vehicle from near-continuous flow to
  free molecular flow, Advances in Aerodynamics 3 (2021) 1--10.

\bibitem{xu2015}
K.~Xu, Direct modeling for computational fluid dynamics: construction and
  application of unified gas-kinetic schemes, Vol.~4, World Scientific, 2014.

\bibitem{liu2020unified}
C.~Liu, Y.~Zhu, K.~Xu, Unified gas-kinetic wave-particle methods{ I: }continuum
  and rarefied gas flow, Journal of Computational Physics 401 (2020) 108977.

\bibitem{zhu2019unified}
Y.~Zhu, C.~Liu, C.~Zhong, K.~Xu, Unified gas-kinetic wave-particle {methods.
  II. Multiscale} simulation on unstructured mesh, Physics of Fluids 31~(6)
  (2019).

\bibitem{xu2001gas}
K.~Xu, A {gas-kinetic BGK scheme for the Navier--Stokes equations and its
  connection with artificial dissipation and Godunov method}, Journal of
  Computational Physics 171~(1) (2001) 289--335.

\bibitem{wei2022unified}
Y.~Wei, Y.~Zhu, K.~Xu, Unified gas-kinetic {wave-particle methods VII:
  diatomic} gas with rotational and vibrational nonequilibrium, arXiv preprint
  arXiv:2211.12922 (2022).

\bibitem{li2020unified}
W.~Li, C.~Liu, Y.~Zhu, J.~Zhang, K.~Xu, Unified gas-kinetic wave-particle
  {methods III: Multiscale} photon transport, Journal of Computational Physics
  408 (2020) 109280.

\bibitem{liu2021unified}
C.~Liu, K.~Xu, Unified gas-kinetic {wave-particle methods IV}: multi-species
  gas mixture and plasma transport, Advances in Aerodynamics 3~(1) (2021)
  1--31.

\bibitem{yang2023unified}
X.~Yang, Y.~Wei, W.~Shyy, K.~Xu, Unified gas-kinetic wave-particle method for
  three-dimensional simulation of gas-particle fluidized bed, Chemical
  Engineering Journal 453 (2023) 139541.

\bibitem{venkatakrishnan1995convergence}
V.~Venkatakrishnan, Convergence to steady state solutions of the {Euler}
  equations on unstructured grids with limiters, Journal of Computational
  Physics 118~(1) (1995) 120--130.

\bibitem{george1999simulation}
J.~George, I.~Boyd, Simulation of nozzle plume flows using a combined
  {CFD-DSMC} approach, in: 33rd Thermophysics Conference, 1999, p. 3454.

\bibitem{karpuzcu2022jet}
I.~T. Karpuzcu, D.~A. Levin, D.~Mamrol, L.~N. Wagner, M.~E. Noftz, J.~S.
  Jewell, S.~T. Smith, {Jet Flow-Shockwave Interactions in a Hypersonic Flow
  using Experimental and Kinetic Methods}, in: AIAA SCITECH 2022 Forum, 2022,
  p. 1577.

\bibitem{karpuzcu2023study}
I.~T. Karpuzcu, D.~A. Levin, {Study of Side-Jet Interactions over a Hypersonic
  Cone Flow Using Kinetic Methods}, AIAA Journal (2023) 1--11.

\end{thebibliography}







\end{document}